\documentclass[12pt]{amsbook}
\usepackage{amssymb,amsfonts,latexsym,amscd}

\begin{document}

\renewcommand{\thechapter}{\arabic{chapter}}
\renewcommand{\thesection}
             {\arabic{chapter}.\arabic{section}}
\renewcommand{\thesubsection}
{\arabic{chapter}.\arabic{section}.\arabic{subsection}}

\def\A{{\mathcal A}}
\def\Af{\mathbf{A}_f}
\def\Afmin{\mathbf{A}_{f,-}}
\def\Bf{\mathbf{B}_f}
\def\bPlperp{\bar{P}_1^\perp}
\def\bPl{\bar{P}_1}
\def\Bound{{\mathcal B}}
\def\BdHfTnl{\kappa_{n-l}^{(1)} }
\def\bTnl{\kappa_{n-l}}
\def\C{{\Bbb C}}

\def\Cnl{\theta_{n-l}}
\def\Cp{C_{|p|}}
\def\Cpuppbd{\frac{7}{4}}

\def\derp{\partial_{|p|}}
\def\deal{\frac{\delta W}{\delta a_\lambda(k)}}
\def\dear{\frac{\delta W}{\delta a^*_\lambda(k)}}
\def\dealn{\frac{\delta W[\h^{(n)}]}{\delta a_\lambda(k)}}
\def\dearn{\frac{\delta W[\h^{(n)}]}{\delta a^*_\lambda(k)}}
\def\der{\mathcal{D}}
\def\derWTs{\der_{WT,n_k,\lambda,\sigma}}
\def\derWTsh{\der_{WT,n_k,\lambda,\hat{\sigma}}}
\def\derWTss{\der_{WT,n_k,\lambda,\sigma_0}}
\def\dPpop{\kappa}
\def\dHfFQQ{\eta}
\def\DEn{\Delta E_{n}}

\def\Eb{E_{bare}[|p|]}
\def\epfrac{\frac{17}{18}}
\def\ez{\epsilon_0}
\def\ezfac{2}
\def\Egrd{E_0[|p|,\sigma_0]}
\def\Elin{E_{lin}}

\def\FchltH{F_{\ch_1,\tau}}
\def\FchrtH{F_{\ch_\rho,\tau}}
\def\FchntHn{F_{\chn,\tau_n}}
\def\FchrtHk{F_{\ch_\rho,\tau^{(k)}}}
\def\Fn{F_n}
\def\Fl{F_1}

\def\gm{g}

\def\hlTnl{\hat{\lambda}_{n-l}}
\def\hbTnl{\hat{\kappa}_{n-l}}
\def\hCnl{\hat{\theta}_{n-l}}
\def\H{{\mathcal H}}
\def\Hps{H_p(\sigma_0)}
\def\Hmap{H}
\def\Hm{H_{lin}}
\def\HW{\omega[H]}
\def\HWz{\omega[H[z]]}

\def\J{\mathfrak{J}}
\def\Kop{K}
\def\Lin{\mathcal{L}}
\def\lTnl{\lambda_{n-l}}
\def\mun{\frac{1}{8}}
\def\num{\frac{1}{12}}
\def\N{{\bf N}}
\def\Ns{N_{\ssig}}
\def\nm{{|\!|\!|\,}}
\def\Omn{N}
\def\Omgrd{\Omega[p;\ssig]}
\def\Omegn{\Omega}
\def\Omgrdn{\Omega[p;\ssig]}
\def\omz{\omega}
\def\P{{\bf P}}

\def\piop{\Pi}
\def\piops{\Pi^\sharp}

\def\Qn{Q_{n}}
\def\Qns{Q_{n}^\sharp}
\def\QchrtH{Q_{\ch_\rho,\tau_z}}
\def\QchrtHs{Q^\sharp_{\ch_\rho,\tau_z}}
\def\sqez{\sqrt{\epsilon}_0}
\def\tQ{\tilde{Q}_{\ch,\tau}}
\def\tQs{\tilde{Q}_{\ch,\tau}^\sharp}

\def\puppbd{p_c}
\def\puppbdnum{\frac{1}{20}}

\def\Pauli{\mathbf{S}}
\def\Ppar{\P_f^\parallel}
\def\Pparop{L}
\def\Ppopk{l}
\def\Pperp{\P_f^\perp}
\def\R{{\Bbb R}}
\def\ssig{{\sigma_0}}

\def\tH{\tau[H]}
\def\tzH{\tau[H]}
\def\tHh{\tau[H[\h]]}
\def\tHz{\tau[H[z]]}
\def\tHpsz{\tau_n[\Hps]}

\def\td{\tilde{\delta}}

\def\Ttild{T'}

\def\Tnlbd{\theta_{n-l}}

\def\uvcut{\kappa_\Lambda}
\def\ul{\underline{\lambda}}
\def\vir{r}
\def\vx{{x}}
\def\vy{{y}}
\def\vk{{k}}
\def\vn{{n}}
\def\vp{{p}}
\def\vq{{q}}
\def\vr{{r}}
\def\vv{{v}}

\def\opsp{\underline{{\cal V}}}

\def\Hspace{{\mathfrak H}}
\def\Mspace{{\mathfrak M}}
\def\Polyd{{\mathfrak V}}
\def\Wspace{{\mathfrak W}}
\def\Uspace{{\mathfrak U}}
\def\Tspace{{\mathfrak T}}
\def\Pl{{\mathcal P}}
\def\h{{\mathfrak h}}
\def\alg{{\mathfrak A}}

\def\op{{\mathcal O}}
\def\opT{\op_0}
\def\tildop{\hat{\op}}
\def\sh{ S}
\def\S{\mathfrak{S}}
\def\ren{{\mathcal R}}
\def\dec{{\mathcal D}_\rho}
\def\decl{{\mathcal D}_1}
\def\decP{{\mathcal D}_P}
\def\Dom{\mathfrak{Dom}}
\def\resc{{\mathcal S}_\rho}

\def\DHf{\Delta_{H_f}}
\def\DHfn{\Delta_{H_f,n}}
\def\DHfnn{\Delta_{H_f,n+1}}

\def\Hel{\H_{el}}
\def\Fo{\H_f}
\def\cut{^{(\ssig)}}
\def\vac{\Omega_f}
\def\Hp{\H_p}
\def\Hsob{\H }
\def\Hpsob{\Hsob_p}
\def\bcr{\bar{\chi}_\rho}
\def\bcl{\bar{\chi}_1}
\def\bPr{\bar{\chi}_\rho}
\def\bP1{\bar{\chi}_1}
\def\bcs{\bar{\kappa}_\sigma}
\def\bcss{\bar{\kappa}_{\ssig}}
\def\ch{\chi}
\def\chr{\chi_\rho}
\def\chn{\ch_{\rho^n}}
\def\bch{\bar{\ch}}
\def\bchr{\bar{\chi}_\rho}
\def\bchn{\bch_{\rho^n}}
\def\dth{\delta}
\def\Pch{P}
\def\Pbch{\bar{P}}
\def\Pchr{P_\rho}
\def\Pchrprp{P_\rho^\perp}
\def\Pbchr{\bar{P}_\rho}
\def\sbch{_{\bch}}
\def\sch{_{\ch}}
\def\Fch{F_{\ch}}
\def\FchtH{F_{\ch,\tau}}

\def\Tlin{T_{lin}}
\def\Tnl{T_{n-l}}
\def\Tnlbch{T_{n-l,\bch}}
\def\TW{T'}
\def\cZ{{\mathcal Z}}
\def\elp{\epsilon_\lambda^\parallel}
\def\crb{\chi_{[\rho,1]}}
\def\heav{h}

\def\1{{\bf 1}}

\def\eqnn{\begin{eqnarray*}}
\def\eeqnn{\end{eqnarray*}}
\def\eqn{\begin{eqnarray}}
\def\eeqn{\end{eqnarray}}
\def\bal{\begin{align}}
\def\eal{\end{align}}

\def\prf{{\bf Proof.}$\;$}

\numberwithin{equation}{section}

\newtheorem{thm}{Theorem}[chapter]
\newtheorem{dfi}{Definition}[chapter]
\newtheorem{prp}{Proposition}[chapter]
\newtheorem{hyp}{Hypothesis}[chapter]
\newtheorem{lm}{Lemma}[chapter]
\newtheorem{cor}{Corollary}[chapter]
\newtheorem{rem}{Remark}[chapter]
\def\qed{\vrule height .6em width .6em depth 0pt\bigbreak}

$\;$
\\

\begin{center}
{\bf \Large OPERATOR-THEORETIC \\

INFRARED RENORMALIZATION\\

AND \\

CONSTRUCTION OF \\

DRESSED 1-PARTICLE STATES \\

IN NON-RELATIVISTIC QED \\

$\;$\\}
\end{center}

\begin{center}
Thomas Chen\footnote{This is the author's PhD thesis under Prof.
J\"urg Fr\"ohlich at ETH Z\"urich, ETH-Diss 14203.}
\\
Institut f\"ur Theoretische Physik\footnote{After August 2001:
Courant Institute, New York University, 251 Mercer Street, New
York,
NY 10012-1185, USA.}\\
ETH H\"onggerberg\\
8093 Z\"urich,
Switzerland\\
Mail: chen@itp.phys.ethz.ch\\
$\;$\\
\end{center}

\noindent{\bf Abstract.} We consider the infrared problem in a
model of a freely propagating, nonrelativistic charged particle of
mass $1$ in interaction with the quantized electromagnetic field.
The hamiltonian $H(\ssig)=H_0+g I(\ssig)$ of the system is
regularized by an infrared cutoff $\ssig\ll1$, and an ultraviolet
cutoff $\Lambda\sim1$  in the interaction term, in units of the
mass of the charged particle. Due to translation invariance, it
suffices to study  the hamiltonian
$\Hps:=\left.H(\ssig)\right|_{\Hp}$, where $\Hp$ denotes the fibre
space of the conserved momentum operator associated to total
momentum $p\in\R^3$. Under the condition that the coupling
constant $g$ is sufficiently small,  there exists a constant
$\puppbd\in[\puppbdnum,1)$, such that for all $p$ with
$|p|\leq\puppbd$, the following statements hold:
\\
\\
(1) For every $\ssig>0$, $\Egrd:=$ inf spec $\Hps$ is an
eigenvalue with corresponding eigenvector $\Omega[p,\ssig]\in\Hp$.
\\
\\
(2) For all $\ssig\geq0$,
$\partial_{|p|}^\beta\left(\Egrd-\frac{|p|^2}{2} \right)\leq
O(g^{\frac{1}{6}})$ for $\beta=0,1,2$.
\\
\\
(3) $\Omega[p,\ssig]$ is not an element of the Fock space $\Hp$ in
the limit $\ssig\rightarrow0$,  if $|p|>0$.
\\

Our proofs are based on the operator-theoretic renormalization
group of V. Bach, J. Fr\"ohlich, and I.M. Sigal \cite{bfs1,bfs2}.
The key difficulty in the analysis of this system is connected to
the strictly marginal nature of the leading interaction term, and
a main issue in our exposition is to develop analytic tools to
control its renormalization   flow.

\thispagestyle{empty}

\tableofcontents

\chapter{INTRODUCTION}

We consider a non-relativistic, charged particle, of mass 1 (where
$\hbar=c=1$), propagating in Euclidean $\R^3$, which interacts
with the quantized electromagnetic field. Due to translation
invariance of the system, the total momentum of the particle and
the boson field is conserved. We thus consider the Hilbert space
of states as a direct integral
$$\H\;=\;\int^\oplus_{\R^3}\;d^3p\;\Hp\;,$$
where $\Hp$ is the fibre   space of the conserved momentum
operator, corresponding to total momentum $p$.

Due to the translational invariance of the system, it suffices to
study $\Hps$, the restriction of the hamiltonian of the system to
$\Hp$, for a range of $p$'s of physical interest. Here, $\ssig$
denotes an artificial, arbitrarily small, but fixed infrared
cutoff in the interaction. Every $\Hp$ is isomorphic to the Fock
space of the quantized electromagnetic field, tensored with
$\C^2$, accounting for electron spin.

\section{Physical motivation and main results}

The questions we propose to answer are:
\\

(1) Does $\Hps$, for sufficiently small values of $|p|$, have an
eigenvalue at the bottom of its spectrum, which accounts for a
dressed 1-particle state (or an infraparticle state) ? For values
$|p|\geq1$ (in units of the mass of the charged particle), this
cannot be true, due to the phenomenon of Cherenkov radiation. In
fact, if $|p|$ approaches 1 (corresponding to the rest energy
$mc^2$ of the charged particle, where $m$ is its mass, and $c$ is
the speed of light), the infraparticle increasingly tends to
reduce its kinetic energy by the emission of electromagnetic
(Cherenkov) radiation. Consequently, if $|p|$ becomes too large,
infspec$\Hps$ ceases to be an eigenstate. The eigenvalue $\Egrd$,
considered as a function of $|p|$, conjecturally becomes a {\bf
resonance} if $|p|$ approaches 1. Clarification of the latter is
beyond the scope of this text.
\\

(2) What is the nature of the ground state vector of $\Hps$ ?
\\

(3) Is the ground state energy of $\Hps$ sufficiently smooth in
$p$, that is, of class $C^2$, for all $\ssig\geq0$ ? This degree
of differentiability is, for instance, required for the
construction of scattering theory according to \cite{fr1,fr2,pi},
and for the analysis of the semiclassical motion of charged
particles \cite{tesp}.
\\

The main physical results obtained in this work are to answer
questions (1) and (3) in the affirmative. As to question (2), the
ground state of $\Hps$ is a 'dressed one-particle state', which
is, for $|p|>0$, contained in the Fock space $\Hp$ for all
$\ssig>0$. However, in the limit $\ssig\rightarrow0$, it ceases to
be an element of $\Hp$, unless $|p|=0$. Instead, it is then
contained in a Hilbert space $\mathcal{H}_{IR}$ that carries a
representation of the canonical commutation relations unitarily
inequivalent to the Fock representation, \cite{fr1,fr2}.

The mathematically precise formulation of the main theorems, is,
together with a discussion of their physical interpretation, given
in Chapter {~\ref{statmainthmsect}}.

\section{Operator-theoretic renormalization group}
The analytical method that we choose for the study of the above
problems is the operator-theoretic renormalization group (RG)
method developed by Bach, Fr\"ohlich, and Sigal \cite{bfs1,bfs2}.

The renormalization group iteration produces a sequence of
effective hamiltonians that can be parametrized by points in a
certain Banach space. This Banach space consists of infinite
sequences of 'kernels'.

The renormalization group iteration is based on a decomposition of
the Fock space $\Hp$ into 'shells' associated with  a dyadic
decomposition of the spectrum of the free photon hamiltonian.

In every recursion step, a shell is 'integrated out', or
'decimated', and a rescaling transformation zooms the reduced
problem to a standard size. The decimation is achieved by an
application of the Feshbach map.

In a manner made precise by a variant of Feshbach's theorem for
smooth Feshbach maps developed by V. Bach, J. Fr\"ohlich and I.M.
Sigal, \cite{bfs3}, all effective hamiltonians are isospectral to
one another. In particular, if the ground state vector of any one
of them is known, the ground state vectors of all of them can be
reconstructed.

Thus, the strategy consists of constructing a long, but finite
sequence of effective hamiltonians, for which the last element has
a known solution of the eigenvalue problem. This is possible
because of the infrared cutoff $\ssig$ which has been inserted
into the interaction. Then, by recursively using the Feshbach
theorem, the eigenvalues at the bottom of every effective
hamiltonian, together with the corresponding eigenvectors, can be
reconstructed. This yields the eigenvalue at the bottom of the
spectrum of $\Hps$, and the associated eigenvector $\Omgrd$.
\\

\subsection{Main problems in the operator-theoretic RG}

The problem of renormalizing the physical hamiltonian is more
difficult than in \cite{bfs1,bfs2}, because the boson form factor
is assumed to exhibit the physical
$\frac{1}{\sqrt{|k|}}$-singularity, and the electron is not
confined to a compact region of space. The main difficulties are:
\\

(i) The interaction operator is purely marginal. Analytical
control of its radiative corrections by resumming perturbation
expansions is very subtle and requires considerable effort.
\\

(ii) If the sequence of effective hamiltonians converges it is
given by an effective hamiltonian with non-vanishing interaction
part, corresponding to a fixed point of the renormalization map.
Unfortunately, the Banach contraction principle is insufficient to
prove its existence, and the Schauder or Sch\"afer fixed point
theorems (which are infinite dimensional analogues, and, as a
matter of fact, consequences of the Brouwer fixed point theorem),
whose application requires the verification of certain compactness
properties, imply rather weak results. A solution of this problem
can presumably be obtained by the development of center manifold
theorems in operator-theoretic renormalization group theory, but
this issue is beyond the scope of the present work.
\\

\subsection{New techniques}
Further development of efficient methods in the operator-theoretic
renormalization group is as important for us as the proof of the
above physical results. Two new main insights are reached. They
concern the following issues:
\\

(I) Gauge invariance of the physical system can, if present, be
exploited, in order to show that various marginal quantities are,
in fact,  renormalized identically. To this end, we develop an
operator-theoretic formulation of the {\bf Ward-Takahashi
identities}, which express the $U(1)$-gauge invariance of the
system. The Ward-Takahashi identities heavily reduce the
complexity in the task of simultaneously renormalizing several
marginal quantities. However, they do not, per se, provide
convergent bounds on the latter.
\\

(II) Control  of the renormalization of purely marginal operators
in the operator-theoretic renormalization group method. The main
insight here is that the radiative corrections of strongly
marginal quantities are subject to nearly complete mutual
cancellations. Consequently, the cumulative radiative correction
to a corresponding quantity remains small in the scaling limit.

The cancellations are consequences of important concatenation
properties of the smooth Feshbach map, which are used to produce
identities that interrelate  certain quantities $X_k,X_n$ of key
interest at arbitrary scales $k,n$. To obtain analytical control
of these quantities, a nested recursion argument is applied.
Passing from a given scale $n$ to the scale $n+1$, the identities
between $X_0$ and  $X_{n+1}$ are determined. Analytical control of
$X_{n+1}$ is obtained from a renormalization group-type
sub-iteration at fixed $n$ that 'integrates' the radiative
corrections up to the actual scale, and which only involves the
quantities $X_k$ for $0\leq k\leq n$.
\\

Among the various approaches to the above points that are
conceivable, we have tried to select efficient and simple  methods
that allow for potential generalizations.
\\

\section{Related recent results}

Aspects of infrared divergences   in the massless Nelson model
that are closely related to the results presented here have
recently been studied by J. L\"orinczi, R.A. Minlos and H. Spohn
\cite{losp, losp1}, by use of probabilistic methods. Also see
\cite{ar} and \cite{behilomisp}. A discussion of the dynamics of
infraparticles in the semiclassical classical limit of
nonrelativistic QED is given by S. Teufel and H. Spohn in
\cite{tesp}; the smoothness properties of the ground state energy
are required in their work. There has been progress in the
scattering theory of massless bosons in the Nelson model in work
of A. Pizzo \cite{pi}, which is to a large extent based on the
work of J. Fr\"ohlich in \cite{fr1}. His results require
$C^2$-smoothness of the ground state energy as a function of the
conserved momentum. Furthermore, there are new results on this
problem owing to C. G\'erard, \cite{ge1}.

Furthermore, recent progress in Rayleigh scattering theory has
been obtained by M. Griesemer, B. Schlein and J. Fr\"ohlich,
\cite{frgrsc1,frgrsc2}, in which one is concerned with charged
particles that are, in contrast to the present case, confined to
atoms or molecules. For properties of the ground state energy in
systems of non-relativistic QED, cf. the recent work of M.
Griesemer, E. H. Lieb and M. Loss, ~\cite{grlilo}.

\chapter{THE MODEL HAMILTONIAN}\label{ModHamchapt}

We introduce a regularized model of a non-relativistic electron
propagating in $\R^3$, coupled to the quantized photon field at
zero temperature.

\section{The physical system}
Writing \eqnn\Hel\;=\;L^2(\R^3,d^3\vx)\;\otimes\; \C^2\;\eeqnn for
the Hilbert space of a single non-relativistic electron, and
\eqnn\Fo\;=\;\bigoplus_{n\geq 0}(L^2(\R^3, d^3 \vk)\;\otimes\;
     \C^2)^{\otimes_{s}n}\;\eeqnn
for the photon Fock space, where "$\otimes_{s}$" denotes the fully
symmetrized tensor product, the Hilbert space of the coupled
system is the tensor product space
\eqnn\H\;=\;\Hel\;\otimes\;\Fo\;.\eeqnn The factors $\C^2$ account
for electron spin and photon polarization.

Moreover, writing $k\in\R^3$ for photon momenta, and
$\lambda\in\lbrace +,-\rbrace$ for photon polarizations,
$a_\lambda^*(\vk)$ and $a_\lambda(\vk)$ denote bosonic creation-
and annihilation operators, respectively. These operator-valued
distributions are subject to the canonical commutation relations
\eqnn [a_{\lambda'}(\vk'),a^*_\lambda(\vk)]\;=\;
      (2\pi)^{3}\delta_{\lambda,\lambda'}\delta^{(3)}(\vk-\vk')\;,\;\;
      [a_{\lambda'}^\sharp(\vk'),a_\lambda^\sharp(\vk)]\;=\;0\;.\eeqnn
For $f\in L^2(\R^3,d^3k)$, we define the operators \eqnn
a_\lambda(f)&\equiv&\int_{\R^3}d^{3}\vk\,
      a_\lambda(\vk)  f^*(\vk) \\
      a^*_\lambda(f)&\equiv&\int_{\R^3}d^{3}\vk\,
      a^*_\lambda(\vk)  f(\vk)\;\eeqnn
on $\Fo$. There is a unique unit ray, the vacuum vector
$\vac\in\Fo$, which obeys \eqnn a_\lambda(f)\;\vac\;=\;0\eeqnn for
all $f\in L^2(\R^3,d^3\vk)$. $\Fo$ is the closure of the span of
all vectors of the form \eqnn\prod_{i} a^{*}_{\lambda_{i}}(f_{i})
       \vac\;,\eeqnn
with $f_{i}\in  L^2(\R^3,d^3\vk)$.
\\

\subsection{Momentum operator and Hamiltonian}
The total momentum operator of the system is given by
\eqn\P_{tot}=\P_{el}\otimes\1_f + \1_{el}\otimes \P_f,
    \label{Ptotdef3}\eeqn
where $\P_{el}$ and $\P_f=\sum_\lambda\int d^3\vk \;\vk\;
a^*_\lambda(\vk)a_\lambda(\vk)$ denote the electron- and photon
momentum operators, respectively.
\\

\noindent{\bf Units.} We set $c=\hbar=1$. The energy will
henceforth be measured in units of the bare electron rest mass,
which, from now on, is set to $m=1$.
\\

The hamiltonian of the system consists of the following operators.
\eqn\;H_{f}\;=\;
     \sum_{\lambda=\pm}\int_{\R^3}\;d^{3}\vk\;
          |\vk|\;a_\lambda^*(\vk)a_\lambda(\vk)\; \eeqn
denotes the usual hamiltonian of the noninteracting
electromagnetic field, in the Coulomb gauge.

The electromagnetic field operator contains an ultraviolet cutoff
$\Lambda$, and an infrared cutoff $\ssig$, which is inserted in
order to render the theory mathematically well-defined. $\Lambda$
is a real number in $(2,3)$, which we keep fixed in the entire
discussion. We let $\uvcut$ denote a smooth function with support
in $[0,\Lambda]$, which equals 1 on $[0,1]$, and
\eqn\left|\;\partial_\vx
    \uvcut(\vx)\;\right|\;\leq\;2\,\chi_{[0,\Lambda]}(\vx)\;,\eeqn
which establishes the ultraviolet cutoff (where $\chi_I$ is the
characteristic function of the interval $I$).

Furthermore, let $0<\ssig\ll1$ denote a positive, but arbitrarily
small  real number, which we   let tend to zero at the end of our
discussion; $\bcss$ is defined by
\eqn\bcss[x]\;=\;\left\lbrace\begin{aligned}1&\;\;&x>\ssig\\
    \frac{x}{\ssig}&\;\;&x\leq\ssig\end{aligned}\right.\eeqn
for $x\in\R_+$. As a matter of fact, we could use any function of
order $O(x^\epsilon)$ for $x\leq\ssig$, and $\epsilon>0$, as long
as it is monotonic, and $\bcss[x]>0$ for all $x>0$. In fact, a
zero of $\bcss$ at some $x>0$ would produce problems in the
operator-theoretic formulation of the Ward-Takahashi identities,
and if $\bcss$ were not monotonic, there would be complications in
the discussion of the renormalization group flow at extremely low
scales.

Other possible implementations of the Ward-Takahashi identities
that do not require this restriction, but lead to other
complications, are addressed in remark {~\ref{alternWTformrem}}.

The interaction between the charged particle, and the quantized
electromagnetic field is implemented by the operator
\begin{gather}\Af(\ssig)=
         \left.\left.
         \sum_{\lambda=\pm}\int_{\R^3}
         \frac{d^3 \vk\;\bcss(|\vk|)\uvcut(|\vk|)}
         {\sqrt{|\vk|}}
         \right(
         \epsilon_{1,0}[n_k,\lambda]\S_{-\vk}\otimes
         a_\lambda^*(\vk)+\epsilon_{0,1}[n_k,\lambda]\S_\vk\otimes
         a_\lambda(\vk)
         \right)\;,
\end{gather}
which is linear in the electromagnetic vector potential in the
Coulomb gauge, where $\epsilon_{M,N}[n_k,\lambda]$, with $M+N=1$,
are polarization vectors, and $n_k=\frac{k}{|k|}$. We have only
indicated the dependence of $\Af(\ssig)$ on the infrared cutoff,
since $\Lambda$ is kept fixed. The factor
$\frac{1}{\sqrt{2}(2\pi)^3}$, which usually appears in the
definition of $\Af$, is absorbed into the coupling constant, that
is, the electron charge, (~\ref{coupelchargedef}).

The choice of the Coulomb gauge implies that
$\lbrace\epsilon_{M,N}[n_k,+],
\epsilon_{M,N}[n_k,-],\vn_\vk\rbrace$ is an orthonormal basis in
$\R^3$, for all $\vk$ and $M+N=1$. It thus also implies that
\eqn\P_f\cdot \Af(\ssig)\;=\;\Af(\ssig)\cdot\P_f\;.\eeqn The
operator $\S_\vk=\S_{-\vk}^{*}$, with $\vk\in\R^3$, shifts the
electron momentum by $k$, that is, it acts on
$\P_{el}$-eigenvectors in terms of
$$\S_\vk\,|\vq\rangle_{el}\; =\; |\vq+\vk\rangle_{el}\;.$$
We also define the operator
\begin{gather}   \Bf(\ssig)\;=\;i\;\left.\left.
        \sum_{\lambda=\pm}\int_{\R^3}
        \frac{d^3 \vk\;\bcss(|\vk|)\uvcut(|\vk|)}
         {\sqrt{|\vk|}}
         \right( \;
         \vk\wedge\epsilon_{1,0}[n_k,\lambda]\S_{-\vk}\otimes
         a_\lambda^*(\vk)\;+\;h.c.
         \right)\;.\end{gather}
The cutoffs are incorporated in a manner that preserves the
$U(1)$-gauge invariance of the system. This is of great importance
for our analysis, since we will make extensive use of the
associated Ward-Takahashi identities.

The Hamiltonian that we will consider is given by
\begin{gather}H\cut\;=\;\left.\left.
          \frac{1}{2}\;\right(\P_{el}\otimes
          \1_{f}\;-\;g \Af(\ssig)\right)^{2}
          \;+\;\frac{g}{2}\;\Pauli\cdot \Bf(\ssig)\;+\;
          \1_{el}\otimes H_f\;.
          \label{uvham}\end{gather}
Due to the cutoffs in the interaction, it is mathematically
well-defined. Here, $\Pauli:=(\sigma_{1},\sigma_{2},\sigma_{3})$
stands for the triple of Pauli matrices, and
\eqn\;g\;:=\;\frac{e}{\sqrt{2}\;(2\pi)^3}\;,
    \label{coupelchargedef}\eeqn
where $e$ denotes the bare charge of the electron. Due to the
absence of antiparticle production in the non-relativistic limit,
there is no charge renormalization (no vacuum polarization
diagrams).
\\

The regularized system is translation invariant,
$[\P_{tot},H\cut]=0$. The Hilbert space $\Hsob$ can be written as
a direct integral \eqnn\Hsob\;=\;\int^{\oplus}_{{\rm
spec}\P_{tot}}\,
    d^{3}\vp\;\Hp\;,\eeqnn
where $\Hp$ are fibre Hilbert spaces associated to fixed values
$p$ of the conserved momentum, which are invariant under space-
and time-translations. That is, the actions of the time evolution
map $e^{it H\cut}$, and of the space translation map
$e^{ix\cdot\P_{tot}}$ on $\Hsob$ are fibre-preserving. Every $\Hp$
is isomorphic to $\C^2\otimes\Fo$. For any fixed value $\vp$ of
the total momentum of the system, the restriction of $H\cut$ to
the fibre $\H_{\vp}$ is given by the operator
\begin{gather}
   H_p\cut\equiv H|_{\H_{\vp}}\;=\;H_f\;+\;\left.\left.\frac{1}{2}
   \right(\vp - \P_f -
   g \Af(\ssig)\right)^2 \;+\;\frac{g}{2}\;\Pauli\cdot
   \Bf(\ssig) \;
\end{gather}
on $\C^2\otimes\Fo$. We restrict the value of $p$ to $|\vp|\leq
\puppbd$,  in all that follows.
\\

\section{The simplified model of a spinless charged particle}

In the rest of this work, we will focus on the simplified model
hamiltonian $\Hps$ obtained by omitting the electron spin, that
is, by dropping the term involving the magnetic field operator.
This is justified by the fact that the magnetic field operator
has, in the absence of cutoffs, a scaling dimension 2. Our later
discussion will show that all operators with scaling dimensions
$>1$ are irrelevant from the point of view of the renormalization
group. Thus, $H_\vp(\ssig)$ and $H_\vp\cut$ belong to the same
universality class of microscopic theories, which justifies the
simplification.

As a preparation for later considerations, we write the
hamiltonian in the form \eqn\;\Hps\;=\;\Eb\;+\;
T[\opT]\;+\;W[\op]\;,
    \label{Hpsdef2}\eeqn
where \eqn\label{E0sigpphyshamdef}
    \Eb\;=\;\frac{|p|^2}{2}\;+\;\frac{g^2}{2}\;
    \left\langle\Af^2(\ssig)\right\rangle_{\vac}\; \eeqn
corresponds to the classical expression for the ground state
energy, except for the Wick ordering correction corresponding to
$\Af^2(\ssig)$.

The operator \eqn T[\opT]&:=& H_f\;
     -\;|p|\;\P_f^\parallel\;+\;
     \frac{1}{2}\;\left(|\Ppar|^2\;+\;|\Pperp|^2\right)
     \;,\;\eeqn
where we define \eqn\opT \;:=\;(H_f,\P_f^\parallel, |\P_f^\perp|)
\;,\eeqn commutes with $H_f$ and $\P_f$, and is referred to as the
{\bf noninteracting hamiltonian}.

In the nonrelativistic limit, the condition that only values of
$p$ with $|p|<1$ are considered, is natural. For technical
reasons, we will impose the more restrictive requirement that
\eqn|p|\;\leq\;\puppbd\;, \eeqn
for some sufficiently small $\puppbd\in[\puppbdnum,1)$. This bound
can be improved, but there is a physical phenomenon which prevents
the main results of this work, which are presented in Chapter
{~\ref{summmainressubsec3}}, to be valid for $|p|\geq1$. We have
already mentioned in the introductory chapter that Cherenkov
radiation occurs in this limit. This point will be discussed in
more detail in Chapter {~\ref{summmainressubsec3}}.

The bounds \eqn(1-|p|)H_f\;\leq\;T[\opT]\;\leq\;
    \left(\frac{3}{2}+|p|\right)H_f\,+\,\frac{1}{2}\,
    |\P_f|^2\;,
    \label{gammaGammabounds1}\eeqn
are an obvious consequence of $|\Ppar|,|\Pperp|\leq H_f$.

Furthermore, defining \eqn\op\;:=\;(H_f,\Ppar,\Pperp)\;,\eeqn the
operator \eqn  W[\op] &=&
    \sum_{1\leq M+N\leq 2}\;W_{M,N}[\op]\;\;,\label{physW}\eeqn
is referred to as the interaction hamiltonian. $W_{M,N}$ denote
Wick ordered operators with $M$ creation and $N$ annihilation
operators, given by \eqn\;W_1[\op]&:=&\sum_{M+N=1}\;W_{M,N}\;=\;
     -\;g\,(p-\P_f)\cdot \Af(\ssig)\;\;, \\
     W_2[\op]&:=&\sum_{M+N=2}\;
     W_{M,N}\;=\;\frac{g^2}{2}\;:\;\Af^2(\ssig)
     \;:\;\;,\eeqn
where $\P_f=\Ppar+\Pperp$, and $n_p:=\frac{p}{|p|}$.

The operators $W_{M,N}$ are referred to as Wick monomials of
degree $M+N$. Their general structure is given by
\eqn\;W_{M,N}[\op]&=&\sum_{\lambda^{(M)},\tilde{\lambda}^{(N)}}\;
     \int \,\prod_{i=1}^M\,\prod_{j=1}^N\,\frac{
     d^3k_i\;\bcss
     (|k_i|)}{\sqrt{|k_i|}} \,
     \frac{d^3\tilde{k}_j\;\bcss(|\tilde{k}_j|)}{
     \sqrt{|\tilde{k}_j|}}
     \;\times\nonumber\\
     &&\times\;
     a^*_{\lambda^{(M)}}(k^{(M)})\,
     w_{M,N}\left[\op;K^{(M,N)}\right]\,
     a^\sharp_{\tilde{\lambda}^{(N)}}(\tilde{k}^{(N)})
     \;,\hspace{1cm}\eeqn
where the integral kernels $w_{M,N}$ are operators that commute
with $H_f$.

We have here introduced the following convenient multiindex
notation:
\eqnn\lambda^{(M)}&:=&(\lambda_1,\dots,\lambda_M)\\
     k^{(M)}&:=&(k_1,\dots,k_M),\eeqnn
as we recall, and likewise for $\tilde{\lambda}^{(N)}$ and
$\tilde{k}^{(N)}$. Moreover,
\eqnn\,K^{(M,N)}\;:=\;\left(k^{(M)},\lambda^{(M)};
     \tilde{k}^{(N)},\tilde{\lambda}^{(N)}\right) ,\eeqnn
and \eqnn a^\sharp_{\lambda^{(M)}}(k^{(M)}):=\prod_{i=1}^M
      a^\sharp_{\lambda_i}(k_i).\eeqnn
In cases where we consider the dependence of the interaction
operators with respect to the interaction kernels, the notation
\eqn\;W_{M,N}\left[[\;w_{M,N}[z;\op;\cdot]\;\right]]\;\eeqn will
be used.

For $M+N=1$, we have
\begin{gather}w_{1,0}[\op;z;k,\lambda]\;=\;
     w_{0,1}^*[\op;k,\lambda]\;=\;
     \,-\,g\,\uvcut(|k|)\,
     (p-\P_f)\cdot\epsilon_\lambda(k)\;,
\end{gather}
and for $M+N=2$, by \eqn
     w_{2,0}[\op;k,\lambda,\tilde{k},\tilde{\lambda}]&=&
     w_{0,2}^*[\op;k,\lambda,\tilde{k},\tilde{\lambda}]
     \nonumber\\
     &=&
     g^2\,\uvcut(|k|)\,\uvcut(|\tilde{k}|)
     \,\epsilon_{\lambda}(k)\,\cdot\,
     \epsilon_{\tilde{\lambda}}(\tilde{k})
\eeqn and
\begin{gather}
     w_{1,1}[\op;k,\lambda,\tilde{k},\tilde{\lambda}]\;=\;
     g^2\;\uvcut(|k|)\uvcut(|\tilde{k}|)
     \;\left(
     \epsilon_\lambda(k)^*\,\cdot\,
     \epsilon_{\tilde{\lambda}}(\tilde{k})\,+\,
     \epsilon_\lambda(k)\,\cdot\,
     \epsilon_{\tilde{\lambda}}^*(\tilde{k})\right)\;.
\end{gather}

The interaction kernels and their derivatives with respect to
operators and momenta are controlled by the following bounds.
\\

\begin{lm}
Let $X$ denote either $|p|$, $|k_i|$ or $|\tilde{k}_j|$, and let
$\alpha=(\alpha_i)_{i=1}^3$ be a multiindex with $|\alpha|=\sum
\alpha_i\leq2$. Then, there is a constant $C$ independent of
$g,|p|$, such that if $|\alpha|=1$,
\eqn\left|\,w_{M,N}[\op;k,\lambda]\,\right|&\leq&
    g\,(|p|\,+\,|\P_f|)\,\uvcut(|k|)\nonumber\\
    \left\|\,\partial^\alpha_{\op}\,w_{M,N}[\op;k,\lambda]\,
    \right\|&\leq&C\,
     g\,\uvcut(|k|)\;
     \nonumber\\
    \left\|\,\partial_{X}w_{M,N}[\op;k,\lambda]\,
    \right\|&\leq&C\,
     g\,\chi_{[0,\Lambda]}(|k|)
     \label{W1physhamest1}\eeqn
for $M+N=1$, and \eqn\left\|\,
     w_{M,N}[\op;k,\lambda,\tilde{k},\tilde{\lambda}]\,\right\|
     &\leq&g^2\,\uvcut(|k|)
     \,\uvcut(|\tilde{k}|)\nonumber\\
     \left\|\,\partial^\alpha_{\op}
     w_{M,N}[\op;k,\lambda,\tilde{k},\tilde{\lambda}]\,\right\|
     &\leq&C\,g^2\,\uvcut(|k|)
     \,\uvcut(|\tilde{k}|)\nonumber\\
     \left\|\,\partial_{X}
     w_{M,N}[\op;k,\lambda,\tilde{k},\tilde{\lambda}]\,\right\|
     &\leq&C\,g^2\,\chi_{[0,\Lambda]}(|k|)
     \,\chi_{[0,\Lambda]}(|\tilde{k}|)\label{W2physhamest1}
    \eeqn
for $M+N=2$. Moreover, if $|\alpha|+\beta\geq2$, where
$|\alpha|\leq2$ and $\beta\leq1$,
\eqn\derp^\beta\,\partial_{\op}^\alpha\, w_{M,N}\;=\;0\;,\eeqn for
$1\leq M+N\leq2$.
\end{lm}

\prf These estimates are obvious consequences of the assumptions
that were imposed on $w_{M,N}$ and $\uvcut$. \qed

$\;$

\section{Scaling and scaling dimensions}
\label{scalingdimsubsect}

The issue of scaling has key importance in renormalization group
analysis of Wilson type. The purpose of this subsection is to
introduce the {\bf scaling transformation}.

For $f\in L^2(\R^3,d^3 \vk)$, let us consider the unitary
transformation \eqn (\Gamma_\rho \;f)(\vk)\;=\;\rho^{-\frac{3}{2}}
    f(\rho^{-1}\;\vk)\;.\eeqn
Lifting $\Gamma_\rho$ to the Fock space $\Hp$, we obtain a unitary
operator $U_\rho$.

Let $\op$ denote any operator that is homogenous under rescaling
of photon momenta. The scaling dimension of an operator $A$ is
defined as the number $d({\mathcal O})$, such that \eqn
Ad_{U_\rho}[A]\;\equiv\;
    U_\rho\;A\;U_\rho^*\;=\;\rho^{d(A)}\;A\;,\eeqn
if $A$ scales homogenously under rescaling of the photon momenta.

The photon number operator \eqn N_f\;=\; \sum_{\lambda=\pm}\int
d^3 \vk\;\; a_\lambda^*(\vk)\;
     a_\lambda(\vk)\;\eeqn
is scale invariant, which implies that
$d(a_\lambda^\sharp)=\frac{-3}{2}$. Moreover, $d[\op^r_{(0)}]=1$
for all $r$, and in the limit $\ssig\rightarrow 0$ and
$\Lambda\rightarrow \infty$, $d[\Af]=1$ and $d[\Bf]=2$. The photon
vacuum is scale invariant, \eqn\label{vacscaleinv}
     U_\rho\;\vac\;=\;\vac\;\;\;\forall \;\rho\;\geq\; 0\;.\eeqn
This follows by observing that
\eqn 0&=&U_\rho\;a_\lambda(\vk)\;U_\rho^*\;U_\rho\;\vac\nonumber\\
       &=&\rho^{-\frac{3}{2}}\;a_\lambda(\rho^{-1}\;\vk)\;U_\rho\;\vac\eeqn
holds for all $\lambda$ and $k$. Thus, uniqueness of the photon
vacuum implies (~\ref{vacscaleinv}).
\\

\subsection{Derivatives of scaled operators}
The following brief section concerns the scaling of derivatives
with respect to $\op$. Let us assume $f$ to be at least $C^2$ in
$\op$ on the Hilbert subspace $P_\rho\Hp$ of $\Hp$, where
$$P_\rho\;\equiv\;\chi[0\leq H_f\leq\rho]\;,$$
for $0<\rho<1$, and that
$$\|\partial_{\op}^\alpha f[\op]\|_{P_\rho\Hp}\;<\;c_{\alpha}$$
holds for $0\leq|\alpha|$. The differential operator
$\partial_{\op}^\alpha$ scales like
$$Ad_{U_\rho}[\partial_{\op}^\alpha\,U_\rho^*]\;=\;
    \rho^{-|\alpha|}\partial_{\op}^\alpha\;,$$
in other words, its scaling dimension is given by $-|\alpha|$.

It is clear that
\eqnn\hat{f}[\op]&\equiv&Ad_{U_\rho}[f[\op]]\\
    &=&U_\rho\,f[\op]\,U_\rho^*\\
    &=&f[\rho\,\op]\;\eeqnn
is an operator acting on the Hilbert space $P_1\Hp$, and that
$$\|\hat{f}[\op]\|_{P_1\Hp}\;=\;\|f[\op]\|_{P_\rho\Hp}$$
by unitarity of $U_\rho$. For the derivatives of $\hat{f}$,  we
have
\eqnn\partial_{\op}^\alpha\hat{f}[\op]&=&\partial_{\op}^\alpha
    \,U_\rho\,f[\op]\,U_\rho^*\\
    &=&U_\rho\,\left(U_\rho^*\,\partial_{\op}^\alpha
    U_\rho\right)\,f[\op]\,U_\rho^*\\
    &=&U_\rho\,Ad_{U_\rho}^{-1}[\partial_{\op}^\alpha]
    \,f[\op]\,U_\rho^*\\
    &=&\rho^{|\alpha|}\,U_\rho\left(\partial_{\op}^\alpha
    f[\op]\right)U_\rho^*\\
    &=&\rho^{|\alpha|}\,Ad_{U_\rho}
    [\partial_{\op}^\alpha f[\op]]\;,\eeqnn
because of \eqnn
    U_\rho^*\;\partial_{\op}^\alpha\,U_\rho\;=\;
    Ad^{-1}_{U_\rho}[\partial_{\op}^\alpha]\;=\;
    \rho^{|\alpha|}\partial_{\op}^\alpha\;.\eeqnn
Notice that the inverse of $Ad_{U_\rho}$ appears here. Thus,
\eqn\|\partial_{\op}^\alpha\hat{f}[\op]\|_{P_1\Hp}&\leq&
    \rho^{|\alpha|}\|f[\op]\|_{P_\rho\Hp}\nonumber\\
    &\leq&\rho^{|\alpha|}\,c_\alpha\;,
    \label{deropalphhatffrelbdsaux3}\eeqn
since
$$\|Ad_{U_\rho}[\partial_{\op}^\alpha f[\op]]\|_{P_1\Hp}
    \;=\;\|\partial_{\op}^\alpha f[\op]\|_{P_\rho\Hp}\;,$$
again by unitarity of $U_\rho$. In particular, we emphasize that
$$\partial_{\op}^\alpha\hat{f}[\op]\;\neq\;
     Ad_{U_\rho}[\partial_{\op}^\alpha f[\op]]\;!$$
The norm of the right hand side on $P_1\Hp$ is bounded by
$\rho^{-|\alpha|}\,c_\alpha$, not by $\rho^{|\alpha|}\,c_\alpha$
as in (~\ref{deropalphhatffrelbdsaux3}), which is {\bf larger}
than $c_\alpha$ if $|\alpha|>0$. Notably, a growth of the bounds
on derivatives of this kind would have catastrophic consequences
for our analysis.
\\

\section{The marginal part: An exactly solvable toy model}
\label{toymodsubsec}

For illustrative purposes, mainly in order to give a motivation
for the main results that are presented in the next chapter, we
now briefly describe a toy model obtained by discarding all
operators in $\Hps$ that are nonlinear in $\op$ or $\Af(\ssig)$.

The point is that $\Hm$, by which we denote the marginal part in
$\Hps$ (consisting of operators with scaling dimension 1 in the
absence of cutoffs), can be easily explicitly diagonalized. It
comprises only operators with a scaling dimension 1, and shares
key properties with the full hamiltonian $\Hps$. By definition,
\eqnn\;\Hm\;:=\;H_f\;-\;|p|\;
    \P_f^\parallel\;+\;g\;|p|\;
    \Af^\parallel(\ssig)\;,\eeqnn
where $\Af^\parallel(\ssig)$ denotes the component of the
electromagnetic vector potential in the direction of the conserved
momentum $p$. As before, this operator acts on the Fock space
$\Hp$, with the Fock vacuum given by $\vac$. The creation and
annihilation operators are also as before.

Diagonalization of $\Hm$ is achieved by the Bogoliubov
transformation
$$\Hm\;\longmapsto\;
    e^{S_\ssig}\;\Hm\; e^{-S_\ssig}$$
generated by
\begin{gather}
    S_\ssig\;=\left.\left.\,g\,|p|\,\sum_\lambda\int\,
    \frac{d^3\vk\,\bcss[|\vk|]}
    {\sqrt{|\vk|}\left(|\vk|-|\vp|\vk^\parallel\right)}
    \right(\,w_{1,0}[\vk,\lambda]a^*_\lambda(\vk)\,-\,
    w_{0,1}[\vk,\lambda]a_\lambda(\vk)\,
    \right)\;,\end{gather}
where \eqnn
    w_{M,N}(\vk,\lambda)&=&g\,|p|\,\uvcut(|\vk|)
    \,
    \epsilon_{M,N}^\parallel[n_k,\lambda]\;.\eeqnn
We recall that $\epsilon_{M,N}^\parallel[n_k,\lambda]$ denotes the
projection of the photon polarization vector in the direction of
the conserved momentum $p$. A trivial calculation shows that
\eqn\;e^{S_\ssig}\;\Hm\;e^{-S_\ssig}\;=\;\Elin
    \,+\,
    H_f\,-\,|p|\,\P_f^\parallel\;,\eeqn
where
\begin{gather}
    \Elin\;=\;-\,g^2\,|p|^2\,\sum_\lambda\int\,
    \frac{d^3k\,\bcss^2[|k|]\,\uvcut^2(|k|)}{|k|
    \left(|k|-|p|k^\parallel\right)}\;
    \epsilon_{0,1}^\parallel[n_k,\lambda]
    \,\epsilon_{1,0}^\parallel[n_k,\lambda]\;
\end{gather}
denotes the ground state energy of $\Hm$. One can easily convince
oneself of \eqnn\left|\;\partial^\alpha_{|p|}\Elin
    \;\right|\;\leq
    \;O(g^2)\;\eeqnn
as long as $|p|<1$, for $\alpha=0,1,2$, and all $\ssig\geq0$.

The associated ground state eigenvector is given by
\eqnn\Psi[\ssig,p]\;=\;e^{S_\ssig}\;\vac\;,\eeqnn which can
clearly be written in the form
\eqnn\;e^{S_\ssig}\;\vac&=&\exp\left(\,-\,\frac{g^2\,|p|^2}{2}\,
    \sum_\lambda\int\,
    \frac{d^3k\,\bcss^2[|k|]\,\uvcut^2(|k|)}{|k|
    \left(|k|-|p|k^\parallel\right)^2}\,\right)\,
    \times\nonumber\\
    &&\times\;\exp\left(g\,|p|\,\int\,
    \frac{d^3k\;\bcss[|k|]\,\uvcut(|k|)}
    {\sqrt{|k|}\left(|k|-|p|k^\parallel\right)}
    \,w_{1,0}[k,\lambda]a^*_\lambda(k)\right)\,\vac\;.\eeqnn
Hence, defining
\eqnn\Omega[\ssig,p]\;:=\;\exp\left(\,\frac{g^2\,|p|^2}{2}
    \,\int\,
    \frac{d^3k\,\bcss^2[|k|]\,w_{0,1}[k,\lambda]\,
    w_{1,0}[k,\lambda]}
    {\left(|k|-|p|k^\parallel\right)^2}\,
    \right)\,
    \Psi[\ssig,p]\;,\eeqnn
and fixing the projection
$$\left\langle\;\vac\;,\;\Omega[\ssig,p]\;\right\rangle
    \;=\;1\;$$
for all $\ssig>0$,
\eqnn&&\left\|\;\Omega[\ssig,p]\;\right\|^2\nonumber\\
    &&\hspace{0.5cm}=\,\exp\left(\,g^2\,|p|^2\,\sum_\lambda\int\,
    \frac{d^3k\,\bcss^2[|k|]\,\uvcut^2(|k|)}{|k|
    \left(|k|-|p|k^\parallel\right)^2}\;
    \epsilon_{0,1}^\parallel[n_k,\lambda]
    \,\epsilon_{1,0}^\parallel[n_k,\lambda]\,\right)\;.
    \eeqnn
In the absence of the infrared regularization, the integrand in
the exponent diverges like $\frac{1}{|k|^3}$ in the limit
$|k|\rightarrow0$; due to the infrared regularization, there are
constants $0<c,C<\infty$ such that
\eqnn\left\|\;\Omega[\ssig,p]\;\right\|
    \;\geq\;C\exp\left(c\;g^2\,|p|^2\,|\log\ssig|\right)\;.
    \eeqnn
In the limit $\ssig\rightarrow0$, the vector $\Psi[\ssig,p]$
leaves the Fock space $\Hp$. For every fixed $|p|>0$, the ground
state belongs to a Hilbert space that carries a representation of
the canonical commutation relations that is unitarily inequivalent
to the Fock representation. For different values of $p$, these
representations are mutually inequivalent.

For $|p|=0$, however, it is trivially clear that the ground state
is given by $\vac\in\Hp$. A slightly more detailed exposition of
matters concerning this toy model can be found in \cite{itzu}.

\chapter{STATEMENT AND DISCUSSION OF THE MAIN THEOREMS}
\label{statmainthmsect}

\label{summmainressubsec3} The central results proved in the
present work are the following two main theorems.
\\

\begin{thm}
\label{mainthm1} For any arbitrarily small, but fixed value of the
infrared cutoff $\ssig>0$, and $g$ sufficiently small, there
exists a constant $\puppbd\in[\puppbdnum,1)$, such that for all
$p$ with $|p|\leq\puppbd$,
\eqn\;\Egrd\;:=\;\inf{\rm
    spec}\lbrace \Hps\rbrace\eeqn
is a unique eigenvalue with an associated eigenvector
$\Omega[p;\ssig]$ that lies in $\Hp$. $\Egrd$ only depends on the
absolute value of $p$ because of rotational symmetry. Under the
normalization condition
$$\left\langle\vac\,,\,\Omgrd\right\rangle
    \;=\;1\;,$$
there are constants, $c,C>0$ and $c',C'>0$, such that
$$c\;\exp\left[\,c'\,g^2\,|p|^2\,
    \left|\,\log\,\ssig\,\right|
    \left]\;\leq\;
    \left\|\, \Omgrd\,\right\|^2\;\leq\;C\,
    \exp\left[\,C'\,g^2\,|p|^2\,
    \left|\,\log\,\ssig\,\right|
    \right]\right.\right.\;.$$
In particular, this implies that $\Hps$ possesses a ground state
vector in $\Hp$,  if and only if $|p|=0$.
\end{thm}

Under the condition that the coupling constant $g$, and the
absolute value $|p|$ of the conserved momentum are sufficiently
small, this gives a description of the bottom of the joint
energy-momentum spectrum of the present system, that is, of the
translation invariant model of a spinless, massive charge, which
interacts with the quantized electromagnetic field. In particular,
this theorem provides the infraparticle state $\Omgrd$, and
demonstrates how the infrared catastrophe arises. The next theorem
gives us analytical control of the first and second derivatives of
the infraparticle energy with respect to the conserved momentum,
again under the assumption that both $g$ and $|p|$ are small
enough.
\\

\begin{thm}
\label{mainthm2} Assume that $|p|\leq\puppbd$, and that $g$ is
sufficiently small. The ground state eigenvalue $\Egrd$ is of
class $C^2$ with respect to the conserved momentum $p$, and
\eqn\left|\;\partial_{|p|}^\alpha\left(\;\Egrd-\frac{|p|^2}{2}\;
    \right)\;\right|\;\leq\;O(g^{1-3\nu}) \eeqn
for all $\ssig\geq 0$, with $\alpha=0,1,2$. Furthermore,
$$\,\derp^2\left(\Egrd\,-\,\frac{|p|^2}{2}\right)\;\leq\;0\;,$$
that is, the renormalized mass of the infraparticle is bounded
from below by the bare mass of the charged particle, for all
$\ssig\geq0$.
\end{thm}

Let us next comment on the physical interpretation of these
results.
\\

\section{The joint energy-momentum
spectrum} Theorem {~\ref{mainthm1}} gives the following
description of the nature of the joint energy-momentum spectrum
spec$(H(\ssig),\P_{tot})$ of the translation invariant system of a
massive, spinless charge that interacts with the quantized
electromagnetic field, on the Hilbert space $\H=\int^\oplus
d^3p\Hp$. The model hamiltonian $H(\ssig)$ is obtained by omitting
$\Bf(\ssig)$ and the electron spin in $H\cut$ which is defined in
(~\ref{uvham}), and $\P_{tot}$ is defined in (~\ref{Ptotdef3}).

In the limit $g\rightarrow0$ of vanishing interaction, the
spectrum consists of the parabola $E[|p|]=\frac{|p|^2}{2}$ in the
$(E,|p|)$-plane (where $E$ denotes the energy, and $p$ the
conserved momentum), which borders to absolutely continuous
spectrum. Thus,
$${\rm spec}\left[\left(\lim_{g\rightarrow0}H(\ssig)\;,\;
    \P_{tot}\left)\left]\;=\;
    \left(\frac{|p|^2}{2}\,+\,\R_+\;,\;p\right)_{|p|\in\R_+}
    \right.\right.\right.\right.
    \;,$$
independently of $\ssig$. For every fixed value of $p$, the
infimum of the spectrum, $\frac{|p|^2}{2}$, is an eigenvalue with
an associated eigenvector given by
$$|p\rangle_{el}\,\otimes\,\vac\;,$$
that is, a generalized electron eigenstate of momentum $p$,
tensored with the photon vacuum $\vac$. Its restriction to the
Hilbert fibrespace $\Hp$ is simply $\vac$.

Turning on the interaction (by choosing some $g>0$), at an
arbitrarily small, but fixed value of the infrared cutoff
$\ssig>0$, Theorem {~\ref{mainthm2}} implies that at least for
values $|p|\leq\puppbd$ of the conserved total momentum, \eqnn{\rm
spec}\left[\left(H(\ssig)\;,\;
    \P_{tot}\left)\;\left|\;|\P_{tot}|\leq\puppbd\left]\;=\;
    \left(\frac{|p|^2}{2}\,+\,O(g^{\frac{1}{6}})\,+\,
    \R_+\;,\;p\right)_{|p|\leq\puppbd}
    \right.\right.\right.\right.\right.
    \;,\eeqnn
uniformly in $\ssig$. That is, the spectrum of the interacting
system is, for sufficiently small $g$ and $|p|$, a small
perturbation of that of the non-interacting one. According to
Theorem {~\ref{mainthm1}}, the infimum of the spectrum is, under
these restrictions, still an eigenvalue. The corresponding
eigenvector, however, is {\bf not} a small perturbation of the
eigenvector at $g=0$. Instead, at fixed $p$, with $|p|\leq\puppbd$
and $g$ sufficiently small, its restriction to the Hilbert
fibrespace $\Hp$, given by $\Omgrd$, Theorem {~\ref{mainthm1}}
asserts that there exist constants $c_1,c_2,C_1,C_2>0$, such that
$$c_1\;\exp\left(\,c_2\,g^2\,|p|^2\,
    \left|\,\log\,\ssig\,\right|
    \left)\;\leq\;
    \left\|\, \Omgrd\,\right\|^2\;\leq\;C_1\,
    \exp\left(\,C_2\,g^2\,|p|^2\,
    \left|\,\log\,\ssig\,\right|
    \right)\right.\right.\;,$$
if its projection in the direction of $\vac$, the eigenstate for
$g=0$, is normalized to the value 1. Most importantly, the lower
bound {\bf diverges} in the limit $\ssig\rightarrow0$.
Consequently, either $\Omgrd$ ceases to have a projection in the
direction of $\vac$, or it ceases to be an element of $\Hp$ if the
infrared regularization is removed.

In fact, the latter is the case.
$$\Psi[p;\ssig]\;:=\;\frac{\Omgrd}{\|\Omgrd\|}$$
corresponds to a generalized coherent state, which is contained in
a Hilbert space orthogonal to $\Hp$ that carries a representation
of the canonical commutation relations that is unitarily
inequivalent to the Fock representation. Both Hilbert spaces are
at this point regarded as Hilbert subspaces embedded in some
infinite tensor product Hilbert space whose construction dates
back to J. v. Neumann \cite{vNe}. For more material on this issue,
cf. for instance \cite{ki}.
\\

\section{Dressed 1-particle states}
Physically, the state $\Omgrd$ describes a massive, charged
particle that is inevitably surrounded by a cloud of infinitely
many photons. It is usually referred to as an {\bf infraparticle}
or a {\bf dressed 1-particle state}.

As has already been recognized by Bloch and Nordsieck long ago,
\cite{blno},  the fact that the number of soft bosons is not
relatively bounded by their energy is responsible for the infrared
catastrophe.

To arrive at a better insight into the structure of the problem,
let us add a brief remark about operator algebraic aspects, and
refer to the work of J. Fr\"ohlich \cite{fr1,fr2}. Let $\alg$
denote the $*$-algebra generated by $\lbrace
\1,a_\lambda^*(f),a_\lambda(g)\rbrace$ for $f,g\in
L^2(\R^3,d^3k)$.

A state on $\alg$  is  a linear functional
\eqn\omega\;:\;\alg\;\longrightarrow\; \C\;\eeqn that is positive
and normalized, that is, $\omega(A^*A)\geq 0$ for all $A\in \alg$,
and $\omega(\1)=1$.

For fixed $\ssig$ and $p$, let $\omega_{p,\ssig}$ denote the
vector state defined by
\eqnn\omega_{p,\ssig}\;:\;\alg&\longrightarrow&\C\\
    A&\longrightarrow&\omega_{p,\ssig}(A)\;=\;
    \left\langle\Psi[p;\ssig]\;,\;A\,\Psi[p;\ssig]
    \right\rangle\;.\eeqnn
For the related case of the massless Nelson model, J. Fr\"ohlich
proved in \cite{fr1} that, in a suitable sense, there exists a
well-defined limiting state, $\omega_p(A)=\lim_{\ssig\rightarrow0}
\omega_{p,\ssig}(A)$ for all $A\in\alg$, and every sufficiently
small value of $|p|$. The Gelfand-Naimark-Segal (GNS-)
construction associated to the state $\omega_p$ yields a Hilbert
space $\H_p^{(IR)}$ that carries a non-Fock representation of the
CCR algebra if $|p|>0$ (that is, a representation of the CCR
algebra unitarily inequivalent to the Fock representation),
together with a vacuum that physically describes a dressed
1-particle state. The same fact is expected to hold for the
present system.
\\

\section{Smoothness of the infraparticle energy}
The fact that the first and second derivatives of the ground state
energy of $\Hps$ with respect to $|p|$ is close to that of the
classical expression $E_{cl}[|p|]=\frac{|p|^2}{2}$ is physically
important. For instance, this result is highly relevant for the
construction of scattering theory according to \cite{fr1,fr2,pi},
or for the semiclassical dynamics of charged particles,
\cite{tesp}. In particular, it provides a formula for the {\bf
mass} of the dressed particle.
\\

\section{Cherenkov radiation}
The fact that Theorems {~\ref{mainthm1}} and {~\ref{mainthm2}} are
only valid for sufficiently small values of $|p|$ is (although
improvements on the bounds on $\puppbd$ are certainly possible)
not only due to a lack of mathematical technology, but also a
consequence of the phenomenon of Cherenkov radiation. In fact, if
$|p|$ approaches 1 (corresponding to the rest energy $mc^2$ of the
charged particle, where $m$ is its mass, and $c$ is the speed of
light), the infraparticle increasingly tends to reduce its kinetic
energy by the emission of electromagnetic (Cherenkov) radiation.
One expects that  if $|p|$ approaches $1$, the eigenvalue $\Egrd$
at the bottom of spec$\Hps$ turns into a resonance, in the sense
of \cite{bfs1,bfs2}.  Clarification of this point is beyond the
scope of this text.
\\

\section{The proof idea}

The proof of Theorems {~\ref{mainthm1}} and {~\ref{mainthm2}} uses
the operator-theoretic renormalization group method developed by
V. Bach, J. Fr\"ohlich, and I.M. Sigal in \cite{bfs1,bfs2}. The
latter is based on the recursive application of the {\bf smooth
Feshbach map} and the {\bf Feshbach theorem}. The smooth Feshbach
map associates to an operator $H$ that acts on a given Hilbert
space $\H$ an operator $H'$ that acts on a Hilbert subspace
$\H'\subset\H$, where, in a sense made precise in the Feshbach
theorem, $H$ and $H'$ are mutually isospectral. The mapping
$\mathcal{L}[\H]\rightarrow \mathcal{L}[\H']$, $H\mapsto H'$ is
referred to as '(Feshbach) decimation', since in the language of
Wilson's renormalization group,  it 'decimates' the 'degrees of
freedom' in $\H-\H'$.

The Bach-Fr\"ohlich-Sigal operator-theoretic renormalization group
is a tool aimed at the analysis of spectral problems in quantum
field theories. In our case, the key issue is to study the bottom
of the spectrum of the hamiltonian $\Hps$, which acts on the
Hilbert fibre space $\Hp$ associated to the fixed value $p$ of the
conserved momentum.

The general setup of the method is given as follows. We introduce
a dyadic decomposition of $\Hp$,
$$\Hp\;=\;\H_{-1}\,\oplus\,\left(\bigoplus_{n\geq0}\,
    \Delta_n\H\right)\;,$$
where
$$\Delta_n\H\;:=\;\chi_{[2^{-(n+1)},2^{-n}]}[H_f]\,\Hp\;,$$
and
$$\H_{-1}\;:=\;\chi_{[1,\infty)}[H_f]\,\Hp\;.$$
Then, the strategy is to employ the smooth Feshbach map to
iteratively decimate the degrees of freedom of the theory in
$\H_{-1}\oplus\H_0\oplus\Delta_1\H\oplus\cdots\oplus\Delta_n\H$,
for increasing $n$ in steps of 1. The purpose is to analytically
control the decimation of the degrees of freedom associated to the
Hilbert space $\chi_{[\mu,\infty)}[H_f]\Hp$, in the limit
$\mu\rightarrow0$.

We will in fact follow the setup of the Wilson renormalization
group, hence a {\bf rescaling transformation} is employed between
each application of Feshbach decimation. This has the effect that
all operators $H_n$ obtained in this process act on the fixed
Hilbert space $P_1\Hp=\chi_{[0,1]}[H_f]\Hp$.

All of these operators are written in Wick ordered normal form
$$H_n\;=\;\sum_{j\in\N_0} H_n^{(j)}\;,$$
where  $H_n^{(j)}$ is the sum of Wick monomials in $H_n$
containing $k$ creation- and $j-k$ annihilation operators, for
$k=0,\dots,j$. In the absence of cutoffs,   $H_n^{(j)}$ is
homogenous under pure scaling, and has the integral scaling
dimension $j$.

Along with the rapid reduction of the initial Hilbert space, we
then observe that under the concatenation of 'Feshbach decimation'
and 'rescaling', which is referred to as the operation of '{\bf
renormalization}', the norms of the operators $H_n^{(j)}$ are
reduced by different numerical fractions under the rescaling map,
the latter depending on their scaling dimensions in the absence of
cutoffs. If this fraction is smaller or equal $\sim\frac{1}{2}$
(up to errors of order $O(g)$), we say that $H_n^{(j)}$ is {\bf
irrelevant}. If it is $\sim1$, $H_n^{(j)}$ is {\bf marginal}, and
if the fraction is larger or equal $\sim2$, $H_n^{(j)}$ is called
{\bf relevant}.

To organize the problem, we introduce a Banach space $\Hspace$,
whose points, denoted by $\h$, are in one-to-one correspondence to
the operators $H_n$, and a map $H$ from $\Hspace$ to
$\Bound(P_1\Hp)$, the bounded operators acting on $P_1\Hp$. That
is, associated to each $H_n$, there is a point
$\h^{(n)}\in\Hspace$, such that $H[\h^{(n)}]= H_n$. We refer to
$H[\h^{(n)}]$ as {\bf effective hamiltonians}.

The {\bf renormalization map} $\ren$ is obtained as follows. Let
$H[\h]$ denote an effective hamiltonian at some given scale,
acting on $P_1\Hp$. From an application of the {\bf Feshbach map},
which is introduced in Chapter {~\ref{AlganpropFeshbsect3}} below,
concatenated with a rescaling transformation, we obtain an
operator $H[\hat{\h}]$. Pulling back the map $H[\h]\mapsto
H[\hat{\h}]$ to $\Hspace$ by way of
$H:\Hspace\rightarrow\Bound(P_1\Hp)$, we obtain the
renormalization map $\ren:\Hspace\rightarrow \Hspace$.

The pair $(\Hspace,\ren)$ is interpreted as a discrete dynamical
system, whose orbits are sequences
$\lbrace\h^{(n)}\rbrace_{n\geq0}$, generated by $n$-fold iterating
$\ren$, applied to some initial value $\h^{(0)}$. The index $n$ is
referred to as the {\bf scale}. Thus, the focus shifts away from
the 'elimination' of the Hilbert spaces $\Delta_n\H$, and moves to
the study of a discrete dynamical system on a Banach space.

All effective hamiltonians are, according to Feshbach's theorem in
the next chapter, mutually isospectral at the bottom of their
spectra. In addition, the Feshbach theorem provides a means to
uniquely determine the ground state eigenvector of any effective
hamiltonian provided that the ground state eigenvector is known
for only one of them.

The very first step, in which the components of $\Hps$ in
$\H_{-1}$ are decimated, 'calibrates' the problem by producing an
initial condition $\h^{(0)}$. Furthermore, $H[\h^{(0)}]$ is
isospectral to $\Hps$ in the sense of the Feshbach theorem.
Therefore, if we succeed in finding the solution of the eigenvalue
problem for any of the $H[\h^{(n)}]$'s, we can immediately obtain
the corresponding result for $\Hps$.

The strategy of our analysis indeed consists of finding a finite
scale $n\sim \Ns$, for which the eigenvalue problem reduces to a
case that has a known solution, and from which the quantities of
interest can be reconstructed for  $\Hps$. This is precisely the
r\^ole of the infrared cutoff in the present model; it provides a
finite scale $\Ns<\infty$, at which the effective hamiltonian
$H[\h^{(\Ns)}]$ has a purely irrelevant interaction part. For the
solution of the eigenvalue problem for $H[\h^{(\Ns)}]$, one can
use the results of Bach, Fr\"ohlich and Sigal \cite{bfs1,bfs2}, to
produce a final condition, from which the ground state data of the
physical hamiltonian  can be reconstructed, basically by backwards
iterating the Feshbach theorem.

The key difficulty in the proof of Theorem {~\ref{mainthm1}} is
the fact that there are purely marginal operators. For their
analysis, we first of all invoke an operator-theoretic version of
the Ward-Takahashi identities of non-relativistic QED to reduce
the level of complexity of the problem. Then, a nested
renormalization group construction is invoked to analytically
master their renormalization flow.

\chapter{ALGEBRAIC AND ANALYTIC
PROPERTIES OF FESHBACH TRIPLES, WICK ORDERING AND THE DECIMATION
MAP} \label{AlganpropFeshbsect3}

A mathematical tool of extraordinary importance in our analysis is
the so-called {\bf smooth Feshbach map}. Combined with the {\bf
Feshbach theorem}, it has, in its initial form, been very
successfully used in the construction of an operator-theoretic
renormalization group for the rigorous analysis of quantum field
theories in \cite{bfs1,bfs2}.

The smooth Feshbach map is a variant that uses smooth cutoff
operators (in the sense of the spectral theorem) instead of
projectors. It has been developed by V. Bach, J. Fr\"ohlich and I.
M. Sigal in the presently unpublished work \cite{bfs3}. I thank
these authors for allowing me to use their work for the exposition
at hand.

There are several properties of the smooth Feshbach map of great
analytical power. The first and foremost one is that of
isospectrality, which is the main reason for its use in the
construction of an operator-theoretic renormalization group.
Furthermore, derivations of the smooth Feshbach map have a
particular algebraic structure that enables us to construct an
operator-theoretic generalization of the Ward-Takahashi identities
in $U(1)-$gauge theory.

In the Bach-Fr\"ohlich-Sigal operator-theoretic renormalization
group, one considers the flow on the Banach space $\Hspace$, which
parametrizes the effective hamiltonians of the theory, that is
generated by a renormalization map. To this end, we write all
effective hamiltonians in Wick ordered normal form. This is
discussed in detail in this chapter.
\\

\section{Feshbach triples and the Feshbach theorem}
\label{isospectrsubsect3}

We assume a closed operator  $H$ with dense domain on a separable
Hilbert space $\H$ to be given, together with a spectral parameter
$z\in\C$ . Furthermore, we associate an operator $\tau[H-z]$ to
$H-z$, and define $\omega[H-z]\equiv H-z-\tau[H-z]$, in a manner
that $\Dom\{\omega[H-z]\}\supseteq\Dom\{\tau[H-z]\}$ are dense in
$\H$.

Let $\ch,\bch$ denote a pair of bounded operators acting on $\H$,
constituting a $C^\infty$-partition of unity,
\eqn\ch^2\,+\,\bch^2\;=\;\1\;,\eeqn
and let
\eqnn\Pch&\Leftrightarrow&{\rm Ran}\{\ch\}\;,\\
    \Pbch&\Leftrightarrow&{\rm Ran}\{\bch\}\eeqnn
denote the orthogonal projections onto their respective ranges.
Likewise, let
\eqnn P^\perp\;\equiv\;\1\,-\,P\;\;\;,\;\;\;
    \bar{P}^\perp\;\equiv\;
    \1\,-\,\bar{P}\;\eeqnn
denote their respective complementary orthogonal projections. We
emphasize that $P P^\perp=0$, but $P\bar{P}\neq0$ (and likewise
for $\bar{P}$). Furthermore, it is, in particular, assumed that
any commutator between any pair in $\tH,\ch,\bch,\Pch$ and $\Pbch$
vanishes.

Let us write $H$ for $H-z$, to simplify the notation. Assuming
that the restrictions of $\tH$ and $\tH+\bch \HW\bch$ to
Ran$\{\Pbch\}$ are both bounded invertible, we may define the
restricted resolvent
\eqn\bar{R}\;:=\;(\tzH+\bch \HW\bch)^{-1} \;
    \label{restrresbarPdef}\eeqn
on $\bar{P}\H$. We assume that
\begin{gather}\bar{R}\;\;,\;\;\bar{R}\,\bch\,\HW\,\ch\;,\;\ch \,\HW\,
     \bch\,\bar{R}\;,\;\nonumber\\
     \ch \HW\,\bch\,\bar{R}\,\bch\,\HW\ch\;\;,\;\;\ch\, \HW\,\ch
     \label{Feshboundopasscond}\end{gather}
all extend to bounded operators on $\H$. Hence, the operators
\eqn\;Q&:=&\chi\;-\;\bch\;\bar{R}\;\bch\;\HW\;\chi\;\;,\nonumber\\
     Q^\sharp&:=&\chi\;-\;\chi\;\HW\;\bch\;\bar{R}\;\bch\eeqn
are well-defined. Furthermore,
\eqn\;\FchtH[H]&:=&\tzH\,+\,\ch
     \,\HW\,\ch\,-
     \ch \,\HW\,\bch\,\bar{R}\,\bch\, \HW\,\ch\nonumber\\
     &=&\tzH\,+\,\ch \,\HW\,Q\nonumber\\
     &=&\tzH\,+\,Q^\sharp\,\HW\, \ch\;\eeqn
is a well-defined operator that acts on $P\H$. $\FchtH$ is called
the {\bf smooth Feshbach map}.
\\

\begin{dfi}\label{Feshbtripledef}
The data $(Q^\sharp,Q,\FchtH)$ will be referred to as the {\bf
Feshbach triple} associated to $(\ch,\H)$.
\end{dfi}

$\;$

Feshbach triples possess a number of miraculous properties. First
and foremost, there is the property of isospectrality in the
following sense.
\\

\begin{thm}\label{Feshprop}
The Feshbach triple $(Q^\sharp,Q,\FchtH)$ associated to $(\ch,\H)$
has the following properties.
\\

\noindent{\bf i) Isospectrality:} $H$ is bounded invertible on
$\H$ if and only if $\FchtH[H]$ is bounded invertible on $P\H$.
Then, the formula
\eqn\FchtH[H]^{-1}\;=\;\chi H^{-1}\chi\;+\;\bch\tzH^{-1}\bch\;,
    \label{feshbinverseform1}\eeqn
relates the respective inverses to each other.
\\

\noindent{\bf ii) Projection of eigenvectors:} $0\neq\psi\in\H$
satisfies $H\psi=0$ if and only if $\FchtH[H]\;\ch\;\psi\;=\;0\;.$
\\

\noindent{\bf iii) Reconstruction of eigenvectors:} $0\neq\zeta\in
P\H$ satisfies $\FchtH[H]\zeta=0$ if and only if $0\neq
Q\zeta\in\H$ satisfies $H Q\zeta=0$.
\end{thm}

$\;$

\prf The proof is reproduced from the preprint \cite{bfs3} by V.
Bach, I.M. Sigal, and J. Fr\"ohlich.

Next, we observe that, writing $F\equiv\FchtH[H]$ in brief, the
identities
\eqn H\,Q&=&\ch\,F\;,\nonumber\\
    Q^\sharp\,H&=&F\,\ch\label{HQeqchFform4444}\eeqn
and
\eqnn F&=&\tau[H]\,+\,\ch\,\omega[H]\,Q\nonumber\\
    &=&\tau[H]\,+\,Q^\sharp\,\omega[H]\,\ch\;.\eeqnn
are satisfied, as can be verified by a straightforward computation
using the fact that $\ch^2+\bch^2=\1$.
\\

(I) Let us suppose that $\tau[H]+\omega[H]$ is invertible. Then,
we claim that
\eqnn F^{-1}&\equiv&
    \ch \,H^{-1}\,\ch\,+\,\bch\,\tzH^{-1}\,\bch\;.\eeqnn
To prove this assertion, we observe that
\eqnn&&\left[\tau[H]^{-1}\bar{P}\,+\,
    \left(\bar{P}^\perp\,-\,\tau[H]^{-1}\bar{P}\,\ch\,
    \omega[H]\right)H^{-1}\,\ch\right]\,F\\
    &&\hspace{1cm}=\;
    \bch\tau[H]^{-1}\bch(\tau[H]+\ch\,\omega[H]\,Q)\,
    \\
    &&\hspace{3cm}+\,\ch\,H^{-1}\,H\,Q\\
    &&\hspace{1cm}=\;\bch^2\,+\,\tau[H]^{-1}
    \left[\ch\,\bch\,-\,\bch^2\,\ch\,\omega[H]\,\bar{R}\right]
    \bch\,\omega[H]\,\ch
    \\
    &&\hspace{3cm}+\,\ch^2\,-\,\ch\,\bch\,\bar{R}\,\bch\,
    \omega[H]\,\ch\\
    &&\hspace{1cm}=\;\1\,+\,\tau[H]^{-1}
    \left[\ch\,\bch\,-\,\ch\,\bch\underbrace{
    (\tau[H]+\bch\omega[H]\bch)
    \bar{R}}_{=\;\bar{P}}\,
    +\,\tau[H]\ch\bch\bar{R}\right]
    \bch\,\omega[H]\,\ch
    \\
    &&\hspace{3cm}-\,\ch\,\bch\,\bar{R}\,\bch\,
    \omega[H]\,\ch\\
    &&\hspace{1cm}=\;\1\;,\eeqnn
hence the expression in discussion  is indeed a left inverse of
$F$. Likewise, one proves that it is also a right inverse. Thus,
$F$ is globally and uniquely invertible on $\H$. In particular,
since $PFP^\perp=P^\perp F P=0$, this also implies the
invertibility of $PFP$ on Ran$\{P\}$.

Conversely, let us assume that the restriction of $F$ to
Ran$\{P\}$ is bounded invertible. Then, defining
\eqnn R&\equiv&Q\,F^{-1}\,Q^\sharp\,+\,\bch\,\bar{R}\,
    \bch\;,\eeqnn
it follows that
\eqnn H\,R&=&\ch\,Q^\sharp\,+\,H\,\bch\,\bar{R}\,\bch\\
    &=&\ch^2\,-\,\ch^2\,\omega[H]\,\bch\,\bar{R}\,\bch
    \,+\,\bch\,H\,\bar{R}\,\bch\\
    &&\;+\;\ch^2\,\omega[H]\,\bch\,\bar{R}\,\bch\\
    &=&\ch^2\,+\,\bch^2\\
    &=&\1\;.\eeqnn
Likewise, one can prove that $R\,H=\1$, hence $R=H^{-1}$.
\\

(II) Assume that $\psi\in\H$, $\psi\neq0$, solves $H\psi=0$. Then,
\eqnn F\,\ch\,\psi&=&Q^\sharp\,H\,\psi\\
    &=&0\; \eeqnn
is satisfied.
\\

(III) Suppose that $\phi\in$ Ran$\{P\}$, $\phi\neq0$, solves
$F\phi=0$. Then,
\eqnn H\,Q\,\phi&=&\ch\,F\,\phi\\
    &=&0\;.\eeqnn
This proves the theorem. \qed

In all cases of interest for us, the hamiltonians in discussion
will be of the form $H=T+W$, where $T$ commutes with $\ch$, while
$W$ contains no part that commutes with $\ch$, and has a small
norm bound relative to $T$. For the construction of the smooth
Feshbach map, the operator $\tau[H]$ will be defined in a manner
that it commutes with $\ch$, as it must be, but
$$\omega[H]\;=\;H\,-\,\tau[H]$$
will in general also contain parts of $T$ that commute with $\ch$.
The following expression for the Feshbach map will later turn out
to be highly useful, because it conveniently separates the terms
in $\FchtH[H]$ that depend on $T$ and $W$.
\\

\begin{lm}\label{piopdeflemma}
Let $\Ttild\equiv T-\tau[H]$, and
\eqn\piop&\equiv&\1\,-\,\bch \Ttild\bch \bar{R}_0\nonumber\\
    &=&\Pbch^\perp\,+\,\Pbch\,\tau[H]\,\bar{R}_0\,\;,
    \label{piopdef3333}\eeqn
where
$$\bar{R}_0\;=\;(\tau[H]\,+\,\bch \Ttild\bch)^{-1}$$
on $\Pbch\H$. Then,
\eqn\FchtH[H]&=&\tau[H]\,+\,\ch\,\Ttild\,\piop\,\ch\,+\,\ch\,
    \piop\,(W\,-\,W\,\bar{R}\,W)\,\piop\,\ch\;,\eeqn
where clearly, $[\piop,\ch]=[\piop,\bch]=0$.
\end{lm}

\prf Using the {\bf second resolvent identity}
\eqn\bar{R}&=&\bar{R}_0\,-\,\bar{R}_0\,\bch
    W\bch\bar{R}\nonumber\\
    &=&\bar{R}_0\,-\,\bar{R}\,\bch
    W\bch\bar{R}_0\;,\label{secondresident4444}\eeqn
and
$$\omega[H]\;=\;\Ttild\,+\,W\;,$$
straightforward calculation shows that
\eqn\FchtH[H]&=&\tau[H]\,+\,\ch \Ttild\ch\,+\,
    \ch W\ch\,\nonumber\\
    &&\hspace{1cm}-\,\ch\omega[H]\bch\bar{R}\bch\omega[H]\ch
    \nonumber\\
    &=&\tau[H]\,+\,\ch \Ttild\ch\,-\,\ch T'\bch\bar{R}_0\bch T'\ch
    \nonumber\\
    &&\,+\,
    \ch W\ch\,-\,\ch W\bch\bar{R}\bch W\ch\nonumber\\
    &&\,-\,
    \ch W\bch\bar{R}_0\bch T'\ch\,-\,
    \ch T'\bch\bar{R}_0\bch W\ch
    \nonumber\\
    &&\,+\,\ch T'\bch\bar{R}_0\bch W\bch\bar{R}_0\bch T'\ch
    \nonumber\\
    &&\,+\,\ch T'\bch\bar{R}_0\bch W\bch\bar{R}\bch W\ch\,+\,
    \ch W\bch\bar{R}\bch W\bch\bar{R}_0\bch T'\ch
    \nonumber\\
    &&\,-\,\ch T'\bch\bar{R}_0\bch W\bch\bar{R}\bch W
        \bch\bar{R}_0\bch T'\ch\nonumber\\
    &=&\tau[H]\,+\,\ch T'\ch\,-\,\ch\,
    \Ttild\bch\bar{R}_0\bch \Ttild\ch\nonumber\\
    &&+\,\ch(\1-\bch \Ttild\bch \bar{R}_0)W
    (\1-\bch \Ttild\bch \bar{R}_0)\ch\nonumber\\
    &&-\,\ch(\1-\bch \Ttild\bch \bar{R}_0)W\bar{R}W
    (\1-\bch \Ttild\bch \bar{R}_0)\ch\;.\eeqn
This proves the claim. \qed

\section{Derivations}
\label{derivationssect3}

Next, we consider derivatives of the smooth Feshbach map with
respect to scalar and operator-valued variables, and of its
commutators with respect to operators and operator valued
distributions. These are cases of {\bf derivations}, and are
collectively governed by Lemma {~\ref{derXbarRlemma1}}, and
Theorem {~\ref{derXfPthm}} below.
\\

\begin{lm}\label{derXbarRlemma1}
Let $\der$ denote a derivation on the bounded linear operators on
$\H$.  That is,
\eqn\der [AB]&=&\der[A]\,B\,+\,A\,\der[B]\nonumber\\
    \der[\lambda A\,+\,\mu B]&=&\lambda\,\der[A]\,+\,\mu\,\der[B]
    \nonumber\\
    \der [\1]&=&0\;,\eeqn
for all $A,B\in\mathcal{L}[\H]$, and all $\lambda,\mu\in\C$. Then,
\eqn\der [\FchtH[H-z]]
    &=&
    \der[\tzH]\; +\;\ch\,\HW\,\bch\,\bar{R}\,
    \der[\tzH]\,
    \bar{R}\,\bch\,\HW\,\ch\nonumber\\
    &+&Q^\sharp\,\der[\HW]\,Q\;\nonumber\\
    &+&\der[\ch]\,(H-z)\,Q\;+\;Q^\sharp\,(H-z)\,\der[\ch]\nonumber\\
    &-&2\,\ch\,\HW\,\bch\,\bar{R}\,
    \frac{\der[\bch]}{\bch}\,\tzH \,
    \bar{R}\,\bch\,\HW\,\ch \;\eeqn
if $\der[\bch],\bch$, and $\tH$ are pairwise commutative.
\end{lm}

\prf Let us first of all note that since
\eqn\bar{R}&=&\tzH^{-1}\,-\,\tzH^{-1}\,\bch\,\HW\,\bch\,
    \bar{R}\nonumber\\
    &=&\tzH^{-1}\,-\,\bar{R}\,\bch\,\HW\,\bch\,
    \tzH^{-1}\;\label{smsecresid333}\eeqn
(the {\bf second resolvent identity}), and
$$[\der[\bch],\tH]\;,\;[\der[\bch],\bch]\;=\;0\;,$$
the operator formally written as
$$\bch\,\bar{R}\,\frac{\der[\bch]}{\bch}$$
has a well-defined meaning.

Next, for any invertible operator $A$,
$$\der[A^{-1}]\;=\;-\,A^{-1}\,\der[A]\,A^{-1}\;,$$
from $0\,=\,\der[\1]\,=\,\der[A\,A^{-1}]\,=\,A\,\der[A^{-1}]\,+\,
\der[A]\,A^{-1}$ (and acting from the left with $A^{-1}$).

Thus,
\eqn\der[\bar{R}]&=&-\bar{R}\,\left(\der[\tzH]\,+\,
    \bch\,\der[\HW]\,\bch\right)\,\bar{R}\nonumber\\
    &&+\;\bar{R}\,\left(\der[\bch]\,\HW\,\bch\;+\;
    \bch\,\HW\,\der[\bch]\right)\,\bar{R}\;,\eeqn
and recalling that
$$\FchtH[H]\;=\;\tH\,-\,z\,+\,\ch\,\HW\,\ch\,-\,
    \ch\,\HW\,\bch\,\bar{R}\,\bch\,\HW\,\ch\;,$$
one finds
\eqn\der[\FchtH[H]]&=&\der[\tzH]\,\nonumber\\
    &&+\,\ch\,\HW\,\bch\,\bar{R}\, \der[\tzH]\,
    \bar{R}\,\bch\,\HW\,\ch\nonumber\\
    &&+\,\ch\,\der[\HW]\,\ch
    \nonumber\\
    &&-\;\ch\,\der[\HW]\,\bch\,\bar{R}\,\bch\,\HW\,\ch
    \;-\;\ch\,\HW\,\bch\,\bar{R}\,\bch\,\der[\HW]
    \nonumber\\
    &&+\;\ch\,\HW\,\bch\,\bar{R}\,\bch\,\der[\HW]\,
    \bch\,\bar{R}\,\bch\,\HW\,\ch\nonumber\\
    &&+\;\der[\ch]\,H\,Q\;+\;
    Q^\sharp\,H\,\der[\ch]\nonumber\\
    &&-\;\ch\,\HW\,\left(\der[\bch]\,\bar{R}\,\bch
    \,+\,\bch\,\bar{R}\,\der[\bch]\right)\,\HW\,\ch\nonumber\\
    &&-\,\ch\,\HW\,\bch\,\bar{R}\,
    \left(\der[\bch]\,\HW\,\bch\,+\,
    \bch\,\HW\,\der[\bch]\right)\,\bar{R}\,\bch\,\HW\,\ch
    \;.\label{derFchtHauxform333}\eeqn
The terms involving $\der[\tzH]$ and $\der[\ch]$ are already in
the form asserted in the lemma. The four terms involving
$\der[\HW]$ are easily seen to combine into
$$Q^\sharp\,\der[\HW]\,Q\;.$$
The terms involving $\der[\bch]$ can be treated as follows. The
last line of (~\ref{derFchtHauxform333}) is the sum of two terms,
one of which is given by
\eqnn\ch\,\HW\,\bch\,\bar{R}\,
    \frac{\der[\bch]}{\bch}\,\bch\,
    \HW\,\bch\,\bar{R}\,\bch\,\HW\,\ch&=&
    \ch\,\HW\,\bch\,\bar{R}\,
    \der[\bch]\,\HW\,\ch\nonumber\\
    &-&\ch\,\HW\,\bch\,\bar{R}\,
    \frac{\der[\bch]}{\bch}\,\tzH\,\bar{R}\,\bch\,\HW\,\ch\;.
    \eeqnn
The first term on the right hand side of the equality sign cancels
one of the two summands on the second last line of
(~\ref{derFchtHauxform333}). The same argument applies to the
terms on the two last lines of (~\ref{derFchtHauxform333}) that
have not been considered yet. Therefore, the terms in
(~\ref{derFchtHauxform333}) that involve $\der[\bch]$  combine to
\eqnn-\,2\,\ch\,\HW\,\bch\,\bar{R}\,
    \frac{\der[\bch]}{\bch}\,\tzH \,\bar{R}\,
    \bch \,\HW\,\ch\;.\eeqnn
Applying the second resolvent identity (~\ref{smsecresid333}) to
the operators $\bar{R}$, one readily arrives at the claim of the
lemma. \qed

\begin{thm}\label{derXfPthm}
Let $\der$, $\der'$ denote derivations as in Lemma
{~\ref{derXbarRlemma1}}. Then, if
$$[\der^{(')}[\ch]\,,\,\bch]\;=\;0\;$$
and
$$\der^{(')}[\tzH]\;=\;0\;,$$
then
\eqn\der[\FchtH[H]]\;=\;Q^\sharp\,\der[\HW]\,Q\;
    \;\label{derXfPformula3}\eeqn
on $\Hsob$. Furthermore,
\eqn\der'[\,\der[
    \FchtH[H-z]]]&=&Q^\sharp\;\der'[\der
    [H-z]]\;Q \nonumber\\
    &-&Q^\sharp\,\der'[\HW]\,\bch\,\bar{R}\,\bch\,
    \der[\HW]\,Q\;\nonumber\\
    &-&Q^\sharp\,\der[\HW]\,\bch\,\bar{R}\,\bch\,
    \der'[\HW]\, Q
    \;\label{derXderYfPthm}\eeqn
on $\Hsob$. In particular, if $\der=\partial_{X}$ for $X$ denoting
some (possibly operator valued) variable, satisfying the above
conditions,
\eqn\partial_X[\FchtH[H]]\;=\;Q^\sharp\,\partial_X[H]\,Q\eeqn on
$\Hsob$, and if $H$ is analytic in  $X$ on $\H$, then $\FchtH[H]$
is also analytic in $X$ on $P\Hsob$.
\end{thm}

\prf Formula (~\ref{derXfPformula3}) follows directly from Lemma
{~\ref{derXbarRlemma1}}.

Next, by use of Lemma {~\ref{derXbarRlemma1}}, and
(~\ref{derXfPformula3}), we obtain
\eqnn\der'[\,\der[\FchtH[H-z]]]&=&\der'[\,Q^\sharp\,\der[H]\,Q\,]
    \nonumber\\
    &=&Q^\sharp\;\der'[\der
    [H]]\;Q\,\nonumber\\
    &&+\,\der'[Q^\sharp]\,H\,Q\,+\,Q^\sharp\,H\,\der'[Q]
    \;.\eeqnn
Inserting
\eqnn\der[Q]&=&\bch\,\bar{R}\,\bch\,\der[\HW]\,\bar{R}\,\bch
    \,\HW\,\ch\,-\,\bch\,\bar{R}\,\bch\,\der[\HW]\,\ch\\
    &=&-\,\bch\,\bar{R}\,\bch\,\der[H]\,Q\;,\eeqnn
using that $\der[\HW]=\der[H]$ and $\der[\bch]=\der[\ch]=0$ for
the special case under investigation, and likewise for $\der'$ and
$Q^\sharp$, we arrive at
\eqnn\der'[\,\der[\FchtH[H]]]&=&\der'[\,Q^\sharp\,\der[H]\,Q\,]
    \nonumber\\
    &=&Q^\sharp\;\der'[\der
    [H]]\;Q\,\nonumber\\
    &&-\,Q^\sharp\,\der'[H]\,\bch\,\bar{R}\,\bch\,\der[H]\,Q\,\\
    &&-\,Q^\sharp\,\der[H]\,\bch\,\bar{R}\,\bch\,\der'[H]\,Q
    \;.\eeqnn
This implies (~\ref{derXderYfPthm}).

The assertion about analyticity follows from the following
argument. Let $H$ be analytic in $X=X_R+iX_I$, where the
subscripts 'R' and 'I' denote the real- and imaginary parts,
respectively. That is, with $\bar{X}\equiv X_R-iX_I$, $H$ is
differentiable with respect to $X$, and
$$\partial_{\bar{X}}H\;=\;0\;.$$
Thus, using (~\ref{derXfPformula3}), $\FchtH[H]$ is differentiable
in $X$, since
\eqn\partial_{X}\FchtH[H]&=&Q^\sharp\,
    (\partial_{X}H)\,Q\;,\eeqn
where by assumption, $\partial_{X}H$ exists and is well-defined,
and \eqn\partial_{\bar{X}}\FchtH[H]&=&Q^\sharp\,
    (\partial_{\bar{X}}H)\,Q\nonumber\\
    &=&0\;\eeqn
implies that $\FchtH[H]$ is analytic in $X$. \qed

The following theorem is central for the operator-theoretic
Ward-Takahashi identities considered later.
\\

\begin{thm}\label{WTauxcor}
Let $\der_i$, $i=1,\dots,m$, denote derivations obeying
$$\der_i[\ch]\;,\;\der_i[\bch]\;,\;\der_i[\tH]\;=\;0\; $$
on $\Dom[\der_i[P]]\supseteq\Hsob$, for all $i$. Furthermore, let
$h_j$, $j=0,\dots,m$, denote linear mappings
$\mathcal{L}[\Hsob]\rightarrow\mathcal{L}[\Hsob]$ which leave
$\mathcal{L}[P\Hsob]$ invariant, such that
$$\left.h_j\,\right|_{P\H}[Q^\sharp\,A\,Q]\;=\;
    Q^\sharp\,h_j[A]\,Q\;$$
holds for all $j$ and $A\in\mathcal{L}[\Hsob]$. Then, defining
$\der_0\equiv\1$, the relationship \eqn\;\sum_{i} h_i[\,
    \der_i[H]\,]\;=\;0\;\eeqn
implies that likewise,
\begin{gather}\;
    \sum_{i} h_i[\,\der_i[\FchtH[H]]\,]\;=\;
    0\;\end{gather}
is satisfied.
\end{thm}

\prf Defining $\der_0:=\1$, we have \eqn\;\sum_{i\geq0} h_i [\,
    \der_i[\FchtH[H]]\,]&=&\sum_{i} h_i [\,
    \der_i[\FchtH[H]]\,]\nonumber\\
    &=&\sum_{i\geq0} h_i [\,
    Q^\sharp\,\der_i[\HW]\,Q\,]\nonumber\\
    &=&Q^\sharp\,\left(\sum_{i\geq0} h_i [\,
    \der_i[\HW]\,]\right)\,Q\nonumber\\
    &=&0\;,\eeqn
as claimed, since by assumption, $\der_i[H]=\der_i[\HW]$. \qed

$\,$
\\

\section{Concatenation rules}
\label{concatlawssect}

Let us in this section formulate the concatenation rules for the
consecutive action of the elements of a Feshbach triple.

For two mutually commuting tuples of cutoff operators
$(\ch_i,\bch_i)$, $i=1,2$, with
$$\ch_1\,\ch_2\;=\;\ch_2,$$
the relationship
\eqn\,F_{\ch_2,\tau_2}[F_{\ch_1,\tau_1}[H]]^{-1}&=&
    F_{\ch_2,\tau_2}[H]^{-1}\;\nonumber\\
    &&+\;
    \bch_2\,\left(\tau_2(F_{\ch_1,\tau_1}[H])^{-1}
    \;-\;\tau_2(H)^{-1}\right)\,\bch_2\eeqn
follows immediately from the inversion formula
(~\ref{feshbinverseform1}) for smooth Feshbach maps. Thus, the
following theorem is clear.
\\

\begin{thm}\label{FchtHconcatthm4444} The  concatenation rule
\eqnn
    F_{\ch_2,\tau_2}[F_{\ch_1,\tau_1}[H]]\;=\;
    F_{\ch_2,\tau_2}[H]\eeqnn
holds if \eqnn\tau_2[F_{\ch_1,\tau_1}[H]]\;=\;\tau_2[H]\eeqnn is
satisfied.
\end{thm}

We will use this theorem in the following way. In the course of
the operator-theoretic renormalization group,  the situation will
be that we successively determine $F_{\ch_1,\tau_1}[H]$ and
$F_{\ch_2,\tau_2}[F_{\ch_1,\tau_1}[H]]$, where $\tau_1[H]$ and
$\tau_2[F_{\ch_1,\tau_1}[H]]$ are defined in a prescribed manner.
Thus, {\bf defining} $\tau_2[H]\equiv\tau_2[F_{\ch_1,\tau_1}[H]]$,
we obtain $F_{\ch_2,\tau_2}[H]$, which, according to the above
theorem, corresponds to the concatenation of the two
'intermediate' smooth Feshbach maps. The intention will then be to
prove certain algebraic identities satisfied by
$F_{\ch_2,\tau_2}[H]$ that are important for the renormalization
of the strongly marginal operators. The quintessence is that
$F_{\ch_2,\tau_2}[H]$ is {\bf independent} of the parameters in
$F_{\ch_1,\tau_1}[H]$. Hence by exploiting the concatenation
property, we control the cancellation of the latter.

For the operators $Q^\sharp,Q$, the following composition rules
holds.
\\

\begin{thm}\label{QQscomplawthm333}
The composition laws
\eqn Q_{\ch_1,\tau_1}[H]\,
    Q_{\ch_2,\tau_2}[F_{\ch_1,\tau_1}[H]]&=&
    Q_{\ch_2,\tau_2}[H]\nonumber\\
    Q_{\ch_2,\tau_2}^\sharp[F_{\ch_1,\tau_1}[H]]\,
    Q_{\ch_1,\tau_1}^\sharp[H]&=&
    Q_{\ch_2,\tau_2}^\sharp[H]\eeqn
hold if $\tau_2[F_{\ch_1,\tau_1}[H]]=\tau_2[H]$. In particular,
\eqn A\,Q_{\ch_2,\tau_2}[H]&=&
    A\,Q_{\ch_2,\tau_2}[F_{\ch_1,\tau_1}[H]]\nonumber\\
    Q_{\ch_2,\tau_2}^\sharp[H]\,A&=&
    Q_{\ch_2,\tau_2}^\sharp[F_{\ch_1,\tau_1}[H]]\,A
    \label{AQpullthrform333}\eeqn
for any operator $A$ that satisfies $A\,\bch_1=0$.
\end{thm}

\prf Assume that $H$ is invertible. From the general formula
\eqn H \,Q_{\chi,\tau}[H]&=&\chi \,\FchtH[H]\nonumber\\
     Q_{\chi,\tau}^\sharp[H]\,H&=& \FchtH[H]\,\chi
     \label{HQeqchFgenform4444}\eeqn
follows that
\eqnn H \,Q_{\ch_2,\tau_2}[H]&=&\ch_2\,F_{\ch_2,\tau_2}[H]\\
    &=&\ch_1\,\ch_2\,F_{\ch_2,\tau_2}[F_{\ch_1,\tau_1}[H]]\\
    &=&\ch_1\,F_{\ch_1,\tau_1}[H]\,
    Q_{\ch_2,\tau_2}[F_{\ch_1,\tau_1}[H]]\\
    &=&H\,Q_{\ch_1,\tau_1}[H]\,
    Q_{\ch_2,\tau_2}[F_{\ch_1,\tau_1}[H]]\;,\eeqnn
using $\ch_1\ch_2=\ch_2$, and likewise for $Q^\sharp$. This
implies the assertion. The fact that (~\ref{AQpullthrform333})
holds if $A\,\bch_1=0$ is immediate. \qed

$\,$

\section{The Wick ordered normal form}
\label{WickordFeshdecsubsec}

In the given problem, we will exclusively study operators that can
be represented as a series of Wick ordered monomials. In the case
of the physical hamiltonian,
\eqn\Hps\;-\;z\;=\;T[\opT]\;-\;z\;+\;\sum_{1\leq
    M+N\leq
    2}\,W_{M,N}[\op]\,\eeqn
on the Hilbert space $\Hp$. The significance of the operators on
the right hand side has already been explained in Chapter
{~\ref{ModHamchapt}}. An essential fact is that there is an
estimate
$$(1-|p|)\,H_f\,+\,\frac{1}{2}
    |\P_f|^2\;\leq\;T[\op]\;\leq\;(1+|p|)H_f\,+\,\frac{1}{2}
    |\P_f|^2\;.$$
As will be shown later, the interaction is relatively bounded with
respect to $T[\op]$.

The situation that we will typically encounter in the
renormalization group analysis, is that we are given an operator
of the form
\eqn\,H[z]\;=\;T[z;\opT]\;-\;z\;+\;\sum_{M+N\geq
    1}\ch_1\,W_{M,N}[z;\op]\,\ch_1\; \label{effhamWickexpdef}\eeqn
on the Hilbert space $P_1\Hp=\chi[H_f<1]\Hp$, which we refer to as
an {\bf effective hamiltonian}. (We recall the notations
$\opT=(H_f,\Ppar,|\Pperp|)$, and $\op=(H_f,\Ppar,\Pperp)$).
$\ch_1$ is, in the sense of the spectral theorem, given by a
smooth operator-valued function $\ch_1[H_f]$, with
supp$\{\ch_1\}=[0,1]$,
$$\ch_1^2\;+\;\bch_1^2\;=\;\1\;.$$
In addition, $\ch_1[x]=1$ if $x\in[0,\frac{2}{3}]$, and $\ch_1[x]$
is strictly monotonic if $x\in(\frac{2}{3},1]$. The terms
appearing in (~\ref{effhamWickexpdef}) are defined as follows.
\\

(I) The parameter $z$ is a complex number in the vicinity of the
origin in $\C$ (assuming that $\Eb$ is set to zero), and
appropriately picked so that in all of our intended applications,
the Feshbach map is well-defined.
\\

(II) The operator $T[z;\opT]$ denotes the noninteracting part of
$H$, and commutes with $H_f$. It can   be written as
\eqn\label{Top11}T[z;\opT]\;=\;H_f\;+\;
      a[z]\;\ch_1\,\P_f^\parallel\,\ch_1\;+\;
      \ch_1\,T_{n-l}[z;\opT]\,\ch_1\;,\eeqn
for $a[z]\in\C$.  The operator $T_{n-l}[\opT;z]$ is $O(H_f^2)$ in
the limit $H_f\rightarrow0$, of class $C^2$ in the variable $H_f$,
and analytic in $\P_f$.  In the cases to be considered, the values
of the coefficient $a[z]$ are such that relative bounds
\eqn\gamma H_f\;\leq\;T[z;\opT]\;\leq\;\Gamma H_f\,
    \label{TopTzgammaGammabound11}\eeqn
hold on $P_1\Hp$ for some  $0<\gamma<1$, and $\Gamma>1$, if $z$ is
appropriately picked in $\R$.
\\

(III) The operator $W_{M,N}[z;\op]$ denotes the interaction term
{\bf of degree $M+N\geq 1$} in $H[z]$, as we will say. It is a
Wick monomial of the form \eqn\label{Wmon}W_{M,N}[z;\op]&=&
     \sum_{\lambda^{(M)},\tilde{\lambda}^{(N)}}\;
     \int \,\prod_{i=1}^M\,\prod_{j=1}^N\,\frac{
     d^3k_i\;\bcs
     (|k_i|)}{\sqrt{|k_i|}} \,
     \frac{d^3\tilde{k}_j\;\bcs(|\tilde{k}_j|)}{
     \sqrt{|\tilde{k}_j|}}
     \;\times\nonumber\\
     &&\hspace{1cm}\times\;
     a^*_{\lambda^{(M)}}(k^{(M)})\,
     w_{M,N}\left[z;\op;K^{(M,N)}\right]\,
     a^\sharp_{\tilde{\lambda}^{(N)}}(\tilde{k}^{(N)})
     \;,\eeqn
containing $M$ creation, and $N$ annihilation operators, where
$$\sigma\;\geq\;\ssig$$
is referred to as the {\bf running infrared cutoff}.

We recall the following multiindex notation:
\eqnn\lambda^{(M)}&:=&(\lambda_1,\dots,\lambda_M)\\
     k^{(M)}&:=&(k_1,\dots,k_M),\eeqnn
as we recall, and likewise for $\tilde{\lambda}^{(N)}$ and
$\tilde{k}^{(N)}$. Moreover,
\eqnn\,K^{(M,N)}\;:=\;\left(k^{(M)},\lambda^{(M)};
     \tilde{k}^{(N)},\tilde{\lambda}^{(N)}\right) ,\eeqnn
and \eqnn a^\sharp_{\lambda^{(M)}}(k^{(M)}):=\prod_{i=1}^M
      a^\sharp_{\lambda_i}(k_i).\eeqnn
In cases where we consider the dependence of the interaction
operators with respect to the interaction kernels, the notation
\eqn\;W_{M,N}\left[[\;w_{M,N}[z;\op;\cdot]\;\right]]\;\eeqn will
be   used.

The {\bf integral kernels} ('{\bf interaction kernels}')
$$w_{M,N}\left[z;\op;K^{(M,N)}\right]$$
are operator valued $C^2$-functions of $\op$, and commute with
$H_f$. They are fully symmetric functions with respect to the
photon momenta $k_1,\dots,k_M$ on the one hand, and
$\tilde{k}_1,\dots, \tilde{k}_N$, on the other hand. Furthermore,
they are of class $C^1$ with respect to $|k_i|$, $|\tilde{k}_j|$,
$z$, $|p|$, and {\bf analytic} in $\P_f$.
\\

\subsection{Derivatives with respect to scalars and $\op$}

Derivatives with respect to complex scalars of the Wick ordered
operators $H=T+\ch_1\sum W_{M,N}\ch_1$ are defined in the obvious
manner.

Derivatives with respect to components $\opT$ and $\op$ are given
as follows. The appropriate definition of
$\partial_{\op^r}W_{M,N}[z;\op]$, compatible with the operation of
Wick ordering, is given by \eqnn\partial_{\op^r}W_{M,N}
    \left[[w_{M,N}[z;\op;\cdot]\;\right]]
    \;=\;W_{M,N}
    \left[[\partial_{\op^r}w_{M,N}[z;\op;\cdot]\;\right]]
    \;,\eeqnn
that is, the derivative acts directly on the integral kernel of
$W_{M,N}$. One then straightforwardly verifies that
\begin{gather}\partial_{\op^r}:\;\left(\;
    f_0[z;\op]W_{M_1,N_1}[z;\op]
    \cdots W_{M_L,N_L}[z;\op]f_L[z;\op]\;\right)
    \;:\;\nonumber\\
    \;\;\;\;\;\;=\;
    :\;\partial_{\op^r}\left(\;
    f_0[z;\op]W_{M_1,N_1}[z;\op]
    \cdots W_{M_L,N_L}[z;\op]f_L[z;\op]\;\right)
    \;:
    \end{gather}
holds for all $C^1$-functions $f_i[z;\op]$, which commute with
$H_f$. The expression on the second line is evaluated by using the
standard product rule, under preservation of the order of
operators in the product.
\\

\section{(Smooth) Feshbach decimation}
\label{Neumserexpsubsect3}

It is an important fact for the Bach-Fr\"ohlich-Sigal
operator-theoretic renormalization group that for a given
effective hamiltonian $H[z]$ in the Wick ordered normal form
(~\ref{effhamWickexpdef}), which acts on the Hilbert space $\H$,
its smooth Feshbach image $\FchtH[H[z]]$, which acts on
$P\H\subset\H$, can, for a unique choice of $\tau[H[z]]$, again be
written in Wick ordered normal form, but under a modification of
$T$ and $w_{M,N}$. The process in which this Wick ordered
expression for $\FchtH[H[z]]$ is generated is called {\bf Feshbach
decimation}.
\\

\subsection{Definition of the smooth Feshbach projection}

The first step in the definition of the smooth Feshbach map
consists of determining an appropriate operator $\tzH$. To this
end, we notice that the operator $H_f-z$ differs from the other
operators occurring in $H[z]$ by the fact that it is not cut off
by operators $\ch_1$ from the left and right. To preserve this
property, we assume that
$$\tHz\;=\;(1+\DHf[z])\,H_f\,-\,z\,+\,\Delta E_0[z]$$
for unknown scalars $\DHf[z],\Delta E_0[z]$, which remain to be
determined.

Then, with
$$\bch\;\equiv\;\bch_\rho\,\bch_1\;,$$
we have
\eqn\FchrtH[H[z]]&=&\tHz\,+\,\chr\,\HWz\,\chr\,
    \nonumber\\
    &&-\,\chr\,\HWz\,\bchr
    \,\bar{R}\,\bcr\,\HWz\,\chr\;,\eeqn
where
\eqnn\HWz&=&-\,\DHf[z]\,H_f\,-\,\Delta E_0[z]
    \,\\
    &&+\,\ch_1\,\left[a[z]\Ppar\,+\,T_{n-l}[z;\opT]\,+\,
    W[z;\op]\right]\,\ch_1\;,\eeqnn
and
$$\bar{R}\;=\;(\tHz\,+\,\bchr\,\HWz\,\bchr)^{-1}$$
on $\bar{P}\H$.
\\

\subsubsection{The correction of $z$}
The number $\Delta E_0[z]$ is determined by the condition that
$$\left\langle\FchrtH[H[z]]\right\rangle_{\vac}\;=\;
    \Delta E_0[z]\,-\,z\;.$$
This implies that
$$\chr\,\left[-\,\Delta E_0[z]\,-\,
    \HWz\,\bchr
    \,\bar{R}\,\bcr\,\HWz\,\right]\,\chr$$
has a vanishing $\vac$-expectation value, that is,
\eqn \Delta E_0[z]\;=\;-\;\left\langle W\,\bchr
    \,\bar{R}\,\bcr\,W\right\rangle_{\vac}\;,\eeqn
where the right hand side depends on $\Delta E_0$ via $\bar{R}$.
\\

\subsubsection{The coefficient of $H_f$}
\label{DHFimplrelsubsubsec333} The quantity $\DHf[z]$ is obtained
from
$$\left\langle\partial_{H_f}\FchrtH[H[z]]\right\rangle_{\vac}\;=\;
    1+\DHf[z] \;.$$
This implies that the Wick ordered expression for
$$\chr\,\left[-\,\DHf[z]\,H_f\,-\,
    \HWz\,\bchr
    \,\bar{R}\,\bcr\,\HWz\,\right]\,\chr$$
contains no term proportional to  $\chr H_f\chr$, so that the term
linear in $H_f$ is given by $(1+\DHf[z])H_f$. Let us next derive
the implicit relation that determines $\DHf[z]$.
\\

\begin{lm} Let
$$z'\,\equiv\,\langle\partial_{H_f}\FchrtH[H[z]]
    (z-\Delta E_0)\rangle_{\vac}\;.$$
Then,
\eqn&&\left\langle\partial_{H_f}\FchrtH[H[z]]\right\rangle_{\vac}
    \;=\; \left\langle Q^\sharp(1+\partial_{H_f}(T_{n-l}+W))\,Q
    \right\rangle_{\vac}\;\times\nonumber\\
    &&\hspace{1cm}\times\left(1-\left\langle W\bch\bar{R}
    \ch^2\bar{R}\bch W
    \right\rangle_{\vac}+ \left\langle W \bch\bar{R}
    \frac{\partial_{H_f} \bch}{\bch}
    (H_f-z')\bar{R}\bch W
    \right\rangle_{\vac}\right)^{-1}
    \;.\label{derHfFchrtHform333}\eeqn
\end{lm}

\prf Application of Lemma {~\ref{derXbarRlemma1}} yields
\eqn1+\DHf[z]&=&\left\langle\partial_{H_f}\FchrtH[H[z]]
    \right\rangle_{\vac}\nonumber\\
    &=&(1+\DHf[z])\left(1+\left\langle W\bch\bar{R}^2\bch W
    \right\rangle_{\vac}\right)\nonumber\\
    &&-\;\DHf[z]\,\left(1+\left\langle W\bch\bar{R}
    \bch^2\bar{R}\bch W
    \right\rangle_{\vac}\right)\nonumber\\
    &&+\;\left\langle Q^\sharp(\partial_{H_f}(T_{n-l}+W))Q
    \right\rangle_{\vac}\nonumber\\
    &&-\;\left\langle W \bch\bar{R}
    \frac{\partial_{H_f} \bch}{\bch}
    \tHz \bar{R}\bch W
    \right\rangle_{\vac}\;,
    \eeqn
which can be straightforwardly converted into
\eqn\left\langle\partial_{H_f}\FchrtH[H[z]]\right\rangle_{\vac}
    &=&\frac{ \left\langle Q^\sharp(1+\partial_{H_f}(T_{n-l}+W))Q
    \right\rangle_{\vac}}{1-\left\langle W\bch\bar{R}
    \ch^2\bar{R}\bch W
    \right\rangle_{\vac}}\nonumber\\
    &&-\;\frac{\left\langle W \bch\bar{R}
    \frac{\partial_{H_f} \bch}{\bch}
    \tHz \bar{R}\bch W
    \right\rangle_{\vac}}{1-\left\langle W\bch\bar{R}
    \ch^2\bar{R}\bch W
    \right\rangle_{\vac}}\;.\eeqn
With
$$z'\,=\,\langle\partial_{H_f}\FchrtH[H[z]]\rangle_{\vac}
    (z-\Delta E_0)\;,$$
we thus have
\eqnn&&\left\langle\partial_{H_f}\FchrtH[H[z]]\right\rangle_{\vac}
    \;=\; \left\langle Q^\sharp(1+\partial_{H_f}(T_{n-l}+W))\,Q
    \right\rangle_{\vac}\;\times\nonumber\\
    &&\hspace{1cm}\times\left(1-\left\langle W\bch\bar{R}
    \ch^2\bar{R}\bch W
    \right\rangle_{\vac}+ \left\langle W \bch\bar{R}
    \frac{\partial_{H_f} \bch}{\bch}
    (H_f-z')\bar{R}\bch W
    \right\rangle_{\vac}\right)^{-1}
    \;, \eeqnn
which proves the claim. \qed

\subsection{Neumann series expansion}

Feshbach decimation is achieved in two steps:\\
1)  Expansion of $\bar{R}$ in $\FchtH[H[z]]$
into a Neumann series. \\
2) Application of Wick ordering to the resulting expression.

In Chapter {~\ref{firstdecchap}}, we will apply the smooth
Feshbach map to the physical hamiltonian $\Hps$. In that case, the
Hilbert space is given by $\H=\Hp$, and $\FchltH$ maps it to an
operator acting on the Hilbert subspace $P_1\Hp$. $\ch_1$ is a
smooth function of $H_f$, with spectral support $[0,1]$, and in
possession all properties required for the construction of
$\FchltH[\Hp-z]$. In all subsequent applications of the Feshbach
map, discussed in Chapter {~\ref{analiterstepchap}}, $\H=P_1\Hp$,
and $P=P_\rho$, with $\rho=\frac{1}{2}$. We will here discuss the
decimation map for the latter cases. For $\Hps$, the arguments are
fully analogous, and will be demonstrated in detail in Chapter
{~\ref{firstdecchap}}.

Let us assume an effective hamiltonian $H[z]$ of the form
(~\ref{effhamWickexpdef}) to be given, which shall act on
$P_1\Hp$.

Step 1) of Feshbach decimation is obtained as follows. Let
$\Ttild\equiv T-\tHz$, and
\eqnn\bar{R}_0[z]\;:=\;
    \left(\tHz\,+\,\bchr\,\Ttild\,\bchr\right)^{-1}
     \; \eeqnn
on $\bar{P}\Hp$. Expansion of $\bar{R}[z]$ in $\FchrtH$ into a
Neumann series yields
\eqnn\bar{R}[z]&=& \bar{R}_0^{\frac{1}{2}}[z]\,
    \sum_{L\geq0}\left(\bar{R}_0^{\frac{1}{2}}[z]\,
    \bch\,W[z;\op]\,\bch\,\bar{R}_0^{\frac{1}{2}}[z]\right)^L
    \,\bar{R}_0^{\frac{1}{2}}[z]\;\eeqnn
(for an appropriate branch of the square root if $z$ has a nonzero
imaginary part), which is absolutely convergent if the estimates
\begin{gather}\left\|\;\bar{R}_0^{\frac{1}{2}}[z]\;\bch\,
    W[z;\op]\;\bch\,\bar{R}_0^{\frac{1}{2}}[z]\;\right\|
    \;,\;
    \left\|\;\bar{R}_0^{\frac{1}{2}}[z]\;\bch\,\HWz\;
    \ch\;\right\|\;,\nonumber\\
    \left\|\;\ch\;\HWz\;\bch\,
    \bar{R}_0^{\frac{1}{2}}[z]\;\right\|\;\ll\;
    1\end{gather}
are satisfied (the norms are evaluated on $\H$, not on $P\H$).
Using (~\ref{TopTzgammaGammabound11}), these estimates are, for
appropriate values of $z$, indeed true. This is demonstrated in
detail in  Chapter {~\ref{firstdecchap}} for the first application
of   Feshbach decimation, and in Chapter {~\ref{analiterstepchap}}
for all other applications of it.

It is thus clear that defining
$$\bch\;:=\;\bchr\,\ch_1\;,$$
and using $\chr\,\ch_1=\chr$, one can write
\eqn\label{Neumexp}\FchrtH[H[z]]&=&\tHz\,+\,\chr\,\Ttild\piop\chr\,+\,
    \chr\,\HWz\,\chr\;
    \nonumber\\
    &-&\sum_{L=2}^\infty\, (-1)^L
    \sum_{\stackrel{1\leq M_i+N_i}{i=1,\dots,L}}\,
    \chr\,\piop\,W_{M_1,N_1}\,
    \bch\,\bar{R}_0 \,\bch\,\cdots
    \nonumber\\
    &&\hspace{0.5cm}\cdots\,
    \bch\,W_{M_{L-1},N_{L-1}}\,\bch\,
    \bar{R}_0\,\bch
    W_{M_L,N_L}\,
    \piop\,\chr\,\;,\eeqn
where the operator $\piop$ is defined in (~\ref{piopdef3333}). The
next issue is to Wick order this expression.
\\

\subsection{Wick ordering}
\label{Wickordsubsubsec}

Applying Wick ordering to each summand in the Neumann series
(~\ref{Neumexp}) for every fixed value of the summation index $L$,
and collecting all Wick monomials that are equal in the number of
creation and annihilation operators, we find
\eqn\dec[H[z]]&=&(1+\DHf[z])\,H_f\,-\,z\,+\,
     \Delta
     E_0[z]\;\nonumber\\
     &&+\;(a[z]+\Delta a[z])\,\chr\,\Ppar\,\chr\,
     \nonumber\\
     &&+\;
     \chr\,( T_{n-l}\piop+\Delta T_{n-l})\,\chr
     \nonumber\\
     &&+\;\left.\left.\sum_{M+N\geq 1}\;\chr\,
     \right(\piop\, W_{M,N}\,\piop\,+\,
     \Delta W_{M,N}\right)\,\chr\;
     \;.\label{momspaceconstr1}\eeqn
The operator $\dec$ denotes the concatenation of the Feshbach map
with Neumann series expansion and Wick ordering, and is referred
to as the {\bf Feshbach decimation map}. Pulling $\chr$ from the
left and right through the creation and annihilation operators in
$W_{M,N}+\Delta W_{M,N}$, one observes that
\eqnn
    H_f\;+\;\sum_{i=1}^M\;|k_i|\;&\leq&\rho\;,\nonumber\\
    H_f\;+\;\sum_{j=1}^N\;|\tilde{k}_j|&\leq&\rho\;\eeqnn
is satisfied by the momentum space arguments. Therefore, their
support is contained in the simplicial subset
\begin{gather}\label{momentumspsubset1}
     \left\lbrace(k_1,\dots,k_M)\in
     \R^{3M}\;,\;(\tilde{k}_1,\dots,\tilde{k}_N)\in\R^{3N}
     \left|\;\sum_{i=1}^M\;|k_i|\;,\;
     \sum_{j=1}^N\;|\tilde{k}_j|\leq1\right.\right\rbrace
\end{gather}
of $\R^{3(M+N)}$. In particular, this shows that every photon
momentum appearing in an interaction kernel of any effective
hamiltonian always lies in a unit ball around the origin.

For any partitioning of the set $\lbrace 1,\dots,m+q\rbrace$ into
two disjoint sets consisting of $m$ respectively $p$ elements, and
of the set $\lbrace 1,\dots,n+q\rbrace$ into two disjoint sets
consisting of $n$ respectively $q$ elements, we define the
operator valued function
\eqn\label{Wmnpq}&&W_{p,q}^{m,n}[z;\op;K^{(m,n)}]
     \;:=\;
     \sum_{\mu^{(p)},\tilde{\mu}^{(q)}}\int
     \left(\prod_{i=1}^p d^3x_i\;\bcs(|x_i|)\;
     |x_i|^{-\frac{1}{2}}\right)\;\times\nonumber\\
     &&\hspace{1cm}\times\;\;
     \left(\prod_{j=1}^q\;d^3\tilde{x}_j\;\bcs(|\tilde{x}_j|)\;
     |\tilde{x}_j|^{-\frac{1}{2}}\right)\;\times\nonumber\\
     &&\hspace{2cm}\times\;\;a^*_{\lambda^{(p)}}(x^{(p)})\;w_{m+p,n+q}
     [z;\op;X^{(p,q)},K^{(m,n)}]\;
     a_{\tilde{\lambda}^{(q)}}(\tilde{x}^{(q)})\;.\eeqn
The integral kernels are totally symmetric functions of the photon
momenta. Hence, each of the
$\left(\begin{array}{c}m+p\\p\end{array}\right)
\left(\begin{array}{c}n+q\\q\end{array}\right)$ possible
partitionings produce the same operator $W_{p,q}^{m,n}$.

The integral kernels $\Delta w_{M,N}$ of $\Delta W_{M,N}$ are
given by \eqn\label{DeltawMN} \Delta
      w_{M,N}[\cdots]\;=\;-\;
      \sum_{L=2}^\infty (-1)^L\;
      \sum_{\stackrel{2\geq M_i+N_i\geq 1}{i=1,\dots,L}}
      \delta_{M,\sum_{i=1}^L m_i}\delta_{N,\sum_{i=1}^L n_i}
      \;\times\hspace{1cm}\nonumber\\
      \times\;
      \prod_{i=1}^L
      \left(\begin{array}{c}m_i+p_i\\p_i\end{array}\right)
      \left(\begin{array}{c}n_i+q_i\\q_i\end{array}\right)
      \; \mathcal{E}_{L}
      \left[\{\underline{m},\underline{n},\underline{p},\underline{q}\};
      z;\op;K^{(M,N)}\right]\eeqn
with $M_i:=m_i+p_i$, and $N_i:=n_i+q_i$,
$\underline{m}:=(m_1,\dots,m_L)$, etc., and
$M:=|\underline{m}|:=\sum m_i$, $N=|\underline{n}|=\sum n_i$.
Furthermore,
\begin{gather}
    \mathcal{E}_{L}
    \left[\{\underline{m},\underline{n},\underline{p},\underline{q}\};
    z;\op;K^{(M,N)}\right]
    \;\equiv\;
    \left\langle\,\piop W_{p_1,q_1}^{m_1,n_1}\bch
    \bar{R}_0\bch W_{p_1,q_1}^{m_1,n_1}\cdots \bch
    W^{m_l,n_l}_{p_L,q_L}\piop\,
    \right\rangle_{\op,sym}\;,\label{mthcalELdef3}\end{gather}
is defined as the expression
\eqnn&&\mathrm{Sym}_{\underline{m},\underline{n}}
      \left.\left\langle \;\piop\left[\sh_X^{(m_1,\dots,m_L)}\op;z
      \right] W^{m_1,n_1}_{p_1,q_1}
      \left[\sh_X^{(m_2,\dots,m_L)}\op;z;K^{(m_1,n_1)}\right]\,
      \times\right.\right.\nonumber\\
      &&\hspace{4cm}\times\;\bch\,\bar{R}_0\,\bch\,
      \left[\sh_X^{(m_2,\dots,m_L;n_1)}\opT ;z\right]
      \;\cdots
      \nonumber\\
      &&\hspace{1cm}\left.\left.\cdots \;\bch\,W^{m_L,n_L}_{p_L,q_L}
      \left[\sh_X^{(n_1,\dots,n_{L-1})}\op;z;K^{(m_L,n_L)}\right]
      \piop\left[\sh_X^{(n_1,\dots,n_{L})}\op;z\right]
      \;\right\rangle_{\vac}\right|_{X\rightarrow\tildop}\;,
      \hspace{1cm} \eeqnn
where $\mathrm{Sym}_{\underline{m},\underline{n}}$ fully
symmetrizes its argument in all $M=|\underline{m}|$ momenta
associated to creation, and all $N=\underline{n}$ momenta labeling
annihilation operators, but not among one another.

The shift operator $\sh_X$ is defined as follows. Let \eqnn
X\;=\;(X^1,X^2,{\bf X}^3)&\in&
     {\rm Spec}\lbrace(H_f,\Ppar,
     \Pperp)\rbrace\nonumber\\
     &&=\;[0,1]\times[-1,1]\times D_1(0)\;\eeqnn
denote the spectral variable associated to
$(H_f,\P_f^\parallel,\P_f)$, where $D_1(0)\ni{\bf X}^3$ is the
unit disc in $\R^2$. The shift operator
$\sh_X^{(m_i,\dots,m_{L};n_{1},\dots,n_{i(-1)})}$ is defined by
\eqn
     \sh_X^{(m_i,\dots,m_{L};n_{1},\dots,n_{i(-1)})}\;:
     \;H_f&\longmapsto& H_f\;+\;\sum_{j=i}^L|k_j|\;+\;
     \sum_{l=1}^{i(-1)}|\tilde{k}_l|\;+\;X^1\;,\;\nonumber\\
     \P_f^\parallel&\longmapsto&\P_f^\parallel\;+\;
     \sum_{j=i}^L k_j^\parallel
     \;+\;\sum_{l=1}^{i(-1)}\tilde{k}_l^\parallel\;+\;X^2\;,
    \nonumber\\
     \Pperp&\longmapsto& \Pperp\;+\;\sum_{j=i}^L k_j
     \;+\;\sum_{l=1}^{i(-1)}\tilde{k}_l\;+\;{\bf X}^3\;.
    \label{defshiftop}\eeqn

The operators $\Delta T$ and $\Delta w_{M,N}$ correspond to the
respective radiative corrections of $T$ and $w_{M,N}$ under
Feshbach decimation. For a more detailed exposition of issues
concerning Wick ordering, we refer the reader to the appendix of
\cite{bfs2}.
\\

\begin{dfi}
If the expectation value $\mathcal{E}_L$ in (~\ref{mthcalELdef3})
exhibits an index $i\in\lbrace 1,\dots,L\rbrace$, so that $p_i\neq
0$ or $q_i\neq 0$, it is called a {\bf loop contribution}.
Otherwise, if $p_i=q_i=0$ for all $i$, it is called a {\bf tree
level contribution}. A momentum which labels a creation or
annihilation operator within the expectation value is called an
{\bf inner momentum} (indices $p_i,q_i$). Otherwise it is called
an {\bf external momentum} (indices $m_i,n_i$).
\end{dfi}

$\;$

\subsection{Estimates on $\nm\mathcal{E}_{L}\nm$}
We remark that for any Borel function $f\in L^\infty($spec$\op)$,
\eqn\|\sh_X^{(m_i,\dots,m_{L};n_{1},\dots,n_{i(-1)})}f[\op]\|
    \;\leq\;\|f\|_\infty\;=\;\|f[\op]\|\;,\eeqn
since shifting $\op$ corresponds to a translation of the support
of $f$ in spec$\op$, which, of course, does not increase the
supremum of $|f|$ in spec$\op$.

Therefore, $\nm\mathcal{E}_{L}\nm$ can be estimated by
\eqn\nm\mathcal{E}_{L}\nm\;\leq\;\|\piop\|^2\,
    \|\bar{R}_0\|^{L-1}\;
    \prod_{i=1}^L\nm
    W_{p_i,q_i}^{m_i,n_i}\nm\;,\eeqn
whenever the norms on the right hand side are finite.
\\

\chapter{THE OPERATOR-THEORETIC RENORMALIZATION GROUP}
\label{rgdefsection}

In this chapter, we  introduce the detailed framework of the
operator-theoretic renormalization group developed in
\cite{bfs1,bfs2}.

The setting of any renormalization group construction generally
comprises a space $\Hspace$, whose points, denoted by $\h$,
parametrize the physical system at hand (coupling constants, for
instance, or operators in the present case). There is always a
natural scale that characterizes the given problem (an energy or
momentum cutoff, for instance). Considering the dependence of
$\h\in\Hspace$ with respect to this scale, and varying it in
discrete steps, a dynamical system is obtained on $\Hspace$, whose
flow map is referred to as the renormalization map $\ren$. The
fixed points of this flow correspond to macroscopic universality
classes of microscopic theories.

In the present case, $\Hspace$ is a Banach space of function
sequences, whose points define certain bounded operators on
$P_1\Hp$. The renormalization map $\ren$ is obtained from
concatenating three distinct operations, the decimation map $\dec$
introduced in the previous chapter, a rescaling transformation
$\resc$, and a redefinition $Z$ of the spectral parameter $z$.
$\dec$ removes all photon states with energies in the interval
$[\rho,1]$ by use of the Feshbach map, for some fixed real number
$0<\rho<1$.
\\

\section{The Banach sequence space of Wick monomials}

Let us now give a detailed definition of the parameter space
$\Hspace$ of the present system.

All effective hamiltonians that we consider can be written as a
series of Wick monomials
$$H[z]\;=\;T[z;\opT]\,-\,z\,+\,\sum_{M+N\geq 1}
    \ch_1\,W_{M,N}[z;\op]\,\ch_1\;,$$
where
$$T[z;\opT]\;=\;H_f\,+\,\ch_1\,\left(
    \underbrace{a[z]\,\Ppar\,+\,
    T_{n-l}[z;\opT]}_{\equiv \Ttild[z;\opT]}
    \right)\,\ch_1\;,$$
which is precisely of the form introduced in and after
(~\ref{effhamWickexpdef}), and acts on the Hilbert space $P_1\Hp$.

$z$ is the spectral parameter that emerges in the definition of
the Feshbach map, and is picked in a $O(g)$-vicinity of
$\lbrace0\rbrace\in\C$.

As before, $T[z;\opT]$ denotes the noninteracting part of $H[z]$,
and commutes with $H_f$, and $W_{M,N}[z;\op]$ accounts for the
interaction operator of degree $M+N\geq 1$, which contains $M$
creation, and $N$ annihilation operators. We refer to Section
{~\ref{WickordFeshdecsubsec}} for their precise definition.

All effective hamiltonians $H[z]$ that emerge in the
renormalization group construction are elements of
$\Bound(P_1\Hp)$, the bounded operators on $P_1\Hp$, and exhibit
this structure. Therefore, we introduce a Banach sequence space
\eqnn\Hspace\;=\;\C\;\oplus\;{\mathfrak T}\;\oplus\;
    {\mathfrak W}\eeqnn
with \eqnn {\mathfrak W}\;=\;\bigoplus_{M+N\geq
    1}{\mathfrak W}_{M,N}\;,\eeqnn
whose points shall parametrize operators of the form $H[z]$.
\\

The building blocks of the space $\Hspace$ are thus given as
follows.
\\

(i) $\C$ is the space of spectral parameters $z$. $z$  is picked
in a $O(g)$-vicinity of the origin in $\C$.
\\

(ii) By definition of $\opT=(H_f,\Ppar,|\Pperp|)$,
$${\rm Spec}\lbrace\opT\rbrace\;=\;
     [0,1]\times[-1,1]\times [0,1]\;.$$
The space \eqnn{\mathfrak T}\;=\;C^2({\rm Spec}\{\opT\})
    \times C^1({\C})\;\eeqnn
parametrizes the noninteracting hamiltonians.

Let $\alpha=(\alpha_i)_{i=1}^3$ denote a 3-component multiindex,
with $\alpha_i\in\{0,1,2\}$, and
$$|\alpha|\;:=\;\sum\alpha_i\;.$$
For $Y=(Y^i)_{i=1}^3$, where $Y^i$ can be either a scalar or a
vector, let \eqnn\partial_{Y}^\alpha\;:=\;\prod_i
    \partial_{Y^i}^{\alpha_i}\;.\eeqnn
We introduce the norm
\eqnn\nm\;T\;\nm\;:=\;\sup_{|\alpha|=0,1,2}\;
     \sup_{Y\in{\rm Spec}\lbrace\opT\rbrace}\;
    \left|\;\partial_{Y}^\alpha T[z;Y]\;\right|\eeqnn
on ${\mathfrak T}$, where $T\in{\mathfrak T}$. Evidently,
$({\mathfrak T},\nm\cdot\nm)$ is a Banach space.
\\

(iii) Clearly, for $\op=(H_f,\Ppar,\Pperp)$,
$${\rm Spec}\lbrace\op\rbrace\;=\;
     [0,1]\times[-1,1]\times D_1(0)\;$$
on $P_1\Hp$, where $D_1(0)\ni Y^3$ is the unit disc in $\R^2$.

The space
\begin{gather}
     {\mathfrak W}_{M,N}\;:=\;
     C^2({\rm Spec}\op)\times C^1({\C})\times
     C_{sym}^1(\R^{3M})\times C_{sym}^1(\R^{3N})
     \; \end{gather}
parametrizes integral kernels of degree $M+N\geq1$.
$C^1_{sym}(\R^{3M})$ is the space of continuously differentiable
functions
$f:\underbrace{\R^3\times\cdots\times\R^3}_{M}\rightarrow\C$ which
are invariant under permutations of the $M$ factors.

Let $\alpha$ again denote a 3-component multiindex. We introduce
the norms \eqnn\nm w_{M,N}\nm\;:=\;
     \sup_{Y\in{\rm Spec}\lbrace\opT\rbrace}\;\;
     \sup_{K^{(M,N)}} \;
     \left|\;
     w_{M,N}[z;Y;K^{(M,N)}]\;\right|\eeqnn
and \eqnn\nm w_{M,N}\nm^{(\zeta)}\;:=\;\sum_{|\alpha|\leq\zeta}\;
     \nm\;\partial_{Y}^{\alpha}
     w_{M,N}[z;Y;K^{(M,N)}]\;\nm\eeqnn
for $w_{M,N}\in{\mathfrak W}_{M,N}$. For every fixed value of
$M+N$ and $\zeta$, the pair $({\mathfrak
W}_{M,N},\;\nm\cdot\nm^{(\zeta)})$ is a Banach space. In
particular, $\nm\cdot\nm=\nm\cdot\nm^{(0)}$.
\\

Thus, a point in $\Hspace$ is a sequence
\eqnn\h\;=\;\left(z,T,\left\lbrace w_{M,N}\right\rbrace\right)\;
     \in\;\Hspace\;.\eeqnn
Next, we introduce a map $H:\Hspace\longrightarrow\Lin(P_1\Hp)$,
the linear operators on $P_1\Hp$. Its action on $\h\in\Hspace$ is
given by
\eqnn H\;:\;\h&\longmapsto&T[z;\opT]\,-\,z\,+\,
    \sum_{M+N\geq1}\,\ch_1\,
    W_{M,N}\left[[w_{M,N}[z;\op;\cdot]\,]\right]\,\ch_1\;,\eeqnn
where \eqn&&
    W_{M,N}\left[[w_{M,N}[z;\op;\cdot]\;]\right]
    \;:=\;W_{M,N}[z;\op]\nonumber\\
    &&\hspace{1.5cm}=\;\sum_{\lambda^{(M)},\tilde{\lambda}^{(N)}}\;
     \int \,\prod_{i=1}^M\,\frac{
     d^3k_i\;\bcs
     (|k_i|)}{\sqrt{|k_i|}} \,\prod_{j=1}^N\,
     \frac{d^3\tilde{k}_j\;\bcs(|\tilde{k}_j|)}{
     \sqrt{|\tilde{k}_j|}}
     \;\times\nonumber\\
     &&\hspace{2.5cm}\times\;
     a^*_{\lambda^{(M)}}(k^{(M)})\,
     w_{M,N}\left[z;\op;K^{(M,N)}\right]\,
     a^\sharp_{\tilde{\lambda}^{(N)}}(\tilde{k}^{(N)})
     \;\label{WMNdefintmeassigl12}\eeqn
is the same as the expression (~\ref{Wmon}), if $\sigma\;\leq\;1$
holds for the running infrared cutoff. Otherwise, we let \eqn&&
    W_{M,N}\left[[w_{M,N}[z;\op;\cdot]\;]\right]
    \;:=\;W_{M,N}[z;\op]\nonumber\\
    &&\hspace{1.5cm}=\;\sum_{\lambda^{(M)},\tilde{\lambda}^{(N)}}\;
     \int \,\prod_{i=1}^M\,
     d^3k_i\,\sqrt{|k_i|}  \,
     \prod_{j=1}^N\,d^3\tilde{k}_j\,
     \sqrt{|\tilde{k}_j|}
     \;\times\nonumber\\
     &&\hspace{2.5cm}\times\;
     a^*_{\lambda^{(M)}}(k^{(M)})\,
     w_{M,N}\left[z;\op;K^{(M,N)}\right]\,
     a^\sharp_{\tilde{\lambda}^{(N)}}(\tilde{k}^{(N)})
     \;\label{WMNdefintmeassigg12}\eeqn
if $\sigma\;>\;1$.

Conversely, every interaction operator $W_{M,N}$ of the form
(~\ref{Wmon}) uniquely defines an associated interaction kernel
$w_{M,N}$ in ${\mathfrak W}_{M,N}$. This correspondence is
obtained by simply reading off the respective components of the
Wick expansion of a given effective hamiltonian in $H(\Hspace)$.

Thus, the inverse of $H$ is given by
\begin{gather}
    H^{-1}\left[T[z;\opT]\;-\;z\;+\;\sum_{M+N\geq 1}\,\ch_1
    W_{M,N}[z;\op]\,\ch_1\right]\;=\;
    \left(z,T,\left\lbrace w_{M,N}\right\rbrace\right)\;.
\end{gather}

$\;$\\

\section{A polydisc of effective hamiltonians}
\label{Polyddefsubsect3}

For the spectral analysis of $\Hps$, we consider the flow
generated by $\ren$ on a certain Banach subspace of $\Hspace$,
defined as follows.

Let $\epsilon,\xi$ be small, positive numbers. The {\bf polydisc}
$\Polyd_{\epsilon,\xi}\subset\Hspace$ consists of all
$\h=(z,T,\lbrace w_{M,N}\rbrace)\in\Hspace$ possessing the
following properties.
\\

(i) The spectral parameter $z$ is contained in
$[-\epsilon,0]\subset\R$ (it is picked {\bf real} !).
\\

(ii) The function
$$T[z;\cdot]\;:\;{\rm Spec}\lbrace\opT\rbrace\;\longrightarrow\;\R$$
obeys the bounds \eqnn\nm T\nm\;\leq\;2\;,\eeqnn and can be
decomposed into
\eqnn\;T[z;\opT]\;=\;H_f\;+\;a[z]\,\ch_1\Ppar\ch_1
    \;+\;\ch_1T_{n-l}[z;\opT]\ch_1\;,
    \eeqnn
where the operator $T_{n-l}$ contains all terms in $T$ that are
$O(H_f^2)$ in the limit $H_f\rightarrow0$. In particular,
$T[z;\opT]$ is analytic in $\P_f$.
\\

(iii) For each value of $M+N\geq1$,
\eqnn\nm\,w_{M,N}\,\nm^{(2)}\;\leq\;\epsilon\,\xi^{M+N}\;. \eeqnn

We recall that the definition of $\nm\cdot\nm^{(2)}$ involves
derivatives up to second order with respect to the spectral
parameters associated to $\op$.

$H$ maps $\Polyd_{\epsilon,\xi}$ to the bounded operators on
$P_1\Hp$. This is a consequence of the following lemma.
\\

\begin{lm}\label{WMNwMNrelboundslemma3}
 For fixed $M+N$,
\eqnn\left\|\;W_{M,N}[[\;w_{M,N}[z;\op;\cdot]\;]]\;\right\|\;\leq\;
     (2\sqrt{\pi})^{M+N}\;
     \nm w_{M,N} \nm\;\eeqnn
on $P_1\Hp$.
\end{lm}

\prf Cf. the appendix of \cite{bfs2}. \qed

Hence, for $(z,T,\lbrace w_{M,N}\rbrace)\in\Polyd_{\epsilon,\xi}$,
we have \eqnn\left\|\;H[\h]\;\right\|&\leq&|z|\;+\;\|\;T\;\|\;+\;
     \epsilon\;\sum_{M+N\geq 1}\;
     (2\sqrt{\pi}\xi)^{M+N}\nonumber\\
     &<&3\;.
     \eeqnn
Thus, $H[\h]$ is a bounded operator on $P_1\Hp$ for all
$\h\in\Polyd_{\epsilon,\xi}$.
\\

\section{Definition of the renormalization map}
The renormalization map
\eqnn\ren\;:\;\Polyd_{\epsilon,\xi}&\longrightarrow&
    \Polyd_{\epsilon,\xi}\nonumber\\
    \h&\longmapsto&\hat{\h}
    \;,\eeqnn
is the pullback of $H[\h]\mapsto H[\hat{\h}]$ to $\Hspace$, that
is,
\eqn
    \begin{CD}
    \Polyd_{\epsilon,\xi} @>\ren>> \Polyd_{\epsilon,\xi}\\
    @VV{H}V          @VV{H}V\\
    \Bound(P_1\Hp)@>Z\circ\resc\circ\dec>>\Bound(P_1\Hp)\;\;
    \end{CD}
    \label{rencommdiag3}\eeqn
is a commutative diagram. The map $H[\h]\mapsto H[\hat{\h}]$ is
obtained from concatenating the three distinct operations, which
we discuss in detail below: Feshbach decimation $\dec$, the
rescaling transformation $\resc$, and the shift of the spectral
parameter $Z$.
\\

\noindent{\bf Remark.} The fact that $\ren$ maps
$\Polyd_{\epsilon,\xi}$ into itself is a central issue whose
verification requires much effort in the subsequent analysis.
\\

\subsection{Feshbach decimation}
Let $\rho\in(0,1)$ denote some fixed, real number. In the present
work, our choice will always be $\rho=\frac{1}{2}$.
$P_\rho=\chi[H_f<\rho]$ is a selfadjoint projector
$P_1\Hp\rightarrow P_\rho\Hp$. For
$\h=(z,T,\{w_{M,N}\})\in\Polyd_{\epsilon,\xi}$, the discussion in
Section (~\ref{Wickordsubsubsec}) has shown that the Feshbach
decimation map $\dec$ is well defined. We have seen that the image
of $H[\h]$ under \eqnn\dec\;:\;\Bound(P_1\Hp)&\longrightarrow&
    \Bound(P_\rho\Hp)\eeqnn
can be written as
\eqnn\dec[H[\h]]&=&(1+\DHf[z])H_f\,
    \,+\,
    (\Delta E_0-z)\,+\,\chr (T'\,\piop+\Delta T')\chr
    \;\nonumber\\
    &&+\;\sum_{M+N\geq1}\chr(\piop\,W_{M,N}\,\piop
    +\Delta W_{M,N})\chr\;,
\eeqnn
where $T'=T-z-\tHh$,
$$T'[z;\opT]\;=\; a[z]\;\Ppar\,-\,\DHf[z]\,H_f\,-\,
    \Delta E_0[z]\,+\,T_{n-l}[z;\opT]\;.$$
$T_{n-l}$ comprises all operators in $T$ that are nonlinear with
respect to the components of $\opT$.
\\

\subsection{The rescaling map} The rescaling map
is obtained by concatenating the scaling transformation
$Ad_{U_\rho}$ of Section {~\ref{scalingdimsubsect}} with
multiplication by a specific numerical factor.

It is a map
$$\resc\;:\;\Bound(P_\rho\Hp)\;\longrightarrow\;\Bound(P_1\Hp)\;,$$
which acts on $A\in\Bound(P_\rho\Hp)$ by
\eqnn\resc[A]\;=\;(1+\DHf[z])^{-1}\;\rho^{-1}\;Ad_{U_\rho}[A]\;.\eeqnn
We recall that \eqn (Ad_{U_\rho}a^\sharp)
    (k)\;=\;\rho^{-\frac{3}{2}}\;a^\sharp(\rho^{-1}\;k)
     \;,\label{scdimasharpaux3}\eeqn
and in particular, that
$Ad_{U_\rho}[\op^r_{(0)}]=\rho\,\op^r_{(0)}$ for the components
$\op^r_{(0)}$ of $\op_{(0)}$.

$\resc$ is defined in this manner in order to keep the coefficient
of $H_f$ in $T_{lin}$ fixed at the value 1 in each application of
$\ren$, thus the division by the factor $\rho(1+\DHf[z])$. As a
consequence,
\eqnn \hat{a}[Z[z]]\;:=\;\;
     \frac{a[z]+\Delta{a}[z]}
     {1+\DHf[z]}\; \eeqnn
corresponds to the renormalized coefficient of $\Ppar$.
\\

\subsection{Shift of the spectral parameter}

Under $\resc\circ\dec$, the spectral parameter $z$ is mapped to
\eqnn\;z\;\longmapsto\;\hat{z}&:=&Z[z]\;:=\;
    (1+\DHf[z])^{-1}\;\rho^{-1}\;(z-\Delta E_0[z])
    \;\;.\eeqnn
$\hat{z}$ is the renormalized spectral parameter.
\\

\subsection{Renormalization of the interaction kernels}
\label{renintkerndefsubsubsect3}

>From Feshbach decimation, we have
\eqnn\,w_{M,N}\;\longrightarrow\;\tilde{w}_{M,N}\;:=
     \;\piop[z;\sh^{(M)}\opT]w_{M,N}\piop[z;\sh^{(N)}\opT]\;
     +\;\Delta w_{M,N}\;,\eeqnn
where the kernel on the left hand side of the arrow acts on
$P_1\Hp$, and the corrected kernel on the right hand side acts on
$P_\rho\Hp$. The formula for $\Delta w_{M,N}$ is given in
(~\ref{DeltawMN}). We recall that $\sh^{(M)}$ shifts the
components of $\opT$ by the photon momenta $(k_1,\dots,k_M)$.

Application of rescaling produces the bounded operators
\eqnn\hat{w}_{M,N}&=&\resc[\tilde{w}_{M,N}]\;\nonumber\\
    &=&\left\{\begin{aligned}\rho^{M+N-1}\,(1+\DHf[z])^{-1}
    \,Ad_{U_\rho}[\tilde{w}_{M,N}]
    \hspace{0.5cm}&{\rm if}\; \sigma\leq1\\
    \rho^{2(M+N)-1}\,(1+\DHf[z])^{-1}
    \,Ad_{U_\rho}[\tilde{w}_{M,N}]
    \hspace{0.5cm}&{\rm if}\; \sigma>1\;,\end{aligned}\right.\eeqnn
on $P_1\Hp$. In the case $\sigma\leq1$, the factor $\rho^{M+N-1}$
is obtained in the following manner: The integration measure in
(~\ref{WMNdefintmeassigl12}), given by
$$\prod_{i=1}^N\prod_{j=1}^N dk_i\,dk_j\,|k_i|^{\frac{1}{2}}
    \,|\tilde{k_j}|^{\frac{1}{2}}\;,$$
contributes a factor $\rho^{\frac{5}{2}(M+N)}$ under the action of
$Ad_{U_\rho}$, since the photon momenta are modified by
$k\mapsto\rho k$. Furthermore, since the scaling dimension of a
creation-and an annihilation operator is $-\frac{3}{2}$, cf.
(~\ref{scdimasharpaux3}), there is a factor
$\rho^{-\frac{3}{2}(M+N)}$ from the action of $Ad_{U_\rho}$ on
$a^\sharp(k)$ in (~\ref{WMNdefintmeassigl12}). Finally, an overall
factor $\rho^{-1}(1+\DHf[z])^{-1}$ is contained in the definition
of $\resc$ (to fix the coefficient of $H_f$ in $T_{lin}$ at the
constant value 1, as we recall).

The additional factor $\rho^{M+N}$ in the case $\sigma>1$ arises
because the integration measure now includes a factor $\prod
|k_i|^{\frac{1}{2}}|\tilde{k_j}|^{\frac{1}{2}}$, cf.
(~\ref{WMNdefintmeassigl12}), instead of its inverse $\prod
|k_i|^{-\frac{1}{2}}|\tilde{k_j}|^{-\frac{1}{2}}$ in the case
$\sigma\leq1$ considered above, cf. (~\ref{WMNdefintmeassigg12}).
This is due to the fact that for $\sigma>1$, the infrared
regularization $\bcs(x)$ is linear in $x$, and absorbed into the
integration measure.

Substituting the spectral parameter by $z\rightarrow
Z^{-1}[\hat{z}]$, one arrives at the renormalized interaction
kernels, given by $\hat{w}_{M,N}[\hat{z};\op;K^{(M,N)}]$.

Let $Y=(Y_1,\dots,Y_r)$ denote a tuple of operators and parameters
are homogenous of degree 1 under rescaling. Let
$\alpha=(\alpha_1,\dots,\alpha_r)$ be a multiindex with
$\alpha_i\in\N_0$ and $|\alpha|=\sum\alpha_i$. Then,
\eqnn\partial_Y^\alpha\hat{w}_{M,N}
    &=&\left\{\begin{aligned}\rho^{M+N+|\alpha|-1}\,(1+\DHf[z])^{-1}
    \,Ad_{U_\rho}[\partial_Y^\alpha\,\tilde{w}_{M,N}]
    \hspace{0.5cm}&{\rm if}\; \sigma\leq1\\
    \rho^{2(M+N)+|\alpha|-1}\,(1+\DHf[z])^{-1}
    \,Ad_{U_\rho}[\partial_Y^\alpha\,\tilde{w}_{M,N}]
    \hspace{0.5cm}&{\rm if}\; \sigma>1\;.\end{aligned}
    \right.\eeqnn
The origin of the additional factors $\rho^{|\alpha|}$ has been
explained in the discussion given in Section
{~\ref{scalingdimsubsect}}.

Finally, the running infrared cutoff is renormalized by
$\sigma\rightarrow\hat{\sigma}=\frac{\sigma}{\rho}$.
\\

\subsection{Construction of $\ren$} Substituting $z$ by \eqnn
z[\hat{z}]\;=\;Z^{-1}[\hat{z}]\;\eeqnn in
$\resc\circ\dec[H[\h]-z]$, we arrive at \eqnn
H[\hat{\h}]\;=\;H_f\,-\,\hat{z}\,+\ch_1\,\hat{T}'\,\ch_1\,+\,
    \sum_{M+N\geq1}\ch_1\,\hat{W}_{M,N}\,\ch_1\;.\eeqnn
Applying $H^{-1}$, we obtain the point
$\hat{\h}=(\hat{z},\hat{T},\{\hat{w}_{M,N}\})$ in $\Hspace$ (we
recall once more that the issue is to prove that indeed,
$\hat{\h}\in\Polyd_{\epsilon,\xi}\subset\Hspace$). This completes
the construction of the renormalization map
$\ren:\h\mapsto\hat{\h}$ in the manner exhibited by the diagram
(~\ref{rencommdiag3}).
\\

\chapter{RENORMALIZATION OF STRONGLY MARGINAL OPERATORS}

The circumstance that the system considered in the work at hand
comprises purely marginal operators distinguishes it from the
majority of rigorous renormalization group studies that are
present in the literature, where the marginal relevant and
irrelevant cases are typically considered. The flow of operators
in systems of the latter kind is dominated by the leading order
radiative corrections in perturbation theory. The renormalization
of strongly marginal quantities, however, is characterized by
almost complete mutual cancellations of all radiative corrections.
In the context of smooth Feshbach renormalization, these
cancellations follow from the exact concatenation formulae for
Feshbach triples that were proved in earlier sections. Our
strategy will be to exploit these concatenation laws, together
with other algebraic properties of the smooth Feshbach map, in
order to derive identities that interrelate certain quantities of
central interest between arbitrary scales. The analytic part,
which produces bounds on the quantities of interest, will be
carried out recursively, by use of a nested renormalization group
scheme.
\\

As it will turn out, the renormalization of strongly marginal
operators in the present theory can be reduced to the
renormalization of the coefficient $a[z]$ of $\Ppar$. This task is
particularly subtle, and it is also especially important, for two
main reasons:
\\

(1) In the limit $n\rightarrow\infty$, $a_n[z_n]$  tends - as our
analysis will show - to a finite, non-zero value. The physical
hamiltonian $\Hps$ exhibits an eigenvalue $\Egrd$ at the bottom of
its spectrum.   The first and second derivatives of the ground
state energy $\Egrd$ of the physical hamiltonian $\Hps$ with
respect to $|p|$, for $|p|\leq\puppbd$, are, for $\alpha=0,1$,
given by
\eqnn\derp \Egrd\;=\;-\,
    \lim_{n\rightarrow\infty}
    \left(\left.\dHfFQQ_n[z_n]^{-1}\,
    a_n[z_n]\right|_{z_n=E_{0,n}}\right)\;,
    \eeqnn
where $E_{0,n}$ denotes the ground state eigenvalue of the
effective hamiltonian $H[\h^{(n)}]$, and the quantity
$\dHfFQQ_n[z_n]$ has the form $1+O(g^2)$. This is explained in
detail below.

Letting $\Omgrd$ denote the ground state vector corresponding to
$\Egrd$, normalized by $\langle\Omgrd,\vac\rangle=1$,
\eqn\derp^2 E_0&=&\frac{\left\langle
    \Omgrd\,,\,(\derp^2 \Hps
    )\Omgrd
    \right\rangle_{\vac}}{\left\langle \Omgrd\,,\,
    \Omgrd\right\rangle_{\vac}}\nonumber\\
    &-&
    \frac{\left\langle\derp \Omgrd\,,\, \left(\Hps-
    E_0\right) \derp\Omgrd
    \right\rangle_{\vac}}{\left\langle \Omgrd\,,\,
    \Omgrd\right\rangle_{\vac}}\;.
    \label{derpE0sig0plimitdef1}\eeqn
Satisfactory bounds on these derivatives are of central importance
for various purposes, including the construction of scattering
theory \cite{fr1,fr2,pi}, and the analysis of the semi-classical
dynamics of charged particles \cite{tesp}.
\\

(2) Via the Ward-Takahashi identities that we will present and
discuss in much detail below, $a[z]$ controls $W_{M,N}$, for
$M+N=1$. Thus, $a[z]$ is responsible for all marginal operators in
the theory other than $H_f$ in $T_{lin}[\h]$.
\\

\section{The ground state energy}

The renormalization group produces a sequence of effective
hamiltonians $\{H[\h^{(k)}]\}$, for $k\in\N_0$, where we recall
that
\eqnn H[\h^{(k+1)}]\;=\;(1+\DHf^{(k)}[z_k])\rho^{-1}
    Ad_{U_\rho}\left[\FchrtHk[H[\h^{(k)}]]\right]\;,\eeqnn
with
\eqnn\tau^{(k)}[H[\h^{(k)}]]&=&
    (1+\DHf^{(k)}[z_k])\left(H_f\,-\,z_k\,+\,\Delta
    E_0^{(k)}[z_k]\right)\\
    &=&(1+\DHf^{(k)}[z_k])\left(H_f\,-\,\rho\,z_{k+1}
    \right)\;.\eeqnn
In particular, if we define
\eqnn\tau_n[\Hps]&\equiv&\left(\prod_{i=0}^n
    (1+\DHf^{(i)}[z_i])\right)
    (H_f\,+\,\rho^n\,z_n)\;,\eeqnn
and assume that all $H[\h^{(k)}]$ are well-defined, the operator
\eqnn\;\Fn[z]\;\equiv\;\FchntHn[\Hps-z]\;\eeqnn
is also well-defined, with
$$\chn\;=\;\ch_1[\rho^{-n}H_f]\;,$$
and
\eqn H[\h^{(n)}]\;=\;\rho^{-n}\left(
    \prod_{i=0}^n(1+\DHf^{(i)}[z_i])^{-1}\right)
    \,Ad_{U_\rho^n}\left[\Fn[z]\right]\;,\eeqn
due to Theorem {~\ref{FchtHconcatthm4444}}. Likewise, the
operators
$$\Qn^{(\sharp)}[z]\;\equiv\;Q_{\chn,\tau_n}^{(\sharp)}[\Hps-z]\;$$
are well-defined.
\\

\begin{prp}
\label{anE0neqderpEgrdprp3} Assume that the Feshbach triple
$(\Fn[z],\Qns[z],\Qn[z])$ associated to   $(\chn, \Hp)$,  is
well-defined for all $n\in\N_0$ and $z$ sufficiently close to the
origin. Furthermore, let $\rho=\frac{1}{2}$. For the sequence of
ground state eigenvalues of $H[\h^{(n)}]$, assume that
$$\lim_{n\rightarrow\infty}\|W[h^{(n)}]\|\;=\;0\;.$$
Then, defining $\cZ_n\equiv Z_n\circ\cdots\circ Z_0$,
\eqn E_0\;=\;\lim_{n\rightarrow\infty}\cZ_n^{-1}(z_n=0)\;,
    \label{E0limexpr4444}\eeqn
and
\eqn \Omgrd\;=\;s-\lim_{n\rightarrow \infty}\Qn[E_0]\,
    \vac \;. \label{Omgrdslimexpr4444}\eeqn
In particular,
\eqn\derp \Egrd&=&
    \lim_{n\rightarrow \infty}\frac{\left\langle \Qns[E_0]
    (\partial_{\Ppar}\Hps)\,\Qn[E_0]
    \right\rangle_{\vac}}{\left\langle \,\Qns[E_0]
    \,\Qn[E_0]
    \right\rangle_{\vac}}\;.\eeqn
Furthermore, defining
\eqn\dHfFQQ_n[z_n]\;:=\;\frac{\left\langle \,
    \partial_{H_f}\Fn[z]
    \right\rangle_{\vac}}{\left\langle \,
    \Qns[z]\,\Qn[z]
    \right\rangle_{\vac}}\;,\label{dHfFQQdefaux333}\eeqn
with
\eqn z_n[z]\;=\; \rho^{-n} \;\frac{z\,-\,\left\langle
    \Fn[z]
    \right\rangle_{\vac}}{\left\langle \,
    \partial_{H_f}\Fn[z]
    \right\rangle_{\vac}}\;,\label{znzfctdefaux333}\eeqn
and assuming that $z\rightarrow z_n$ is invertible, the
coefficient of $\Ppar$ in $T_{lin}[\h^{(n)}]$ is given by
\eqn
\left.(a_n[z_n])\right|_{z_n=E_{0,n}}&=&\dHfFQQ_n[E_{0,n}]^{-1}\,
    \frac{\left\langle \Qns[E_0]
    (\partial_{\Ppar}\Hps)\,\Qn[E_0]
    \right\rangle_{\vac}}{\left\langle \,\Qns[E_0]
    \,\Qn[E_0]
    \right\rangle_{\vac}}\;,\eeqn
such that
\eqn\left.\lim_{n\rightarrow\infty}(a_n[z_n]
    \dHfFQQ_n[z_n])\right|_{z_n=E_{0,n}}
    &=&-\,\derp\,\Egrd\;
    \;.
    \label{derpE0sig0azform1}\eeqn
Furthermore,
$$\,\derp^2\left(\Egrd\,-\,\frac{|p|^2}{2}\right)\;\leq\;0\;,$$
that is, the renormalized mass of the infraparticle is bounded
from below by the bare mass of the charged particle, for all
$\ssig\geq0$.
\end{prp}

\prf The condition that the Feshbach triple $(\Fn,\Qns,\Qn)$ is
well-defined for all $n\in\N_0$ will be proved in the
renormalization group analysis of Chapter {~\ref{RGfloweqssect3}}.
The fact that the limits in the asserted expressions exist is also
proved there.

Let us thus assume that the limits exist. In order to fix the
coefficient of $H_f$ in the Wick ordered expression for $\Fn[E_0]$
at the value 1, we must divide by $\langle
\partial_{H_f}\Fn[E_0]\rangle_{\vac}$. We therefore have
\eqnn H[\h^{(n)}_0]\;=\;\frac{\rho^{-n}\,}
    {\left\langle \partial_{H_f}\Fn[E_0]
    \right\rangle_{\vac}}\,Ad_{U_{\rho^n}}
    \left[ \Fn[E_0]\right]\;,\eeqnn
again by Theorem  {~\ref{derXfPthm}}.

Let us first prove that the expressions given in
(~\ref{E0limexpr4444}) and (~\ref{Omgrdslimexpr4444}) satisfy the
eigenvalue equation
$$(\Hps-E_0)\Omgrd\;=\;0$$
for $\Hps$. To this end, the general formula
(~\ref{HQeqchFgenform4444}) implies
\eqnn&&(\Hps-\cZ_n^{-1}(0))\,\Qn[\cZ_n^{-1}(0)]\\
    &&\hspace{1cm}=\;
    \chi_{\rho^{n}}\Fn[\cZ_n^{-1}(0)]\\
    &&\hspace{1cm}=\;
    \rho^{n}\chi_{\rho^{n}}\langle\partial_{H_f}
    \Fn[\cZ_n^{-1}(0)]\rangle_{\vac}Ad_{U_{\rho^n}^*}
    \left[H[\h^{(n)}]|_{z_n=0}\right]\\
    &&\hspace{1cm}=\;
    \rho^{n}\chi_{\rho^{n}}\langle\partial_{H_f}
    \Fn[\cZ_n^{-1}(0)]\rangle_{\vac}Ad_{U_{\rho^n}^*}
    \left[H_f+\ch_1(T'[\h^{(n)}]+ W[\h^{(n)}])|_{z_n=0}
    \ch_1\right]\;.\eeqnn
Thus, if $\lim_{n\rightarrow\infty}\|W[\h^{(n)}]\|=0$, it is
evident that
\eqnn 0&=&s-\lim_{n\rightarrow\infty}
    \left(\Hps-\cZ_n^{-1}(0)\,\Qn[\cZ_n^{-1}(0)]\vac\right)\\
    &=&\left(\Hps-\lim_{n\rightarrow\infty}\cZ_n^{-1}(0)\right)
    \left(s-\lim_{n\rightarrow\infty}
    \Qn[\cZ_n^{-1}(0)]\vac\right)\;.
    \eeqnn
This proves the claim.

The coefficient $a_n[z_n]|_{z_n=E_{0,n}}$ of $\Ppar$ in
$T_{lin}[\h^{(n)}]$ is obtained from multiplying
\eqnn\left\langle \partial_{\Ppar}\Fn[E_0]
    \right\rangle_{\vac}
    \;=\;\left\langle \Qns[E_0]
    (\partial_{\Ppar}\Hps)\Qn[E_0]
    \right\rangle_{\vac}\;\eeqnn
by
$\left\langle\partial_{H_f}\FchntHn[\Hps]\right\rangle_{\vac}^{-1}$.
Therefore,
\eqnn\left.a_n[z_n]\right|_{z_n=E_{0,n}}&=&
    \frac{\left\langle \Qns[E_0]
    (\partial_{\Ppar}\Hps)\,\Qn[E_0]
    \right\rangle_{\vac}}{\left\langle \,
    \partial_{H_f}\Fn[E_0]
    \right\rangle_{\vac}}\\
    &=&\dHfFQQ_n[E_{0,n}]^{-1}\,
    \frac{\left\langle \Qns[E_0]
    (\partial_{\Ppar}\Hps)\,\Qn[E_0]
    \right\rangle_{\vac}}{\left\langle \,\Qns[E_0]
    \,\Qn[E_0]
    \right\rangle_{\vac}}\;,\eeqnn
as claimed.

Next, we demonstrate that
\eqnn \derp \Egrd&=&
    \lim_{n\rightarrow \infty}\frac{\left\langle \Qns[E_0]
    (\partial_{\Ppar}\Hps)\,\Qn[E_0]
    \right\rangle_{\vac}}{\left\langle \,\Qns[E_0]
    \,\Qn[E_0]
    \right\rangle_{\vac}}\;\nonumber\\
    &=&-\,\lim_{n\rightarrow \infty}\frac{\left\langle \Qns[E_0]
    (\derp\Hps)\Qn[E_0]
    \right\rangle_{\vac}}{\left\langle \Qns[E_0]
    \,\Qn[E_0]
    \right\rangle_{\vac}}\;,\eeqnn
where, passing to the second line, we have used
$\derp\Hps=-\partial_{\Ppar}\Hps$.

To this end, let us consider the eigenvalue problem for the ground
state energy of $\Hps$. We recall that
$$s-\lim_{n\rightarrow\infty}\Qn[E_0]\,\vac\;=\;\Omgrd\;.$$
Hence, defining
$$\Psi[p;\ssig]\;:=\;\frac{\Omgrd}{\|\Omgrd\|}\; $$
on $\Hp$, we see that
$$\lim_{n\rightarrow \infty}\frac{\left\langle \Qns
    (\derp\Hps)\Qn[E_0]
    \right\rangle_{\vac}}{\left\langle \Qns[E_0]
    \,\Qn[E_0]
    \right\rangle_{\vac}}
    \;=\;-\,\left\langle \Psi[p;\ssig]\;,\;
    (\derp\Hps)\Psi[p;\ssig]
    \right\rangle \;.$$
On the other hand, differentiating the eigenvalue equation
\eqn\left(\Hps\,-\,\Egrd\right)\,\Psi[p;\ssig]\;=\;0\;\eeqn
with respect to $|p|$, we find
\eqnn\left(\Hps\,-\,\Egrd\right)\,\derp\,\Psi[p;\ssig]\;=\;-\,
    \left(\derp\left(\Hps\,-\,\Egrd\right)
    \right)\,\Psi[p;\ssig]\;.\eeqnn
Taking the scalar product with respect to $\Psi[p;\ssig]$, we
obtain
$$\left\langle \Psi[p;\ssig]\;,\;(\derp\Hps)\,\Psi[p;\ssig]
    \right\rangle\;=\;
    \derp\Egrd\;,$$
using that
\eqnn\left\langle \derp\Psi[p;\ssig]\;,\;\Psi[p;\ssig]
    \right\rangle&=&\frac{1}{2}\,\derp\,\underbrace{\left\langle
    \Psi[p;\ssig]\;,\;\Psi[p;\ssig]
    \right\rangle}_{=\;1}\\
    &=&0\;.\eeqnn
Finally, the formula
\eqnn\derp^2 E_0&=&\lim_{n\rightarrow\infty}\frac{\left\langle
    \Qns[E_0](\derp^2 \Hps
    )\Qn[E_0]
    \right\rangle_{\vac}}{\left\langle \Qns[E_0]
    \Qn[E_0]\right\rangle_{\vac}}\nonumber\\
    &-&\lim_{n\rightarrow\infty}
    \frac{\left\langle(\derp \Qns[E_0]) \left(\Hps-
    E_0\right) \derp\Qn[E_0]
    \right\rangle_{\vac}}{\left\langle \Qns[E_0]
    \Qn[E_0]\right\rangle_{\vac}}\;,
    \eeqnn
together with $\derp^2 \Hps=1$, immediately implies that
$$\derp^2\left(\Egrd\,-\,\frac{|p|^2}{2}\right)\,\leq\,0\;.$$
This proves the claim. \qed

To estimate the quotient
\eqnn\dHfFQQ_n[z]\;=\;\frac{\left\langle \,
    \partial_{H_f}\Fn[z]
    \right\rangle_{\vac}}{\left\langle \,
    \Qns[z]\,\Qn[z]
    \right\rangle_{\vac}}\;,\eeqnn
we employ formula ({~\ref{derHfFchrtHform333}}). Using the fact
that $\partial_{H_f}(T_{n-l}+W)=0$ for $\Hps=T_{lin}+T_{n-l}+W$,
one straightforwardly obtains
\eqn&& \left\langle\partial_{H_f}\Fn[z]
    \right\rangle_{\vac}\;=\;
    \frac{ \left\langle
    \Qns[z]\,\Qn[z]
    \right\rangle_{\vac}}{1-\left\langle W\bchn\bar{R}_n
    \chn^2\bar{R}_n\bchn W
    \right\rangle_{\vac}}\;\times\nonumber\\
    &&\hspace{2cm}\times
    \left(1\,-\,\frac{\left\langle W\bchn\bar{R}_n
    \frac{\partial_{H_f}\bchn}{\bchn}
    \tHpsz \bar{R}_n\bchn W
    \right\rangle_{\vac}}{\left\langle
    \Qns[z]\,\Qn[z]
    \right\rangle_{\vac}}\right)\;,\eeqn
hence
\eqn\dHfFQQ_n[z]&=&\frac{1}{1-\left\langle W\bchn\bar{R}_n
    \chn^2\bar{R}_n\bchn W
    \right\rangle_{\vac}}\;\times\nonumber\\
    &&\times
    \left(1\,-\,\frac{\left\langle W\bchn\bar{R}_n
    \frac{\partial_{H_f}\bchn}{\bchn}\,
    \tHpsz \,\bar{R}_n\bchn W
    \right\rangle_{\vac}}{\left\langle
    \Qns[z]\,\Qn[z]
    \right\rangle_{\vac}}\right)\;.
    \label{dHfFQQnformimplaux333}\eeqn
A very important element of the analysis in subsequent chapters
will be to prove, by means of the renormalization group iteration,
that
\eqnn\left|\left\langle W\bchn\bar{R}_n
    \chn^2\bar{R}_n\bchn W
    \right\rangle_{\vac}\right|\;\leq\;O(\ez )\;,\eeqnn
and that
\eqnn\left|\frac{\left\langle W\bchn\bar{R}_n
    \frac{\partial_{H_f}\bchn}{\bchn}
    \tHpsz \bar{R}_n\bchn W
    \right\rangle_{\vac}}{
    \left\langle \,
    \Qns[z]\,\Qn[z]
    \right\rangle_{\vac}}\right|\;\leq\;O(\ez)\;,\eeqnn
both uniformly in $n$. This implies that $\dHfFQQ_n[z]=1+O(\ez)$,
uniformly in $n$. For the proof of these estimates, the
concatenation laws of section {~\ref{concatlawssect}} have central
importance.
\\

\section{Analytical control of $\dHfFQQ_n[z]$ and
cancellation phenomena}

One of the key insights that makes the subsequent analysis
possible is that the quantity
\eqnn\dHfFQQ_n[z]\;=\;\frac{\left\langle \,
    \partial_{H_f}\Fn[z]
    \right\rangle_{\vac}}{\left\langle \,
    \Qns[z]\,\Qn[z]
    \right\rangle_{\vac}}\;,\eeqnn
does in fact  solely depend on the effective hamiltonian of the
scale $n-1$, as stated in the following proposition.
\\

\begin{prp}
\label{dHfFQQformprop333} Let $H[\h]$ denote the effective
hamiltonian at some arbitrary, but fixed scale, acting on the
Hilbert space $P_1\Hp$, and define
\eqn\dHfFQQ[\h]&\equiv&\left(1\,+\,
    \left\langle W[\h]\,
    \bch\,\bar{R}[\h]\,
    \ch_\rho^2\,\bar{R}[\h]\,\bch \,W[\h]
    \right\rangle_{\vac}\right.\nonumber\\
    &&\hspace{0.5cm}+\;\left.\left\langle W[\h]\,
    \bch\,\bar{R}[\h]\,
    \frac{\partial_{H_f}\bch_\rho}{\bch_\rho}\,
    (H_f-\rho\hat{z})
    \,\bar{R}[\h]\,\bch \,W[\h]
    \right\rangle_{\vac}\right)^{-1}\;,\eeqn
where $\bch:=\bch_\rho\ch_1$, and where $\hat{z}$ is the spectral
parameter associated with $\hat{\h}=\ren[\h]$. Then,
\eqnn\dHfFQQ[\h^{(n-1)}]&=&\dHfFQQ_n[z]\\
    &=&\frac{\left\langle \,
    \partial_{H_f}\Fn[z]
    \right\rangle_{\vac}}{\left\langle \,
    \Qns[z]\,\Qn[z]
    \right\rangle_{\vac}}\;.\eeqnn
where the relationship between the spectral parameters $z_{n-1}$
(for $\h^{(n-1)}$), $z_n$ and $z$ is determined by
(~\ref{znzfctdefaux333}).
\end{prp}

\prf Let us separately discuss the overlap term (due to $\chn^2
\bar{P}_n$)
$$\left\langle W\,\bchn\,\bar{R}_n\,
\chn^2\,\bar{R}_n\,\bchn\, W \right\rangle_{\vac}\;,$$
and the term
$$\frac{\left\langle W\bchn\bar{R}_n
    \frac{\partial_{H_f}\bchn}{\bchn}
    \tHpsz \bar{R}_n\bchn W
    \right\rangle_{\vac}}{\left\langle
    \Qns[z]\,\Qn[z]
    \right\rangle_{\vac}}\;,$$
comprising the $\bchn$-derivative, which occur in
$\dHfFQQ_n[z_n]$.
\\

\begin{lm}
\label{Ovlaptermglobloclemma333} Let
$H[\h^{(k)}]=T[\h^{(k)}]-z_k+\ch_1 W[\h^{(k)}]\ch_1$ denote the
effective hamiltonian at the scale $k$, acting on $P_1\Hp$, for
$0\leq k<n$. Then,
\eqn&&\left\langle W\,\bchn\,\bar{R}_n\,
    \chn^2\,\bar{R}_n\,\bchn\, W
    \right\rangle_{\vac}\;\nonumber\\
    &&\hspace{1cm}=\;
    \left\langle W[\h^{(n-1)}]\,
    \bch\,\bar{R}[\h^{(n-1)}]\,
    \ch_\rho^2\,\bar{R}[\h^{(n-1)}]\,\bch \,W[\h^{(n-1)}]
    \right\rangle_{\vac}\;,\label{Ovlaptermglobloc333}
    \eeqn
where $\bch:=\bch_\rho\ch_1$, and where the expectation value on
the left hand side of the equality sign is determined by $\Hps$
and $\bchn[H_f]=\bch[\rho^{-n}H_f]$.
\end{lm}

\prf Clearly,
\eqnn&&\left\langle W\,\bchn\,\bar{R}_n\,
    \chn^2\,\bar{R}_n\,\bchn \,W
    \right\rangle_{\vac}\\
    &&\hspace{1cm}=\;\lim_{\delta\rightarrow0}
    \left\langle \Qns[z]\,P_n\bar{P}_n\,\frac{\chn^2}{\bchn^2+\delta}
    \,P_n\bar{P}_n\,\Qn[z]
    \right\rangle_{\vac}\;,\eeqnn
where $P_n,\bar{P}_n,$ are the projectors onto the ranges of
$\chn,\bchn$ respectively. Now, by the concatenation law for
$Q^\sharp,Q$ of Theorem {~\ref{QQscomplawthm333}},
\eqnn&&\Qn[z]\;=\;Q_{\ch_1,\tau_0}[\Hps-z]\,
    \\
    &&\hspace{2cm}\,Q_{\ch_\rho,\tau_1}[
    [F_{\ch_0,\tau_0}[\Hps-z]] \,\cdots\,\\
    &&\hspace{2cm}\cdots
    \,Q_{\ch_{\rho^n},\tau_n}[
     F_{\ch_{\rho^{n-1}},\tau_{n-1}}[\,\cdots\,
    F_{\ch_1,\tau_0}[\Hps-z]\,
    \cdots\,] ]\,,\eeqnn
thus by the second part of Theorem {~\ref{QQscomplawthm333}},
\begin{gather} P_n \,\bar{P}_n\,\Qn[z]\;=\;P_n \,\bar{P}_n
    \,Q_{\chn,\tau_n}[F_{_{\ch_{\rho^{n-1}},\tau_{n-1}}}
    [\,\cdots\,
    F_{_{\ch_0,\tau_0}}[\Hps-z]\,
    \cdots\,]]\;,\label{PbPQnauxform333}\end{gather}
since $P_n \,\bar{P}_n\bch_k=0$ for all $k<n$. In this notation,
the operator $\tau_k[H[\h^{(k-1)}]]$ determines the Feshbach
decimation map passing from $H[\h^{(k-1)}]$ to $H[\h^{(k)}]$ in
the renormalization step.

Now clearly,
\eqnn H[\h^{(n-1)}]&=&\frac{\rho^{-n}}{\left\langle\partial_{H_f}
    F_{n-1}[z]
    \right\rangle_{\vac}}\,Ad_{U_{\rho^n}}
    \left[F_{\chi_{\rho^{n-1}},\tau_{n-1}}
    [\,\cdots\,
    F_{\chi_1,\tau_0}[\Hps-z]\,
    \cdots\,]\right]\;,\eeqnn
thus
\eqnn (~\ref{PbPQnauxform333})&=&P_n \,\bar{P}_n
    \,Q_{\chn,\tau_n}\left[
    \rho^{n} \left\langle\partial_{H_f}
    F_{n-1}[z]\right\rangle_{\vac}\,Ad_{U_{\rho^{n-1}}^*}
    [H[\h^{(n)}]]
    \right]\\
    &=&
    Ad_{U_{\rho^{n-1}}^*}\left[P_\rho\,\bar{P}_\rho\,
    Q_{\ch_\rho,\tau}
    [H[\h^{(n-1)}]]\right]\;,\eeqnn
since the $Q^\sharp,Q$-operators are {\bf homogenous of degree 0}
with respect to the arguments in square brackets.

Due to
$$Ad_{U_{\rho^{n-1}}^*}\left[\frac{\chn^2}{\bchn^2+\delta}
    \right]\;=\;
    \frac{\ch_\rho^2}{\bch_\rho^2+\delta}\;,$$
we have
\eqnn&&\lim_{\delta\rightarrow0}
    \left\langle \Qns[z]
    \,P_n\bar{P}_n\,\frac{\chn^2}{\bchn^2+\delta}
    \,P_n\bar{P}_n\,\Qn[z]
    \right\rangle_{\vac}\\
    &=&\lim_{\delta\rightarrow0}
    \left\langle Q_{\ch_\rho,\tau}^\sharp
    [H[\h^{(n)}]]\,P_\rho\bar{P}_\rho\,
    \frac{\ch_\rho^2}{\bch_\rho^2+\delta}
    \,P_\rho\bar{P}_\rho\,Q_{\ch_\rho,\tau}
    [H[\h^{(n)}]]
    \right\rangle_{\vac}\\
    &=&\left\langle W[\h^{(n)}]\,
    \bch\,\bar{R}[\h^{(n)}]\,
    \ch_\rho^2\,\bar{R}[\h^{(n)}]\,\bch \,W[\h^{(n)}]
    \right\rangle_{\vac}\;.\eeqnn
This proves the lemma. \qed

The identity expressed by this lemma is based on algebraic
properties of Feshbach triples. To obtain bounds on the right hand
side of (~\ref{Ovlaptermglobloc333}), we must invoke the
renormalization group iteration.
\\

Let us next consider the term
\eqn\frac{\left\langle W\bchn\bar{R}_n
    \frac{\partial_{H_f}\bchn}{\bchn}
    \tHpsz \bar{R}_n\bchn W
    \right\rangle_{\vac}}{
    \left\langle \,
    \Qns[z]\,\Qn[z]
    \right\rangle_{\vac}}\;,\eeqn
which occurs in $\dHfFQQ_n[z]$.
\\

\begin{lm}
Using the notations of the previous lemma,  the identity
\eqn&&\left\langle W\bchn\bar{R}_n
    \frac{\partial_{H_f}\bchn}{\bchn}
    \tHpsz \bar{R}_n\bchn W
    \right\rangle_{\vac}\nonumber\\
    &&\hspace{1cm}=\;\left\langle\partial_{H_f}
    F_{n}[z]\right\rangle_{\vac}\,\times
    \nonumber\\
    &&\hspace{1.5cm}\times\;
    \left\langle W[\h^{(n-1)}]\,\bch \,\bar{R}[\h^{(n-1)}]
    \frac{\partial_{H_f}\bch_\rho}{\bch_\rho}\,
    (H_f-\rho z_{n})\,
    \bar{R}[\h^{(n-1)}]\,\bch \,W[\h^{(n-1)}]
    \right\rangle_{\vac}\label{etandeltterm2expvalidaux333}
    \eeqn
is satisfied, where $\bch=\ch_\rho\bch_\rho$, and where
$H[\h^{(n)}]$ denotes the effective hamiltonian at the scale $n$,
acting on $P_1\Hp$.
\end{lm}

\prf The left hand side of the equality sign in
(~\ref{etandeltterm2expvalidaux333}) can be written as
$$\lim_{\delta\rightarrow0}\left\langle \Qns[z]\,
    \frac{\partial_{H_f}\bchn }{\bchn^3+\delta}\;
    \tHpsz^{-1}\,P_n\,\bar{P}_n\,\Qn[z]
    \right\rangle_{\vac}\;,$$
where,  as in the proof of Lemma
{~\ref{Ovlaptermglobloclemma333}}, $P_n,\bar{P}_n,$ are the
projectors onto the ranges of $\chn,\bchn$ respectively. Notice
that all operators apart from $Q^\sharp,Q$ in the expectation
value are mutually commutative. By the arguments demonstrated in
the first part of the proof of Lemma
{~\ref{Ovlaptermglobloclemma333}}, based on the concatenation
rules for $Q^\sharp,Q$, and
$$Ad_{U_{\rho^{n-1}}^*}\left[
    \frac{\partial_{H_f}\bchn}{\bchn^3+\delta}
    \right]\;=\;\rho^{-(n-1)}\,
    \frac{\partial_{H_f}\bch_\rho}{\bch_\rho^3+\delta}\;,$$
as well as
\eqn Ad_{U_{\rho^{-(n-1)}}}\left[\tHpsz
    \right]&=&\left\langle\partial_{H_f}
    F_n[z]\right\rangle_{\vac}\,\times\nonumber\\
    &&\hspace{1cm}\times\;\left(\rho^{n-1}\,
    H_f\,-\,z\,+\,\left\langle\FchntHn[\Hps-z]\right\rangle_{\vac}
    \right)\nonumber\\
    &=&\rho^{n-1}\,\left\langle\partial_{H_f}
    F_n[z]\right\rangle_{\vac}\,
    (H_f-\rho z_n)\;,
    \label{AdUrhotHpszauxform333}\eeqn
cf. (~\ref{znzfctdefaux333}), the left hand side of
(~\ref{etandeltterm2expvalidaux333}) equals
\eqnn&& \lim_{\delta\rightarrow0}
    \left\langle Q_{\ch_\rho,\tau}^\sharp
    [H[\h^{(n-1)}]]\,
    \frac{\partial_{H_f}\bch}{\bch^3+\delta}\;
    (H_f-\rho z_n)
    \,P_\rho\bar{P}_\rho\,Q_{\ch_\rho,\tau}
    [H[\h^{(n-1)}]]
    \right\rangle_{\vac}\\
    &&\hspace{1cm}=\;\left\langle\partial_{H_f}
    F_n[z]\right\rangle_{\vac}\,\times\\
    &&\hspace{1.5cm}\times\;\left\langle W[\h^{(n-1)}]\,
    \bch\,\bar{R}[\h^{(n)}]\,
    \frac{\partial_{H_f}\bch_\rho}{\bch_\rho}\,
    (H_f-\rho z_n)
    \,\bar{R}[\h^{(n)}]\,\bch \,W[\h^{(n-1)}]
    \right\rangle_{\vac}\;,\eeqnn
as claimed. \qed

Thus,
\eqn&&\frac{\left\langle W\bchn\bar{R}_n
    \frac{\partial_{H_f}\bchn}{\bchn}
    \tHpsz \bar{R}_n\bchn W
    \right\rangle_{\vac}}{
    \left\langle \,
    \Qns[z]\,\Qn[z]
    \right\rangle_{\vac}}\nonumber\\
    &&\hspace{1cm}=\;\dHfFQQ_n[z]\,
    \left\langle W[\h^{(n)}]\,\bch \,\bar{R}[\h^{(n-1)}]
    \frac{\partial_{H_f}\bch_\rho}{\bch_\rho}\,
    (H_f-\rho z_n)\,
    \bar{R}[\h^{(n-1)}]\,\bch \,W[\h^{(n-1)}]
    \right\rangle_{\vac}\;.\eeqn
Inserting this expression into the formula
(~\ref{dHfFQQnformimplaux333}), and solving it for $\dHfFQQ_n[z]$,
we arrive at the asserted formula. This proves the proposition.
\qed

\subsection{Cancellation phenomena in the renormalization
of purely marginal quantities} The quantity $\dHfFQQ_n[z]$ does
is, for $n>0$, independent of $\h^{(k)}$, for all $0\leq k\leq
n-2$. The task of  controlling it analytically is therefore almost
trivial. From the point of view of the 'global' argument in the
proof of the above proposition, by which $H[\h^{(n)}]$ is directly
related to $\Hps$ via the concatenation properties of the smooth
Feshbach map, this is easy to understand.

However, from the 'local' renormalization group point of view, in
which $H[\h^{(k)}]$ is only compared to $H[\h^{k+1}]$, for all
$k=0,\dots,n-1$, this result looks far from obvious. Passing from
the scale $k$ to the scale $k+1$,   the ratio
$$\frac{\dHfFQQ_{k+1}[z]}{\dHfFQQ_{k}[z]}\;=\;
    \frac{\langle\partial_{H_f}\FchrtH[H[\h^{(k)}]]\rangle_{\vac}}
    {\langle Q^\sharp[\h^{(k)}](\partial_{\Ppar}H[\h^{(k)}])Q[\h^{(k)}]
    \rangle_{\vac}}$$
determines the renormalized value of $a[z]$. Looking at the right
hand side of this expression, it seems far from evident that the
product
$$\dHfFQQ_{n}[z]\;=\;\prod_{k=0}^{n-1}
    \frac{\langle\partial_{H_f}\FchrtH[H[\h^{(k)}]]\rangle_{\vac}}
    {\langle Q^\sharp[\h^{(k)}](\partial_{\Ppar}H[\h^{(k)}])Q[\h^{(k)}]
    \rangle_{\vac}}$$
should only depend  on $\h^{(n-1)}$. The fact that all
contributions from intermediate scales mutually cancel each other
in this product is absolutely crucial in the renormalization of
the purely marginal quantity $a[z]$.

This insight in fact suggests that these cancellation phenomena
are typical for the renormalization of purely marginal
interactions in any quantum field theory. To explain this
statement, let us for simplicity consider the renormalization of a
marginal coupling $g_k\geq0$ in some given quantum field theory,
where $k\in\N$ labels the scale. The correction of the running
coupling constant $g_k$ under a renormalization step is at most of
order $O(g_k^2)$. The familiar scenario where
$$g_{k+1}\;=\;g_k(1\pm c_k\epsilon_k^2)$$
holds for some $c_k>0$ (i.e. a   definite sign ), appears in the
cases of marginal relevance and marginal irrelevance. One easily
concludes there that either $g_k\rightarrow\infty$, or
$g_k\rightarrow0$ (UV/IR asymptotic freedom). In particular, it is
not possible that $g_n\,=\,g_0(1+O(g_0))$ for all $n\in\N$.
However, the latter condition defines purely marginal
interactions.

In a theory exhibiting purely marginal interactions, the mechanism
at work must thus necessarily be different. The fact that, as in
the above cases of marginal relevance and irrelevance,
$$g_{k+1}\;=\;g_k\,+\,\Delta g_k\; $$
is given, where $|\Delta g_k|=O(g_k^2)$, cannot be circumvented.
But in contrast to the situation there, $\Delta g_k$  does not
have a definite sign !  This makes it possible that
\eqnn g_n&=&g_0\,+\,\sum_{k=0}^{n-1}\Delta g_k\\
    &=&g_0(1+O(g_0^2))\; \eeqnn
holds, due to lots of mutual cancellations among the $\Delta
g_k$'s. It is clear that perturbation theory to any arbitrary
finite order is insufficient to analytically control the
cancellations in these sums. If the error term were $O(g_k^\nu)$
for some $\nu\geq2$, one would have an error of order $\geq
c\sum_{k=0}^{n-1}g_k^\nu$ at the scale $n$, which diverges as
$n\rightarrow\infty$.

Thus, one can putatively only rely on finding 'global identities'
that govern the cancellations, as is the case in the present work.
\\

\section{Renormalization of $a[z]$}
\label{Renazstrategysubsec}

As has been remarked, it suffices, due to the Ward-Takahashi
identities, to bound the parameter $a[z]$ and its derivative with
respect to $|p|$ in order to analytically control all marginal
operators of the theory apart from $H_f$ in $T[z;\opT]$ (which we
choose as the 'reference operator'). Thus, proving sufficiently
small upper bounds on $|\derp^\beta a[z]|$, for $\beta=0,1$, is
the key point of the present renormalization group study.

Under the assumption that $\dHfFQQ[\h]=1+O(\ez^2)$ can be proved
by the renormalization group analysis, bounds on $\derp\Egrd$ are
implied by bounds on $a_n[z_n]$. We will now explain our strategy
to determine
\eqn a_n[z_n]&=&\frac{\left\langle \Qns[z[z_n]]\,
    (\partial_{\Ppar}\Hps)\,\Qn[z[z_n]]
    \right\rangle_{\vac}}{\left\langle \,\partial_{H_f}
    \Fn[z[z_n]]\right\rangle_{\vac}}\nonumber\\
    &=&|p|\,\dHfFQQ_n[z[z_n]]^{-1}\,\nonumber\\
    &&-\,\dHfFQQ_n[z[z_n]]^{-1}
    \frac{\left\langle \Qns[z[z_n]]
    (\Ppar+g\Af^\parallel(\ssig))\,\Qn[z[z_n]]
    \right\rangle_{\vac}}{\left\langle
    \Qns[z[z_n]]\,\Qn[z[z_n]]
    \right\rangle_{\vac}}\;,\label{anzndefexpr4444}\eeqn
where we recall the shorthand
\eqnn\Qn^{(\sharp)}[z]&\equiv&Q_{\chn,\tau_n}^{(\sharp)}[\Hps-z]\;,\\
    \Fn[z]&\equiv&\FchntHn[\Hps-z]\;,\eeqnn
in order to analytically control the coefficient $a_n[z_n]$ of
$\Ppar$ in $T_{lin}[\h^{(n)}]$. We recall that
$(a_n[z_n]\dHfFQQ[\h^{(n-1)}]^{-1})|_{z_n=E_{0,n}}$ tends to
$\derp\Egrd$ in the limit $n\rightarrow\infty$.
\\

\begin{prp}\label{anznprp3333}
Assume that there exist constants $C,C'<\infty$ that are
independent of $n$, such that
\eqn\left\langle\Qns[z] \,H_f\,\Qn[z]\right\rangle_{\vac}
    \;\leq\;C\,\ez^2\,\left\langle\Qns[z] \,\Qn[z]\right\rangle_{\vac}
    \;,\label{dthhnboundaux333}\eeqn
and
\eqn\left|\dHfFQQ_n[z[z_n]]^{-1}-1\right|\;\leq\;
    C'\,\ez^2\eeqn
hold. Then,
$$|a_n[z_n]+|p||\;\leq\; \sqez\; $$
is satisfied for any $n\geq0$.
\end{prp}

\prf Clearly,
\eqn\left|\left\langle \Qns[z[z_n]]\, \Ppar \,\Qn[z[z_n]]
    \right\rangle_{\vac}\right|&\leq&
    \left\langle \Qns[z[z_n]]\,
    \left|\Ppar\right| \,\Qn[z[z_n]]
    \right\rangle_{\vac}\nonumber\\
    &\leq&\left\langle \Qns[z[z_n]]\,
    H_f\,\Qn[z[z_n]]
    \right\rangle_{\vac}\;.\eeqn
Furthermore,
\eqn\left|\left\langle \Qns[z[z_n]]\, \Af^\parallel(\ssig)
    \,\Qn[z[z_n]]
    \right\rangle_{\vac}\right|&\leq&2
    \left\langle \Qns[z[z_n]]\,
    \Afmin^\parallel(\ssig) \,\Qn[z[z_n]]
    \right\rangle_{\vac}\nonumber\\
    &\leq&2\,b\, \left\langle \Qns[z[z_n]]\,
    \,\Qn[z[z_n]]
    \right\rangle_{\vac}^{\frac{1}{2}}\,\times\nonumber\\
    &&\hspace{1cm}\times\,
    \left\langle \Qns[z[z_n]]\,
    H_f\,\Qn[z[z_n]]
    \right\rangle_{\vac}^{\frac{1}{2}}\;,\eeqn
where $\Afmin^\parallel(\ssig)$ is the part in
$\Af^\parallel(\ssig)$ linear in annihilation operators, and the
constant $b$ is bounded by
$$b\;\leq\;\left(\int_{|k|\leq\Lambda}\,d^3k\,\frac{1}{|k|}
    \right)^{\frac{1}{2}}\;,$$
where $\Lambda$ denotes the ultraviolet cutoff introduced into the
interaction in $\Hps$, as we recall. This is due to the standard
inequality
$$\left(\phi\,,\,a_\lambda(f)\,\phi\right)\;\leq\;
    \left(\int\,d^3k\,\frac{ |f(k)|^2}{|k|}\right)^{\frac{1}{2}}\,
    \|\phi\|\,\left(\phi\,,\,H_f\,\phi\right)^{\frac{1}{2}}$$
for any $\phi\in\Hp$. Comparison with (~\ref{anzndefexpr4444})
shows that this immediately implies the claim. \qed

The key point in the construction thus is to prove that
(~\ref{dthhnboundaux333}) is indeed satisfied. To this end, we
invoke a renormalization group iteration, which corresponds to an
induction argument in $n$. Let
\eqn Q_0^{(\sharp)}[z]\;\equiv\;
    Q_{\ch_1,\tau_0}^{(\sharp)}[\Hps-z]\;,\eeqn
and let us introduce the notation
\begin{gather}\langle Q^\sharp A Q \rangle_{(j;k)}\;:=\;\left\langle
    Q^\sharp[\h^{(k-1)}]Ad_{U_\rho}\left[
    \cdots Ad_{U_\rho}\left[ Q^\sharp[\h^{(j)}]A Q[\h^{(j)}]
    \right]\cdots \right]Q[\h^{(k-1)}]
    \right\rangle_{\vac}\;\label{langleQnsQndefaux3333} \end{gather}
for any operator $A$ acting on $\Hp$, and $0\leq j<k$.
\\

\begin{lm}\label{QnsQnQksQkboundlemmaaux3333}
Assume that for any arbitrary, but fixed scale $n$,
$$\epsilon_k\;\leq\;2\,\epsilon_0 \;, $$
and
\eqn\left|
    \frac{\left\langle
    Q^\sharp \,Q
    \right\rangle_{(k;n)}}{\left\langle
    \Qns[z[z_n]]\Qn[z[z_n]]
    \right\rangle_{\vac}}\right|&\leq&2\;,
    \eeqn
for $0\leq k< n$. In addition, assume that there exists a constant
$B<\infty$ that is independent of $n$ and $\epsilon_0$, such that
\eqn \left\|Q^\sharp[\h^{(0)}]\,\left(Q_0^{\sharp}[z]\,H_f\,
    Q_0[z]\,-\,H_f\right)\,Q[\h^{(0)}]\right\|&\leq&
    B\,\epsilon_0^2 \eeqn
and
\eqn\left\|Q^\sharp[\h^{(k+1)}]\,
    U_\rho\,\left(Q^{\sharp}[\h^{(k)}]\,H_f\,
    Q[\h^{(k)}]\,-\,H_f\right)\,U_\rho^*\,
    Q[\h^{(k+1)}]\right\|&\leq&
    B\,\epsilon_0^2 \label{QksQkboundauxform3333}\eeqn
for all $0\leq k< n$. Then,
$$\left\langle\Qns[z] \,H_f\,\Qn[z]\right\rangle_{\vac}
    \;\leq\;4\,B\,\epsilon_0^2\,\left\langle \Qns[z]\, \Qn[z]
    \right\rangle_{\vac}\;.$$
\end{lm}

\prf Clearly,
\eqn\left\langle\Qns[z] \,H_f\,\Qn[z]\right\rangle_{\vac}
    \;=\;\left\langle Q^\sharp \,\left(Q_0^{\sharp}[z] \,H_f\,
    Q_0[z]\right)\,Q\right\rangle_{(0;n)}\;.
    \eeqn
The renormalization group recursion is defined as follows. Due to
the condition
\eqn \left\|Q^\sharp[\h^{(0)}]\,\left(Q_0^{\sharp}[z]\,H_f\,
    Q_0[z]\,-\,H_f P_1\right)\,Q[\h^{(0)}]\right\|&\leq&
    B\,\epsilon_0^2 \; \eeqn
which is assumed to be satisified, one has
\eqn\left\langle\Qns[z] \,H_f\,\Qn[z]\right\rangle_{\vac}
    \;\leq\;\left\langle Q^\sharp \,H_f
    \,Q\right\rangle_{(0;n)}\,+\,B\,\epsilon_0^2
    \,\left\langle Q^\sharp
    \,Q\right\rangle_{(1;n)}\;
    \eeqn
for a constant $B$ that is independent of $n$. The assumption that
$$\epsilon_k\;\leq\;2\,\epsilon_0 \;,$$
for general $0\leq k\leq n$, is contained in a 'global' induction
argument of which the statement of the lemma is only a part. Let
us then consider the quantity
$$\left\langle Q^\sharp \,H_f\,Q
    \right\rangle_{(k;n)}$$
for $k\leq n-3$, which equals
\eqnn&&
    \left\langle Q^\sharp \,U_\rho\,H_f\,U_\rho^*\,Q
    \right\rangle_{(k+1;n)}\,\nonumber\\
    &&+\,
    \left\langle Q^\sharp\,\left(
    Q^\sharp[\h^{(k+1)}]U_\rho\,\left(
    Q^\sharp[\h^{(k)}]H_f Q[\h^{(k)}]-H_f P_\rho
    \right)U_\rho^*Q[\h^{(k+1)}]\right)Q
    \right\rangle_{(k+2;n)}\;.
    \eeqnn
By (~\ref{QksQkboundauxform3333}) and $U_\rho H_f U_\rho^*=\rho\,
H_f $, this is bounded from above by
\eqn\rho\,
    \left\langle Q^\sharp \,H_f\,Q
    \right\rangle_{(k+1;n)}\;+\;B\,\epsilon_0^2
    \left\langle Q^\sharp \, Q
    \right\rangle_{(k+2;n)}\;.\eeqn
Using the 'final condition'
\eqn\left\langle Q^\sharp \,H_f\,Q
    \right\rangle_{(n-1;n)}&\leq&\rho\,
    \left\langle Q^\sharp[\h^{(n-1)}] \,H_f\,Q[\h^{(n-1)}]
    \right\rangle_{\vac}\nonumber\\
    &\leq&B\,\epsilon_0^2\;,\eeqn
and backwards iterating the above estimates according to
$n\rightarrow n-1\rightarrow\cdots\rightarrow
k\rightarrow\cdots\rightarrow0$, one obtains
\eqn\left\langle \Qns[z]\,H_f\,\Qn[z]
    \right\rangle_{\vac}&\leq&\sum_{k=0}^n\rho^{n-k}
    B\,\epsilon_0^2\,
    2\,\left\langle \Qns[z]\, \Qn[z]
    \right\rangle_{\vac}\nonumber\\
    &=&4\,B\,\epsilon_0^2\,\left\langle \Qns[z]\, \Qn[z]
    \right\rangle_{\vac}\; \eeqn
for $\rho=\frac{1}{2}$, which is the asserted result. \qed

\section{Renormalization of $\derp a[z]$}

In order to prove $C^2$-smoothness of the ground state energy
$\Egrd$, it is necessary to analytically control first order
derivatives of $H[\h]$ with respect to $|p|$. Due to the pure
marginality of $a[z]$, this task is complicated.

To begin with, let us recall that the smooth Feshbach map is
defined by
\eqn\Fn[z]&=&\tau[\Hps]\,+\,\chn\omega[\Hps]\chn\,
    \nonumber\\
    &&-\,\chn\,\omega[\Hps]\,\bchn\,\bar{R}\,
    \bchn\,\omega[\Hps]\,\chn\eeqn
with
\eqn\tau[\Hps]&=&\langle\partial_{H_f}\Fn[z]\rangle_{\vac}
    \,H_f\,-\,z'\;\eeqn
and
\eqn\omega[\Hps]&=&\Hps\,-\,\left(\langle\partial_{H_f}
    \Fn[z]\rangle_{\vac}-1\right)\,H_f\,-\,
    \DEn[z]\;,\eeqn
where
\eqn\DEn[z]\;\equiv\;\left\langle\Fn[z]+z\right
    \rangle_{\vac}\eeqn
and
$$z'\;=\;z\,-\,\DEn[z]\;.$$
Furthermore, defining
\eqn z_n\;\equiv\;\langle\partial_{H_f}
    \Fn[z]\rangle_{\vac}^{-1}\,\rho^{-n}\,z'\;,\eeqn
and assuming that $z\rightarrow z_n$ is invertible, we recall that
\eqn a_n[z_n]&=& \dHfFQQ_n[z[z_n]]^{-1}\,
    \frac{\left\langle
    \Qns[z[z_n]](\partial_{\Ppar} \Hps)\Qn[z[z_n]]
    \right\rangle_{\vac}}{\left\langle
    \Qns[z[z_n]]\Qn[z[z_n]]
    \right\rangle_{\vac}}\,.
    \label{anzndefrecallaux3333}\eeqn
The explicit expression for $\derp a_n[z_n]$  is given in the
following proposition.
\\

\begin{prp}\label{derpanznprp4444}
Let $n$ denote a fixed scale, and
$z_n\in[-\epsilon,0]\subset\R_-$. Then, $\derp a_n[z_n]$ is
explicitly given by
\eqn\derp
    a_n[z_n]&=&-\,
    \frac{\derp\dHfFQQ_n[z_n]}{\dHfFQQ_n[z_n]}
    \,a_n[z_n] \,+\,\dHfFQQ_n[z[z_n]]^{-1}\nonumber\\
    &&-\,2\,\dHfFQQ_n[z[z_n]]^{-1}\,\frac{\left\langle
    (\derp\Qns[z[z_n]])\left(\partial_{\Ppar} \Hps
    \right)\derp\Qn[z[z_n]]
    \right\rangle_{\vac}}{\left\langle
    \Qns[z[z_n]]\Qn[z[z_n]]
    \right\rangle_{\vac}}\nonumber\\
    &&+\,\rho^{n-1}\,
    \left(\left\langle(\derp Q^\sharp[\h^{(n-1)}])
    \,\bar{P}_\rho\frac{(H_f-\,\rho z_n)\chr^2}{\bchr^2}\bar{P}_\rho\,
    Q[\h^{(n-1)}]\right\rangle_{\vac}+\,h.c.\right)
    \;\times\nonumber\\
    &&\hspace{0.5cm} \times\,\dHfFQQ_n[z[z_n]]^{-1}\,
    \frac{\langle\derp\partial_{H_f}\Fn[z[z_n]]\rangle_{\vac}}
    {\left\langle\,\Qns[z[z_n]]
    \Qn[z[z_n]]\right\rangle_{\vac}}\,
    \frac{\left\langle
    (\derp\Qns[z[z_n]]) \, \Qn[z[z_n]]
    \right\rangle_{\vac}}{\left\langle
    \Qns[z[z_n]]\Qn[z[z_n]]
    \right\rangle_{\vac}}\nonumber\\
    &&+\,\rho^{n-1}\, \left(\left\langle
    (\derp Q^\sharp[\h^{(n-1)}])\chr\derp H[\h^{(n)}]
    \right\rangle_{\vac}+\,h.c.\right)\;.
    \eeqn
In addition,  for $|p|\leq\puppbd$, let us assume that there exist
constants $C,C',C''<\infty$ independent of $n$ and $\epsilon_0$,
such that
\eqn\epsilon_k&\leq&2\,\epsilon_0\nonumber\\
    \left|\frac{\langle\derp\partial_{H_f}
    \Fn[z[z_n]]\rangle_{\vac}}
    {\left\langle\,\Qns[z[z_n]]
    \Qn[z[z_n]]\right\rangle_{\vac}}\right|
    &\leq&C\,\epsilon_0^2\,n
    \nonumber\\
    \left|
    \frac{\left\langle
    (\derp\Qns[z[z_n]]) \, \Qn[z[z_n]]
    \right\rangle_{\vac}}{\left\langle
    \Qns[z[z_n]]\Qn[z[z_n]]
    \right\rangle_{\vac}}\right|&\leq&C\,\epsilon_0^2\,n
    \nonumber\\
    \left|
    \frac{\left\langle
    Q^\sharp \,Q
    \right\rangle_{(k;n)}}{\left\langle
    \Qns[z[z_n]]\Qn[z[z_n]]
    \right\rangle_{\vac}}\right|&\leq&2
    \nonumber\\
    \left\|(\derp Q[\h_0^{(k)}])
    U_\rho^*Q[\h_0^{(k+1)}]\right\|&\leq&C'\,\ez\;,\nonumber\\
    \left|\derp^\beta\dHfFQQ_n[z[z_n]]\right|&\leq&
    C''\,\epsilon_0^2\nonumber\\
    Q^\sharp[\h^{(k)}]H[\h^{(k)}]Q[\h^{(k)}]&\leq&
    \frac{7}{4}\,\FchrtH[H[\h^{(k)}]]\eeqn
for $\beta=0,1$, and all $0\leq k\leq n$. Then, if $\left|\derp
a_k[z_k]+1\right|\leq \sqrt{\epsilon}_0$ holds for all $0\leq
k<n$, it follows that
\eqn\left|\derp a_n[z_n]+1\right|\;\leq\;\sqrt{\epsilon}_0
    \eeqn
is satisfied for $k=n$.
\end{prp}

\prf Taking the derivative with respect to $|p|$ of
(~\ref{anzndefrecallaux3333}),
\eqn\derp
    a_n[z_n]&=&-\,
    \frac{\derp\dHfFQQ_n[z_n]}{\dHfFQQ_n[z_n]}
    \,a_n[z_n]\nonumber\\
    &&+\,\dHfFQQ_n[z_n]^{-1}\,\frac{\left\langle
    \Qns[z[z_n]](\derp\partial_{\Ppar} \Hps)\Qn[z[z_n]]
    \right\rangle_{\vac}}{\left\langle
    \Qns[z[z_n]]\Qn[z[z_n]]
    \right\rangle_{\vac}}\nonumber\\
    &&+\,\dHfFQQ_n[z_n]^{-1}\,\frac{\left\langle
    (\derp\Qns[z[z_n]])\left(\partial_{\Ppar} \Hps-
    a_n[z_n]\dHfFQQ_n[z_n]\right)\Qn[z[z_n]]
    \right\rangle_{\vac}}{\left\langle
    \Qns[z[z_n]]\Qn[z[z_n]]
    \right\rangle_{\vac}}\nonumber\\
    &&+\,\dHfFQQ_n[z_n]^{-1}\,\frac{\left\langle
    \Qns[z[z_n]]\left(\partial_{\Ppar} \Hps-
    a_n[z_n]\dHfFQQ_n[z_n]\right)(\derp\Qn[z[z_n]])
    \right\rangle_{\vac}}{\left\langle
    \Qns[z[z_n]]\Qn[z[z_n]]
    \right\rangle_{\vac}}\label{derpanznformaux3333}\eeqn
The first two terms on the right hand side are, due to
$$\derp\partial_{\Ppar} \Hps\;=\;1\;,$$
determined by $\derp^\beta\dHfFQQ_n[z[z_n]]$ alone, for
$\beta=0,1$. The main problem consists of estimating the two last
terms in (~\ref{derpanznformaux3333}).

Since both of them can be treated in the precise same manner, we
will only focus on one of them, say the expression on the second
last line. It will shown in the course of the renormalization
group analysis in the subsequent chapters that for fixed $n$, all
maps $z_{k}\mapsto z_{k+1}$, with $k=0,\dots,n-1$, are invertible
(for all $z_k$ in suitable domains). Hence, the map $z\rightarrow
z_n$ is invertible, and
$$z[z_n]\;=\;\DEn[z[z_n]]\,+\,
    \langle \partial_{H_f}\Fn[z[z_n]]\rangle_{\vac}\,
    \rho^n\,z_n\;$$
is well-defined. Taking the derivative with respect to $|p|$ on
both sides of
$$\left(\Hps\,-\,z[z_n]
    \right)\Qn[z_n]\;=\;
    \chn\Fn[z[z_n]]\;,$$
and recalling that
$$\derp\Hps\;=\;-\,\partial_{\Ppar}\Hps\;,$$
we find
\eqnn\left(\partial_{\Ppar}\Hps \right)\Qn[z[z_n]]&=&-\,
    \left(\Hps-z[z_n]\right)
    \derp\Qn[z[z_n]]\nonumber\\
    &&+\,\langle \derp\Fn[z[z_n]]\rangle_{\vac}\Qn[z[z_n]]\,
    \nonumber\\
    &&+\,\chn\derp\Fn[z[z_n]]\;.\eeqnn
Thus,
\eqn&&\dHfFQQ_n[z_n]^{-1}\,\frac{\left\langle
    (\derp\Qns[z[z_n]])\left(\partial_{\Ppar} \Hps-
    a_n[z_n]\dHfFQQ_n[z_n]\right)\Qn[z[z_n]]
    \right\rangle_{\vac}}{\left\langle
    \Qns[z[z_n]]\Qn[z[z_n]]
    \right\rangle_{\vac}}\nonumber\\
    &&\hspace{1cm}=\,-\,
    \dHfFQQ_n[z_n]^{-1}\,\frac{\left\langle
    (\derp\Qns[z[z_n]])\left( \Hps-z[z_n]
    \right)\derp\Qn[z[z_n]]
    \right\rangle_{\vac}}{\left\langle
    \Qns[z[z_n]]\Qn[z[z_n]]
    \right\rangle_{\vac}}\label{derpanzncritterm1aux3333}\\
    &&\hspace{1.5cm}+\,\dHfFQQ_n[z_n]^{-1}\,\frac{\left\langle
    (\derp\Qns[z_n])\left(  \derp\DEn[z[z_n]] -
    a_n[z_n]\dHfFQQ_n[z_n]
    \right) \Qn[z[z_n]]
    \right\rangle_{\vac}}{\left\langle
    \Qns[z[z_n]]\Qn[z[z_n]]
    \right\rangle_{\vac}}
    \label{derpanzncritterm2aux3333}\\
    &&\hspace{1.5cm}+\,\dHfFQQ_n[z_n]^{-1}\,\frac{\left\langle
    (\derp\Qns[z[z_n]])\chn\derp\Fn[z[z_n]]
    \right\rangle_{\vac}}{\left\langle
    \Qns[z[z_n]]\Qn[z[z_n]]
    \right\rangle_{\vac}}\label{derpanzncritterm3aux3333}
    \eeqn
Let us first focus on the term (~\ref{derpanzncritterm2aux3333}).

It is straightforward to deduce by means of Lemma
{~\ref{derXbarRlemma1}}, that
\eqn\derp\DEn[z[z_n]] &=&
    \langle\derp\partial_{H_f}\Fn[z[z_n]]\rangle_{\vac}
    \left\langle\Qns[z[z_n]]\,\bar{P}
    \frac{(H_f-\rho^n\,z_n)}{\bchn^2}\bar{P}\,
    \Qn[z[z_n]]\right\rangle_{\vac}\nonumber\\
    &&-\,\left\langle\Qns[z[z_n]]\,(\partial_{\Ppar}\Hps)\,
    \Qn[z[z_n]]\right\rangle_{\vac}\nonumber\\
    &&-\,(\derp\DEn[z[z_n]])
    \left\langle\Qns[z[z_n]]\,\bar{P}\,
    \Qn[z[z_n]]\right\rangle_{\vac}\;.\eeqn
Therefore, using
$$\left\langle\Qns[z[z_n]]\,
    \Qn[z[z_n]]\right\rangle_{\vac}\;=\;1\,+\,
    \left\langle\Qns[z[z_n]]\,\bar{P}\,
    \Qn[z[z_n]]\right\rangle_{\vac}\;,$$
we have
\eqn&&\derp\DEn[z[z_n]]\;=\;-\,
    \frac{\left\langle\Qns[z[z_n]]\,(\partial_{\Ppar}\Hps)\,
    \Qn[z[z_n]]\right\rangle_{\vac}}{\left\langle\Qns[z[z_n]]\,
    \Qn[z[z_n]]\right\rangle_{\vac}}\nonumber\\
    &&\hspace{1cm}
    +\,\langle\derp\partial_{H_f}\Fn[z[z_n]]\rangle_{\vac}\,
    \frac{
    \left\langle\Qns[z[z_n]]\,\bar{P}
    \frac{(H_f-\rho^n\,z_n)\chn^2}{\bchn^2}\bar{P}\,
    \Qn[z[z_n]]\right\rangle_{\vac}}{\left\langle\Qns[z[z_n]]\,
    \Qn[z[z_n]]\right\rangle_{\vac}}\;,\eeqn
so that by (~\ref{anzndefrecallaux3333}),
\eqn(~\ref{derpanzncritterm2aux3333})&=&
    \frac{
    \left\langle(\derp\Qns[z[z_n]])
    \,\bar{P}\frac{(H_f-\rho^n\,z_n)\chn^2}{\bchn^2}\bar{P}\,
    \Qn[z[z_n]]\right\rangle_{\vac}}{\left\langle\,\Qns[z[z_n]]
    \Qn[z[z_n]]\right\rangle_{\vac}}\,\times\nonumber\\
    &&
    \times\,\dHfFQQ_n[z_n]^{-1}\,
    \langle\derp\partial_{H_f}\Fn[z[z_n]]\rangle_{\vac}
    \,\frac{\left\langle
    (\derp\Qns[z[z_n]]) \, \Qn[z[z_n]]
    \right\rangle_{\vac}}{\left\langle
    \Qns[z[z_n]]\Qn[z[z_n]]
    \right\rangle_{\vac}}\nonumber\\
    &=&\rho^{n-1}\left\langle(\derp Q^\sharp[\h^{(n-1)}])
    \,\bar{P}_\rho\frac{(H_f-\,\rho z_n)\chr^2}{\bchr^2}\bar{P}_\rho\,
    Q[\h^{(n-1)}]\right\rangle_{\vac}\;\times\nonumber\\
    && \times\,\dHfFQQ_n[z[z_n]]^{-1}\,
    \frac{\langle\derp\partial_{H_f}\Fn[z[z_n]]\rangle_{\vac}}
    {\left\langle\,\Qns[z[z_n]]
    \Qn[z[z_n]]\right\rangle_{\vac}}\,
    \frac{\left\langle
    (\derp\Qns[z[z_n]]) \, \Qn[z[z_n]]
    \right\rangle_{\vac}}{\left\langle
    \Qns[z[z_n]]\Qn[z[z_n]]
    \right\rangle_{\vac}}\;.\nonumber\eeqn
Finally,
\eqn(~\ref{derpanzncritterm3aux3333})&=&\rho^{n-1}\,
    \dHfFQQ_n[z[z_n]]^{-1}\,
    \langle \partial_{H_f}\Fn[z[z_n]]\rangle_{\vac}
    \,\times\nonumber\\
    &&\hspace{1cm}\times
    \,\frac{\left\langle
    (\derp Q^\sharp[\h^{(n-1)}])\chr\derp H[\h^{(n)}]
    \right\rangle_{\vac}}{\left\langle
    \Qns[z[z_n]]\Qn[z[z_n]]
    \right\rangle_{\vac}}\nonumber\\
    &=&\rho^{n-1}\, \left\langle
    (\derp Q^\sharp[\h^{(n-1)}])\chr\derp H[\h^{(n)}]
    \right\rangle_{\vac}\;.
     \eeqn
This proves the asserted expression for $\derp a_n[z_n]$.

The renormalization group analysis in the subsequent chapters will
show that the bound
$$\left|\frac{\left\langle
    \langle\derp\partial_{H_f}\Fn[z[z_n]]\rangle_{\vac}
    \right\rangle_{\vac}}{\left\langle
    \Qns[z[z_n]]\Qn[z[z_n]] \right\rangle_{\vac}}\right|\;,\;
    \left|\frac{\left\langle
    (\derp\Qns[z[z_n]]) \, \Qn[z[z_n]] \right\rangle_{\vac}}{\left\langle
    \Qns[z[z_n]]\Qn[z[z_n]] \right\rangle_{\vac}}\right|\leq C\,
    \epsilon_0\,n\;,
    $$
indeed holds, where the constant is  $O(1)$ with respect to $n$
and $g$. Due to
$$\left\langle(\derp Q^\sharp[\h^{(n-1)}])
    \,\bar{P}_\rho\frac{(H_f-\,\rho z_n)\chr^2}{\bchn^2}\bar{P}_\rho\,
    Q[\h^{(n-1)}]\right\rangle_{\vac}
    \;\leq\;O(\epsilon_0)\;,$$
one has
$$|(~\ref{derpanzncritterm2aux3333})|\;\leq\;
    C\,\epsilon_0\,\rho^n\,n^2\;
    \;,$$
which tends to 0 in the limit $n\rightarrow\infty$. Furthermore,
one can straightforwardly show that
$$(~\ref{derpanzncritterm3aux3333})\;\leq\;\rho^n\,
    C\,\epsilon_0\;.$$
In order to estimate (~\ref{derpanzncritterm1aux3333}), we invoke
a renormalization group iteration. We define
\eqnn A_0^{(n)} \;:=\;\left(\frac{\left\langle
    (\derp\Qns[z[z_n]])\left( \Hps-z[z_n]
    \right)\derp\Qn[z[z_n]]
    \right\rangle_{\vac}}{\left\langle
    \Qns[z[z_n]]\Qn[z[z_n]]
    \right\rangle_{\vac}}\right)^{\frac{1}{2}}\;,\eeqnn
and for   general $n$,
\eqnn A_k^{(n)}
    \;:=\;\left(\frac{\left\langle(\derp Q^\sharp )
    H[\h^{(k)}]
    \derp Q
    \right\rangle_{(k;n)}}{\left\langle
    \Qns[z[z_n]]\Qn[z[z_n]]
    \right\rangle_{\vac}}\right)^{\frac{1}{2}}\;,\eeqnn
using the notation introduced in (~\ref{langleQnsQndefaux3333}),
and
$$\h=(z,T,\{w_{M,N}\})\;.$$
To relate subsequent $A_k^{(n)}$'s for $k\rightarrow k+1$, for
$0\leq k<n$, to one another, we exploit the concatenation rules
for $Q^\sharp,Q$,
\eqnn\Omega[\h_0]\;=\;Q[\h_0]\,U_\rho^*\,\Omega[\hat{\h}_0]
    \;,\eeqnn
where $U_\rho^*$ reverts the action of rescaling, and
$\hat{\h}_0=\ren[\h_0]$. Next, we exploit that
$$\derp \Omega[\h]\;=\;\Qn[E_0]\,U_\rho^*\,\derp
    \Omega[\hat{\h}_0]\,+\,(\derp\Qn[E_0])\,
    U_\rho^*\,\Omega[\hat{\h}_0] ,$$
and
\eqnn Q^\sharp[\h]H[\h]Q[\h]&\leq&(1+\mu)\FchtH[H[\h]]\\
    &=&(1+\mu)\langle \partial_{H_f}\FchtH[H[\h]]
    \rangle_{\vac}\rho\,Ad_{U_\rho^*}[H[\hat{\h}_0]]
    \eeqnn
for some $\mu<\frac{3}{4}$, and $|\langle
\partial_{H_f}\FchtH[H[\h]] \rangle_{\vac}|=1+ O(\epsilon^2)$. Thus,
we find through the Cauchy-Schwarz inequality that
\eqnn A_k^{(n)}&\leq&\sqrt{\rho(1+\mu)\langle
    \partial_{H_f}\FchtH[H[\h]]
    \rangle_{\vac}}\,A_{k+1}^{(n)}
    \nonumber\\
    &&+\,\left\| \sqrt{H[\h^{(k)}]}(\derp Q[\h^{(k)}])
    U_\rho^*Q[\h^{(k+1)}]\right\|\,
    \sqrt{\frac{\left\langle  Q^\sharp Q
    \right\rangle_{(k+2;n)}}{\left\langle
    \Qns[z[z_n]]\Qn[z[z_n]]
    \right\rangle_{\vac}} }\\
    &\leq&\frac{9}{10}\,A_{k+1}^{(n)}\,+\,
    C\,\ez\,
    \; \eeqnn
for $\rho=\frac{1}{2}$. This follows from the assumption that
$$\left\|(\derp Q[\h^{(k)}])
    U_\rho^*Q[\h^{(k+1)}]\right\|\;\leq\;O(\ez)\;,$$
and
\eqnn\frac{\left\langle  Q^\sharp  Q
    \right\rangle_{(k;n)}}{\left\langle
    \Qns[z[z_n]]\Qn[z[z_n]]
    \right\rangle_{\vac}}\;\leq\;2\eeqnn
for all $0\leq k<n$. Therefore,
\eqnn A_0^{(n)}\;\leq\;
    C'\,\ez\;,
    \eeqnn
for a constant $C'$ that is independent of $\epsilon$ and $n$. By
assumption, $\epsilon_k\leq 2\epsilon_0$ for all $k\geq0$.

Collecting the above results, one arrives at an estimate of the
form
\eqnn|\derp a_n[z_n]|\;\leq\;(1+C_1\epsilon_0 )
    \left(1\,+\,C_2\,\ez
    \right)\;.\eeqnn
Thus, we obtain
$$|\derp a_n[z_n]|\;\leq\; 1\,+\,\sqrt{\epsilon_0}\;,$$
for $n\geq0$.

This proves the proposition. \qed

An immediate consequence is that due to
$$\derp^2\Egrd\;=\;\lim_{n\rightarrow\infty}
    \derp\left(\dHfFQQ_n[z[z_n]]
    \,a_n[z_n] \right) \;,$$
the bounds on $\derp a_n[z_n]$ obtained above immediately imply
$$\left|\derp^2\left(\Egrd-\frac{|p|^2}{2}\right)\right|\;
    \leq\;\sqrt{\epsilon_0}\; $$
for the ground state energy of $\Hps$.
\\

\section{The Ward-Takahashi identities}

Analytically controlling the renormalization flow of purely
marginal interactions is usually a very difficult task. The key to
the solution of this problem here is to exploit the $U(1)$-gauge
invariance of the effective hamiltonians $H[\h^{(n)}]$. This
reduces the number of marginal operators to merely one in the
present system, given by $a[z]$ (since the coefficient of the
purely marginal operator $H_f$ is kept fixed at the value 1). As
has been demonstrated in the preceding sections, our approach to
renormalizing $a[z]$ heavily uses the concatenation identities
fulfilled by Feshbach triples.

On the quantum level, gauge invariance imposes certain constraints
on the correlation functions of the theory. They manifest
themselves in the form of relationships between the fully
renormalized $n$-point functions of the theory, for varying $n$,
which are in the context of QED referred to as Ward-Takahashi
(W-T) identities. A standard textbook exposition can for instance
be found in \cite{itzu}.

In the present operator-theoretic framework, our focus will not be
on $n$-point functions; they do in fact not enter the discussion
at all. Instead, we will prove a relationship satisfied by the
effective hamiltonians at arbitrary scales, which represent (a
generalization of) the W-T identities.
\\

\section{Ward-Takahashi identities in non-relativistic QED}

The structure of the W-T identities in non-relativistic QED is the
same as in the special relativistic theory. However, due to the
lack of antiparticle production in the non-relativistic limit, the
"spinor-algebraic" aspects can be simplified, and for the spinless
model at hand, they are trivial. The derivation of the W-T
identities of non-relativistic QED in the standard textbook way is
accomplished in full analogy to the relativistic case, for which
we refer for instance to \cite{itzu}. They can, of course, also be
obtained by properly taking the non-relativistic limit of the
relativistic W-T identities in QED. We omit an exposition of these
matters, and simply state the final result, since our discussion
yields a different, Feshbach-theoretic proof.

To this end, let the scalar $$\Sigma(p)$$ denote the {\bf
renormalized 2-point function} of the spinless fermion, and
$$\Gamma^i(p,k,p+k)$$ the {\bf renormalized vertex function} in
non-relativistic, zero-fermion-spin QED. The vertex function
describes an incoming fermion of momentum $p$ that absorbs a
transverse photon of momentum $k$, after which it propagates on
with a different momentum $p+k$. The index $'i'$ couples it to the
components of the quantized electromagnetic vector potential in
Coulomb gauge.

The Ward-Takahashi identities \eqnn
k_i\Gamma^i(p,k,p+k)\;=\;\Sigma(p)-\Sigma(p+k)\;. \eeqnn are fully
analogous to the relativistic case. The renormalized vertex
function is thus determined by the renormalized charge propagator.
More relevant for us is its differential expression,
\eqn\Gamma^i(p,0,p)\,=\,-\,\partial_{p_i}\Sigma(p)\;,
    \label{WTidtextbookrel3}\eeqn
obtained in the limit $|k|\rightarrow0$, which will occur in this
form in the operator-theoretic context.

In the context of non-relativistic QED, it is clear that
$$\Sigma(p)\;=\;\Egrd\;,$$
the {\bf renormalized ground state energy} of $\Hps$, and
$g\Gamma^i(p,0,p)$ turns out to be the renormalized coefficient of
the marginal interaction term. Thus,
$$\Gamma^i(p,0,p)\;=\;n_p^i\,
    \lim_{n\rightarrow\infty}a_n[E_{0,n}]\;,$$
and we observe that the Ward-Takahashi identities are equivalent
to the statement of Proposition {~\ref{anE0neqderpEgrdprp3}},
given that its assumptions are satisfied.

Likewise, there are higher order Ward-Takahashi identities which
relate $n$-point functions of the theory to $n'$-point functions,
with $n'<n$, which we will not discuss.

In the present operator-theoretic formulation, we will verify that
a particular relationship is fulfilled on the operator level. This
is to say, in contrast to relations between particular kinds of
expectation values, for instance. They are satisfied by all
effective hamiltonians, and will be  interpreted  as a generalized
form of Ward-Takahashi identities.
\\

\section{Ward-Takahashi identities in
Feshbach renormalization} \label{WTidentFeshbrensubsec}

We will now return to the functional analytic setting introduced
in the previous chapters, and formulate a variant of the W-T
identities that is applicable in the operator theoretical
renormalization scheme. The relations between $n$-point functions
in the above discussion of the functional integral formulation of
the problem are now translated into relations between partial
derivatives of the effective hamiltonians with respect to
operators, and commutators with creation and annihilation
operators.

The strategy is based on the recursive application of Theorem
{~\ref{WTauxcor}}. We first prove that the physical hamiltonian
$\Hps$ obeys a relationship of the form stated there. The
consequence is that the effective hamiltonian $H[\h^{(0)}]$ obeys
the same relationship, thus also $H[\h^{(1)}]$, and so on.

The starting point is the following fact.
\\

\begin{lm}
The operator
$$\der^\sharp_{comm}\;:=\;\lim_{\stackrel{|k|
    \rightarrow0}{n_k\;{\rm fixed}}}
    \bcs^{-1}[|k|]\,\sqrt{|k|}\;
    [a_\lambda^\sharp(k)\,,\,\cdot]\; $$
satisfies $\der_{comm}^\sharp[f[H_f]]=0$ for all $f$ of class
$C^1$.
\end{lm}

\prf The derivation rules are obviously satisfied by commutators,
thus $\der_{comm}^\sharp$ is a derivation.

Next, for any $C^1$-function $f$,
\eqnn\lim_{\stackrel{|k|\rightarrow0}{n_k\;{\rm fixed
    }}}\bcs^{-1}[|k|]\,|k|^{\frac{1}{2}}\,
    [a_\lambda(k)\,,\,f[H_f]]&=&0\;,\eeqnn
This is because the expression on the left hand side, of which the
limit is taken, can be written as
\eqnn\sigma\,|k|^{-\frac{1}{2}}\,
    \left(f[H_f+|k|]\,-\,f[H_f]\right)\,a_\lambda(k)\;,\eeqnn
since $\bcs[|k|]=\frac{1}{\sigma}|k|$ for small $|k|$.
$C^1$-smoothness of $f$ implies the claim. \qed

Next, we define the two following special derivations, which we
will in the sequel refer to as the '{\bf Ward-Takahashi
derivations}':
\eqn\derWTs^*&:=&g\,\epsilon_{1,0}[n_k;\lambda]\,
    \cdot\,\partial_{\P_f}\,+\,
    \lim_{\stackrel{|k|\rightarrow0}{n_k\;{\rm fixed
    }}}\,
    \bcs^{-1}[|k|]\sqrt{|k|}\,\left[
    \cdot,a_\lambda^*(k)\right]\nonumber\\
    \derWTs&:=&g\,\epsilon_{0,1}[n_k;\lambda]\,
    \cdot\,\partial_{\P_f}\,+\,
    \lim_{\stackrel{|k|\rightarrow0}{n_k\;{\rm fixed
    }}}\,
    \bcs^{-1}[|k|]\sqrt{|k|}\,\left[
    a_\lambda(k),\cdot\right]\;,\eeqn
where $\bcs$ denotes the running infrared regulator. Then, the
following proposition holds.
\\

\begin{prp}
The physical hamiltonian $\Hps$ satisfies
\begin{gather}
    \derWTss^\sharp[\Hps]\;=\;0\;.
\end{gather}
\end{prp}

\prf The cutoff regularized hamiltonian of the system in
discussion is, as we recall, given by
\begin{gather}\Hps\;=\;\Eb+\;H_f\,+\,
    \left.\left.\frac{1}{2}\,
    :\right(p\,-\,\P_f+g \Af(\ssig)\right)^2:\,\;,
\end{gather}
with \eqnn\Eb\;=\;\frac{|p|^2}{2}\;+\;\frac{g^2}{2}\;
    \left\langle \Af(\ssig)^2
    \right\rangle_{\vac}\;,\eeqnn
and acts on the Hilbert space $\Hp$, where $p$ denotes the
conserved total momentum.

It can be easily checked that
\eqn\partial_{\P_f}\Hps\;=\;-\;p\;+\;\P_f\;+\;
    g\;\Af(\ssig)\;,\eeqn
and from
\begin{gather}\Af(\ssig)\;=\;\left.\left.
    \sum_\lambda\int
    \frac{d^3k\;\bcss[|k|]\;\uvcut[|k|]}{\sqrt{|k|}}
    \right(\epsilon_{1,0}[n_k,\lambda]a^*_\lambda(k)+
    \epsilon_{0,1}[n_k,\lambda]a_\lambda(k)\right)\;,
\end{gather}
that \eqnn\left[
    \Hps,a_\lambda^*(k)\right]&=&a_\lambda^*(k)
    \left(\frac{1}{2}
    \left((\P_f+k)^2-\P_f^2\right)+|k|-|p|k^\parallel\right)\;
    \nonumber\\
    &&+\;g\;a_\lambda^*(k)\; k\cdot\Af(\ssig)\;+\;
    \frac{1}{2}a_\lambda^*(k)\;\nonumber\\
    &&+\;g\;\frac{\bcss[|k|]}{\sqrt{|k|}}\;
    \epsilon_{1,0}[n_k;\lambda]\cdot
    \left(p\;-\;\P_f\;-\;
    g\;\Af(\ssig)\right)\;.\eeqnn
Therefore,
\begin{gather}
    \lim_{\stackrel{|k|
    \rightarrow0}{n_k\;{\rm fixed}}}
    \bcss^{-1}[|k|]\sqrt{|k|}\;\;\left[
    \Hps,a_\lambda^*(k)\right]\;=\;-\;
    g\;\epsilon_{1,0}[n_k;\lambda]\;
    \cdot\;\partial_{\P_f}\Hps\;,
\end{gather}
and likewise,
\begin{gather}
    \lim_{\stackrel{|k|
    \rightarrow0}{n_k\;{\rm fixed}}}
    \bcss^{-1}[|k|]\sqrt{|k|}\;\;\left[
    a_\lambda(k),\Hps\right]\;=\;-\;
    g\;\epsilon_{0,1}[n_k;\lambda]\;
    \cdot\;\partial_{\P_f}\Hps\;,
\end{gather}
for the commutator with $a_\lambda(k)$. \qed

We notice that the limit $|k|\rightarrow0$ is in full analogy to
the corresponding limit of vanishing external photon momentum
which was used to obtain (~\ref{WTidtextbookrel3}).
\\

According to the notation used in Theorem {~\ref{WTauxcor}}, let
\eqnn\der_1\;:=\;\partial_{\P_f}\;,\eeqnn and \eqnn
h_1\;:=\;-\,g\,\epsilon_{0,1}[n_k,\lambda]\,\1\;.\eeqnn
Furthermore, let \eqnn\der_2\;:=\;\lim_{\stackrel{|k|
    \rightarrow0}{n_k\;{\rm fixed}}}
    \bcss^{-1}[|k|]\sqrt{|k|}\;[a_\lambda(k)\,,\,\cdot]\;,\eeqnn
and \eqnn h_2\;:=\;\1\;.\eeqnn Theorem {~\ref{WTauxcor}}
straightforwardly implies that these relations also hold for
$$H[\h^{(0)}]\;=\;\mathcal{D}_1[\Hps-z]\;.$$
That is,
\begin{gather}
    \derWTss^\sharp[H[\h^{(0)}]]\;=\;0\;.
\end{gather}

\section{The Ward-Takahashi identities in inductive form}
\label{WTindformsubsec}

Arguing by induction, this relationship can be extended to all
effective hamiltonians produced in the renormalization group
iteration. The following proposition establishes this.
\\

\begin{prp}\label{WaTaidprop2}
Let  $H[\h]$, with $\h\in\Polyd_{\epsilon,\xi}$ for $\epsilon,\xi$
sufficiently small, be a given effective hamiltonian, which
satisfies
\eqnn
    \derWTs^\sharp[H[\h]]\;=\;0\;,
\eeqnn
$\sigma$ denotes the running infrared cutoff. Then, its
renormalized counterpart $H[\hat{\h}]$, with $\hat{\h}=\ren[\h]$,
satisfies
$$\derWTsh^\sharp[H[\hat{h}]]\;=\;0\;,$$
where $\hat{\sigma}=\frac{\sigma}{\rho}$.
\end{prp}

\prf This follows directly from Theorem {~\ref{WTauxcor}}, and the
fact that $T$ and $w_{M,N}$ are of class $C^1$, for all
$M+N\geq1$, with respect to the components of $\op$, since
$\h\in\Polyd_{\epsilon,\xi}$. \qed

The Ward-Takahashi identities in the operator-theoretic
renormalization group are used in Chapters {~\ref{firstdecchap}},
{~\ref{analiterstepchap}}, for the purpose of proving that the
marginal interaction kernels $w_{M,N}$, for $M+N=1$, are fully
determined by $a[z]$, the coefficient of the operator $\Ppar$ in
$T_{lin}[\h]$. This reduces the number of independent marginal
operators, whose renormalization must be controlled, by one.
\\

\section{Remark on alternative formulations}
\label{alternWTformrem}

Other  operator-theoretic versions of the Ward-Takahashi
identities  certainly deserve consideration. For instance, one
could consider coupling the charged particle to a classical
external electromagnetic vector potential $\Af^{(cl)}$, and study
derivatives of the effective hamiltonians with respect to
$\Af^{(cl)}$. Likewise, one could have thought of comparing
derivatives with respect to $|p|$, of effective hamiltonians at
subsequent scales to each other.

The problem here is that in order to achieve maximal simplicity of
the method, it is desirable only to involve operations into the
formulation of the W-T identities which {\bf commute with
rescaling}. Neither of the two aforementioned tentative
alternatives have this property, because the factor
$(1+\DHf[z])^{-1}$ is a function both of $|p|$ and $\Af^{(cl)}$,
in case the latter is not zero.

In our favored approach presented in Section
{~\ref{WTidentFeshbrensubsec}}, however, the operations
$\partial_{\Ppar}$, and $[a_\lambda^\sharp(k)\,,\,\cdot]$
(commutation with creation- and annihilation operators) both
commute with the rescaling transformation $\resc$. The price that
we pay for this convenient  property is that the infrared
regularization $\bcs[|k|]$ cannot be picked entirely arbitrarily.
However, since the latter is an artificial item of the given
theory anyway, we shall accept this mild restriction.

\chapter{ANALYSIS OF THE FIRST DECIMATION STEP}
\label{firstdecchap}

\section{The main Theorem}

The purpose of this chapter is to determine the initial condition
$\h^{(0)}\in\Polyd_{\epsilon,\xi}$ (the polydisc defined in
Section {~\ref{Polyddefsubsect3}}), for the renormalization group
recursion. We recall the selfadjoint that the smooth cutoff
operators are given by
\eqnn \ch_1&=&\ch_1[H_f]\nonumber\\
     \ch_1^2\,+\,\bch_1^2&=&\1\;,\eeqnn
and that the projectors onto the respective ranges are given by
$P_1,\bar{P}_1$. Furthermore, $P_1^\perp=\1-P_1$, and
$\bar{P}^\perp_1=\1-\bar{P}_1$.

For the spectral parameter, we choose
$$z\;\in\;\left[\Eb-g\;,\;\Eb\right]\;,$$
where $\Eb$ is the bare ground state energy defined in
(~\ref{E0sigpphyshamdef}). We assume that the conserved total
momentum satisfies
$$|p|\;\leq\;\puppbd\;,$$
for some constant $\puppbd\in[\puppbdnum,1)$, which is required to
be sufficiently small for the results of this and the next chapter
to hold. Cf. also the remark on Cherenkov radiation in the
introductory chapter.

The Feshbach triple $(Q_1,Q_1^\sharp,\FchltH)$ associated to
$(\ch_1,\Hp)$ (cf. definition {~\ref{Feshbtripledef}}) is given by
\eqn\;Q_1[z]&:=&\ch_1\;-\;\bch_1\bar{R}[z;\opT]\bch_1
    \omz[\Hps]\ch_1\nonumber\\
    Q_1^\sharp[z]&:=&\ch_1\;-\;\ch_1 \;\omz[\Hps]\; \bch_1
    \bar{R}[z;\opT]\bch_1
    \nonumber\\
    \Fl[z]&\equiv&\FchltH[\Hps]\nonumber\\
    &=&\tau[\Hps]\,+\,\ch_1 \omega[\Hps]\ch_1\,
    \nonumber\\
    &&\hspace{2cm}-\,
    \ch_1\omz[\Hps]\bch \bar{R}[z;\op]
    \bch_1\omz[\Hps]\ch_1\;,\eeqn
where
\eqnn\bar{R}[z;\op]\;=\;\left(\tau[\Hps]+\bch_1
    \omz[\Hps]\bch_1 \right)^{-1} \; \eeqnn
on $\bar{P}_1\Hp$, and
\eqnn\tau[\Hps]&=&(1+\DHf[z])\,H_f\,-\,z\,+\,\Delta
    E_0[z]\\
    \omz[\Hps]&=&\Hps\,-\,z\,-\,\tau[\Hps]\;,\eeqnn
cf. the discussion in Section {~\ref{Neumserexpsubsect3}}. The
quantity $\DHf[z]$ is defined by the implicit relation
\eqnn1+\DHf[z]\;:=\;\left\langle\partial_{H_f}
    \FchltH[\Hps-z]\right\rangle_{\vac}\;.
    \label{DHfzfirstdecstepfirstdefaux3}\eeqnn
Furthermore,
\eqnn T'[\h]&\equiv&T[\h]\,-\,\tau[H[\h]]\nonumber\\
    &=&-\,\DHf[z]H_f\,+\,\Delta E_0[z]\,
    +\,\ch_1\left(-|p|\Ppar
    \,+\,\lTnl|\P_f|^2\right)\ch_1\;,\eeqnn
where
$$\lTnl\;=\;\frac{1}{2}$$
(we introduce this parameter for later convenience). The element
$\h^{(0)}\in\Polyd_{\epsilon,\xi}$ is determined by
\eqnn
    H[\h^{(0)}]\;:=\;(1+\DHf[z])^{-1}\mathcal{D}_{1}
    [\Hps-z]\;,\eeqnn
where we recall that $\mathcal{D}_{1}$ is the decimation operator,
given by the concatenation of the Feshbach map $\Fl[z]$ with
Neumann series expansion and Wick ordering, cf. Sections
{~\ref{Neumserexpsubsect3}} and {~\ref{Wickordsubsubsec}}.
$H[\h^{(0)}]$ is referred to as the effective hamiltonian at the
scale 0.

Neumann series expansion of $\bar{R}[z;\opT]$ in $\Fl[z]$ yields
\eqn\label{Neumexpfirststep}
    \Fl[z]&=&\tau[\Hps]\,+\,
    \ch_1\,T'\,\piop\,\ch_1\,+\,
    \ch_1\,\piop\,W\,\piop\,\ch_1\nonumber\\
    &-&\sum_{L=2}^\infty\, (-1)^L
    \sum_{\stackrel{1\leq M_i+N_i\leq 2}{i=2,\dots,L-1}}\,
    \ch_1\,\piop\,\left( \sum_{1\leq M_1+N_1\leq 2}W_{M_1,N_1}
    \,\right)\,
    \bch_1\,\bar{R}_0 \,\bch_1\,\cdots
    \nonumber\\
    &&\hspace{0.5cm}\cdots\,
    \bch\,W_{M_{L-1},N_{L-1}}\,\bch\,
    \bar{R}_0\,\bch_1
    \left( \sum_{1\leq M_L+N_L\leq 2}W_{M_L,N_L}\,\right)\,
    \piop\,
    \ch_1\;,\eeqn
where
\eqn\piop&=&\1\,-\,\bch_1 \Ttild\bch_1 \bar{R}_0\nonumber\\
    &=&\Pbch_1^\perp\,+\,\Pbch_1\,\tau[H]\,\bar{R}_0\,\;,
    \label{piopldef4444}\eeqn
cf. Lemma {~\ref{piopdeflemma}}, and
$$T'\,=\,T\,-\,\tau[\Hps]\;.$$
Wick ordering each of the summands in the Neumann series, and
collecting all Wick monomials that are equal in the number of
creation and annihilation operators, we find
\eqn\mathcal{D}_1[\Hps-z]&=&(1+\DHf[z'])H_f\,+\,\Delta
     E_0[z']\;-\;z' \nonumber\\
     &&+\,
     \chi_1( T'\,\piop\,+\,
     \Delta T')[z';\opT]\,\chi_1\;\nonumber\\
     &&+\;\left.\left.\sum_{M+N\geq 1}\;\;\chi_1\;
     \right( \piop\,W_{M,N}\,\piop
     +\Delta W_{M,N}\right)[z';\op]\;
     \chi_1\;,\label{Decfirststeppiop4444}\eeqn
where again, $z'=z-\Eb$.

The main result of this chapter is the following theorem.
\\

\begin{thm}
\label{firstdecstepmainthm3} The point $\h^{(0)}\in\Hspace$ lies
in a polydisc $\Polyd_{\epsilon_0,\xi}$ (for definitions, cf.
Section {~\ref{Polyddefsubsect3}}), where
\eqnn\epsilon_0&=&\xi^2\nonumber\\
    \epsilon_0\,\xi&=&2\,g\;,
    \eeqnn
for $g$ sufficiently small, and it is assumed that
$$|p|\;\leq\;\puppbd\; $$
holds for the conserved momentum.

The function $\dHfFQQ[\Hps]$ (cf. Proposition
{~\ref{dHfFQQformprop333}} for its definition) obeys the bound
\eqnn\left|\derp^\beta\left(\dHfFQQ[\Hps]-1\right)\right|\;\leq\;
    O(g^{2})\;, \eeqnn
with $\beta=0,1$, and the renormalized coefficient $\hat{a}[Z[z]]$
of $\Ppar$ in $T_{lin}[\hat{\h}]$ satisfies
\eqnn\lim_{|p|\rightarrow0}\hat{a}[Z[z]]&=&0\;,\nonumber\\
    \left|\,\derp^\beta(\hat{a}[Z[z]]+|p|) \,\right|&\leq&
    O(g^2)\,
    \;.\eeqnn
For the   operator $T_{n-l}[\h^{(0)}]$, there exists a constant
$\bTnl<\infty$ independent of $g$, such that
\eqnn\left|
    \derp^\beta\partial_{\op}^\alpha\left(
    T_{n-l}[\h^{(0)}]\,-\,\hlTnl\,|\P_f|^2\right)
    \right|&\leq&O(g^2)\,+\,
    \bTnl\bar{P}_1
    \;,
    \eeqnn
for $0\leq|\alpha|\leq2$, $\beta=0,1$, where
\eqnn\hlTnl&=&\frac{1}{2}\;,\\
    |\derp\hlTnl|\,,\,|\partial_z\hlTnl|&=&
    0\;.\eeqnn
In particular,
\eqn\left|
    \left(T_{n-l}[\h^{(0)}]\,-\,\hlTnl|\P_f|^2\right)
    \right|&\leq&O(g^2)\,+\,\left(\frac{|p|+\hlTnl}{2}\,
    +\,O(\sqez)\right)\bar{P}_1 \eeqn
and
\eqn\left|
    \partial_{H_f}
    T_{n-l}[\h^{(0)}]
    \right|&\leq&O(\sqez)\;+\;(|p|\,+\,\hlTnl)\,\times\nonumber\\
    &&\;\times\,\left(2\,+\,
    3\,|\partial_{H_f}(\ch_1^2\bch_1^2)|\,
    +\,2(|p|+\hlTnl)\,|\partial_{H_f}
    \bch_1^2|\right)\;. \label{partHfTnlfirststepboudn4444}\eeqn
For $M+N=1$,
\eqnn u_{M,N}[\h^{(0)}]&=&
    \lim_{\stackrel{|k|\rightarrow0}{n_k\;{\rm fixed}}}
    w_{M,N}[\h^{(0)}]|_{\op\rightarrow0}\\
    &=&g\;\hat{a}[Z[z]]\;
    \epsilon_{M,N}^\parallel[n_k,\lambda]\;,\eeqnn
and for $Y$ denoting $|k|$, $z$, $H_f$ or $\Pperp$,
\eqnn \nm\,\bPlperp[H_f]\,
    \,\partial_{X}w_{M,N}[\h^{(0)}]\,\bPlperp[H_f+|k|]\,\nm
    &\leq&O(g^2)\;,\\
    \nm\,\bPlperp[H_f]\,
    \,\partial_{\Ppar}
    w_{M,N}[\h^{(0)}]\,\bPlperp[H_f+|k|]\,\nm
    &\leq&2\,g\,\hlTnl\,+\,O(g^2)\;
    \;.\eeqnn
More generally, the following  estimates hold for
$0\leq|\alpha|\leq2$, $0\leq\beta\leq1$:
\eqnn
    \nm\,\derp^\beta\,\partial_{X}
    \,w_{M,N}[\h^{(0)}]\,\nm\;,\;
    \nm\,\derp^\beta
    \,\partial_{\op}^\alpha
    w_{M,N}[\h^{(0)}]\,\nm&\leq&
    \epsilon_0 \xi\;\;\;\;{\rm if}\; M+N=1\;,
    \eeqnn
while for  $M+N\geq2$,
\eqnn\nm\,\derp^\beta\,\partial_{\op}^\alpha\,w_{M,N}[\h^{(0)}]\,\nm
    &\leq&
    \epsilon_0^{\frac{7}{4}}\,\xi^{M+N}\;,
    \\
    \nm\,\derp^\beta\,\partial_X\,w_{M,N}[\h^{(0)}]\,\nm&\leq&
    \epsilon_0^{\frac{7}{4}}\,\xi^{M+N}\;.
    \eeqnn
All bounds are uniform in $|p|\leq\puppbd$, and for sufficiently
small $\puppbd\in[\puppbdnum,1)$.
\end{thm}

\section{Well-definedness of the Feshbach triple}
\label{FirstdecWelldefFbtripsect3}

First of all, it must be verified that for the given choice of
$z$, the Feshbach triple $(Q_1,Q_1^\sharp,\FchltH)$ is
well-defined.

For notational convenience, let us make a redefinition of the
spectral parameter,
\eqn\;z\;\longrightarrow\;z'\;:=\;z\;-\;\Eb\;\in\;\R_-\;,\eeqn so
that $z'$ lies in the interval $[-g,0]$. In all of the above
expressions, $z$ can be replaced by $z'$ if $\Eb$ is set to zero.

We will now prove that the quantities (~\ref{Feshboundopasscond})
needed for the definition of the Feshbach triple extend to bounded
operators on $\Hp$, and start by showing that $\bar{R}[z;\op]$ is
well-defined.
\\

\begin{lm}\label{R0Ruppbdfirstdecsteplemma4444}
For $z'=z-\Eb\in[-g,0]$,
\eqn\left\|\,\bar{R}[z;\opT]\,\right\|\;,\;
    \left\|\,\bar{R}_0[z;\opT]\,\right\|\;\leq\;
    \frac{1}{1-|p|+O(g^2)}\;\leq\;2\eeqn
for all $|p|\leq\puppbd$, with $\puppbd\in[\puppbdnum,1)$
sufficiently small.
\end{lm}

\prf Let, for $T'\equiv
T-z-\tau[\Hps]=\omz[\Hps]|_{g\rightarrow0}$,
\eqnn\bar{R}_0[z';\opT]\;=\;\bP1\left(\tau[\Hps]+\bP1
    T'\bP1
    \right)^{-1}\bP1\eeqnn
denote the free resolvent, obtained from $\bar{R}[z;\opT]$ by
setting the coupling constant $g$ equal to zero. Clearly,
\eqn\left\|\;\bar{R}_0[z';\opT]\;\right\|\;\leq\;
    (1-|p|)^{-1}\;\leq\;O(1)\;,\eeqn
since
\eqnn\tau[\Hps]+\bP1
    T'\bP1
    &\geq&(1-|p|)H_f\;.\eeqnn
>From Lemma {~\ref{elhighenrelW1W2bounds1}} below follows that
\eqnn\left\|\bar{R}_0^{\frac{1}{2}}\,\ch_1\,W  \,\ch_1\,
      \bar{R}_0^{\frac{1}{2}}\right\|\;
      \leq\;O(g )\;\ll\;1\eeqnn
if $|p|\leq \puppbd$, hence $\bar{R}[z;\op]$ can be expanded into
an absolutely convergent Neumann series, such that
\eqnn\|\bar{R}\|&\leq&\|\bar{R}_0^{\frac{1}{2}}\|
    \,\sum_{j\geq0}\left(\|\bar{R}_0^{\frac{1}{2}}\,
    \ch_1\,W\,\ch_1\,
    \bar{R}_0^{\frac{1}{2}}\|\right)^j\,
    \|\bar{R}_0^{\frac{1}{2}}\|\\
    &\leq& \sum_{j\geq0}O(g^{ j})\\
    &=&O(1)\; \eeqnn
is satisfied for the given range of $z$, resp. $z'$. This proves
the lemma. \qed

\begin{lm}\label{elhighenrelW1W2bounds1}
For the interaction operators in $\Hps$, cf. (~\ref{physW}),
\eqn\left\|\;\bar{R}_0^{\frac{1}{2}}[z;\opT]
    \bch_1\;W_1[\op]\;\bch_1
    \bar{R}_0^{\frac{1}{2}}[z;\opT]\;\right\|&\leq&
    O(g)\;\nonumber\\
    \left\|\;\bar{R}_0^{\frac{1}{2}}[z;\opT]\;\bch_1 W_2[\op]\;
    \bch_1\bar{R}_0^{\frac{1}{2}}[z;\opT]\;
    \right\|&\leq& O(g^{2})\;\;\eeqn
holds on $\Hp$, for $|p|\leq \puppbd$.
\end{lm}

\prf The proof is given in the Appendix in Chapter
{~\ref{firstdecappendix}}. \qed

\begin{lm}\label{elhighenrelW1W2bounds2}
The interaction operators in $\Hps$, cf. (~\ref{physW}), obey
\eqn\left\|\;\bar{R}_0^{\frac{1}{2}}[z;\opT]\;
    \bch_1 W_1[\op]\;
    \ch_1\;\right\|&\leq&O(g)\nonumber\\
    \left\|\;\bar{R}_0^{\frac{1}{2}}[z;\opT]\;\bch_1 W_2[\op]\;
    \ch_1\;\right\|&\leq&O(g^{2})\;
\eeqn
on $\Hp$, for $|p|\leq \puppbd$.
\end{lm}

\prf Cf. the Appendix in Chapter {~\ref{firstdecappendix}}. \qed

These bounds imply that the Feshbach triple
$(Q_1,Q_1^\sharp,\FchltH)$ is well-defined, cf. Section
{~\ref{isospectrsubsect3}}.
\\

\begin{lm}\label{piopderboundsfirstdeclemma4444}
There is a constant $C$ that is independent of $g$, such that
\eqn\|\derp^\beta\partial_{\op}^\alpha\piop\|&\leq&C
    \label{derpbetapartopalphpiopleqCfirstdec4444}\eeqn
holds on $P_1\Hp$, for $0\leq |\alpha|\leq2$ and $\beta=0,1$. In
particular, with
\eqnn
    \|\partial_{H_f}\bch_1^2\|\;,\;
    \|\partial_{H_f}\bch_1^2\|&\leq&4 \;,\eeqnn
the estimates
\eqn\|\piop \|&\leq&
    \frac{4}{3}\nonumber\\
    \|\ch_1\left(\piop\,-\,\1\right)\ch_1\|&\leq&
    \frac{1}{3}\label{piopmin1firstdecest4444}\\
    \|\derp\piop \|&\leq&2\label{derppiopfirstdecest4444}\\
    |\partial_{H_f}(\ch_1\piop\ch_1)|&\leq&
    2\,+\,
    3\,|\partial_{H_f}(\ch_1^2\bch_1^2)|\,
    \nonumber\\
    &&\hspace{1cm}+\,2(|p|+\lTnl)\,|\partial_{H_f}
    \bch_1^2|
    \label{partialHfpiopfirstdecest4444}\eeqn
hold for $|p|\leq\puppbd$ with $\puppbd\in[\puppbdnum,1)$
sufficiently small.
\end{lm}

\prf From
\eqnn\left\|P_1(\derp^\beta\partial_{\op}^\alpha X)
    P_1\right\|\;
    \leq\;c\eeqnn
for
$$X\;=\;\ch_1^2\bch_1^2\;,\;\bar{R}_0\;{\rm and}\;T'[\h]$$
(using $|\derp\DHf[z]|,|\derp\Delta E_0[z]|\leq O(g^2)$, as will
be proved below, and Lemma {~\ref{R0Ruppbdfirstdecsteplemma4444}})
follows straightforwardly that there exists a constant $C<\infty$
such that (~\ref{derpbetapartopalphpiopleqCfirstdec4444}) holds.

To prove (~\ref{piopmin1firstdecest4444}), where (on $P_1\Hp$)
$$\piop[\h]\,-\,\1\;=\;-\,\bch_1 \Ttild
    \bch_1 \bar{R}_0\;,$$
we use
\eqnn\left\|P_1 \,\bar{R}_0[\h]\,P_1\right\|
    &\leq&
    \left(1-|p|+O(g^2)\right)^{-1}\;,\eeqnn
from Lemma {~\ref{R0Ruppbdfirstdecsteplemma4444}}. Furthermore,
\eqn\|\ch_1^2\bch_1^2\|\;\leq\;\frac{1}{4}\;,\eeqn
since $\max_{x\in[0,1]}\{x^2(1-x^2)\}=\frac{1}{4}$. Moreover, by
definition of $T' $ and $|\DHf[z]|,|\Delta E_0[z]| \leq O(g^2)$,
\eqnn\|P_1 T' P_1\|&\leq&
    |p|\,+\,\lTnl\,+\,O(g^2)\;.\eeqnn
Combining these estimates with $\lTnl=\frac{1}{2}$, we find that
\eqnn\|\ch_1\left(\piop \,-\,\1\right)\ch_1\|&\leq&
    \frac{1}{4}\frac{|p|+\frac{1}{2}+
    O(g^2)}
    {1-|p|+O(g^2)}\\
    &\leq&\frac{1}{3}\;,\eeqnn
and from
\eqnn\piop &=&\bPlperp\,+\,\bPl\,\tau[\Hps]\,
    \bar{R}_0 \eeqnn
we have
\eqnn\|\piop \|&\leq&\max\left\{1\;,\;
    \|\tau[\Hps]\,\bPl\bar{R}_0 \|\right\}
    \\
    &\leq&(1+O(g^2))\max_{x\geq 1}\frac{x}
    {(1-|p|)x+O(g^2)}\\
    &\leq&\frac{4}{3}\;,\eeqnn
for all $|p|\leq\puppbd$ if  $\puppbd\in[\puppbdnum,1)$ is picked
sufficiently small, as claimed.
\\

As to (~\ref{derppiopfirstdecest4444}), we have
\eqnn\|\derp\piop \|&\leq&
    \|(\derp \tau[\Hps])\,\bar{R}_0[\h]\|
     \,\nonumber\\
    &&+\,\|\tau[\Hps]\bar{R}_0^2
    \derp(\tau[\Hps]+\bch_1 T'\bch_1)\|\\
    &\leq&\max_{x\geq1}
    \frac{(1+O(g^2))\,x^2}
    {((1-|p|)x+O(g^2))^2}\,
    +\,O(g^2)\\
    &\leq&2\eeqnn
on $P_\rho\Hp$. This establishes the claim.
\\

To prove (~\ref{partialHfpiopfirstdecest4444}), we use
\eqnn\left\|P_1(\partial_{H_f}T'[\h])P_1
    \right\|\;\leq\;
    O(g^2)\;,\eeqnn
using the fact that $P_1\,(\partial_{H_f}\bch_1)\,P_1 =0$ since
the spectral supports of $\partial_{H_f}\bch_1$ and $P_1$ are
mutually disjoint, as well as
\eqnn\left|P_1\left(\partial_{H_f}\bar{R}_0 \right)
    P_1\right|&\leq&\left\|P_1 \bar{R}_0
    P_1\right\|^2\left|P_1\left(\partial_{H_f}
    (\tau[\Hps]+\bch_1 T' \bch_1)\right)
    P_1\right|\\
    &\leq&\frac{\left(1+O(g^2)+
    \|P_1 T' P_1\|\,|\partial_{H_f}
    \bchr^2|\right)}
    {(1-|p|+O(g^2))^2}\\
    &\leq&\frac{ \left(1+(|p|+\lTnl)|\partial_{H_f}
    \bch_1^2|+O(g^2)\right)}
    {(1-|p|+O(g^2))^2}\;.
    \eeqnn

Therefore,
\eqnn\left|\partial_{H_f}
    (\ch_1\,\piop[\h]\,\ch_1)\right|&\leq&
    |\partial_{H_f}(\ch_1^2\bch_1^2)|\,\|P_1 T' P_1\|\,
    \|P_1\,\bar{R}_0 \,P_1\|\\
    &&\;+\;\|\ch_1^2\bch_1^2\|\,\|P_1 (\partial_{H_f}T' )
    P_1\|\,
    \|P_1\,\bar{R}_0 \,P_1\|\\
    &&\;+\;\|\ch_1^2\bch_1^2\|\,\|P_1  T'
    P_1\|\,
    |P_1\,(\partial_{H_f}\bar{R}_0 )\,P_1|\\
    &\leq&\frac{(1+|p|+\lTnl+O(g^2))}{1-|p|+O(g^2)}
    |\partial_{H_f}(\ch_1^2\bch_1^2)|\,+\,O(g^2)\,\\
    &&\;+\;\frac{1}{4\,(1-|p|+O(g^2))^2}
    (1+O(g^2))\\
    &&\;+\;\frac{1}{4}
    \frac{|p|+\lTnl}
    {(1-|p|+O(g^2))^2}\,|\partial_{H_f}
    \bch_1^2|\\
    &\leq&2\,+\,3\,|\partial_{H_f}(\ch_1^2\bch_1^2)|\,
    +\,2(|p|+\lTnl)\,|\partial_{H_f}
    \bch_1^2|\eeqnn
for $|p|\leq\puppbd$ with $\puppbd\in[\puppbdnum,1)$ sufficiently
small. \qed

\section{Two technical Lemmata}

The following two   technical lemmata will be very helpful in the
subsequent analysis given in this chapter.
\\

\begin{lm}\label{firststepoverallbdslm}
Let  $Y$ denote $|p|,z$ or any component of $\op$. Then,
\eqn\left\|\;\bP1\,\partial_Y\Hps\,Q_1\;\right\|&\leq&
    O(g)\;,\nonumber\\
    \left\|\;\bar{R}\,Q_1\;\right\|&\leq&O(g)\;.\eeqn
\end{lm}

\prf These estimates are straightforward consequences of Lemmata
{~\ref{elhighenrelW1W2bounds1}} and
{~\ref{elhighenrelW1W2bounds2}}. \qed

\begin{lm}\label{firststepderopalphaDeltwMNlemma1}
Let $M+N\geq0$, and let stands for $z$ or $|k_r|,|\tilde{k}_s|$.
Then, are  constants $C_{\alpha,\beta}$ and $C_{X,\beta}$
independent of $\epsilon_0,\xi$, such that
\eqnn\nm\;\derp^\beta\partial_{\op}^\alpha\Delta
      w_{M,N}\;\nm\;\leq\;
      C_{\alpha,\beta}\,
      \epsilon_0^{2}\;
      \xi^{M+N}\;,\eeqnn
for $0\leq|\alpha|\leq2$ and $\beta=0,1$, and
\eqnn\nm\;\derp^\beta\partial_{X}\Delta
      w_{M,N}\;\nm\;\leq\;
      C_{X,\beta}\,
      \epsilon_0^{2}\;
      \xi^{M+N}\;,\eeqnn
for all $M+N\geq0$. In particular, $\Delta w_{M,N}$ are {\bf
analytic} in $\P_f$.
\end{lm}

\prf Cf. the Appendix of Chapter {~\ref{firstdecappendix}}. \qed

\section{Shift of the spectral parameter}
By a redefinition of the spectral parameter, we eliminate the
quantity
\eqnn\Delta
     E_0[z]&=&\left\langle\FchltH[\Hps]+z
     \right\rangle_{\vac}\\
     &=&-\;\left\langle \;
     W[\op]\,\bch_1\,\bar{R}[z;\op]\,\bch_1\,
     W[\op]\right\rangle_{\vac}\;.
    \eeqnn
Due to Lemma {~\ref{elhighenrelW1W2bounds2}}, \eqnn\left|\;\Delta
    E_0[z]\;\right|&\leq&
    \left\|\;\bar{R}^{\frac{1}{2}}
    [z;\op]\,\bch_1\,W[\op]\,P_1\;\right\|^2\;
    \nonumber\\
    &\leq&O(g^{2})\;.\eeqnn
We define the map
$$Z\;:\;z\;\rightarrow\;\hat{z}\,:=\,Z[z]$$
by \eqnn\left.\left.\hat{z}\;=\;(1+\DHf[z])^{-1}\,
    \right(z\,-\,\Eb\,-\,\Delta
    E_0[z]\right)\;,\eeqnn
which corresponds to a translation of the origin in $\C$ by
$\Eb+\Delta E_0[z]$. The quantity $\DHf[z]$ is defined in
(~\ref{DHfzfirstdecstepfirstdefaux3}), and discussed below in
Section {~\ref{DHfzsubsubsectfirststep3}}.
\\

\begin{prp}\label{ZC1firstdecstepprp3}
The map $Z:z\rightarrow\hat{z}$ is of class $C^1$, and
$$\partial_z Z[z]\;=\;\rho^{-1}(1+\DHf[z])^{-1}\;+\;
    O(\epsilon_0^{2 }\,\xi^2)\;.$$
Thus, for any function $h$ of class $C^1$ in $z$,
\eqn\left|\partial_{\hat{z}}h[Z^{-1}[\hat{z}]]\right|\;\leq\;
    (1\;+\;O(\epsilon))\,\left|\partial_z h[z]\right|
    \;.\label{derhatzderzfirstdecstepcomp1}\eeqn
In addition,
$$\left|\;\partial_z^\alpha\Delta E_0[z]\;\right|
    \;\leq\;O(\epsilon_0^{2}\,\xi^2)$$
for $\alpha=0,1$.
\end{prp}

\prf Cf. the proof of Proposition {~\ref{ZC1indstepprp3}} below,
but employing Lemmata {~\ref{elhighenrelW1W2bounds1}} and
{~\ref{elhighenrelW1W2bounds2}} instead of the estimates used
there. \qed

\section{Correction of the noninteracting hamiltonian}
Next, let us focus on $(T'+\Delta T')[z;\opT]$, which we separate
into
\eqnn-\DHf[z] H_f\,-\,\Delta E_0[z]\,+\,
    (-|p|+\Delta a[z])\Ppar\,+\,
    \left(\frac{|\P_f^2|}{2}\,+\,
    \Delta T_{n-l}[z;\opT]\right)\;,\eeqnn
where $T_{n-l}+\Delta T_{n-l}$ comprises all operators in $T'$ of
order $O(H_f^2)$ in the limit $H_f\rightarrow0$. We recall that
$\opT=(H_f,\Ppar,|\Pperp|)$, and that $T'+\Delta T'$ depends on
$\Pperp$ only via $|\Pperp|$, due to the $O(2)$-rotation and
reflection symmetry of the model, which is preserved under
Feshbach decimation. There is no term linear in $|\Pperp|$ because
$\mathcal{D}_1[\Hps-z]$ is analytic in $\op$ by Theorem
{~\ref{derXfPthm}}. $|\Pperp|$ is not analytic in $\Pperp$, hence
only even powers of $|\Pperp|$ (which are, of course, analytic)
appear in $\Delta T$.

Let us next discuss  $\DHf[z]$ and $\Delta a[z]$.
\\

\subsection{The coefficient of $H_f$}
\label{DHfzsubsubsectfirststep3} By Proposition
{~\ref{dHfFQQformprop333}} (i.e. setting $n=0$), the coefficient 1
of $H_f$ in $\Hps$ is modified into
\eqnn1+\DHf[z]&=&\left\langle \partial_{H_f}\FchltH[\Hps]
    \right\rangle_{\vac}\\
    &=&\dHfFQQ_0[z]\,\left\langle Q_1^\sharp
    Q_1\right\rangle_{\vac}\;,\eeqnn
where
\eqnn\dHfFQQ_0[z]&=&\left(1\,+\,
    \left\langle W\,
    \bch_1\,\bar{R} \,
    \ch_1^2\,\bar{R} \,\bch_1 \,W
    \right\rangle_{\vac}\right.\nonumber\\
    &&\hspace{0.5cm}+\;\left.\left\langle W \,
    \bch_1\,\bar{R} \,
    \frac{\partial_{H_f}\bch_1}{\bch_1}\,
    (H_f- \hat{z})
    \,\bar{R} \,\bch_1 \,W
    \right\rangle_{\vac}\right)^{-1}\;,\eeqnn
as we recall.
\\

\begin{lm}\label{derpbetaDHfzboundlemma4444}
The quantity $\DHf[z]$ satisfies $|\derp^\beta\DHf[z]|\leq
O(g^2)$, for $\beta=0,1$.
\end{lm}

\prf This is easily obtained along the lines demonstrated above,
using the fact that $\|\partial_{H_f}\bch_1\|\leq4$. \qed

We require that the fully renormalized coefficient of $H_f$ shall
again be 1. Therefore, we divide $\FchltH[\Hps]$ by $(1+\DHf[z])$.
\\

\subsection{The coefficient of $\Ppar$}
The renormalized expression for the coefficient of $\Ppar$ in
$T_{lin}[\h^{(0)}]$ is given by
\eqn\hat{a}[Z[z]]\;=\;\frac{a[z]\,+\,\Delta
a[z]}{1+\DHf[z]}\;.\eeqn This is obtained from Feshbach
decimation, \eqn\;a[z]+\Delta
a[z]&=&\left.\left.\left\langle\partial_{\Ppar}
    \right(\FchltH
    [\Hps-z]\right)
    \right\rangle_{\vac}\nonumber\\
    &=&\left\langle Q_1^\sharp\left(\partial_{\Ppar}
    \Hps\right)Q_1\right\rangle_{\vac}\;
    \nonumber\\
    &=&-|p|\left\langle Q_1^\sharp Q_1\right\rangle_{\vac}
    \;+\;
    \left\langle Q_1^\sharp\left(\Ppar+g\Af^\parallel(\ssig)
    \right) Q_1\right\rangle_{\vac} \;,
    \label{azplDeltazeq3}\eeqn
and subsequent application of rescaling, yielding
\eqn\hat{a}[Z[z]]\;=\;-|p|\,\frac{\left\langle Q_1^\sharp
    Q_1\right\rangle_{\vac}}{1+\DHf[z]}\,+\,
    \frac{\left\langle Q_1^\sharp\left(\Ppar+g\Af^\parallel(\ssig)
    \right) Q_1\right\rangle_{\vac}}{1+\DHf[z]}\;.
    \label{renazfirstdecaux3}\eeqn

$\;$

\begin{prp}\label{firststepazcorrprp3}
The renormalized coefficient of $\Ppar$ in $H[\h^{(0)}]$ is
$O(|p|)$ in the limit $|p|\rightarrow0$, satisfying
$$|\derp(\hat{a}[Z[z]]+|p|)|\;\leq\;O(g^{2})\;,$$
uniformly in $|p|$ for $|p|\leq\puppbd$.
\end{prp}

\prf The projection of the vector
$$\left\langle Q_1^\sharp\left(\P_f+g\Af(\ssig)
    \right) Q_1\right\rangle_{\vac}\;$$
in the direction of $p$ defines the second term in
(~\ref{renazfirstdecaux3}).

In the limit $|p|\rightarrow0$, $Q^{(\sharp)}_1$ is
$O(3)$-symmetric with respect to spatial rotations. Hence
$$\lim_{|p|\rightarrow0}
    \left\langle Q_1^\sharp\left(\P_f+g\Af(\ssig)
    \right) Q_1\right\rangle_{\vac}\;$$
transforms like a vector, but is invariant under all linear
$O(3)$-actions on $\R^3$. Therefore, it must be zero. This implies
that $\hat{a}[Z[z]]\rightarrow0$  in the limit $|p|\rightarrow0$,
as claimed.

Straightforward calculation shows that
\eqn\derp(\hat{a}[Z[z]]\,+\,|p|)&=&-\;
    \frac{\derp\DHf[z]}{1+\DHf[z]}\,(\hat{a}[Z[z]]\,+\,|p|)\;
    \nonumber\\
    &&-\;
    \frac{\left\langle Q_1^\sharp\,(\derp \Hps)\,\bar{R}\,
    (\partial_{\Ppar}\Hps)\, Q_1\right\rangle_{\vac}\,+\,h.c.}
    {1+\DHf[z]}\,.\label{derpdeltazpaux3}\eeqn
Let us first control the first term on the right hand side of the
equality sign, and bound $|(\hat{a}[Z[z]]+|p|)|$. Because of
$\Ppar\vac=0$, \eqnn\left|\left\langle Q_1^\sharp\;\Ppar\;
Q_1\right\rangle
    \right|&=&\left|\left\langle W[\op]\,\bch_1\,\bar{R}[z;\op]
    \,\bch_1\,
    \Ppar\,\bch_1\,\bar{R}[z;\op]\,\bch_1\,W[\op]\right\rangle
    \right|\nonumber\\
    &\leq&O(g^{2})\;,\eeqnn
using
$$\left\|\Ppar\,\bch_1\, \bar{R}[z;\op]\right\|\;\leq\; O(1)$$
for $|p|\leq\puppbd$.

Moreover, \eqnn g\left|\left\langle
Q_1^\sharp\;\Af^\parallel(\ssig)\;
    Q_1\right\rangle
    \right|&\leq&g\left|\left\langle
    W[\op]\,\bch_1\,\bar{R}[z;\op]\,
    \bch_1\,
    \Af^\parallel(\ssig) \right\rangle\;+\;h.c.\right|
    \nonumber\\
    &&+\;g\left|\left\langle
    W[\op]\,\bch_1\,\bar{R}[z;\op]\,\bch_1\,
    \Af^\parallel(\ssig)\,\bch_1\, \bar{R}[z;\op]\,\bch_1\,
    W[\op]\right\rangle
    \right|\nonumber\\
    &\leq&O(g^{2})\;,\eeqnn
using the estimates \eqnn
    \left\|\;\bar{R}^{\frac{1}{2}}[z;\op]\,\bch_1\,
    \Af^\parallel(\ssig)\,\bch_1\,
    \bar{R}^{\frac{1}{2}}[z;\op]\;\right\|
    &\leq&O(1)\;,\nonumber\\
    \left\|\;\bar{R}^{\frac{1}{2}}[z;\op]\,\bch_1\,
    \Af^\parallel(\ssig)\;\ch_1\right\|&\leq&
    O(1)\;,
\eeqnn which are proved in a manner fully analogous to Lemmata
{~\ref{elhighenrelW1W2bounds1}} and
{~\ref{elhighenrelW1W2bounds2}}.

Hence, we arrive at
\eqnn\left|\,\hat{a}[Z[z]]\,+\,|p|\,\left|\;\leq\;
    O(g^{2})\right.\right.\;,\eeqnn
uniformly in $|p|\leq\puppbd$, for $\puppbd\in[\puppbdnum,1)$
sufficiently small.

To control the first term on the right hand side of the equality
sign in (~\ref{derpdeltazpaux3}), we use Lemma
{~\ref{derpbetaDHfzboundlemma4444}}. Collecting all estimates,
\eqn\left|\derp(\hat{a}[Z[z]]\,+\,|p|)\right|&\leq&
    O(g^2)\;,
    \label{Deltazfirstdecres}\eeqn
uniformly in $|p|$ for $|p|\leq\puppbd$.  Taylor's theorem thus
implies
$$|\hat{\delta}_a[Z[z]]|\;\leq\;
    O(g^{2}) \;.$$
This proves the proposition. \qed

\section{Correction of $T_{n-l}[\opT]$}
Let us next consider the renormalized expression for the operator
$T_{n-l}$.
\\

\begin{prp}
The renormalized operator $T_{n-l}[\h^{(0)}]$ is $O(H_f^2)$ in the
limit $H_f\rightarrow0$, and there exists a constant
$\bTnl<\infty$ independent of $g$, such that
\eqn\left|
    \derp^\beta\partial_{\op}^\alpha\left(
    T_{n-l}[\h^{(0)}]\,-\,\hlTnl\,|\P_f|^2\right)
    \right|&\leq&O(g^2)\,+\,
    \bTnl\bar{P}_1
    \;,\label{Tnlfirststepthm33}
    \eeqn
for $0\leq|\alpha|\leq2$, $\beta=0,1$, where
\eqnn\hlTnl&=&\frac{1}{2}\;,\\
    |\derp\hlTnl|\,,\,|\partial_z\hlTnl|&=&
    0\;.\eeqnn
In particular,
\eqn\left|
    \left(T_{n-l}[\h^{(0)}]\,-\,\hlTnl|\P_f|^2\right)
    \right|&\leq&O(g^2)\,+\,\left(\frac{|p|+\hlTnl}{2}\,
    +\,O(\sqez)\right)\bar{P}_1\label{Tnlprpfirstdecformest4444}
    \eeqn
and
\eqn\left|
    \partial_{H_f}
    T_{n-l}[\h^{(0)}]
    \right|&\leq&O(\sqez)\;+\;(|p|\,+\,\hlTnl)\,\times\nonumber\\
    &&\;\times\,\left(2\,+\,
    3\,|\partial_{H_f}(\ch_1^2\bch_1^2)|\,
    +\,2(|p|+\hlTnl)\,|\partial_{H_f}
    \bch_1^2|\right)\;.  \eeqn
are satisfied.
\end{prp}

\prf

>From smooth Feshbach decimation, $\dec :T_{n-l}\rightarrow
T_{n-l}+\Delta T_{n-l}$, where $\Delta T_{n-l}$ is given as
follows. By Lemma {~\ref{piopdeflemma}},
\eqnn \decl\;:\;T_{n-l}&\longrightarrow&P_1 T_{n-l}
    P_1
     \,+\,T'\,(\piop-\1)
     \,\\
    &&+\,(\Delta
     w_{0,0}-\Delta
     w_{0,0}|_{\op\rightarrow0}-\DHf[z]H_f-\Delta a[z]\Ppar)\\
     &=&P_1 T_{n-l} P_1
     \,-\,\bch_1^2 (T')^2\bar{R}_0\,\\
    &&+\,(\Delta
     w_{0,0}-\Delta
     w_{0,0}|_{\op\rightarrow0}-\DHf[z]H_f-\Delta a[z]
    \Ppar)\;\\
    &\equiv&P_1 \left(T_{n-l}
    \,+\,\Delta T_{n-l}\right)P_\rho\eeqnn
on $P_1\Hp$, since
$$\piop\;=\;\1\,-\,\bch_1 T'\bch_1\bar{R}_0\;,$$
as we recall. The contribution to $\Delta T_{n-l}$ from $\Delta
w_{0,0}$ can be estimated by
$$\frac{1}{2!}\;\sum_{|\alpha|=2}
    \|\,\partial_{\opT^r}^\alpha\,\Delta w_{0,0}\|
    H_f^2\;\leq\;O(\epsilon^2)\,H_f^2
    \;,$$
which is immediate from Lemma
{~\ref{firststepderopalphaDeltwMNlemma1}} and Taylor's theorem.

It thus remains to bound the operator
\eqn T'(\piop-\1)\;=\;-\,\bch_1^2(T' )^2\bar{R}_0 \;,
    \label{Tnlfirstdecoverlcorr4444}\eeqn
together with its derivatives with respect to $\op$.

It is clear that
\eqnn&&\left|P_1\,\left(T_{n-l} \,-\,\lTnl|\P_f|^2\,+\,
    \Delta T_{n-l}
    \right)\,P_1\right|\\
    &&\hspace{2cm}\leq\;O(\epsilon)\;+\;
    \|P_1 T'[\h]P_1 \|\,|P_1(\piop[\h]-\1)P_1|
    \nonumber\\
    &&\hspace{2cm}\leq\;O(\epsilon)\,+\,
    \frac{|p|+\hlTnl+O(\sqez)}{2}\bar{P}_1\;.
    \eeqnn
This is obtained by using Lemma
{~\ref{piopderboundsfirstdeclemma4444}}, and the fact that
\eqnn\|P_1 T' P_1 \|&\leq&O(\sqez)\,+\,
    |p|\|P_1 \Ppar P_1 \|\,+\,\lTnl
    \|P_\rho |\P_f|^2 P_\rho \|\\
    &\leq&O(\sqez)\,+\,
    |p|\, +\,\hlTnl \;,\eeqnn
where by definition, $\hlTnl=\lTnl=\frac{1}{2}$. Hence, rescaling
by multiplication with $(1+\DHf[z])^{-1}$ yields
(~\ref{Tnlprpfirstdecformest4444}).
\\

Let us next consider the derivative with respect to $H_f$. We have
\eqnn&&\left|\partial_{H_f}\left(\ch_1\left(
    T_{n-l}\,-\,\lTnl|\P_f|^2\,+\,
    \Delta T_{n-l}
    \right)\ch_1\right)\right|\\
    &&\hspace{1cm}\leq\;O(\sqez)\,+\,
    \|P_1 (\partial_{H_f}T') P_1\|\,\|\piop\|\\
    &&\hspace{1.5cm}+\,
    \|P_1 T' P_1\|\,
    |\partial_{H_f}(\ch_1\piop \ch_1)|\\
    &&\hspace{1cm}\leq\;O(\sqez)\,+\,
    (|p|+\hlTnl)\,
    |\partial_{H_f}(\ch_1\piop \ch_1)|\;,\eeqnn
since $\|P_1 (\partial_{H_f}T' ) P_1\|\leq O(\epsilon^2)$.
Applying Lemma {~\ref{piopderboundsfirstdeclemma4444}}, and
rescaling, one obtains the asserted bound.
\\

For general $|\alpha|+\beta\geq1$, the part in
$\derp^\beta\partial_{\op}^\alpha\Delta T_{n-l}$ stemming $\Delta
w_{0,0}$ has an operator norm bounded by $O(\epsilon^2)$, again by
Lemma {~\ref{firststepderopalphaDeltwMNlemma1}} and Taylor's
theorem. Since
\eqnn\|\bar{R}_0^{\frac{1}{2}}
    (\derp^\beta\partial_{\op}^\alpha T)
    \bar{R}_0^{\frac{1}{2}}\|&\leq&O(1)\;,\\
    \|\bar{R}_0^{\frac{1}{2}}
    (\derp^\beta\partial_{\op}^\alpha W)
    \bar{R}_0^{\frac{1}{2}}\|&\leq&
    O(\epsilon_0^\beta\epsilon^{1-\beta}\xi)\;,\\
    \|\derp^\beta\partial_{\op}^\alpha\ch_1^2\|&\leq&O(1)\;,\eeqnn
where the implicit constants only depend on $|p|$, it follows
straightforwardly that there exists a bound of the asserted form
(~\ref{Tnlfirststepthm33}). \qed

\section{The interaction kernels of degree 1}
The modification of the interaction kernels $w_{0,1}$ and
$w_{1,0}$ due to Feshbach decimation is in part controlled by the
operator-theoretic Ward-Takahashi identities derived in Section
{~\ref{WTidentFeshbrensubsec}}. The result is that $\Delta
w_{M,N}$, for $M+N=1$, are corrected in the same manner as the
coefficient of $\Ppar$.
\\

\subsection{The marginal interaction and the
Ward-Takahashi identities}

Let \eqnn\;u_{M,N}[z;k,\lambda]&:=&
    \lim_{\stackrel{|k|\rightarrow0}{n_k\;fixed}}
    \;w_{M,N}[\op\rightarrow0;z;k,\lambda]\nonumber\\
    &=&g\;|p|\;\epsilon_{M,N}^\parallel[n_k,\lambda]\eeqnn
for $M+N=1$, where we recall that $\epsilon_{M,N}[n_k,\lambda]$
denote  the photon polarization vectors. For its image under the
renormalizion map, we have the following proposition.
\\

\begin{prp} Let $M+N=1$.
The renormalized expression for $u_{M,N}$ is given by
\eqn\;\hat{u}_{M,N}[Z[z];n_k,\lambda]
    &=&-\;g\;\hat{a}[Z[z]]
    \;\epsilon_{M,N}^\parallel[n_k,\lambda]\;,
    \label{hatuMNeqauxprp3}\eeqn
where $\hat{a}[Z[z]]$ is the coefficient of the operator
$\ch_1\Ppar\ch_1$ in $T[\h^{(0)}]$. Furthermore, for
$0\leq|\alpha|\leq2$ and $\beta=0,1$, the renormalized interaction
kernels satisfy
\eqn\nm\,\derp^\beta\partial_{\op}^\alpha
    w_{M,N}[\h^{(0)}]\,\nm\;\leq\;\epsilon_0\,\xi\;.\eeqn
In particular,
\eqnn \nm\,\bPlperp[H_f]\,
    \,\partial_{X}w_{M,N}[\h^{(0)}]\,\bPlperp[H_f+|k|]\,\nm
    &\leq&O(g^2)\\
    \nm\,\bPlperp[H_f]\,
    \,\partial_{\Ppar}
    w_{M,N}[\h^{(0)}]\,\bPlperp[H_f+|k|]\,\nm
    &\leq&2\,g\,\lTnl\,+\,O(g^2)\;
    \;.\eeqnn
for $X$ denoting either the absolute value of the photon momentum
$|k|$, $z$, $H_f$ or $\Pperp$.
\end{prp}
\prf

(I) Let us first prove (~\ref{hatuMNeqauxprp3}). According to the
discussion of the iterative Ward-Takahashi identities in Section
{~\ref{WTidentFeshbrensubsec}}, we focus on the relationship
\eqnn\lim_{\stackrel{|k|\rightarrow0}{n_k\;fixed}}
    \bcs^{-1}[|k|]\sqrt{|k|}\;\;\left[
    \Hps,a_\lambda^*(k)\right]&=&-\;g\;\epsilon_{1,0}[n_k;z]
    \cdot\partial_{\P_f}\Hps\;,
    \nonumber\\
    \lim_{\stackrel{|k|\rightarrow0}{n_k\;fixed}}
    \bcs^{-1}[|k|]\sqrt{|k|}\;\;\left[a_\lambda(k),
    \Hps\right]&=&-\;g\;\epsilon_{0,1}[n_k;z]
    \cdot\partial_{\P_f}\Hps\;,\eeqnn
which can be straightforwardly checked by inserting the definition
of $\Hps$.

Next, we consider the correction of $u_{M,N}$ under Feshbach
decimation $\mathcal{D}_1$, that is, \eqnn
    \Delta u_{M,N}[z;n_k,\lambda]&=&
    \lim_{\stackrel{|k|\rightarrow0}{n_k\;fixed}}
    \bcs^{-1}[|k|]\sqrt{|k|}\;\Delta w_{M,N}
    [z;\op\rightarrow0;k,\lambda]\;,
\eeqnn where $\Delta w_{M,N}$ is obtained from the Wick monomial
of degree $M+N=1$ in $\mathcal{D}_1[\Hps-z]$ (its interaction
kernel is given by $w_{M,N}+\Delta w_{M,N}$). In particular,
\eqnn\Delta u_{1,0}[z;n_k,\lambda]
    &=&\lim_{\stackrel{|k|\rightarrow0}{n_k\;fixed}}
    \,\bcs^{-1}[|k|] |k|^{\frac{1}{2}}\;
    \left[\Delta W_{1,0}
    [z;\op\rightarrow0]\,,\,a_\lambda(k)\right]\;,\eeqnn
and likewise for $\Delta u_{0,1}[z;n_k,\lambda]$.

In order to exploit the O(2)-symmetry of the system, we write this
in the form
\eqnn\lim_{\stackrel{|k|\rightarrow0}{n_k\;fixed}}\,
    \bcs^{-1}[|k|]\sqrt{|k|}\;\;\left[a_\lambda(k),
    \Hps\right]&=&-\;
    g\;\epsilon^\parallel_{0,1}[n_k,\lambda]
    \partial_{\Ppar}\Hps\;\nonumber\\
    &&-\;
    g\;\epsilon^\perp_{0,1}[n_k;\lambda]
    \cdot\partial_{\Pperp}\Hps\;,\eeqnn
where the right hand side is split into contributions from
directions parallel to $p$, and contributions perpendicular to it.
Likewise, the same applies to the commutator with
$a_\lambda^*(k)$.

The Ward-Takahashi identities are a consequence of Theorem
{~\ref{WTauxcor}}, which implies that these relations also hold
for $H[\h^{(0)}]= \mathcal{D}_1[\Hps-z]$. That is,
\eqnn&&\lim_{\stackrel{|k|\rightarrow0}{n_k\;fixed}}\,
    \bcs^{-1}[|k|]\sqrt{|k|}\;\left[a_\lambda(k),
    \mathcal{D}_1[\Hps-z]\right]\\
    &&\hspace{1cm}=\;-\;
    g\;\epsilon_{0,1}^\parallel[n_k,\lambda]
    \partial_{\Ppar}\mathcal{D}_1[\Hps-z]
    \nonumber\\
    &&\hspace{1.5cm}-\;g\;\epsilon_{0,1}^\perp[n_k,\lambda]\cdot
    \partial_{\Pperp}\mathcal{D}_1[\Hps-z]\;.\eeqnn
Next, we take the vacuum expectation value with respect to the
Fock vacuum,
\eqnn\lim_{\stackrel{|k|\rightarrow0}{n_k\;fixed}}
    \left(w_{0,1}+\Delta w_{0,1}\right)
    [\op\rightarrow0;z;k,\lambda]&=&-\;
    g\;\epsilon_{0,1}^\parallel[n_k,\lambda]
    \left\langle\partial_{\Ppar}\mathcal{D}_1[\Hps-z]
    \right\rangle_{\vac}\nonumber\\
    &&-\;g\;\epsilon_{0,1}^\perp[n_k,\lambda]\cdot
    \left\langle\partial_{\Pperp}\mathcal{D}_1[\Hps-z]
    \right\rangle_{\vac}\;.\eeqnn
The expectation value in the first term on the right hand side of
the equality sign equals the corrected coefficient  of $\Ppar$ in
$T[z;\op]$ under Feshbach decimation,
\eqnn\left\langle
    \partial_{\Ppar}\mathcal{D}_1[\Hps-z]
    \right\rangle_{\vac}\;=\;-\,|p|\;+\;\Delta a[z]
    \;,\eeqnn
as one observes from comparison with (~\ref{azplDeltazeq3}). On
the other hand, the expectation value in the second term,
\eqn\left\langle\partial_{\Pperp}\mathcal{D}_1[\Hps-z]
    \right\rangle_{\vac}\;,\eeqn
is a vector in the 1-plane perpendicular to $p$. Since both
$\mathcal{D}_1[\Hps-z]$ and $\vac$ are invariant under
O(2)-actions on $\R^3$ which keep $p$ fixed, it is necessarily
O(2)-invariant, and must thus be zero.

Finally, the rescaling transformation acts on $u_{M,N}+\Delta
u_{M,N}$ by division by $(1+\DHf[z])$, which gives
\eqnn\hat{u}_{M,N}[Z[z];n_k,\lambda]&=&g\;\hat{a}[Z[z]]\;
    \epsilon^\parallel_{M,N}[n_k,\lambda]
    \;.\eeqnn
This proves the assertion about $\hat{u}_{M,N}$.
\\

(II) Next, let us verify the inequality
$$\nm\,\hat{w}_{M,N}\,\nm\;\leq\;\epsilon_0\,\xi\;.$$
The left hand side is clearly bounded by
\eqnn\nm\,\piop\,w_{M,N}\,\piop\,\nm\,+\,\nm\,\Delta w_{M,N}\,\nm
    &\leq&\|\piop\|^2\,g\,(|p|+1)\;+\;
    \nm\,\Delta w_{M,N}\,\nm\\
    &\leq&\left(\frac{4}{3}\right)^2\,g\,(|p|+1)\,+\,
    O(g^{2})\nonumber\\
    &\leq&2\,g\nonumber\\
    &=&\epsilon_0\,\xi\;,\eeqnn
for $|p|\leq\puppbd$ for $\puppbd\in[\puppbdnum,1)$ sufficient
small, since Lemma {~\ref{firststepderopalphaDeltwMNlemma1}}
implies
$$\nm\Delta w_{M,N}\nm\;\leq\;O(g^{2})\;,$$
and $\|\piop\|\leq\frac{4}{3}$ by Lemma
{~\ref{piopderboundsfirstdeclemma4444}}.
\\

(III) Next,  considering that
$\bar{P}_\rho^\perp\piop=\bar{P}_\rho^\perp$, we have
\eqnn&&\nm\,\bar{P}_1^\perp[H_f]\left(
    \partial_{Y}
    \hat{w}_{M,N} \right)\bar{P}_1^\perp[H_f+|k|]\,\nm\\
    &&\hspace{1cm}\leq\;
    (1+\DHf[z])^{-1}\nm\bar{P}_1^\perp[H_f]
    \left(\partial_{Y} \left(
    \piop[z;\op]w_{M,N}\piop[z;\sh_k\op]\right)\right)
    \bar{P}_1^\perp[H_f+|k|]\nm
    \,\nonumber\\
    &&\hspace{1.5cm}+\,(1+\DHf[z])^{-1}\nm\partial_{Y}
    \Delta w_{M,N}\nm\\
    &&\hspace{1cm}\leq\;
    (1+\DHf[z])^{-1}\nm\bar{P}_1^\perp[H_f]
    \left(\partial_{Y} w_{M,N}\right)
    \bar{P}_1^\perp[H_f+|k|]\nm
    \,\nonumber\\
    &&\hspace{1.5cm}+\,(1+\DHf[z])^{-1}\nm\partial_{Y}
    \Delta w_{M,N}\nm\\
    &&\hspace{1cm}\leq\; \nm\bar{P}_1^\perp[H_f]
    \left(\partial_{Y} w_{M,N}\right)
    \bar{P}_1^\perp[H_f+|k|]\nm\,+\,
    O(g^2)\;.
    \eeqnn
The asserted bounds follow immediately, since $\nm\partial_{\Ppar}
w_{M,N}\nm=g=2\,\lTnl\,g$ for $Y=\Ppar$, and $\nm\partial_{Y}
w_{M,N}\nm\leq O(g^2)$ for $Y$ denoting any of the quantities
accounted for by the variable $X$. \qed

\section{The interaction kernels of degree 2 and higher}

\begin{prp}
\label{firststepderopalphatildwMNprp1} Let $X$ stand for $z$ or
$|k_r|,|\tilde{k}_s|$. Then, for $0\leq\beta\leq1$ and
$0\leq|\alpha|\leq2$, the renormalized integral kernels
$\{\hat{w}_{M,N}\}$ in $\hat{\h^{(0)}}$ satisfy
\eqnn\nm\;\partial_{|p|}^\beta
    \partial^\alpha_{\op}\hat{w}_{M,N}\;\nm
    &\leq&
    \epsilon_0^{\frac{7}{4}}\,\xi^{M+N}\;,
    \eeqnn
and
\eqnn\nm\;\partial_{|p|}^\beta
    \partial_X\hat{w}_{M,N}\;\nm
    &\leq&
    \epsilon_0^{\frac{7}{4}}\,\xi^{M+N}\;,
    \eeqnn
uniformly in $|p|$, for $|p|\leq\puppbd$, and $M+N\geq1$. In
particular, $\hat{w}_{M,N}$ is {\bf analytic} in $\P_f$.
\end{prp}

\prf This follows from \eqnn\nm\;\partial_{|p|}^\beta\,
    \partial^\alpha_{\op}\,\hat{w}_{M,N}\;\nm
    &\leq&
    \rho^{M+N-1+|\alpha|}\,\sum_{\alpha'+\alpha''=\alpha}
    \left(\begin{array}{c}|\alpha|\\|\alpha'|,
    |\alpha'' |\end{array}\right)\,\times
    \nonumber\\
    &&\hspace{1cm}\times\,\left(
    2^\beta\|\derp^\beta
    \partial_\op^{\alpha'}\piop\| \,
    \|\partial_\op^{\alpha''}\piop\|\,
    \nm
    \partial_\op^{\alpha-\alpha'-\alpha''}\,
    w_{M,N}\nm\;\right.\nonumber\\
    &&\hspace{2cm}+\,\left.
    \|\partial_\op^{\alpha'}\piop\| \,
    \|\partial_\op^{\alpha''}\piop\|\,
    \nm\derp^\beta
    \partial_\op^{\alpha-\alpha'-\alpha''}\,
    w_{M,N}\nm\;\right)\nonumber\\
    &&+\;C_{\alpha,\beta}\,
    g^{M+N+2}\;,\eeqnn
and \eqnn\nm\;\partial_{|p|}^\beta\,
    \partial_X\,\hat{w}_{M,N}\;\nm
    &\leq&\rho^{M+N}\,\nm\partial_{|p|}^\beta\,
    \partial_X\,w_{M,N}\nm\;\nonumber\\
    &&+\;C_{X,\beta}\,
    g^{M+N+2}\;,\eeqnn
which is a straightforward consequence of Lemma
{~\ref{firststepderopalphaDeltwMNlemma1}}. The constants
$C_{\alpha,\beta}$, $C_{X,\beta}$ do not depend on $\epsilon,\xi$.
We remark that it is trivially, $\nm w_{M,N}\nm=0$ if $M+N\geq3$.
\qed

\chapter{ANALYSIS OF THE INDUCTION STEP}
\label{analiterstepchap}

\section{General framework}

The purpose of this chapter is to study the renormalization map
\eqnn\ren\;:\;\Polyd_{2\epsilon_0,\xi}&\longrightarrow&
    \Polyd_{2\epsilon_0,\xi}\\
    \h&\longmapsto&\hat{\h}\;,\eeqnn
where we recall from the previous chapter that
\eqnn\epsilon_0\;\xi&=&2\,g\\
    \epsilon_0&=&\xi^2\;,\eeqnn
such that $\xi \ll 1$ for $g\ll1$, which we assume to be
sufficiently small. The detailed constructions of $\ren$ and of
the polydisc $\Polyd_{\epsilon,\xi}$ are given in Chapter
{~\ref{rgdefsection}}.

Let us pick a number
$$\epsilon\;\leq\;2\epsilon_0\;,$$
together with an arbitrary element
\eqn\h\;\in\;\Polyd_{\epsilon,\xi}\;\subseteq\;
    \Polyd_{2\epsilon_0,\xi}\;.\eeqn
The main purpose is to demonstrate that
\eqn\hat{\h}\;\equiv\;\ren[\h]\;\in\;\Polyd_{\hat{\epsilon},\xi}
    \label{hathPolyddef3}\eeqn
for a renormalized value $\hat{\epsilon}\leq\epsilon$.

The value of $\xi$ is kept fixed ($\xi=(2g)^{\frac{1}{3}}$), while
the number $\epsilon$ is a measure for the renormalization group
flow of the electron charge $g$. Notably it does a priori {\bf not
depend on $|p|$}, a fact that plays a r\^ole in connection to the
derivation of uniform bounds in $|p|$ at various places in the
sequel.
\\

\section{General induction hypotheses for $H[\h]$}

Let us present the list of induction hypotheses imposed on
$H[\h]$. To begin with, the general structure of $H[\h]$ is
assumed to be of the form
$$H[\h]\;=\;T[\h]\,+\,
    \ch_1\,W[\h]\,\ch_1\;,$$
with
$$T[\h]\;=\;H_f\,-\,z\,+\,\ch_1\,(a[z]\Ppar\,+\,T_{n-l}[\h])\,
    \ch_1\;,$$
where $T_{n-l}[\h]$ comprises all terms of order $O(H_f^2)$ in the
limit $H_f\rightarrow0$ that commute with $H_f$, and where $a[z]$
is a real scalar for $z\in\R$. Furthermore,
$$W[\h]\;=\;\sum_{M+N\geq1}W_{M,N}[\h]\;,$$
where the Wick monomials $W_{M,N}[\h]$ have the form given in
(~\ref{WMNdefintmeassigl12}) and (~\ref{WMNdefintmeassigg12}).
\\

Next, it is assumed that
$$H[\h]\;\;{\rm is\;\; analytic\;\; in}\;\;\Ppar\;,\;\Pperp\;.$$
This condition is immediately seen to be preserved by the
induction step, due to consequence of Theorem {~\ref{derXfPthm}}.
It is needed to determine the functional structure of
$T_{lin}[\h]$, the linear part of the noninteracting hamiltonian.
\\

The absolute value of the conserved momentum $p$ is assumed to be
restricted to the values
\eqn\,|p|\;\leq\;\puppbd\;,
    \label{pabsboundindhyp2}\eeqn
with some constant
$$\puppbd\;\in\;[\puppbdnum,1)\;,$$
which must be picked sufficiently small for the results in this
chapter to hold. In the subsequent discussion, bounds that are
{\bf uniform in $|p|$} will always be implicitly understood only
for the range of values specified in (~\ref{pabsboundindhyp2}).

The induction assumption on the symmetry properties of $H[\h]$ are
that for $|p|>0$, $H[\h]$ is $O(2)$-invariant with respect to
rotations and reflections in $\R^3$ that leave $p$ fixed. In the
special case $|p|=0$, \eqn\lim_{|p|\rightarrow0}H[\h]\eeqn is
fully $O(3)$-rotation and reflection invariant.
\\

The smooth cutoff operators are assumed to obey the bounds
\eqn\|\partial_{H_f}(\chr^2\bchr^2)\|\;,\;
    \|\partial_{H_f}\bchr^2\|&\leq&8\;,\;\nonumber\\
    \|\partial_{H_f}(\ch_1^2\bch_1^2)\|\;,\;
    \|\partial_{H_f}\bch_1^2\|&\leq&4 \;,
    \label{derch1derchrboundsindstep4444}\eeqn
where the spectral support of $\ch_1$ is assumed to be given by
$[\frac{2}{3},1]$.
\\

\section{Induction hypotheses for the spectral parameter}

For the spectral parameter, we assume that
$$z\;\in\; [-\epsilon,0]\;.$$
This ensures well-definedness of the Feshbach map. Furthermore,
the interval is chosen appropriately such that its interior
contains the infimum of the spectrum of $H[\h]$. This is essential
for the application of the reconstruction part of the Feshbach
theorem, which will be invoked to determine the ground state
eigenvalue of $\Hps$, and the corresponding eigenvector.
\\

\section{Induction hypotheses for $T[\h]$}
\label{IndhypThsect33}

For $|p|>0$, the noninteracting hamiltonian $T[\h]$ is invariant
under linear $O(2)$-actions on $\R^3$ which leave $p$ invariant.
Therefore, it can only depend on $\Pperp$ only via $|\Pperp|$, it
is thus a function of $\opT=(H_f,\Ppar,|\Pperp|)$. In addition, it
is assumed to be {\bf analytic} in $\P_f$. Hence, it only depends
on even powers of $|\Pperp|$. We introduce the decomposition
\eqn\;T[\h]\;=\;H_f\,-\,z\,+\,\ch_1\left(a[z]\,\Ppar\,+
    \,T_{n-l}[\h]\right)\ch_1\;,\eeqn
where we assume that
\eqn\left|
    \derp^\beta\partial_{\op}^\alpha\left(
    T_{n-l}[\h]\,-\,\lTnl\,|\P_f|^2\right)
    \right|&\leq&\epsilon\,+\,
    \bTnl\,\bar{P}_1
    \;,\label{Tnllowhighboundindhyp}
    \eeqn
on $P_1\Hp$, for $0\leq|\alpha|\leq2$, $\beta=0,1$, where
\eqnn0&\leq&\lTnl\;\leq\;\frac{1}{2} \;,\\
    |\derp\lTnl|\,,\,|\partial_z\lTnl|&=&
    0\;,\eeqnn
and where the constant $\bTnl$ is independent of $\epsilon$. In
particular,
\eqn\left|
    \left(T_{n-l}[\h]\,-\,\lTnl|\P_f|^2\right)
    \right|&\leq&\epsilon\,+\,\left(\frac{|p|+\lTnl}{2}\,
    +\,O(\sqez)\right)\bar{P}_1 \label{Tnlindhypboudn4444}\eeqn
and
\eqn\left|
    \partial_{H_f}
    T_{n-l}[\h]
    \right|&\leq&O(\sqez)\;+\;(|p|\,+\,\lTnl)\,\times\nonumber\\
    &&\;\times\,\left(2\,+\,
    3\,|\partial_{H_f}(\ch_1^2\bch_1^2)|\,
    +\,2(|p|+\lTnl)\,|\partial_{H_f}
    \bch_1^2|\right)\;. \label{partHfTnlindhypboudn4444}\eeqn
A direct consequence is that
\eqn |( T_{n-l}[\h]-\lTnl|\P_f|^2)|
    &\leq& \frac{3}{2}\,\left(\frac{|p|+\lTnl}{2}\,
    +\,O(\sqez)\right)\,H_f\;.\label{TnlHfbound4444}\eeqn
To show this, we observe that
\eqn P_1\bar{P_1}&\leq&\frac{3}{2}\,H_f\;,\eeqn
since the spectral support of $P_1\bar{P_1}$ (as a function of
$H_f$) is the interval $[\frac{2}{3},1]$, so
\eqnn|\bar{P}_1 ( T_{n-l}[\h]-\lTnl|\P_f|^2)\bar{P}_1 |&\leq&
    \|\bar{P}_1 ( T_{n-l}[\h]-\lTnl|\P_f|^2)\bar{P}_1 \|\,\bar{P}_1\\
    &\leq&\frac{3}{2}\,
    \|\bar{P}_1 ( T_{n-l}-\lTnl|\P_f|^2)\bar{P}_1 \|\,H_f \;.\eeqnn
Furthermore, since  $|\Ppar|,|\Pperp|\leq H_f$,
\eqnn|\bar{P}_1^\perp  ( T_{n-l}[\h]-\lTnl|\P_f|^2)
    \bar{P}_1^\perp |&\leq&\sum_{|\alpha|=1}
    \|\bar{P}_1^\perp
    (\partial_\op^\alpha ( T_{n-l}[\h]-\lTnl|\P_f|^2))
    \bar{P}_1^\perp \|\,H_f\,
    \bar{P}_1^\perp\\
    &\leq&O(\epsilon)\,H_f\,\bar{P}_1^\perp \;.\eeqnn
This immediately implies (~\ref{TnlHfbound4444}).
\\

The induction assumption on $a[z]$ is  that
\eqn\left|\,\derp^\beta(a[z]\,+\,|p|)\,\right|&\leq&
    \sqrt{\epsilon}_0\nonumber\\
    \label{derpalphaazconscondindhyp}\eeqn
uniformly in $|p|$, for $|p|\leq\puppbd$, and $\beta=0,1$. Due to
the strong marginality of the operator $a[z]\Ppar$, our method to
bound the left hand side of (~\ref{derpalphaazconscondindhyp}), is
to 'integrate' the renormalization group flow equations up to the
present scale in each renormalization iteration. This is done in
Proposition {~\ref{hatazminazestkappaprp}}, by which one ensures
the validity of (~\ref{derpalphaazconscondindhyp}) in every
iteration step. Since we prefer to solve the flow equations in the
end, this approach via a 'nested induction argument' is
convenient.
\\

Furthermore, in the special case $|p|=0$,
\eqn\lim_{|p|\rightarrow0}T[\h]\eeqn is fully $O(3)$-rotation and
reflection invariant, and thus only depends on the operators $H_f$
and $|\P_f|^2=|\Ppar|^2+|\Pperp|^2$. Therefore,
\eqn\lim_{|p|\rightarrow 0}a[z]\;=\;0\;.\eeqn
This, together with (~\ref{derpalphaazconscondindhyp}), implies
that, by Taylor's theorem, $a[z]$ can be written in the form
$$a[z]\;=\;-|p|(1+\delta_a[z])\;,$$
where $|\delta_a[z]|\leq\sqrt{\epsilon}_0$, uniformly in $|p|$,
for $|p|\leq\puppbd$, and $\puppbd\in[\puppbdnum,1)$ sufficiently
small.

Furthermore, it is assumed that
\eqn\|\partial_z T[\h]\|\;\leq\;C\;,\eeqn
for a constant $C$ that is independent of $\epsilon$.

Under the induction hypotheses formulated above for $T[\h]$, we
obtain the upper and lower relative bounds
\eqn\frac{1}{3}\;H_f\;\leq\;
    T[\h]\;\leq\;
    3\;H_f\;,
    \label{TgammaGammabd}\eeqn
uniformly in $|p|$ for $|p|\leq\puppbd$. The upper bound is a
consequence of \eqn T[\h]&\leq&(1\,+\,|a[z]|)H_f\,+\,
    \lTnl H_f^2\,+\,\frac{3}{2}\,
    \left(\frac{|p|+\lTnl}{2}\,
    +\,O(\sqez)\right)\,H_f\;
    \nonumber\\
    &\leq&\left( 1+\frac{7|p|}{4}+\frac{7}{8}\,
    +\,O(\sqrt{\epsilon}_0)\right)
    H_f\,\nonumber\\
    &\leq&3\,H_f
    \;,\label{Gammaindhypbound4444}\eeqn
using (~\ref{TnlHfbound4444}) and $\lTnl\leq\frac{1}{2}$, for
$|p|\leq \puppbd$ and $\puppbd\in[\puppbdnum,1)$ sufficiently
small.

The lower bound follows from
\eqn T[\h]&\geq&(1\,-\,|a[z]|)H_f\,-\,\frac{3}{2}\,
    \left(\frac{|p|+\lTnl}{2}\,
    +\,O(\sqez)\right)\,H_f\;
    \nonumber\\
    &\geq&\left( 1-\frac{7|p|}{4}-\frac{3}{8}\,
    +\,O(\sqrt{\epsilon}_0)\right)
    H_f\,\nonumber\\
    &\geq&\frac{1}{3}\,H_f
    \;,\label{gammaindhypbound2}\eeqn
again using (~\ref{TnlHfbound4444}) and $\lTnl\leq\frac{1}{2}$,
for $|p|\leq \puppbd$ and $\puppbd\in[\puppbdnum,1)$ sufficiently
small. This completes the list of induction hypotheses for the
non-interacting effective hamiltonian $T[\h]$.
\\

\section{Induction hypotheses for interaction kernels
of degree 1} It is, to begin with, assumed that the interaction
kernels $w_{M,N}[z;\op;k,\lambda]$, $M+N=1$, are {\bf analytic} in
$\P_f$.

They comprise the only marginal interaction kernels in the theory,
namely
\eqn\;u_{M,N}[\h]&=&u_{M,N}[z;n_k,\lambda]\nonumber\\
    &=&
    \lim_{\stackrel{|k|\rightarrow0}{n_k\;{\rm fixed}}}
    w_{M,N}[\op\rightarrow0;z;k,\lambda]\;.\eeqn
The induction hypothesis on $u_{M,N}[\h]$ is that
\eqn\;u_{M,N}[\h]\;=\;g\;a[z]\;
    \epsilon_{M,N}^\parallel[n_k,\lambda]\;,\eeqn
where $a[z]$ is the coefficient of $\Ppar$ in $T_{lin}[\h]$, and
$\epsilon_{M,N}^\parallel[n_k,\lambda]$ is the projection of a
photon polarization vector in the direction of the conserved
momentum $p$.

Moreover, it is assumed that the lower bound
\eqn\epsilon\;\geq\;\ezfac\,\nm
u_{M,N}\nm\;=\;\ezfac\,g\,|a[z]|\;,
    \label{wMN1epsxibdindhyp31}\eeqn
holds for $\epsilon$, and that the auxiliary bound
\eqnn\nm w_{M,N}[\h]\nm&\leq&\frac{16}{9}\,g\,|p|\,(1+\sqez)
    +\,\frac{32}{9}\,g\,\lTnl\,+\,O(\epsilon^{\frac{3}{2}}\xi)\\
    &\leq&\epsilon\;,\eeqnn
holds. Furthermore, letting $X$ denote the absolute value of a
photon momentum $|k|$, the spectral parameter $z$, $H_f$, or
$\Pperp$, that
\eqn\nm\,\bPlperp[H_f]\left(\partial_{\Ppar}
    w_{M,N}\right)\bPlperp[H_f+|k|]\,\nm&\leq&
    2\,g\,\lTnl\,+\,\epsilon^{\frac{3}{2}}\,\xi\;,
    \nonumber\\
    \nm\,\bPlperp[H_f]\left(\partial_{X}
    w_{M,N}\right)\bPlperp[H_f+|k|]\,\nm
    &\leq&
    \epsilon^{\frac{3}{2}}\,\xi\;,
    \label{derpalphwMNminuMNindhypbd3}\eeqn
for $\bPlperp[|k|]=\bPlperp [\sh_k H_f]|_{H_f\rightarrow0}$. These
induction assumptions are supplemented by additional requirements
below, which can be formulated in a more general manner together
with the conditions imposed on the cases $M+N\geq2$.
\\

\subsection{Ward-Takahashi identities}
The fact that $u_{M,N}[\h]$ is determined by $a[z]$, the
coefficient of $\Ppar$ in $T[\h]$, is a consequence of the
Ward-Takahashi identities. Formulated in the form of an induction
hypothesis, they express that $H[\h]$ satisfies \eqnn
    \derWTs^\sharp[H[\h]]&=&0\;,
\eeqnn in the manner explained in Section
{~\ref{WTindformsubsec}}.
\\

\section{General induction hypotheses for $w_{M,N}[\h]$}
\label{wMNindhypsubsect3}

All $w_{M,N}$ are assumed to be {\bf analytic} in $\P_f$. Let $X$
denote photon momenta $|k_i|,|\tilde{k}_j|$, for $i=1,\dots,M$,
and $j=1,\dots,N$, or the spectral parameter $z$. Then, the
induction hypotheses are that for $0\leq|\alpha|\leq2$,
$0\leq\beta\leq1$,
\eqn\nm\,\derp^\beta\,\partial_{\op}^\alpha
    \,w_{M,N}[\h]\,\nm&\leq&
    (2\ez)^\beta\,
    \epsilon^{1-\beta}\xi\;\;\;,\nonumber\\
    \nm\,\derp^\beta\,\partial_X\,w_{M,N}[\h]\,\nm&\leq&
    (2\ez)^\beta\,\epsilon^{1
    -\beta}\,\xi\;,\;\;\; M+N=1\;,
    \label{wMN1epsxibdindhyp32}\eeqn
while for $M+N\geq2$,
\eqnn\nm\,\derp^\beta\,\partial_{\op}^\alpha\,w_{M,N}[\h]\,\nm
    &\leq&
    (2\ez)^\beta \,\epsilon^{\frac{7}{4}
    -\frac{|\alpha|}{4}-\beta}\,\xi^{M+N}\;,
    \\
    \nm\,\derp^\beta\,\partial_X\,w_{M,N}[\h]\,\nm&\leq&
    (2\ez)^\beta\,\epsilon^{\frac{3}{2}
    -\beta}\,\xi^{M+N}\;.
    \eeqnn
All bounds are assumed to be uniform in $|p|$, for
$|p|\leq\puppbd$, with $\puppbd\in[\puppbdnum,1)$ sufficiently
small.
\\

\section{The main Theorem} The following Theorem is the main
result of this chapter, and fully controls the renormalization
group iteration.
\\

\begin{thm}
\label{indstepmainthm3} Let $\rho=\frac{1}{2}$, and let $X$ denote
photon momenta $|k_i|,|\tilde{k}_j|$, or the renormalized spectral
parameter $\hat{z}=Z[z]$. There is a constant
$\puppbd\in[\puppbdnum,1)$, such that for all $p$ with
$|p|\leq\puppbd$, the following results hold. Assuming that
$\epsilon_0,\xi$ are sufficiently small, the renormalized values
of $\epsilon$ and $\dPpop$ are defined by
\eqnn\hat{\epsilon}\;=\;\left\{
    \begin{aligned}
    \max\left\lbrace\epfrac\;\epsilon\;,\;
    \ezfac\,(1+\sqrt{\epsilon}_0)\,|p|\,\ez\right\rbrace
    &\;\;\;\;{\rm if }\;\sigma\leq1\\
    \epfrac\;\epsilon\hspace{2cm}&\;\;\;\;{\rm if }\;
    \sigma>1\;\;,\;\end{aligned}\right.\eeqnn
with $\epsilon\leq2\epsilon_0$, and
$$\sigma\;\longrightarrow\;
    \hat{\sigma}\;=\;\frac{\sigma}{\rho}\;=\;2\,\sigma\;.$$
accounts for the renormalization of the running infrared cutoff.
The renormalized effective hamiltonian $H[\hat{\h}]$ is analytic
in $\op$, and $\lim_{|p|\rightarrow}H[\hat{\h}]$ is fully
$O(3)$-symmetric.

The function $\dHfFQQ[\h]$ (cf. Proposition
{~\ref{dHfFQQformprop333}} for its definition) obeys the bound
\eqnn\left|\derp^\beta\left(\dHfFQQ[\h]-1\right)\right|\;\leq\;
    O(\ez^\beta\hat{\epsilon}^{2-\beta})\;, \eeqnn
with $\beta=0,1$, and the renormalized coefficient $\hat{a}[Z[z]]$
of $\Ppar$ in $T_{lin}[\hat{\h}]$ satisfies
\eqnn\lim_{|p|\rightarrow0}\hat{a}[Z[z]]&=&0\;,\nonumber\\
    \left|\,\derp^\beta(\hat{a}[Z[z]]+|p|) \,\right|&\leq&
    \sqrt{\epsilon}_0\,
    \;.\eeqnn
For the renormalized operator $T_{n-l}[\hat{\h}]$,
\eqnn\left|
    \derp^\beta\partial_{\op}^\alpha\left(\ch_1\left(
    T_{n-l}[\hat{\h}]\,-\,\hlTnl\,|\P_f|^2\right)\ch_1\right)
    \right|&\leq&\epsilon\,+\,
    \hbTnl\,\bar{P}_1
    \;,
    \eeqnn
for $0\leq|\alpha|\leq2$, $\beta=0,1$, and
\eqnn0\;\leq\;\hlTnl&=&\rho\,\lTnl\;,\\
    |\derp\hlTnl|\,,\,|\partial_z\hlTnl|&=&
    0\;.\eeqnn
The constant $\hbTnl<\infty$ is independent of $\hat{\epsilon}$.
In particular,
\eqnn\left|
    \left(T_{n-l}[\h]\,-\,\lTnl|\P_f|^2\right)
    \right|&\leq&\epsilon\,+\,\left(\frac{|p|+\lTnl}{2}\,
    +\,O(\sqez)\right)\bar{P}_1 \eeqnn
and
\eqnn\left|
    \partial_{H_f}
    T_{n-l}[\h]
    \right|&\leq&O(\sqez)\;+\;(|p|\,+\,\lTnl)\,\times\nonumber\\
    &&\;\times\,\left(2\,+\,
    3\,|\partial_{H_f}(\ch_1^2\bch_1^2)|\,
    +\,2(|p|+\lTnl)\,|\partial_{H_f}
    \bch_1^2|\right)\;.\eeqnn
For $M+N=1$,
\eqnn u_{M,N}[\hat{\h}]&=&
    \lim_{\stackrel{|k|\rightarrow0}{n_k\;{\rm fixed}}}
    w_{M,N}[\hat{\h}]\\
    &=&g\;\hat{a}[Z[z]]\;
    \epsilon_{M,N}^\parallel[n_k,\lambda]\;,\eeqnn
and in particular, for $|\alpha|=1$,
\eqnn\nm\,\bPlperp[H_f]\left(\partial_{\op}^\alpha
    w_{M,N}[\hat{\h}]\right)\bPlperp[H_f+|k|]\,\nm&\leq&
    2\,g\,\lTnl\,+\,\hat{\epsilon}^{\frac{3}{2}}\,\xi\;,
    \nonumber\\
    \nm\,\bPlperp[H_f]\left(\partial_{Y}
    w_{M,N}[\hat{\h}]\right)\bPlperp[H_f+|k|]\,\nm
    &\leq&
    \hat{\epsilon}^{\frac{3}{2}}\,\xi\;, \eeqnn
where $Y$ denotes the absolute value of a photon momentum $|k|$,
$z$, $H_f$ or $\Pperp$. Furthermore, for $0\leq|\alpha|\leq2$ and
$0\leq\beta\leq1$,
\eqnn
    \nm\,\derp^\beta\,\partial_X\,w_{M,N}[\hat{\h}]\,\nm\;
    ,\;\nm\,\derp^\beta\,\partial_\op^\alpha
    \,w_{M,N}[\hat{\h}]\,\nm&\leq&
    (2\ez)^\beta\,
    \hat{\epsilon}^{1-\beta}\xi\;\;\;,\;\;{\rm if}\; M+N=1\;,
    \eeqnn
while for $M+N\geq2$,
\eqnn\nm\,\derp^\beta\,\partial_{\op}^\alpha\,\hat{w}_{M,N}\,\nm
    &\leq&
    (2\ez)^\beta\,\hat{\epsilon}^{\frac{7}{4}
    -\frac{|\alpha|}{4}-\beta}\,\xi^{M+N}\;,
    \\
    \nm\,\derp^\beta\,\partial_X\,\hat{w}_{M,N}\,\nm&\leq&
    (2\ez)^\beta\,\hat{\epsilon}^{\frac{3}{2}
    -\beta}\,\xi^{M+N}\;.
    \eeqnn
All bounds are uniform in $|p|$, for $|p|\leq\puppbd$.
\end{thm}

\section{Two key technical Lemmata}

The auxiliary results presented in this section will be very
helpful for the rest of the discussion.
\\

\begin{lm}\label{overallbdslm}
Let
$$K[\h]\,=\,a\,+\,\sum_{M+N\geq0}K_{M,N}[\h]$$
denote a Wick ordered operator on $P_1\Hp$ (for instance
$\partial_{|p|}^\beta\partial_{\op}^\alpha W[\h]$ or
$\partial_{|p|}^\beta\partial_{\op}^\alpha H[\h]$), where
$$a\;:=\;\langle K[\h]\rangle_{\vac}\;.$$
Furthermore, assume that the integral kernels are subject to
bounds
\eqnn\|\,k_{0,0}\,\|&\leq&b\\
    \nm\,k_{M,N}\,\nm&\leq&c\,\xi^{M+N}\;.\eeqnn
Then, \eqnn
    \|\,K[\h]\,\vac\,\|&\leq&a\,+\,4\,\sqrt{\pi}\,c\,\xi\\
    \|\,\bPr\,K[\h]\,\vac\,\|&\leq&4\,\sqrt{\pi}\,c\,\xi
    \eeqnn
and \eqnn\|\,K[\h]\,\|&\leq&a\,+\,b\,+\,4\,\sqrt{\pi}\,
    c\,\xi\;.\eeqnn
\end{lm}

\prf

We have
$$\|K_{M,N}[\h]\|\,\leq\,
    (2\sqrt{\pi})^{M+N}\nm k_{M,N}\nm\;,$$
cf. Lemma {~\ref{WMNwMNrelboundslemma3}}, thus
\eqnn\|\sum_{M+N\geq1}K_{M,N}[\h]\|&\leq&c\sum_{M+N\geq1}
    (2\sqrt{\pi}\xi)^{M+N}\\
   &\leq&4\sqrt{\pi}\,c\;\xi\;.\eeqnn
The asserted bounds follow immediately from $\|K_{0,0}\|\leq b$
and $K_{0,0}\vac=0$. \qed

The following {\bf main lemma} provides analytical control of the
corrections of the interaction kernels and their derivatives under
Feshbach decimation.
\\

\begin{lm}\label{deropalphaDeltwMNlemma1}
Let $X$ denote $z$  or $|k_r|,|\tilde{k}_s|$. We assume that
$M+N\geq0$, and set $\rho=\frac{1}{2}$. Then, there are constants
$C_{\alpha,\beta}$, $C_{X,\beta}<\infty$ independent of
$\epsilon,\xi$, such that
\eqnn\nm\;\derp^\beta\partial_{\op}^\alpha \Delta
      w_{M,N} \;\nm\;\leq\;\ez^\beta\,
      C_{\alpha,\beta}\,\epsilon^{2-\beta}\,2^{M+N}\,
      \xi^{M+N}\;,\eeqnn
for $\beta=0,1$ and $0\leq|\alpha|\leq2$, and
\eqnn\nm\;\derp^\beta\partial_X \Delta
      w_{M,N}\;\nm\;\leq\;\ez^\beta\,
      C_{X,\beta}\,\epsilon^{2-\beta}\,2^{M+N}\,
      \xi^{M+N}\;,\eeqnn
for  $|p|\leq\puppbd$. In particular, the kernels $\Delta w_{M,N}$
are {\bf analytic} in $\P_f$ for all $M+N\geq0$.
\end{lm}

\prf The proof is given in the Appendix of Chapter
{~\ref{iteratstepappendsect}}. \qed

\section{Well-definedness of the Feshbach triple}

A preliminary task in the discussion of the renormalization group
iteration step is to verify that the Feshbach triple
$(Q^\sharp_{P_\rho}[\h],Q_{P_\rho}[\h],\FchrtH[H[\h]])$ associated
to $(P_\rho, P_1\Hp)$ is well-defined. In addition, it must be
demonstrated that $\FchrtH[H[\h]]$ can be expanded into an
absolutely convergent Neumann series for the operation of Feshbach
decimation to be applicable.

Let
\eqn\tau[H[\h]]&=&(1+\DHf[z])H_f\,-\,(z-\Delta E_0[z])
    \nonumber\\
    \omega[H[\h]]&=&H[\h]\,-\,\tau[H[\h]]
    \nonumber\\
    &=&-\,\DHf[z]H_f\,-\,\Delta E_0[z]\,+\,
    \ch_1\,W[\h]\,\ch_1\;,\eeqn
where the definitions of $\DHf[z]$ and $\Delta E_0[z]$ are given
in Section {~\ref{Neumserexpsubsect3}}. Then,
\eqnn\bar{R}_0[\h]&=&\left(\tau[H[\h]]\,+\,
    \bchr\,T'[\h]\,\bchr\right)^{-1}\,
    \;,\eeqnn
denotes the 'free resolvent' on $\bar{P}_\rho\Hp$, with
\eqnn T'[\h]&=&T[\h]\,-\,\tau[H[\h]]\nonumber\\
    &=&-\,\DHf[z]H_f\,+\,\Delta E_0[z]\,
    +\,\ch_1\left(a[z]\Ppar
    \,+\,T_{n-l}[\h]\right)\ch_1\;,\eeqnn
and writing
$$\bch\;\equiv\;\bchr\ch_1\;,$$
the operator
\eqnn\bar{R}[\h]&=&\left(\tau[H[\h]]\,+\,
    \bch\,\omega[H[\h]]\,
    \bch\right)^{-1}\;,\eeqnn
denotes the 'full resolvent'.

>From the general discussion of the Feshbach map, we recall that it
must be verified that
\begin{gather}\bar{R}[\h]\;,\;\bchr\,\bar{R}\,\bchr\,
     \omega[H[\h]]\, \chr\;,\nonumber\\
     \chr \,\omega[H[\h]]\,\bchr
     \bar{R}[\h]\,\bchr\;,\;\nonumber\\
     \chr\,\omega[H[\h]]\,\bchr\,\bar{R}[\h]\,\bchr\,
     \omega[H[\h]]\,\bchr
     \;,\nonumber\\
     \chr\, \omega[H[\h]]\,\chr\label{boundopwelldefFbtripaux33}
     \end{gather}
extend to bounded operators on $P_1\Hp$ in order for
\eqnn
    Q^\sharp_{\chr,\tau}[\h]&=&\chr\,-\,\chr\,\omega[H[\h]]\,\bchr\,
    \bar{R}\,\bchr\;,\\
    Q_{\chr,\tau}[\h]&=&\chr\,-\,\bchr\,\bar{R}\,\bchr\,\omega[H[\h]]
    \,\chr\;,\\
    \FchrtH[H[\h]]&=&\tau[H[\h]]\,+\,\chr\,\omega[H[\h]]\,\chr
    \,\nonumber\\
    &&-\,\chr\,\omega[H[\h]]\,\bchr\,\bar{R}[\h]\,\bchr\,
     \omega[H[\h]]\,\bchr\; \eeqnn
to be well-defined.

>From Lemma {~\ref{piopdeflemma}}, Neumann series expansion of the
resolvent in the Feshbach map yields
\eqn\label{Neumexpindstep}
    \FchrtH[H[\h]]&=&\tau[H[\h]]\,+\,
    \chr\,T'[\h]\,\piop[\h]\,\chr\,+\,
    \chr\,\piop[\h]\,W[\h]\,\piop[\h]\,\chr\nonumber\\
    &-&\sum_{L=2}^\infty\, (-1)^L
    \sum_{\stackrel{1\leq M_i+N_i}{i=2,\dots,L}}\,
    \chr\,\piop[\h]\,\left( \sum_{1\leq M_1+N_1 }
    W_{M_1,N_1}[\h]\,\right)\,
    \bch\,\bar{R}_0[\h] \,\bch\,\cdots
    \nonumber\\
    &&\hspace{0.5cm}\cdots\,
    \bch\,W_{M_{L-1},N_{L-1}}[\h]\,\bch\,
    \bar{R}_0[\h]\,\bch
    \left( \sum_{1\leq M_L+N_L }W_{M_L,N_L}[\h]\,\right)\,
    \piop[\h]\,
    \chr\;,\eeqn
where
\eqn\piop[\h]&=&\1\,-\,\bchr \Ttild[\h]
    \bchr \bar{R}_0[\h]\nonumber\\
    &=&\Pbchr^\perp\,+\,\Pbchr\,\tau[H[\h]]\,\bar{R}_0[\h]\,\;.
    \label{piopindstepdef4444}\eeqn
Wick ordering each of the summands in the Neumann series, and
collecting all Wick monomials that are equal in the number of
creation and annihilation operators, we find the image of $H[\h]$
under smooth Feshbach decimation,
\eqn\dec[H[\h]]&=&(1+\DHf[z])H_f\,-\,(z\,-\,\Delta
     E_0[z]) \nonumber\\
     &&+\,
     \chr( T'[\h]\,\piop[\h]\,+\,
     \Delta T'[z';\opT])\,\chr\;\nonumber\\
     &&+\;\left.\left.\sum_{M+N\geq 1}\;\;\chr\;
     \right( \piop[\h]\,W_{M,N}[\h]\,\piop[\h]
     +\Delta W_{M,N}[z;\op]\right)\;
     \chr\;,\label{Decindsteppiop4444}\eeqn
in full analogy to the expression obtained in the first decimation
step.
\\

\begin{lm}\label{barRnmboundlemma1}
For $z\in[-\epsilon,0]$ and $|p|\leq\puppbd$ as required in the
induction hypothesis,
\eqnn \frac{1}{4} &\leq&\bar{R}_0[\h]\,,\,\bar{R}[\h]\; \leq\;
    7 \;,\eeqnn
and in particular,
\eqnn\|P_\rho\bar{R}_0[\h] P_\rho\|&\leq&\rho^{-1}
    \left(1-|p|+O(\sqrt{\epsilon}_0)\right)^{-1}\eeqnn
with $\rho=\frac{1}{2}$. Furthermore,
\eqnn\|\,\derp^\beta\partial_\op^\alpha\,\bar{R}_0[\h]\,\|
    \;,\;\|\,\partial_z\,\bar{R}_0[\h]\,\|&\leq&
    C\\
    \|\,\derp^\beta\partial_\op^\alpha\,\bar{R}[\h]\,\|
    \;,\;\|\,\partial_z\,\bar{R}[\h]\,\|&\leq&
    C\eeqnn
and
\eqnn\|\derp^\beta\partial_\op^\alpha
    W[\h]\|&\leq&C'\,\ez^\beta
    \epsilon^{1-\beta}\,\xi\\
    \|\partial_z W[\h]\|&\leq&C\,
    \epsilon \,\xi\eeqnn
for some constants $C,C'$ that are independent of $\epsilon$.
\end{lm}

\prf From $\frac{1}{3} H_f\leq T\leq 3 H_f$ in
(~\ref{TgammaGammabd}) of the induction hypothesis immediately
follows that
$$\frac{1}{3}\;\leq\;\bar{R}_0\;\leq\;\frac{3}{\,\rho}\;.$$
Expanding $\bar{R}$ into a Neumann series, we have
$$\bar{R}\;=\;\bar{R}_0\,+\,\;
    \sum_{j\geq1} \bar{R}_0\left(\bch W\bch\bar{R}_0\right)^j\;,$$
where the sum can be estimated by
\eqnn\sum_{j\geq1}\|\bar{R}_0\|\,\|\left(\bch
    W\bch\bar{R}_0\right)\|^j
    &\leq&\|\bar{R}_0\|\,\sum_{j\geq1}O(\epsilon^{j}\,\xi^j)\\
    &=&\|\bar{R}_0\|\,
    O(\epsilon\,\xi)\;,\eeqnn
using $\|W[\h]\|\leq O(\epsilon\,\xi)$ on $P_1\Hp$ from choosing
$K[\h]=W[\h]$ in Lemma {~\ref{overallbdslm}}. We thus obtain the
asserted bounds, for $\epsilon$ sufficiently small.

Next, we note that
\eqnn\left\|P_\rho \,\bar{R}_0[\h]\,P_\rho\right\|&\leq&\rho^{-1}
    \left(1-|a[z]|-\|P_\rho \,(T_{n-l}[\h]-\lTnl|\P_f|^2)\,
    P_\rho\|\right)^{-1}\\
    &\leq&\rho^{-1}
    \left(1-|p|+O(\sqrt{\epsilon}_0)\right)^{-1}\;,\eeqnn
using (~\ref{Tnllowhighboundindhyp}) of the induction hypothesis.

The inequalities asserted for the derivatives of $\bar{R}_0[\h]$
and $\bar{R}[\h]$ and $W[\h]$ are straightforward consequences of
Lemma {~\ref{derXbarRlemma1}}, and the induction assumptions on
derivatives of $T[\h]$ and $W[\h]$, in addition to the fact that
$\|\partial_{\op}^\alpha\chi\|,\|\partial_{\op}^\alpha\bch\|\leq
O(1)$. \qed

Hence, all operators in (~\ref{boundopwelldefFbtripaux33}) are
indeed bounded on $P_1\Hp$. This result also implies that
$\FchrtH[H[\h]]$ can be expanded into an absolutely convergent
Neumann series, cf. the detailed discussion in Section
{~\ref{FirstdecWelldefFbtripsect3}}.
\\

\begin{lm}\label{piopderboundslemma4444}
Let $\rho=\frac{1}{2}$. Then, there is a constant $C$ that is
independent of $\epsilon$, such that
\eqn\|\derp^\beta\partial_{\op}^\alpha\piop\|&\leq&C
    \label{derpbetapartopalphpiopleqC4444}\eeqn
holds on $P_1\Hp$, for $0\leq |\alpha|\leq2$ and $\beta=0,1$. In
particular, with
\eqnn\|\partial_{H_f}(\chr^2\bchr^2)\|\;,\;
    \|\partial_{H_f}\bchr^2\|\;,\;
    \|\partial_{H_f}\bch_1^2\|&\leq&10 \;,\eeqnn
the estimates
\eqn\|\piop[\h]\|&\leq&
    \frac{4}{3}\nonumber\\
    \|\chr\left(\piop\,-\,\1\right)\chr\|&\leq&
    \frac{1}{3}\label{piopmin1est4444}\\
    \|\derp\piop[\h]\|&\leq&2\label{derppiopest4444}\\
    |\partial_{H_f}(\chr\piop\chr)|&\leq&
    2\,+\,
    3\,|\partial_{H_f}(\chr^2\bchr^2)|\,
    \nonumber\\
    &&\hspace{1cm}+\,2(|p|+\lTnl)\,|\partial_{H_f}
    \bchr^2|
    \label{partialHfpiopest4444}\eeqn
hold for $|p|\leq\puppbd$ with $\puppbd\in[\puppbdnum,1)$
sufficiently small.
\end{lm}

\prf From
\eqnn\left\|\derp^\beta\partial_{\op}^\alpha X\right\|\;
    \leq\;c\eeqnn
for
$$X\;=\;\chr^2\bchr^2\;,\;\bar{R}_0\;{\rm and}\;T'[\h]$$
(since $|\derp\DHf[z]|,|\derp\Delta E_0[z]|\leq O(\epsilon_0)$, as
will be proved below, and using Lemma {~\ref{barRnmboundlemma1}})
follows immediately that there exists a constant $C<\infty$ such
that (~\ref{derpbetapartopalphpiopleqC4444}) holds.

To prove (~\ref{piopmin1est4444}), where (on $P_\rho\Hp$)
$$\piop[\h]\,-\,\1\;=\;-\,\bchr \Ttild[\h]
    \bchr \bar{R}_0[\h]\;,$$
we first recall that
\eqnn\left\|P_\rho \,\bar{R}_0[\h]\,P_\rho\right\|
    &\leq&\rho^{-1}
    \left(1-|p|+O(\sqrt{\epsilon}_0)\right)^{-1}\;,\eeqnn
from Lemma {~\ref{barRnmboundlemma1}}. Furthermore,
\eqn\|\chr^2\bchr^2\|\;\leq\;\frac{1}{4}\;,\eeqn
because of $\max_{x\in[0,1]}\{x^2(1-x^2)\}=\frac{1}{4}$. Finally,
by definition of $T'[\h]$ and $|\DHf[z]|,|\Delta E_0[z]| \leq
O(\epsilon^2)$,
\eqnn\|T'[\h]\|&\leq&|a[z]|\,+\,\lTnl\,+\,O(\epsilon^2)\\
    &\leq&|p|\,+\,\lTnl\,+\,O(\sqrt{\epsilon}_0)\;.\eeqnn
Combining these estimates with $\rho=\frac{1}{2}$ and
$\lTnl\leq\frac{1}{2}$, we find that
\eqnn\|\chr\left(\piop[\h]\,-\,\1\right)\chr\|&\leq&
    \frac{\rho^{-1}}{4}\frac{|p|+\frac{1}{2}+
    O(\sqrt{\epsilon}_0)}
    {1-|p|+O(\sqrt{\epsilon}_0)}\\
    &\leq&\frac{1}{3}\;,\eeqnn
and from
\eqnn\piop[\h]&=&\Pbchr^\perp\,+\,\Pbchr\,\tau[H[\h]]\,
    \bar{R}_0[\h]\eeqnn
we have
\eqnn\|\piop[\h]\|&\leq&\max\left\{1\;,\;
    \|\tau[H[\h]]\,\Pbchr\bar{R}_0[\h]\|\right\}
    \\
    &\leq&(1+O(\epsilon^2))\max_{x\in[\frac{1}{2},1]}\frac{x}
    {(1-|p|)x+O(\sqrt{\epsilon}_0)}\\
    &\leq&\frac{4}{3}\;,\eeqnn
for sufficiently small $|p|\leq\puppbd$ with
$\puppbd\in[\puppbdnum,1)$, as claimed.
\\

As to (~\ref{derppiopest4444}), we have
\eqnn\|\derp\piop[\h]\|&\leq&
    \|(\derp \tau[H[\h]])\,\bar{R}_0[\h]\|
     \,\nonumber\\
    &&+\,\|\tau[H[\h]]\bar{R}_0^2[\h]
    \derp(\tau[H[\h]]+\bchr T'\bchr)\|\\
    &\leq&\max_{x\in[\frac{1}{2},1]}
    \frac{(1+O(\epsilon^2))\,|\derp a[z]|\,x^2}
    {((1-|p|)x+O(\sqez))^2}\,
    +\,O(\epsilon_0)\\
    &\leq&2\eeqnn
on $P_\rho\Hp$. This establishes the claim.
\\

To prove (~\ref{partialHfpiopest4444}), we use
\eqnn\left\|P_\rho\left(\partial_{H_f}T'[\h]\right)
    P_\rho\right\|\;\leq\;
    O(\epsilon^2)\;,\eeqnn
using the fact that $P_\rho\,(\partial_{H_f}\bch_1)\,P_\rho =0$
since the spectral supports of $\partial_{H_f}\bch_1$ and $P_\rho$
are mutually disjoint, as well as
\eqnn\left|P_\rho\left(\partial_{H_f}\bar{R}_0[\h]\right)
    P_\rho\right|&\leq&\left\|P_\rho \bar{R}_0[\h]
    P_\rho\right\|^2\left|P_\rho\left(\partial_{H_f}
    (\tau[H[\h]]+\bchr T'[\h]\bchr)\right)
    P_\rho\right|\\
    &\leq&\frac{\rho^{-2}\left(1+O(\epsilon^2)+
    \|P_\rho T'[\h]P_\rho\|\,|\partial_{H_f}
    \bchr^2|\right)}
    {(1-|p|+O(\sqrt{\epsilon}_0))^2}\\
    &\leq&\frac{\rho^{-2}\left(1+(|p|+\lTnl)|\partial_{H_f}
    \bchr^2|+O(\sqez)\right)}
    {(1-|p|+O(\sqrt{\epsilon}_0))^2}\;.
    \eeqnn

Therefore,
\eqnn\left|\partial(\chr\,\piop[\h]\,\chr)\right|&\leq&
    |\partial_{H_f}(\chr^2\bchr^2)|\,\|P_\rho T'[\h]P_\rho\|\,
    \|P_\rho\,\bar{R}_0[\h]\,P_\rho\|\\
    &&\;+\;\|\chr^2\bchr^2\|\,\|P_\rho (\partial_{H_f}T'[\h])
    P_\rho\|\,
    \|P_\rho\,\bar{R}_0[\h]\,P_\rho\|\\
    &&\;+\;\|\chr^2\bchr^2\|\,\|P_\rho  T'[\h]
    P_\rho\|\,
    |P_\rho\,(\partial_{H_f}\bar{R}_0[\h])\,P_\rho|\\
    &\leq&\rho^{-1}\,\frac{(1+|p|+\lTnl+O(\sqez))}{1-|p|+O(\sqez)}
    |\partial_{H_f}(\chr^2\bchr^2)|\,+\,O(\epsilon)\,\\
    &&\;+\;\frac{\rho^{-2}}{4\,(1-|p|+O(\sqrt{\epsilon}_0))^2}
    (1+O(\sqez))\\
    &&\;+\;\frac{\rho^{-2}}{4}
    \frac{|p|+\lTnl}
    {(1-|p|+O(\sqrt{\epsilon}_0))^2}\,|\partial_{H_f}
    \bchr^2|\\
    &\leq&2\,+\,3\,|\partial_{H_f}(\chr^2\bchr^2)|\,
    +\,2(|p|+\lTnl)\,|\partial_{H_f}
    \bchr^2|\eeqnn
for $|p|\leq\puppbd$ with $\puppbd\in[\puppbdnum,1)$ sufficiently
small. \qed

\section{Renormalization of the spectral parameter}

We absorb the quantity
\eqn\Delta E_0[z]\;:=\;\left\langle \FchrtH
    \left[H[\h]\right]+z\right\rangle_{\vac}
    \label{defE01}\eeqn
into a shift of the spectral parameter,
\eqn\;z\;\longrightarrow\;z\;-\;\Delta E_0[z]\;.\eeqn
Application of the rescaling transformation yields
\eqn\hat{z}\;=\;Z[z]\;:=\;(1+\DHf[z])^{-1}\;\rho^{-1}\;
    \left(z\;-\;\Delta E_0[z]\right)\;,\eeqn
the renormalized spectral parameter.
\\

\begin{prp}\label{ZC1indstepprp3}
The map $Z:z\rightarrow\hat{z}$ is of class $C^1$, and its first
derivative is given by
$$\partial_z Z[z]\;=\;\rho^{-1}(1+\DHf[z])^{-1}\;+\;
    O(\epsilon^{2}\,\xi^2)\;,$$
for $z$ as required in the induction hypothesis. Thus, for any
function $f$ of class $C^1$ in $z$,
\eqn\left|\partial_{\hat{z}}f[Z^{-1}[\hat{z}]]\right|\;\leq\;
    (\rho\;+\;O(\epsilon))\,\left|\partial_z f[z]\right|
    \;.\label{derhatzderzcomp1}\eeqn
In addition,
$$\left|\;\partial_z^\alpha\Delta E_0[z]\;\right|
    \;\leq\;O(\epsilon^{2}\,\xi^2)$$
for $\alpha=0,1$.
\end{prp}

\prf From straightforward calculation,
$$\partial_z Z[z]\;=\;\rho^{-1}(1+\DHf[z])^{-1}
    \left(1\;-\;\partial_z\Delta E_0[z]\left)
    \;-\;\frac{\partial_z\DHf[z]}{1+\DHf[z]}\,Z[z]
    \right.\right.\;.$$
We have
\eqnn|\partial_z \Delta E_0[z]|&=&\left|\;\partial_z\left\langle
W[\h]\,
    \bar{R}[\h]\,W[\h]\right\rangle_{\vac}\;\right|
    \\
    &\leq&O(\epsilon^{2}\,\xi^2)\;, \eeqnn
using
$$\|\partial_z^\beta W[\h]\|\;\leq \;O(\epsilon\,\xi)$$
and
$$\|\partial_z^\beta\bar{R}\|\;\leq\; O(1)\;,$$
for $\beta=0,1$, from Lemma {~\ref{barRnmboundlemma1}}, and
$$|\partial_z^\beta\DHf[z]|\;\leq\; O(\epsilon^{2}\,\xi^2)$$
from Lemma {~\ref{DHfupperlowerbdlemma3}} below. In the same
manner, one shows that $|\Delta E_0[z]|\leq
O(\epsilon^{2}\,\xi^2)$.

Therefore,
$$\partial_z Z[z]\;=\;\rho^{-1}(1+\DHf[z])^{-1}\;+\;
    O(\epsilon^{2}\,\xi^2)\;,$$
and (~\ref{derhatzderzcomp1}) follows immediately. \qed

The renormalized spectral parameter in
$\hat{\h}=\ren[\h]\in\Hspace$ is obtained by the substitution
$z\rightarrow Z^{-1}[\hat{z}]$.
\\

\section{Renormalization of $T[\h]$}

Let us discuss the renormalization of the operators $H_f$,
$a[z]\Ppar$ and $T_{n-l}[\h]$ in $T[\h]$.
\\

\subsection{The coefficient of $H_f$}

Under Feshbach decimation, the coefficient 1 of $H_f$ in
$T_{lin}[\h]$ is modified into
\eqn1+\DHf[z]&=&\left\langle\partial_{H_f}\FchrtH
    \left[H[\h]\right]\right\rangle_{\vac}\nonumber\\
    &=&
    \left\langle \QchrtHs[\h]\;(1+\partial_{H_f}(T_{n-l}[\h]
    +W[\h]))\;
    \QchrtH[\h]\right\rangle_{\vac}\,\times\nonumber\\
    &&\times\, \left(
     1-\left\langle
    W[\h]\bch\bar{R}[\h] \chr^2\bar{R}[\h]\bch
    W[\h]\right\rangle_{\vac}\right.\nonumber\\
    &&\hspace{2cm}\left.
    +\,\left\langle \QchrtHs[\h]\;
    \frac{\partial_{H_f}\bchr}{\bchr^3+\delta}\,\tau[H[\h]]\;
    \QchrtH[\h]\right\rangle_{\vac}\right)^{-1}\;.
    \label{oneplusDHfzauxform3333}\eeqn
Since we have decided to keep the coefficient of $H_f$ fixed at
the constant value 1 in the renormalization group iteration, the
rescaling transformation must act by
\eqn\resc[A]\;=\;(1+\DHf[z])^{-1}\;\rho^{-1}\;
    Ad_{U_\rho}[A]\eeqn
on operators $A:P_\rho\Hp\rightarrow P_\rho\Hp$.
\\

\begin{lm}\label{DHfupperlowerbdlemma3}
For $|p|\leq\puppbd$, with $\puppbd\in[\puppbdnum,1)$ sufficiently
small, and under the assumptions of the induction hypothesis,
$$\;\frac{1}{10}\,g^2\,|p|^2\,+\,\lTnl\,O(\epsilon^2\xi^2)\,+\,
    O(\epsilon^{\frac{5}{2}}\xi^2)\;\leq\;\DHf[z]
    \;\leq\;O(\epsilon^{2}\xi^2)\;.$$
Furthermore,
$$\left|\derp\DHf[z]\right|\;\leq\;C\,\epsilon_0^2\; $$
holds for a constant $C$ that is independent of $\epsilon$ and
$g$.
\end{lm}

\prf First of all, the denominator  in
(~\ref{oneplusDHfzauxform3333}) can be estimated by use of
\eqnn\lim_{\delta\rightarrow0}\left|\left\langle
    \QchrtHs[\h]\bar{P}_\rho
    \frac{\chr^2}{\bchr^2+\delta}\bar{P}_\rho
    \QchrtH[\h]\right\rangle_{\vac}\right|&\leq&
    \left\|\chr\bar{R}[\h]\bchr\right\|^2\,\left\|W[\h]\right\|^2
    \\
    &\leq&
    O(\epsilon^2\xi^2)\;.\eeqnn
and
\eqnn&&\lim_{\delta\rightarrow0}\left|\left\langle
    \QchrtHs[\h]\bar{P}_\rho
    \frac{\partial_{H_f}\chr}{\bchr^3+\delta}(H_f-\hat{z})\bar{P}_\rho
    \QchrtH[\h]\right\rangle_{\vac}\right|\\
    &&\hspace{1cm}=\,
    \left|\left\langle
    W[\h]\bch\bar{R}(\partial_{H_f}\chr)(H_f-\hat{z})\bar{R}_0W[\h]
    \right\rangle_{\vac}\right|\\
    &&\hspace{1.5cm}+\,\left|\left\langle
    W[\h]\bch\bar{R} (\partial_{H_f}\chr)(H_f-\hat{z})\bar{R}_0W[\h]
    \bch\bar{R}\bch W[\h]
    \right\rangle_{\vac}\right|\\
    &&\hspace{1cm}\leq\,
    O(\epsilon^2\xi^2)\;,\eeqnn
by the second resolvent identity for $\bar{R}[\h]$.
\\

(I) The lower bound. Using the abbreviated notation
$$Q^{(\sharp)}[\h]\;\equiv\;\QchrtH^{(\sharp)}[\h]$$
it is clear that
\eqn&&\left\langle
    Q^{\sharp}[\h]\;(1+\partial_{H_f}(T_{n-l}[\h]
    +W[\h]))\;
    Q[\h]\right\rangle_{\vac}\nonumber\\
    &&\hspace{1cm}=\;1\;+\;\left\langle Q^\sharp[\h]\;\bar{P}_\rho\;
    Q[\h]\right\rangle_{\vac}\;\nonumber\\
    &&\hspace{1cm}\;\;\;\;+\;
    \left\langle Q^\sharp[\h]\bar{P}_\rho
    \left(\partial_{H_f}T_{n-l}[\h]\right)\bar{P}_\rho
    Q[\h]\right\rangle_{\vac}
    \;\nonumber\\
    &&\hspace{1cm}\;\;\;\;+\;\left\langle Q^\sharp[\h]
    \left(\partial_{H_f}W[\h]\right)
    Q[\h]\right\rangle_{\vac}\;.\label{DHfzformula4}\eeqn
The leading term in (~\ref{DHfzformula4}) is given by
\eqn\left\langle Q^\sharp[\h]\;\bar{P}_\rho\;
    Q[\h]\right\rangle_{\vac}&=&
    \left\langle\, W \,\bchr\,\bar{R}\,\bchr^2\,
    \,\bar{R}\,\bchr\, W\,
    \right\rangle_{\vac}\nonumber\\
    &=&\left\langle  W_1\,\bchr\,\bar{R}\,\bchr^2\,
    \,\bar{R}\,\bchr\,W_1
    \right\rangle_{\vac}\;\\
    &&+\;\left(\;
    \left\langle \;W_1\,\bchr\,\bar{R}\,\bchr^2\,
    \,\bar{R}\,\bchr\,W_{\geq2}\;
    \right\rangle_{\vac}\;+\;{\rm h.c.}\;\right)
    \label{DHferrterm11}\\
    &&+\;\left\langle \;W_{\geq2}\,\bchr\,\bar{R}\,\bchr^2\,
    \,\bar{R}\,\bchr\,W_{\geq2}\;
    \right\rangle_{\vac}\;,\label{DHferrterm12}\eeqn
where $W_1=W_{0,1}+W_{1,0}$, and $W_{\geq2}=W-W_1$.
\\

(I.1) We claim that the leading term satisfies
\eqn\left\langle
    W_1\,\bchr\,\bar{R}\,\bchr^2\,
    \,\bar{R}\,\bchr\,W_1\;
    \right\rangle_{\vac}&\geq&\;
    \mathfrak{I}[z]\;
    +\;O(\epsilon^{\frac{5}{2}}\xi^2)\;,
    \label{leadtermDHfzest3}\eeqn
where
\begin{gather}\mathfrak{I}[z]\;:=\;|a[z]|^2\,\sum_\lambda
    \int_{\rho\leq|k|\leq\frac{2}{3}}
    \frac{d^3k}{|k|}\left.\left.
    \left|\,\epsilon_\lambda^\parallel(n_k)\,\right|^2
    \,\bchr^2[|k|]\,\right\langle
    \bar{R}_0^2[z;\sh_k\opT]\right\rangle_{\vac}
    \;.\label{mfrakIdef3}\end{gather}
This can be seen as follows.

Application of the second resolvent identity to $\bar{R}$ yields
$$\bar{R}\;=\;\bar{R}_0\;-\;\bar{R}_0\bch
    W\bch\bar{R}\;,$$
where $\|W\|=O(\epsilon\xi)$ and $\|\bar{R}\|,\|\bar{R}_0\|=O(1)$.
Therefore,
\eqnn\left\langle
    W_1\,\bch\,\bar{R}\,\bch^2\,
    \,\bar{R}\,\bchr\,W_1\;
    \right\rangle_{\vac}&=&\left\langle
    W_1\,\bch\,\bar{R}_0\,\bch^2\,
    \,\bar{R}_0\,\bch\,W_1\;
    \right\rangle_{\vac}\,+\,O(\epsilon^3\xi^3)\;\eeqnn
is immediate. Furthermore, it is clear that
\eqnn\left\langle
    W_1\,\bch\,\bar{R}_0\,\bchr^2\,
    \,\bar{R}_0\,\bch\,W_1\;
    \right\rangle_{\vac}&=&\left\langle
    W_1\,\bch^2\,\bPlperp\,\bar{R}_0^2\,\bPlperp\,\bch^2\,W_1\;
    \right\rangle_{\vac}\\
    &&+\;\left\langle
    W_1\,\bch^2\,\bPl\,\bar{R}_0^2\,\bPl\,\bch^2\,W_1\;
    \right\rangle_{\vac}\\
    &\geq&\left\langle
    W_1\,\bch^2\,\bPlperp\,\bar{R}_0^2\,\bPlperp\,\bch^2\,W_1\;
    \right\rangle_{\vac}\;,\eeqnn
and that
\eqnn\mathfrak{I}[z]&=&\left\langle
\;U_1\,\bPlperp\,\bch^2\,\bar{R}_0^2
    \,\bch^2\,\bPlperp\,U_1\;
    \right\rangle_{\vac}\;,\eeqnn
where $U_1$ is defined in the same manner as $W_1$, but with the
interaction kernels $w_{M,N}$ replaced by $u_{M,N}$ ($M+N=1$).
Defining the error term
\eqn \mathfrak{R}&:=&\left\langle \;W_1\,\bch^2\,\bPlperp\,
    \bar{R}^2_0\,\bPlperp\,\bch^2\,W_1\;
    \right\rangle_{\vac}\,\nonumber\\
    &&-\;
    \left\langle \;U_1\,\bch^2\,\bPlperp\,\bar{R}^2_0\,
    \bPlperp\,\bch^2\,U_1\;
    \right\rangle_{\vac}\;,\eeqn
we have that, due to $W_1\vac=W_1|_{\op\rightarrow0}\vac$ and
$$\|\bPlperp(W_1|_{\op\rightarrow0}\,-\,U_1)\bPlperp\|\;
    \leq\;
    O(\epsilon^{\frac{3}{2}})\,\xi\;,$$
which is a consequence of (~\ref{derpalphwMNminuMNindhypbd3}) of
the induction hypothesis,
\eqnn \mathfrak{R}&=&\left\langle \;(W_1-U_1)\bPlperp
    \,\bchr^2\,\bar{R}^2_0\,\bchr^2\,W_1\;
    \right\rangle_{\vac}\,\nonumber\\
    &&-\;
    \left\langle \;W_1\,\bchr^2\,\bar{R}^2_0
    \,\bchr^2\,\bPlperp(W_1-U_1)\;
    \right\rangle_{\vac}\;\nonumber\\
    &&+\;
    \left\langle \;(W_1-U_1)\bPlperp
    \,\bchr^2\,\bar{R}^2_0\,\bchr^2\,\bPlperp(W_1+U_1)\;
    \right\rangle_{\vac}\; \eeqnn
is bounded by
$$|\mathfrak{R}|\;\leq\;
    O(\epsilon^{\frac{5}{2}}\,\xi^2)\;.$$
Finally, an explicit calculation shows that for
$\puppbd\in[\puppbdnum,1)$ sufficiently small, and
$\rho=\frac{1}{2}$, there is an estimate
$$\mathfrak{I}[z]\,\geq\,\frac{2}{5}\,g^2\,|p|^2\;,$$
for all $|p|\leq\puppbd$ with $\puppbd\in[\puppbdnum,1)$
sufficiently small, and for $z$ as required in the induction
hypothesis. Furthermore, we have used the fact that
$|a[z]|\leq|p|(1+O(\sqrt{\epsilon}_0))$, which is a direct
consequence of the induction assumption on $a[z]$.
\\

(I.2) The error term (~\ref{DHferrterm11}) can be estimated by
$$\left|
    \left\langle \;W_1\,\bch^2\,\bar{R}^2\,\bch^2\,W_{\geq2}\;
    \right\rangle_{\vac}\;+\;{\rm h.c.}\right|\;\leq\;
    O(\epsilon^{\frac{5}{2}}\,\xi^3)\;,$$
since $\|W_{\geq2}\|\leq O(\epsilon^{\frac{3}{2}}\,\xi^2)$, and
using Lemma {~\ref{barRnmboundlemma1}}.
\\

(I.3) The error term (~\ref{DHferrterm12}) can be bounded by
$$\left|\left\langle \;W_{\geq2}\,\bch^2\,\bar{R}^2
    \,\bch^2\,W_{\geq2}\;
    \right\rangle_{\vac}\right|\;\leq\;
    O(\epsilon^{3}\,\xi^4)\;,$$
for the same reasons as in the previous case under point (2).
\\

(I.4) There are two additional error terms in
(~\ref{DHfzformula4}) which have to be considered. The second
vacuum expectation value appearing there can be estimated by
\eqn&&\left|\left\langle Q^\sharp[\h]\bar{P}_\rho
    \left(\partial_{H_f}T_{n-l}[\h]\right)\bar{P}_\rho
    Q[\h]\right\rangle_{\vac}\right|\nonumber\\
    &&\hspace{1cm}=\;
    \left|\left\langle W\bch\bar{R}\bch
    \left(\partial_{H_f}T_{n-l}\right)\bch\bar{R}\bch
    \,W\;\right\rangle_{\vac}\right|\nonumber\\
    &&\hspace{1cm}\leq\;(2(|p|+\lTnl)+O(\sqez))
    \left\langle W\bch\bar{R}\bch^2
    \bar{R}\bch
    W\;\right\rangle_{\vac}\nonumber\\
    &&\hspace{1.5cm}\;+\;3\,(|p|\,+\,\lTnl)
    \left\langle W\bch\bar{R}\bch
    |\partial_{H_f}(\ch_1^2\bch_1^2)|
    \bch\bar{R}\bch
    W\;\right\rangle_{\vac}\nonumber\\
    &&\hspace{1.5cm}\;+\;2(|p|+\lTnl)^2
    \left\langle W\bch\bar{R}\bch
    |\partial_{H_f}\bch_1^2|
    \bch\bar{R}\bch
    W\;\right\rangle_{\vac}
    \nonumber\\
    &&\hspace{1cm}=\;2\,(|p|+O(\sqez))\,\left\langle W\bch\bar{R}
    \bch^2\bar{R}\bch
    W\;\right\rangle_{\vac}\nonumber\\
    &&\hspace{1.5cm}\;+\;3\,|p|\,
    \left\langle W\bch\bar{R}\bch
    |\partial_{H_f}(\ch_1^2\bch_1^2)|
    \bch\bar{R}\bch
    W\;\right\rangle_{\vac}\nonumber\\
    &&\hspace{1.5cm}\;+\;2|p|^2
    \left\langle W\bch\bar{R}\bch
    |\partial_{H_f}\bch_1^2|
    \bch\bar{R}\bch
    W\;\right\rangle_{\vac}
    \nonumber\\
    &&\hspace{1.5cm}\;+\;\lTnl\,O(\epsilon^2\xi^2)\;, \eeqn
using (~\ref{partHfTnlindhypboudn4444}) of the induction
hypothesis, and the fact that
$\left(\partial_{H_f}T_{n-l}\right)\vac=0$. Thus, the bounds
(~\ref{derch1derchrboundsindstep4444}) on the derivatives of
$\chr,\ch_1$ imply that
\eqn\left|\left\langle Q^\sharp[\h]\bar{P}_\rho
    \left(\partial_{H_f}T_{n-l}[\h]\right)\bar{P}_\rho
    Q[\h]\right\rangle_{\vac}\right|&\leq&
    (14+8|p|)\,|p|\,\left\langle W\bch\bar{R}\bch^2\bar{R}
    \bch
    W\;\right\rangle_{\vac}\nonumber\\
    &&\;+\;\lTnl\,O(\epsilon^2\xi^2)\nonumber\\
    &\leq&\frac{3}{4}\left\langle W\bch\bar{R}\bch^2\bar{R}
    \bch
    W\;\right\rangle_{\vac}\nonumber\\
    &&\;+\;\lTnl\,O(\epsilon^2\xi^2)\eeqn
for $|p|\leq\puppbd$ with $\puppbd\in[\puppbdnum,1)$ small enough.
\\

(I.5) The last error term in (~\ref{DHfzformula4}) can be
estimated by \eqn\left|\;\left\langle Q^\sharp[\h]
    \left(\partial_{H_f}W[\h]\right)
    Q[\h]\right\rangle_{\vac}\;\right|&\leq&\|\partial_{H_f}W[\h]\|
    \left\langle W\bch\bar{R}\bch^2\bar{R}\bch
    W\;\right\rangle_{\vac}\nonumber\\
    &&+\;2\,\left|\;\left\langle Q^\sharp[\h]\,\bar{P}_\rho\,
    \left(\partial_{H_f}W[\h]\right)
    \right\rangle_{\vac}\;\right|
    \nonumber\\
    &\leq&
    O(\epsilon^{\frac{5}{2}}\xi^2)\;,\eeqn
since
$$\|\bPlperp\partial_{H_f}W\vac\|\;\leq\;
    O(\epsilon^{\frac{3}{2}}\xi)\;,$$
due to (~\ref{derpalphwMNminuMNindhypbd3}) of the induction
hypothesis, and
$$\|Q\,\vac\|\;\leq\; O(\epsilon\,\xi)\;.$$

$\;$

(I.6) In total, we have, in view of (~\ref{leadtermDHfzest3}) and
the estimates derived in (I.1) $\sim$ (I.5), the lower bound
\eqnn\DHf[z]&\geq&(1-\frac{3}{4})\left\langle
    U_1\bch^2\bar{R}_0^2\bch^2
    U_1\;\right\rangle_{\vac}\,\nonumber\\
    &&\;+\,\lTnl\,O(\epsilon^2\xi^2)\,
    +\,O(\epsilon^{\frac{5}{3}}\xi^2)
    \;,\eeqnn
which implies the claim.
\\

(II) The upper bound. The estimate
\eqnn|\DHf[z]|&\leq&
     O(\epsilon^{2}\,\xi^2)
    \eeqnn
follows from
$$\left\langle W\bch\bar{R}\bch^2\bar{R}\bch
    W\;\right\rangle_{\vac}\;\leq\;O(\epsilon^2\xi^2)\;,$$
which is evident from $\|\bar{R}[\h]\|\leq O(1)$, $\| W[\h]\|\leq
O(\epsilon\xi)$.
\\

(III) The estimate
\eqnn|\derp\DHf[z]|&\leq&
     O(\epsilon_0^2)
    \eeqnn
follows from $\|\derp\bar{R}[\h]\|\leq O(1)$, $\|\derp W[\h]\|\leq
O(g)$, $g\ll\epsilon_0$, $\epsilon\leq 2\epsilon_0$, and
straightforwardly taking the derivative of
(~\ref{oneplusDHfzauxform3333}) with respect to $|p|$. \qed

\subsection{Renormalization of the coefficient of $\Ppar$}

The strategy in renormalizing $a[z]$ is to employ a nested
induction argument. Assuming that Theorem {~\ref{indstepmainthm3}}
and the bound (~\ref{derpalphaazconscondindhyp}) hold for all
$\h^{(k)}$ with $0\leq k<n$, it is proved that the assumptions of
Propositions {~\ref{anznprp3333}} and {~\ref{dHfFQQformprop333}}
are fulfilled. Consequently, the bound
(~\ref{derpalphaazconscondindhyp}) also holds for $k=n$.
\\

\begin{prp}\label{hatazminazestkappaprp}
Let $\h^{(n)}$ denote the point in $\Polyd_{2\epsilon_0,\xi}$
obtained from $\h^{(0)}$ by $n$-fold iterating $\ren$, and let
$\rho=\frac{1}{2}$.  Assume that both Theorem
{~\ref{indstepmainthm3}} and the bound
(~\ref{derpalphaazconscondindhyp}) hold  for
$\h^{(j-1)}\rightarrow\h^{(j)}$, for all $j<n$. Then,
\eqn\left| \derp^\beta(a_n[z_n]\,+\,|p|) \right|
    \;\leq\;\sqrt{\epsilon}_0 \; \eeqn
for $|p|\leq\puppbd$.
\end{prp}

\prf

Lemmata {~\ref{overallbdslm}} and {~\ref{deropalphaDeltwMNlemma1}}
imply that
\eqnn\|\derp^\beta\bar{R}[\h^{(k)}]\|&\leq&C\\
    \|\derp^\beta W[\h^{(k)}]\|&\leq&\ez^\beta\,
    C'\,\epsilon^{1-\beta}\,\xi\eeqnn
for constants $C,C'$ that are independent of $\epsilon$,
$\beta=0,1$, and all $0\leq k<n$. Furthermore,
$$\left|\derp^\beta\left(\dHfFQQ[\h^{(n-1)}]-1\right)\right|\;
    \leq\;C\,\ez^\beta
    \epsilon^{2-\beta}\;,$$
for $\beta=0,1$, where $\dHfFQQ[\h]$ is defined in Propositions
{~\ref{anE0neqderpEgrdprp3}} and {~\ref{dHfFQQformprop333}}.
\\

(I) The assertion for the case $\beta=0$ is a consequence of
Proposition {~\ref{anznprp3333}} under the condition that the main
theorem holds for all $\h^{(k)}$ with $0\leq k<n$.

Lemma {~\ref{QnsQnQksQkboundlemmaaux3333}} implies that for
Proposition {~\ref{anznprp3333}} to hold, one must show
\eqn\left|
    \frac{\left\langle
    Q^\sharp \,Q
    \right\rangle_{(k;n)}}{\left\langle
    \Qns[z[z_n]]\Qn[z[z_n]]
    \right\rangle_{\vac}}\right|&\leq&2\;,
    \label{anznprpcond13333}\eeqn
for all $0\leq k\leq n$, and that there exists a constant
$B<\infty$ independent of $n$ and $\epsilon_0$, such that
\eqn \left\|Q^\sharp[\h^{(0)}]\,\left(Q_0^{\sharp}[z]\,H_f\,
    Q_0[z]\,-\,H_f\right)\,Q[\h^{(0)}]\right\|&\leq&
    B\,\epsilon_0^2 \label{QksQkboundauxform03333}\eeqn
and
\eqn\left\|Q^\sharp[\h^{(k+1)}]\,
    U_\rho\,\left(Q^{\sharp}[\h^{(n)}]\,H_f\,
    Q[\h^{(k)}]\,-\,H_f\right)\,U_\rho^*\,
    Q[\h^{(k+1)}]\right\|&\leq&
    B\,\epsilon_0^2 \label{QksQkboundauxformk3333}\eeqn
for all $0\leq k\leq n$.
\\

(I.1) In order to prove (~\ref{anznprpcond13333}), let us
introduce the notation
\eqn F_{(k;n)}[H[\h^{(k)}]]&:=&
    F_{\ch_{\rho^{n-k}},\tau_n}[H[\h^{(k)}]]\;,\eeqn
where
\eqnn H[\h^{(n)}]&=&\frac{\rho^{-(n-k)}}{\langle
    \partial_{H_f}F_{(k;n)}[H[\h^{(k)}]]\rangle_{\vac}}
    Ad_{U_{\rho^{n-k}}}
    \left[F_{(k;n)}[H[\h^{(k)}]]\right]\eeqnn

Next, we observe that according to
(~\ref{oneplusDHfzauxform3333}),
\eqnn \left\langle\partial_{H_f}F_{(k;n)}[H[\h]]
    \right\rangle_{\vac}&=&
    \left\langle Q^\sharp (1+\partial_{H_f}
    (\ch_1(T_{n-l}+W)[\h^{(k)}])\ch_1)
    Q\right\rangle_{(k;n)}\,\kappa[\h^{(n-1)}]\;,\eeqnn
with
\eqnn\kappa^{-1}[\h^{(n-1)}]&=& 1-\left\langle
    Q^\sharp\,
    \frac{\ch_{\rho^{n-k}}^2}{\bch_{\rho^{n-k}}^2}\, Q
    \right\rangle_{(k;n)}\nonumber\\
    &&\;+\, \left\langle Q^\sharp\,
    \frac{\partial_{H_f} \bch_{\rho^{n-k}}}{\bch_{\rho^{n-k}}^3}
    (H_f-\rho z_n)\,Q
    \right\rangle_{(k;n)}\nonumber\\
    &=&\dHfFQQ_n^{-1}[z[z_n]]\;.\eeqnn
Thus,
\eqn\kappa[\h^{(n-1)}]&=&1\,+\,O(\epsilon_{n-1}^2)\;.\eeqn
It is clear that
\eqn&&\left|\left\langle Q^\sharp
    (\partial_{H_f}(\ch_1(T_{n-l}+W)\ch_1)[\h^{(k)}])
    Q\right\rangle_{(k;n)}\right|\nonumber\\
    &&\hspace{1cm}\leq\;\left(
    \|\partial_{H_f}(\ch_1 T_{n-l}\ch_1)\|+
    \|\partial_{H_f}(\ch_1 W[\h^{(k)}]\ch_1)\|\right)
    \left\langle Q^\sharp
    Q\right\rangle_{(k;n)}\nonumber\\
    &&\hspace{1cm}\leq\;\left(
    \bTnl+\sqrt{\epsilon}_0+
    O(\epsilon_k\xi)\right)\,
    \left\langle Q^\sharp
    Q\right\rangle_{(k;n)}\nonumber\\
    &&\hspace{1cm}\leq\;\frac{1}{3}\left\langle Q^\sharp
    Q\right\rangle_{(k;n)}\;,\eeqn
for $|p|\leq\puppbd$ if $\puppbd\in[\puppbdnum,1)$ is sufficiently
small.

Hence, we find that
\eqn \left\langle
    Q^\sharp
    Q\right\rangle_{(k;n)}&\leq&
    \left(\frac{3}{2}\,+\,O(\epsilon_{n-1}^2)\right)
    \left\langle\partial_{H_f}F_{(k;n)}[H[\h]]
    \right\rangle_{\vac}\nonumber\\
    &=&\left(\frac{3}{2}\,+\,O(\epsilon_{n-1}^2)\right)
    \prod_{i=k}^n(1+\DHf^{(i)}[z_i])\;.\eeqn
Consequently,
\eqn\frac{\left\langle
    Q^\sharp
    Q\right\rangle_{(k;n)}}{\left\langle
    \Qns
    \Qn\right\rangle_{\vac}}&=&\dHfFQQ_n[z[z_n]]\,
    \frac{\left\langle
    Q^\sharp
    Q\right\rangle_{(k;n)}}{\left\langle\partial_{H_f}
    \Fn[z]
    \right\rangle_{\vac}}\nonumber\\
    &=&\dHfFQQ_n[z[z_n]]\,
    \frac{\left\langle
    Q^\sharp
    Q\right\rangle_{(k;n)}}{\left\langle
    \partial_{H_f}F_{(k;n)}[H[\h]]
    \right\rangle_{\vac}}\,
    \frac{\left\langle\partial_{H_f}F_{(k;n)}[H[\h]]
    \right\rangle_{\vac}}{\left\langle\partial_{H_f}
    \Fn[z]
    \right\rangle_{\vac}}\nonumber\\
    &\leq&\left(\frac{3}{2}\,+\,O(\epsilon_{n-1}^2)\right)
    \prod_{i=0}^{k-1}(1+\DHf^{(i)}[z_i])^{-1}\nonumber\\
    &\leq&2\;,\eeqn
as claimed, since $0<\DHf^{(i)}[z_i]\leq C\epsilon_i^2$, and
$\epsilon_k\leq2\epsilon_0$ for $0\leq k\leq n$.
\\

(I.2) Next, let us prove (~\ref{QksQkboundauxform03333}). A
straightforward calculation leads to
\eqnn &&Q_0^{\sharp}[z]\,H_f\,
    Q_0[z]\,-\,H_f P_1\;=\;\ch_1 \omega[\Hps]\bch_1\bar{R} H_f
    \bar{R}\bch_1 \omega[\Hps]\ch_1
    \\
    &&\hspace{3cm}
    -\,(\bch_1+\ch_1 \omega[\Hps]\bch_1\bar{R}\ch_1)H_f
    (\bch_1+\ch_1\bar{R}\bch_1 \omega[\Hps]\ch_1)\;,\eeqnn
where we recall that
\eqnn\omega[\Hps]&=&\underbrace{-\,\Delta
    E_0[z]\,-\,\DHf[z]H_f\,+\,|p|\,\Ppar\,+\,
    \frac{1}{2}|\P_f|^2}_{=T'}\,+\,W\;,\eeqnn
with $|\Delta E_0[z]|,|\DHf[z]|\leq O(g^2)$. We claim that
\eqn\left\|H_f^{\frac{1}{2}}\bar{R}\bch_1
    \omega[\Hps]\ch_1\,Q[\h^{(0)}]\right\|
    &\leq&C\,\epsilon_0\label{relbdsHfaux4444}\\
    \left\|H_f^{\frac{1}{2}}\bch_1 \,Q[\h^{(0)}]\right\|
    &\leq&C\,\epsilon_0\label{relbdsHfaux24444}\eeqn
which implies (~\ref{QksQkboundauxform03333}). To prove
(~\ref{relbdsHfaux4444}), we note that since
\eqn Q[\h^{(0)}]&=&\chr\,-\,\bchr\bar{R}[\h^{(0)}]
    \bchr\omega[H[\h^{(0)}]]\chr\;,\eeqn
we have
\eqnn&&\left\|H_f^{\frac{1}{2}}\bar{R}\bch_1
    \omega[\Hps]\ch_1\,Q[\h^{(0)}]\right\|
    \;\leq\;\left\|H_f^{\frac{1}{2}}\bar{R}\bch_1
    W\chr \right\|\,\\
    &&\hspace{2cm}+\,
    \left\|H_f^{\frac{1}{2}}\bar{R}\bch_1
    \omega[\Hps]\ch_1\bchr\bar{R}[\h^{(0)}]
    \bchr\omega[H[\h^{(0)}]]\chr \right\|\;,\eeqnn
since $\bch_1 T'[\Hps]\bchr=0$. As a consequence of Lemma
{~\ref{elhighenrelW1W2bounds2}},
\eqnn\left\|H_f^{\frac{1}{2}}\bar{R}\bch_1
    W\chr \right\|&\leq&
    \left\|H_f^{\frac{1}{2}}\bar{R}^{\frac{1}{2}}\right\|\,
    \left\|\bar{R}^{\frac{1}{2}}\bch_1
    W\chr \right\|\\
    &\leq&O(g)\;.\eeqnn
Next, we apply the second resolvent formula to
$\bar{R}[\h^{(0)}]$, which yields
\eqnn&&\left\|H_f^{\frac{1}{2}}\bar{R}\bch_1
    \omega[\Hps]\ch_1\,\bchr\bar{R}[\h^{(0)}]
    \bchr\omega[H[\h^{(0)}]]\chr \right\|\\
    &&\hspace{1cm}\leq\;\left\|H_f^{\frac{1}{2}}\bar{R}\bch_1
    \omega[\Hps]\ch_1\,\bchr\bar{R}_0[\h^{(0)}]
    \bchr\omega[H[\h^{(0)}]]\chr \right\|\\
    &&\hspace{1.5cm}+\;\left\|H_f^{\frac{1}{2}}\bar{R}\bch_1
    \omega[\Hps]\ch_1\,\bchr\bar{R}_0[\h^{(0)}]
    \bch W[\h^{(0)}]\bch
    \bar{R}[\h^{(0)}]
    \bchr\omega[H[\h^{(0)}]]\chr \right\|\;.\eeqnn
The second term on the right hand side of the inequality sign is
bounded by
\eqnn\underbrace{\left\|H_f^{\frac{1}{2}}\bar{R}^{\frac{1}{2}}
    \right\|}_{\leq\;O(1)}\,
    \underbrace{\left\|\bar{R}^{\frac{1}{2}}\bch_1
    \omega[\Hps]\ch_1\right\|}_{\leq\;O(1)}\,
    \underbrace{\left\|\bchr\bar{R}_0^{\frac{1}{2}}[\h^{(0)}]
    \right\|}_{\leq\;O(1)}\,
    \underbrace{\left\|\bar{R}_0^{\frac{1}{2}}[\h^{(0)}]
    \bch W[\h^{(0)}]\bch
    \bar{R}^{\frac{1}{2}}[\h^{(0)}]\right\|}_{\leq\;O(\epsilon_0)}
    \,\times\\
    \times\,\underbrace{\left\|
    \bar{R}^{\frac{1}{2}}[\h^{(0)}]
    \bchr\omega[H[\h^{(0)}]]\chr \right\|}_{\leq\;O(1)}
    \;\leq\;O(\epsilon_0)\;,\eeqnn
as a consequence of Lemmata {~\ref{elhighenrelW1W2bounds1}},
{~\ref{elhighenrelW1W2bounds2}} and {~\ref{barRnmboundlemma1}}.
The first term is bounded by
\eqnn&&\left\|\underbrace{H_f^{\frac{1}{2}}\bar{R}\bch_1
    T'[\Hps]\ch_1\,\bchr\bar{R}_0[\h^{(0)}]
    \bchr T'[H[\h^{(0)}]]\chr}_{=\;0} \right\|\\
    &&+\,\underbrace{\left\|H_f^{\frac{1}{2}}\bar{R}\bch_1
    W\ch_1\right\|}_{\leq\;O(g)}\,
    \underbrace{\left\|\bchr\bar{R}_0[\h^{(0)}]
    \bchr T'[H[\h^{(0)}]]\chr \right\|}_{\leq\;O(1)}\\
    &&+\,\underbrace{\left\|H_f^{\frac{1}{2}}\bar{R}\bch_1
    T'[\Hps]\ch_1\right\|}_{\leq\;O(1)}\,
    \underbrace{\left\|\bchr\bar{R}_0[\h^{(0)}]
    \bchr W[H[\h^{(0)}]]\chr \right\|}_{\leq\;O(\epsilon_0)}\\
    &&+\,\underbrace{\left\|H_f^{\frac{1}{2}}\bar{R}\bch_1
    W[\Hps]\ch_1\right\|}_{\leq\;O(g)}\,
    \underbrace{\left\|\bchr\bar{R}_0[\h^{(0)}]
    \bchr W[H[\h^{(0)}]]\chr \right\|}_{\leq\;O(\epsilon_0)}
    \;\leq\;O(\epsilon_0)\;.\eeqnn
The first term here vanishes because $H_f,T'[\Hps],
\bar{R}_0[\h^{(0)}]$ and $T'[H[\h^{(0)}]]$ mutually commute, and
$\bch_1\chr=0$.

Likewise, (~\ref{relbdsHfaux24444}) follows from
\eqnn\left\|H_f^{\frac{1}{2}}\bch_1 \,Q[\h^{(0)}]\right\|
    &=&\left\|H_f^{\frac{1}{2}}\bch_1  \bar{R}[\h^{(0)}]
    \bchr\omega[H[\h^{(0)}]]\chr \right\|\\
    &\leq&\underbrace{\left\|H_f^{\frac{1}{2}}\bch_1
    \bar{R}_0^{\frac{1}{2}}[\h^{(0)}]\right\|}_{\leq\;O(1)}\,
    \underbrace{\left\|
    \bar{R}_0^{\frac{1}{2}}[\h^{(0)}]
    \bchr W[H[\h^{(0)}]]\chr \right\|}_{\leq\;O(\epsilon_0)}\\
    &&+\,\underbrace{\left\|H_f^{\frac{1}{2}}\bch_1
    \bar{R}_0^{\frac{1}{2}}[\h^{(0)}]\right\|}_{\leq\;O(1)}\,
    \underbrace{\left\|\bar{R}_0^{\frac{1}{2}}[\h^{(0)}]
    \bch W[\h^{(0)}]\bch\bar{R}^{\frac{1}{2}}[\h^{(0)}]
    \right\|}_{\leq\;O(\epsilon_0)}\,\times\\
    &&\hspace{2cm}\times
    \underbrace{\left\|\bar{R}^{\frac{1}{2}}[\h^{(0)}]
    \bchr\omega[H[\h^{(0)}]]\chr \right\|}_{\leq\;O(1)}\\
    &\leq&
    O(\epsilon_0)\eeqnn
by the second resolvent identity, applied $\bar{R}[\h^{(0)}]$.
This proves  (~\ref{QksQkboundauxform03333}).
\\

(I.3) The proof of (~\ref{QksQkboundauxformk3333}) is completely
analogous to the case (I.2). This is because the operator
$$U_\rho\,\left(Q^{\sharp}[\h^{(n)}]\,H_f\,
    Q[\h^{(k)}]\,-\,H_f\right)\,U_\rho^*$$
has (up to a numerical factor) precisely the  same structure as
$$Q_0^{\sharp}[z]\,H_f\,
    Q_0[z]\,-\,H_f \;,$$
which has been studied above. We note that in the present case,
$H[\h^{(k)}]$ is, in contrast to $\Hps$, a bounded operator, and
in comparison to the above case concerning the first decimation
map, one must substitute $\Hps\rightarrow H[\h^{(k)}]$,
$H[\h^{(0)}]\rightarrow H[\h^{(n+1)}]$ in the expressions
appearing under (I.2). (~\ref{QksQkboundauxformk3333}) follows
straightforwardly by applying Lemmata {~\ref{overallbdslm}} and
{~\ref{deropalphaDeltwMNlemma1}}. Due to the similarity to the
case (I.2), this will not be carried out explicitly here.
\\

This proves the claim for the case $\beta=0$.
\\

(II) The case $\beta=1$. Here, we need to verify the assumptions
of Proposition  {~\ref{dHfFQQformprop333}} under the condition
that both Theorem {~\ref{indstepmainthm3}} and the main hypothesis
are correct for all scales $0\leq k<n$.

The statement that
\eqn\epsilon_k&\leq&2\,\epsilon_0\eeqn
and
\eqn\|\derp^\beta T[\h^{(k)}]\|\;\leq\;C\;,\;\|
    \derp^\beta W[\h^{(k)}]\|&\leq&
    C\,\ez^\beta\epsilon_k^{1-\beta}\xi\eeqn
straightforwardly follow from Theorem {~\ref{indstepmainthm3}} and
Lemma {~\ref{overallbdslm}}, and the induction hypothesis that
\eqn \left|\derp a_k[z_k]+1\right|&\leq&
    \sqrt{\epsilon}_0 \label{derpakzkauxform3333}\eeqn
holds for all $0\leq k<n$.

For Proposition  {~\ref{dHfFQQformprop333}} to hold, we must show
that
\eqn\left|\frac{\langle\derp\partial_{H_f}
    \Fn[z[z_n]]\rangle_{\vac}}
    {\left\langle\,\Qns[z[z_n]]
    \Qn[z[z_n]]\right\rangle_{\vac}}\right|
    &\leq&\epsilon_0\,n
    \label{PrpdHfFQQformpropcond13333}\\
    \left|
    \frac{\derp\left\langle
    \Qns[z[z_n]]\, \Qn[z[z_n]]
    \right\rangle_{\vac}}{\left\langle
    \Qns[z[z_n]]\Qn[z[z_n]]
    \right\rangle_{\vac}}\right|&\leq&\epsilon_0\,n
    \label{PrpdHfFQQformpropcond23333}\\
    \left|
    \frac{\left\langle
    Q^\sharp \,Q
    \right\rangle_{(k;n)}}{\left\langle
    \Qns[z[z_n]]\Qn[z[z_n]]
    \right\rangle_{\vac}}\right|&\leq&2
    \label{PrpdHfFQQformpropcond33333} \\
    \left\|(\derp Q[\h^{(k)}])
    U_\rho^*Q[\h^{(k+1)}]\right\|&\leq&C\,\ez
    \label{PrpdHfFQQformpropcond53333} \\
    Q^\sharp[\h^{(k)}]H[\h^{(k)}]Q[\h^{(k)}]&\leq&
    (1+\mu)\FchrtH[H[\h^{(k)}]]
    \label{PrpdHfFQQformpropcond43333}\eeqn
for $\beta=0,1$, and all $0\leq k\leq n$, and some constant
$\mu\leq\frac{3}{4}$.
\\

(II.1) To verify (~\ref{PrpdHfFQQformpropcond13333}), we observe
that
\eqnn\left| \langle\derp\partial_{H_f}
    \Fn[z[z_n]]\rangle_{\vac} \right|
    &=&\left|\derp\prod_{i=0}^{n}(1+\DHf^{(i)}[z_i])\right|
    \\
    &\leq&\left(\sum_{i=0}^n\left|\derp \DHf^{(i)}[z_i]
    \right|\right)\left|\prod_{i=0}^{n}
    (1+\DHf^{(i)}[z_i])\right|\\
    &=&\left(\sum_{i=0}^n\left|\derp \DHf^{(i)}[z_i]
    \right|\right)\left|\langle\partial_{H_f}
    \Fn[z[z_n]]\rangle_{\vac}\right|\;,\eeqnn
since $\DHf^{(i)}[z_i]\geq0$. Therefore,
\eqnn\rm{l.h.s.\;of\;}(~\ref{PrpdHfFQQformpropcond13333})
    &\leq&\left(\sum_{i=0}^n
    \left|\derp \DHf^{(i)}[z_i]
    \right|\right)\left|\dHfFQQ_n[z[z_n]]\right|\\
    &\leq&C\,g\,\epsilon_0\,n\,(1+O(\epsilon_0^2))\;,
    \eeqnn
since by Lemma {~\ref{DHfupperlowerbdlemma3}},
$$\left|\derp \DHf^{(i)}[z_i]
    \right|\;\leq\;O(g\,\epsilon_i)\;,$$
which implies the claim.
\\

(II.2) In the case of  (~\ref{PrpdHfFQQformpropcond23333}), we
have
\eqnn\derp\left\langle
    \Qns[z[z_n]]\, \Qn[z[z_n]]
    \right\rangle_{\vac}&=&\derp\left(\langle\partial_{H_f}
    \Fn[z[z_n]]\rangle_{\vac}\dHfFQQ_n^{-1}[z[z_n]]\right)\\
    &=&\dHfFQQ_n^{-1}[z[z_n]]\, \derp\langle\partial_{H_f}
    \Fn[z[z_n]]\rangle_{\vac}\\
    &&\hspace{1cm}-\;
    \langle\partial_{H_f}
    \Fn[z[z_n]]\rangle_{\vac}\,
    \dHfFQQ_n^{-2}[z[z_n]]\,\derp\dHfFQQ_n[z_n]\;,\eeqnn
by definition of $\dHfFQQ_n[z[z_n]]$. Therefore,
\eqnn\left|\rm{l.h.s.\;of\;}(~\ref{PrpdHfFQQformpropcond23333})
    \right|&=&\left|
    \dHfFQQ_n^{-1}[z[z_n]]\,\frac{\langle\derp\partial_{H_f}
    \Fn[z[z_n]]\rangle_{\vac}}
    {\left\langle\,\Qns[z[z_n]]
    \Qn[z[z_n]]\right\rangle_{\vac}}\right|\\
    &&\hspace{1cm}+\,
    \left|\dHfFQQ_n^{-1}[z[z_n]]\,\derp\dHfFQQ_n[z[z_n]]\right|\\
    &\leq&(1+O(\epsilon_0^2))C\, \epsilon_0^2\,n\,\\
    &&\hspace{1cm}+\,
    (1+O(\epsilon_0^2))C'\, \epsilon_0^2\;,
    \eeqnn
using the above estimate on the l.h.s. of
(~\ref{PrpdHfFQQformpropcond13333}).
\\

(II.3) Inequality  (~\ref{PrpdHfFQQformpropcond33333}) has already
been proved in the case $\beta=0$.
\\

(II.4) The bound (~\ref{PrpdHfFQQformpropcond53333}) follows from
the following argument. Since
$$\omega[\h]\;=\;a[z]\Ppar\,+\,O(\epsilon)\;,$$
(where $O(\epsilon)$ is with respect to the operator norm, of
course) we have, for $0\leq k<n$,
\eqnn&&\left\|(\derp Q[\h^{(k)}])
    U_\rho^*Q[\h^{(k+1)}]\right\|\;\leq\;O(\ez)\,\\
    &&\hspace{1cm}+\,|a^{(k)}[z_k]|\,|a^{(k+1)}[z_{k+1}]|
    \left\|\bchr(\derp \bar{R}[\h^{(k)}])\bchr\Ppar
    \chr
    U_\rho^*\bchr\bar{R}[\h^{(k+1)}]\bchr\Ppar\chr
    \right\| \\
    &&\hspace{1cm}+\,
    |\derp a^{(k)}[z_k]|\,|a^{(k+1)}[z_{k+1}]|\,
    \left\|\bchr\bar{R}[\h^{(k)}]\bchr\Ppar
    \chr
    U_\rho^*\bchr\bar{R}[\h^{(k+1)}]\bchr\Ppar\chr
    \right\| \;.\\
    \eeqnn
Expanding $\bar{R}=\bar{R}_0+O(\epsilon)$ according to the second
resolvent identity (~\ref{secondresident4444}), and using that
$\chr U_{\rho}\chr =0$, we arrive at the claim.
\\

(II.5) To prove (~\ref{PrpdHfFQQformpropcond43333}), we first of
all observe that for $\h\equiv\h^{(k)}$ and any $0\leq k<n$,
\eqnn &&Q^\sharp[\h]\,H[\h]\,Q[\h]\;=\;
    \chr^2 \,\FchrtH[H[\h]]\,\chr^2\,
    +\,\tau[H[\h]]\bchr^2\chr^2\;,
    \eeqnn
which is obtained from straightforward calculation. Let us then
find a sufficiently small number $\mu$ such that an estimate
\eqn\chr^2 \,\FchrtH[H[\h]]\,\chr^2\,
    +\,\tau[H[\h]]\bchr^2\chr^2\;\leq\;(1+\mu)\FchrtH[H[\h]]
    \,\label{mudefrelaux3333} \eeqn
holds, and show that one can pick $\mu\leq\frac{3}{4}$. For
$|p|\leq\puppbd$, and $\puppbd\in[\puppbdnum,1)$ sufficiently
small, there is a bound
\eqnn\FchrtH[H[\h]]&\geq&C_{|p|}\,\tau[H[\h]]\,+\,
    O(\sqrt{\epsilon}_0)\;,
    \eeqnn
for a constant
$$C_{|p|}\;\geq\;\frac{4}{9}\;,$$
since
$$(1+\DHf[z])H_f\,+\,(a[z]+\Delta a[z])\Ppar\;\geq\;(1-|p|)
    \tau[H[\h]]\,+\,O(\sqrt{\epsilon}_0)\;,$$
and
$$\|T_{n-l}[\h]\,\piop[\h]+\Delta T_{n-l}-\lTnl|\P_f|^2\|\;
    \leq\;\frac{1}{2} \;,$$
cf. (~\ref{Neumexp}) and Proposition {~\ref{Tnlrenhathprop4444}}
below, while
$$\|\piop[\h] W[\h]\piop[\h]+\Delta W\|\;\leq\;O(\epsilon)\;.$$
Thus,
\eqnn\tau[H[\h]]\bchr^2\chr^2\;\leq\;C_{|p|}^{-1}
    \bchr\chr \FchrtH[H[\h]]
    \bchr\chr\,+\,O(\sqrt{\epsilon}_0)\bchr^2\chr^2\; ,\eeqnn
and since
$$\bchr^2\chr^2 \;\geq\;\frac{1}{2}\tau[H[\h]]\bchr^2\chr^2\;,$$
it follows that
\eqnn\tau[H[\h]]\bchr^2\chr^2&\leq&(1+O(\sqrt{\epsilon}_0))C_{|p|}^{-1}
    \bchr\chr \FchrtH[H[\h]]
    \bchr\chr\;.\eeqnn
Since $\|\bchr\chr\|\leq\frac{1}{2}$, this implies that
\eqnn\tau[H[\h]]\bchr^2\chr^2&\leq&\frac{3}{4}
    \FchrtH[H[\h]]\;,\eeqnn
thus (~\ref{mudefrelaux3333}) holds for $\mu\leq\frac{3}{4}$.
\\

(III) By (I) and (II), Proposition  {~\ref{dHfFQQformprop333}}
implies that (~\ref{derpakzkauxform3333}) holds for $k=n$. This
proves the claim of the proposition. \qed

It is thus a trivial consequence of the above proposition that
\eqn\hat{a}[Z[z]]\;=\;-\;|p|(1\,+\,\hat{\delta}_a[Z[z]])\;,
    \label{hatazeq2}\eeqn
with
$$|\hat{\delta}_a[Z[z]]|\;\leq\;\sqrt{\epsilon}_0
    \;,$$
uniformly in $|p|$, for $|p|\leq\puppbd$, with
$\puppbd\in[\puppbdnum,1)$ small enough.
\\

\section{Renormalization of $T_{n-l}$}

The following proposition controls the behaviour of $T_{n-l}$
under $\ren$.
\\

\begin{prp}\label{Tnlrenhathprop4444}
Let $\rho=\frac{1}{2}$, and $\beta=0,1$. The renormalized
nonlinear part $\hat{T}_{n-l}\equiv T_{n-l}[\hat{\h}]$ of
$T[\hat{\h}]$ is $O(H_f^2)$ in the limit $H_f\rightarrow0$, and
satisfies
\eqn\left|
    \derp^\beta\partial_{\op}^\alpha\left(\ch_1\left(
    T_{n-l}[\hat{\h}]\,-\,\hlTnl\,|\P_f|^2\right)\ch_1\right)
    \right|&\leq&\hat{\epsilon}\,+\,
    \bTnl\,\bar{P}_1
    \;,\label{Tnlrenindstepbound1}
    \eeqn
for $0\leq|\alpha|\leq2$, $\beta=0,1$, where
\eqnn0\;\leq\;\hlTnl&\leq&\rho\,\lTnl\;,\\
    |\derp\hlTnl|\,,\,|\partial_z\hlTnl|&=&
    0\;.\eeqnn
The constant $\hbTnl<\infty$ is independent of $\hat{\epsilon}$.
In particular,
\eqn\left|
    \left(T_{n-l}[\hat{\h}]\,-\,\hlTnl|\P_f|^2\right)
    \right|&\leq&\hat{\epsilon}\,+\,\left(\frac{|p|+\hlTnl}{2}\,
    +\,O(\sqez)\right)\bar{P}_1
    \label{Tnlprpformest4444}\eeqn
and
\eqn\left|
    \partial_{H_f}
    T_{n-l}[\hat{\h}]
    \right|&\leq&O(\sqez)\;+\;(|p|\,+\,\hlTnl)\,\times\nonumber\\
    &&\;\times\,\left(2\,+\,
    3\,|\partial_{H_f}(\ch_1^2\bch_1^2)|\,
    +\,2(|p|+\hlTnl)\,|\partial_{H_f}
    \bch_1^2|\right)\;. \label{partHfTnlprpform4444}\eeqn
The quantity $\hat{\epsilon}$ is defined in Proposition
{~\ref{deropalphatildwMNprp1}} below.
\end{prp}

\prf

>From smooth Feshbach decimation, $\dec :T_{n-l}\rightarrow
T_{n-l}+\Delta T_{n-l}$, where $\Delta T_{n-l}$ is given as
follows. By Lemma {~\ref{piopdeflemma}},
\eqnn \decl\;:\;T_{n-l}[\h]&\longrightarrow&P_\rho T_{n-l}[\h]
    P_\rho
     \,+\,T'[\h]\,(\piop[\h]-\1)
     \,\\
    &&+\,(\Delta
     w_{0,0}-\Delta
     w_{0,0}|_{\op\rightarrow0}-\DHf[z]H_f-\Delta a[z]\Ppar)\\
     &=&P_\rho T_{n-l}[\h]P_\rho
     \,-\,\bchr^2 (T')^2\bar{R}_0\,\\
    &&+\,(\Delta
     w_{0,0}-\Delta
     w_{0,0}|_{\op\rightarrow0}-\DHf[z]H_f-\Delta a[z]
    \Ppar)\;\\
    &\equiv&P_\rho \left(T_{n-l}[\h]
    \,+\,\Delta T_{n-l}\right)P_\rho\eeqnn
on $P_\rho\Hp$, since
$$\piop[\h]\;=\;\1\,-\,\bchr T'[\h]\bchr\bar{R}_0[\h]\;,$$
as we recall. The contribution to $\Delta T_{n-l}$ from $\Delta
w_{0,0}$ can be estimated by
$$\frac{1}{2!}\;\sum_{|\alpha|=2}
    \|\,\partial_{\opT^r}^\alpha\,\Delta w_{0,0}\|
    H_f^2\;\leq\;O(\epsilon^2)\,H_f^2
    \;,$$
which is immediate from Lemma {~\ref{deropalphaDeltwMNlemma1}} and
Taylor's theorem.

It thus remains to bound the operator
\eqn T'[\h](\piop[\h]-\1)\;=\;-\,\bchr^2(T'[\h])^2\bar{R}_0[\h]\;,
    \label{Tnlindstepoverlcorr4444}\eeqn
together with its derivatives with respect to $\op$.

By (~\ref{Tnlindhypboudn4444}) of the induction hypothesis, it is
clear that
\eqnn&&\left|P_\rho\,\left(T_{n-l}[\h]\,-\,\lTnl|\P_f|^2\,+\,
    \Delta T_{n-l}
    \right)\,P_\rho\right|\\
    &&\hspace{2cm}\leq\;O(\epsilon)\;+\;
    \|P_\rho T'[\h]P_\rho \|\,|P_\rho(\piop[\h]-\1)P_\rho|
    \nonumber\\
    &&\hspace{2cm}\leq\;O(\epsilon)\,+\,\rho\,
    \frac{|p|+\hlTnl+O(\sqez)}{2}\bar{P}_\rho\;.
    \eeqnn
This is obtained by using Lemma {~\ref{piopderboundslemma4444}},
and the fact that
\eqnn\|P_\rho T'[\h]P_\rho \|&\leq&O(\epsilon)\,+\,
    |a[z]|\|P_\rho \Ppar P_\rho \|\,+\,\lTnl
    \|P_\rho |\P_f|^2 P_\rho \|\\
    &\leq&O(\sqez)\,+\,
    |p|\,\rho\,+\,\hlTnl\,\rho\;,\eeqnn
where by definition, $\hlTnl=\rho\lTnl$. Hence, rescaling by
$\resc=\rho^{-1}(1+\DHf[z])^{-1}Ad_{U_\rho}$ yields
(~\ref{Tnlprpformest4444}).
\\

Next, let us consider the derivative with respect to $H_f$. By
(~\ref{partHfTnlindhypboudn4444}) of the induction hypothesis,
\eqnn&&\left|\partial_{H_f}\left(\chr\left(
    T_{n-l}[\h]\,-\,\lTnl|\P_f|^2\,+\,
    \Delta T_{n-l}
    \right)\chr\right)\right|\\
    &&\hspace{1cm}\leq\;O(\sqez)\,+\,
    \|P_\rho (\partial_{H_f}T'[\h]) P_\rho\|\,\|\piop[\h]\|\\
    &&\hspace{1.5cm}+\,
    \|P_\rho T'[\h] P_\rho\|\,
    |\partial_{H_f}(\chr\piop[\h]\chr)|\\
    &&\hspace{1cm}\leq\;O(\sqez)\,+\,
    (|p|+\hlTnl)\,
    |\partial_{H_f}(\chr\piop[\h]\chr)|\;,\eeqnn
since $\|P_\rho (\partial_{H_f}T'[\h]) P_\rho\|\leq
O(\epsilon^2)$. Applying Lemma {~\ref{piopderboundslemma4444}},
and rescaling, one obtains the asserted bound.
\\

For general $|\alpha|+\beta\geq1$, the part in
$\derp^\beta\partial_{\op}^\alpha\Delta T_{n-l}$ stemming $\Delta
w_{0,0}$ has an operator norm bounded by $O(\epsilon^2)$, again by
Lemma {~\ref{deropalphaDeltwMNlemma1}} and Taylor's theorem. Since
\eqnn\|\derp^\beta\partial_{\op}^\alpha T[\h]\|&\leq&O(1)\;,\\
    \|\derp^\beta\partial_{\op}^\alpha W[\h]\|&\leq&
    O(\epsilon_0^\beta\epsilon^{1-\beta}\xi)\;,\\
    \|\derp^\beta\partial_{\op}^\alpha\ch_1^2\|&\leq&O(1)\;,\eeqnn
where the implicit constants only depend on $\rho$, $|p|$, it
follows straightforwardly that there exists a bound of the
asserted form (~\ref{Tnlrenindstepbound1}).   \qed

\section{Renormalization of the marginal kernels of
degree 1} The interaction kernels of degree $M+N=1$ are marginal,
but their their renormalization group flow is determined by the
marginal quantity $a[z]$ that has already been considered before.
This is due to the operator-theoretic formulation of the
Ward-Takahashi identities derived in Section
{~\ref{WTidentFeshbrensubsec}}.

We use the notation
$$\hat{w}_{M,N}\;=\;w_{M,N}[\hat{\h}]$$
for brevity, and recall the shorthand
\eqnn\hat{u}_{M,N}&\equiv&u_{M,N}[\hat{\h}]\;=\;
    \lim_{\stackrel{|k|\rightarrow0}{n_k\;{\rm fixed}}}
    w_{M,N}[\hat{\h}]\;,\eeqnn
from our previous discussion.
\\

\begin{prp}
The renormalized expression for $u_{M,N}[\h]$ is given by
\eqn\hat{u}_{M,N}[Z[z];n_k,\lambda]\;=\;-\;g\;\hat{a}[Z[z]]\;
    \epsilon_{M,N}^\parallel[n_k;\lambda]\;,\eeqn
where $\hat{a}[Z[z]]$ is the coefficient of $\Ppar$ in
$T_{lin}[\hat{\h}]$. Furthermore, picking $\rho=\frac{1}{2}$,
\eqn\nm\,\bar{P}_1^\perp[H_f]\left(
    \partial_{\Ppar}\hat{w}_{M,N} \right)
    \bar{P}_1^\perp[H_f+|k|]\,\nm
    &\leq&2\,g\,\hlTnl\,+\,\hat{\epsilon}^{\frac{3}{2}}\,\xi\;,
    \nonumber\\
    \nm\,\bar{P}_1^\perp[H_f]\left(
    \partial_{X}
    \hat{w}_{M,N} \right)\bar{P}_1^\perp[H_f+|k|]\,\nm
    &\leq&
    \hat{\epsilon}^{\frac{3}{2}}\,\xi\;.
    \eeqn
for  $X$ denoting either the absolute value $|k|$ of a photon
momentum, $H_f$, $\Pperp$ or $z$. The quantity $\hat{\epsilon}$ is
defined in Proposition {~\ref{deropalphatildwMNprp1}} below.
\end{prp}

\prf

(I) By induction hypothesis,
\eqn\;u_{M,N}[z;n_k,\lambda]\;=\;g\;a[z]\;
    \epsilon_{M,N}^\parallel[n_k;\lambda]\;.\eeqn
We claim that the correction of $u_{M,N}[z;n_k,\lambda]$ under
$\dec$ is given by \eqn\;\Delta u_{M,N}[z;n_k,\lambda]\;=\;-\;g\;
    \Delta a[z]\;\epsilon^\parallel_{M,N}[n_k,\lambda]\;,\eeqn
where $\Delta a[z]$ denotes its correction under Feshbach
decimation $\dec$.

From
\eqn u_{0,1}[z;n_k,\lambda]&=&
    \lim_{\stackrel{|k|\rightarrow0}{n_k\;{\rm fixed}}}\,
    \bcs^{-1}[|k|]\sqrt{|k|} \left[
    H[\h],a_\lambda^*(k)\right]\nonumber\\
    &=&
    -\;g\;\epsilon_{0,1}[n_k;\lambda]\cdot
    \partial_{\P_f}H[\h]
    \;,\eeqn
and
\eqn u_{1,0}[z;n_k,\lambda]&=&
    \lim_{\stackrel{|k|\rightarrow0}{n_k\;{\rm fixed}}}\,
    \bcs^{-1}[|k|]\sqrt{|k|}
    \left[a_\lambda(k),H[\h]\right]\nonumber\\
    &=&
    -\;g\;\epsilon_{1,0}[n_k;\lambda]\cdot
    \partial_{\P_f}H[\h]\;,
\eeqn as required by the induction hypothesis, follows that
\eqn\;(u_{0,1}+\Delta u_{0,1})[z;n_k,\lambda]
    &=&\lim_{\stackrel{|k|\rightarrow0}{n_k\;{\rm fixed}}}\,
    \bcs^{-1}[|k|]\sqrt{|k|} \left[
    \FchrtH[H[\h]],a_\lambda^*(k)\right]
    \nonumber\\
    &=&-\;g\;
    \epsilon^\parallel_{0,1}[n_k;\lambda]\;\left\langle
    \partial_{\Ppar}\FchrtH[H[\h]]\right\rangle
    \nonumber\\
    &=&-\;g\;(a[z]+\Delta a[z])\;
    \epsilon^\parallel_{0,1}[n_k;\lambda]\;\eeqn
due to Proposition {~\ref{WaTaidprop2}}, and likewise for
$u_{1,0}+\Delta u_{1,0}$. Thus, by definition of $\hat{a}[Z[z]]$,
\eqn\left(u_{0,1}+
    \Delta u_{0,1}\right)[z;n_k,\lambda]&=&-\,
    g\,\hat{a}[Z[z]]\,
    (1+\DHf[z])\;\epsilon^\parallel_{0,1}[n_k,\lambda]
    \;.\eeqn
Hence, application of the rescaling transformation $\resc$ yields
\eqn\hat{u}_{0,1}[Z[z];n_k,\lambda]&=&g\;
    \hat{a}[Z[z]]\;\epsilon^\parallel_{0,1}[n_k,\lambda]
    \;\eeqn
for the renormalized expression.
\\

(II) Next, for $|\alpha|=1$, and considering that
$\bar{P}_\rho^\perp\piop=\bar{P}_\rho^\perp$, we have
\eqnn&&\nm\,\bar{P}_1^\perp[H_f]\left(
    \partial_{Y}
    \hat{w}_{M,N} \right)\bar{P}_1^\perp[H_f+|k|]\,\nm\\
    &&\hspace{1cm}\leq\;\rho\,
    (1+\DHf[z])^{-1}\nm\bar{P}_\rho^\perp[H_f]
    \left(\partial_{Y} \left(
    \piop[z;\op]w_{M,N}\piop[z;\sh_k\op]\right)\right)
    \bar{P}_\rho^\perp[H_f+|k|]\nm
    \,\nonumber\\
    &&\hspace{1.5cm}+\,(1+\DHf[z])^{-1}\nm\partial_{Y}
    \Delta w_{M,N}\nm\\
    &&\hspace{1cm}\leq\;\rho\,
    (1+\DHf[z])^{-1}\nm\bar{P}_\rho^\perp[H_f]
    \left(\partial_{Y}  w_{M,N}\right)
    \bar{P}_\rho^\perp[H_f+|k|]\nm
    \,\nonumber\\
    &&\hspace{1.5cm}+\,(1+\DHf[z])^{-1}\nm\partial_{Y}
    \Delta w_{M,N}\nm\\
    &&\hspace{1cm}\leq\;\rho\,\nm\bar{P}_\rho^\perp[H_f]
    \left(\partial_{Y}  w_{M,N}\right)
    \bar{P}_\rho^\perp[H_f+|k|]\nm\,+\,
    O(\epsilon^{2}\xi)\;,
    \eeqnn
by unitarity of $Ad_{U_\rho}$, and for $Y$ denoting $\Ppar$ or any
of the quantities accounted for by $X$. Inserting
(~\ref{derpalphwMNminuMNindhypbd3}) of the induction hypothesis,
one immediately arrives at the claim. \qed

\section{Renormalization of kernels of degree 1 and higher}
\label{RenwMNsubsect2}

The following proposition controls general properties of the
renormalized kernels
$$\hat{w}_{M,N}\;=\;w_{M,N}[\hat{\h}]\;,$$
for all $M+N\geq1$.
\\

\begin{prp}
\label{deropalphatildwMNprp1} Let $X$ stand for $\hat{z}=Z[z]$, or
$|k_r|,|\tilde{k}_s|$. Moreover, let $\rho=\frac{1}{2}$. Then, the
renormalized integral kernels $\{\hat{w}_{M,N}\}$, for $M+N\geq1$,
satisfy
\eqn\nm \hat{w}_{M,N}\nm&\leq&\frac{16}{9}\,g\,|p|\,(1+\sqez)
    +\,\frac{32}{9}\,g\,\hlTnl\,+\,O(\hat{\epsilon}^{\frac{3}{2}}\xi)
    \nonumber\\
    &\leq&\hat{\epsilon}\,\xi\;,
    \label{indstepwMNmainprp1aux4444}\eeqn
and generally, for $0\leq|\alpha|\leq2$, $0\leq\beta\leq1$,
\eqn\nm\,\derp\,
    \partial_{\op}^\alpha \,\hat{w}_{M,N}\,\nm&\leq&
    (2\ez)^\beta\,\epsilon^{1-\beta}\,\xi\;,\nonumber\\
    \nm\,\derp\,
    \partial_{X} \,\hat{w}_{M,N}\,\nm&\leq&
    (2\ez)^\beta\,\epsilon^{1-\beta}\,\xi\;,\;\;\; M+N=1\;,
    \label{indstepwMNmainprp1}
    \eeqn
while for  $M+N\geq2$,
\eqn\nm\,\derp^\beta\,\partial_{\op}^\alpha\,
    \hat{w}_{M,N}\,\nm
    &\leq&
    (2\ez)^\beta\,\hat{\epsilon}^{\frac{7}{4}
    -\frac{|\alpha|}{4}-\beta}\,\xi^{M+N}\;,
    \label{indstepwMNmainprp2}\\
    \nm\,\derp^\beta\,\partial_X\,
    \hat{w}_{M,N}\,\nm&\leq&
    (2\ez)^\beta
    \,\hat{\epsilon}^{\frac{3}{2}
    -\beta}\,\xi^{M+N}\;.
    \label{indstepwMNmainprp3}\eeqn
All bounds are uniform in $|p|$, for $|p|\leq\puppbd$. The
renormalized value of $\epsilon$ is given by
\eqn\hat{\epsilon}\;=\;\left\{
    \begin{aligned}
    \max\left\lbrace\epfrac\;\epsilon\;,\;
    \ezfac\,(1+\sqrt{\epsilon}_0)\,|p|\,\ez\right\rbrace
    &\;\;\;\;{\rm if }\;\sigma\leq1\\
    \epfrac\;\epsilon\hspace{2cm}&\;\;\;\;{\rm if }\;
    \sigma>1\;\;,\end{aligned}\right.\;,
    \label{indstepwMNmainprp6}\eeqn
with $\hat{\epsilon}<2\epsilon_0$. The running infrared cutoff
$\sigma$ is renormalized via
\eqn\sigma\;\longrightarrow\;\hat{\sigma}\;=\;\frac{\sigma}{\rho}
    \;.\eeqn
In particular, $\hat{w}_{M,N}$ are, for all $M+N\geq1$, {\bf
analytic} in the components of $\P_f$.
\end{prp}

\prf We recall, to begin with, that
\eqnn \hat{w}_{M,N}&=&\frac{\rho^{M+N-1}}{1+\DHf[z]}
    Ad_{U_\rho}\left[\piop[z;\sh^{(M)}\op]\,w_{M,N}[\h]\,
    \piop[z;\sh^{(N)}\op]\,\right.\\
    &&\hspace{1cm}\left.+\,
    \Delta w_{M,N}[\op;z;K^{(M,N)}]\right]\;,\eeqnn
where $\sh^{(M)}$ shifts $\op$ by $(k_1,\dots,k_M)$, that is,
\eqnn\sh^{(M)}H_f\,=\,H_f\,+\,\sum_{i=1}^M |k_i|\;\;\;,\;\;\;
    \sh^{(M)}\P_f\,=\,\P_f\,+\,\sum_{i=1}^M k_i\;,\eeqnn
and likewise for $\sh^{(N)}$. Furhermore, we recall the induction
hypotheses on $w_{M,N}$ of Section {~\ref{wMNindhypsubsect3}}.
\\

(I) Let us first consider (~\ref{indstepwMNmainprp1}). For the
cases $M+N\geq2$, we have \eqn\nm\,\hat{w}_{M,N}\,\nm
    &\leq&\rho\,\nm \,w_{M,N}\,\nm\,+\,
    O(\epsilon^{2 })\,
    \xi^{M+N}\nonumber\\
    &\leq&\left(\frac{1}{2}\,+\,O(\epsilon^{\frac{1}{4}})
    \right)\,\epsilon^{\frac{7}{4}}\,\xi^{M+N}\nonumber\\
    &\leq&\hat{\epsilon}^{\frac{7}{4}}\,\xi^{M+N}\;,
    \eeqn
using the induction hypothesis $\nm w_{M,N}\nm\leq
\epsilon^{\frac{7}{4}}\,\xi^{M+N}$, and for $\hat{\epsilon}$ as
defined in (~\ref{indstepwMNmainprp6}).

For the case $M+N=1$,
\eqn\nm\,\hat{w}_{M,N}\,\nm&\leq&(1+\DHf[z])^{-1}
    \left(\nm Ad_{U_\rho}\left[\piop[z;\sh_k\op]
    u_{M,N}\piop[z;\op]\right]\nm\,\right.\nonumber\\
    &&+\,
    \nm Ad_{U_\rho}\left[\piop[z;\sh_k\op]
    (w_{M,N}-u_{M,N})\piop[z;\op]\right]\nm\,\nonumber\\
    &&+\,
    \left.\nm\Delta w_{M,N}\nm\right)\nonumber\\
    &\leq&\|\piop\|^2
    \nm u_{M,N}\nm\,\nonumber\\
    &&+\,\rho\,\|\piop\|^2
    \left(\nm P_\rho(\partial_{\Ppar} w_{M,N})P_\rho\nm\,+\,
    \sum_{Y=|k|,H_f,\Pperp}
    \nm P_\rho(\partial_Y w_{M,N})P_\rho\nm\right)
    \,\nonumber\\
    &&+\,O(\epsilon^{2}\xi)\nonumber\\
    &\leq&\frac{16}{9}(\nm u_{M,N}\nm\,+\,2\,g\,\hlTnl)\,
    +\,O(\epsilon^{\frac{3}{2}}\xi)
    \nonumber\\
    &=&\frac{16}{9}\,g\,(|a[Z^{-1}[\hat{z}]]|
    +\,2\,\hlTnl)\,+\,O(\epsilon^{\frac{3}{2}}\xi)\;,
    \label{hatepsrenest4444}
\eeqn
where we have used Lemma {~\ref{deropalphaDeltwMNlemma1}} and
(~\ref{wMN1epsxibdindhyp32}) of the induction hypothesis, together
with the fact that $\|\piop\|\leq\frac{4}{3}$, by Lemma
{~\ref{piopderboundslemma4444}}. This implies
(~\ref{indstepwMNmainprp1aux4444}). In particular, since
$\hlTnl=\rho\lTnl\leq\frac{1}{4}$, this is bounded by
\eqnn\nm\,\hat{w}_{M,N}\,\nm&\leq&\frac{16}{9}\,g\,|p|\,(1+O(\sqez))\,+\,
    \frac{8}{9}\,g\,+\,
    O(\epsilon^{\frac{3}{2}}\xi)\nonumber\\
    &\leq&2\,g\,\nonumber\\
    &=&\ez\,\xi\;,
    \eeqnn
for $|p|\leq\puppbd$ with $\puppbd\in[\puppbdnum,1)$.

Inequality (~\ref{hatepsrenest4444}) implies that if we choose
\eqnn\hat{\epsilon}&=& \max\left\{\epfrac\,\epsilon\,,\,
    \ezfac(1+\sqrt{\epsilon}_0)\,|p|\,g
    \right\}\;,\eeqnn
it is always true that
\eqnn\hat{\epsilon}\;\geq\;
    \ezfac\nm\hat{u}_{M,N}\nm\;.\eeqnn
Furthermore, it follows from a straightforward calculation that
under the induction assumption
$$\frac{16}{9}\,g\,|p|\,(1+\sqez)
    +\,\frac{32}{9}\,g\,\lTnl\,+\,O(\epsilon^{\frac{3}{2}}\xi)
    \;\leq\;\epsilon\,\xi\;,$$
the estimate
$$\frac{16}{9}\,g\,|p|\,(1+\sqez)
    +\,\frac{32}{9}\,g\,\hlTnl\,+\,O(\hat{\epsilon}^{\frac{3}{2}}\xi)
    \;\leq\;\hat{\epsilon}\,\xi$$
holds for this definition of $\hat{\epsilon}$. Therefore, we
conclude that
\eqnn\nm\hat{w}_{M,N}\nm\;\leq\;\hat{\epsilon}\,\xi\;,\eeqnn
with $\hat{\epsilon}\leq\ez$.
\\

(II) For (~\ref{indstepwMNmainprp2}), we first consider the case
$M+N=1$, $|\alpha|=0$ and $\beta=1$. We have
\eqn\nm\,\derp\,\hat{w}_{M,N}\,\nm&\leq&
    (1+\DHf[z])^{-1}\left(\frac{|\derp\DHf[z]|}{1+\DHf[z]}\,
     \nm \hat{w}_{M,N}\nm\,\right.\nonumber\\
    &&\hspace{1cm}+\,\underbrace{
    2
    \|\derp\piop\|\,\|\piop\|}_{\leq\;\frac{16}{3}\;
    ,\;{\rm cf.\;Lemma\;} {~\ref{piopderboundslemma4444}}.}\,
    \nm u_{M,N}\nm\,\nonumber\\
    &&\hspace{1cm}+\,\|\piop\|^2
    \left(\underbrace{\nm\,\derp\,
    u_{M,N}\,\nm}_{\leq\;1+\sqez}
    \,+\,\nm\,P_\rho(\derp\,\partial_{|k|}
    w_{M,N})P_\rho\,\nm\,\right)\nonumber\\
    &&\hspace{1cm}+\,\left.\|\piop\|^2
    \sum_{|\alpha|=1}\nm\,P_\rho(\derp\,\partial_{\op}^\alpha\,
    w_{M,N})P_\rho\,\nm\,+\,
    \nm\derp\Delta w_{M,N}\nm\right)\nonumber\\
    &\leq&\frac{16}{3}\,|p|\,g\,+\,\frac{16}{9}\,g\,(1+O(\sqez))\,
    +\,O(\ez\epsilon^{\frac{1}{2}}\xi) \nonumber\\
    &\leq&2\,\ez\,
    \xi\;,\eeqn
for $|p|\leq\puppbd$, with $\puppbd\in[\puppbdnum,1)$ sufficiently
small, where we used $2g=\epsilon_0\xi$, cf.
(~\ref{indstepwMNmainprp3}), and Lemma
{~\ref{piopderboundslemma4444}}.

Next, we consider the cases $M+N\geq2$ and $0\leq|\alpha|\leq2$,
$0\leq\beta\leq1$.
\\
Here, we have
\eqn\nm\;\derp^\beta
    \partial^\alpha_{\op}\hat{w}_{M,N}\;\nm
    &\leq&
    \rho^{M+N-1+|\alpha|}\,\sum_{\alpha'+\alpha''=\alpha}
    \left(\begin{array}{c}|\alpha|\\|\alpha'|,
    |\alpha'' |\end{array}\right)\,\times
    \nonumber\\
    &&\hspace{1cm}\times\,\left(
    2^\beta\|\derp^\beta
    \partial_\op^{\alpha'}\piop\| \,
    \|\partial_\op^{\alpha''}\piop\|\,
    \nm
    \partial_\op^{\alpha-\alpha'-\alpha''}\,
    w_{M,N}\nm\;\right.\nonumber\\
    &&\hspace{2cm}+\,\left.
    \|\partial_\op^{\alpha'}\piop\| \,
    \|\partial_\op^{\alpha''}\piop\|\,
    \nm\derp^\beta
    \partial_\op^{\alpha-\alpha'-\alpha''}\,
    w_{M,N}\nm\;\right)\nonumber\\
    &&+\;\ez^\beta\,C_{\alpha,\beta}\,
    \epsilon^{2-\beta}\,(2\xi)^{M+N}\nonumber\\
    &\leq&
    \rho\,\|\piop\|^2\,\nm\partial_{|p|}
    \partial^\alpha_{\op} w_{M,N}\nm\,+\,g^\beta
    \,
    C\,\epsilon^{\frac{7}{4}-\frac{|\alpha|-1}{4}-\beta}\,
    \xi^{M+N}\nonumber\\
    &\leq&\ez^\beta
    \,\left(\frac{3}{4}\,+\,
    O(\epsilon^{\frac{1}{4}})\right)\,
    \epsilon^{\frac{7}{4}-\frac{|\alpha|}{4}-\beta}
    \,\xi^{M+N}\nonumber\\
    &\leq&\ez^\beta
    \left(\frac{4}{5}\right)^{\frac{7}{4}-\frac{|\alpha|}{4}-\beta}\,
    \epsilon^{\frac{7}{4}-\frac{|\alpha|}{4}-\beta}\,
    \xi^{M+N}\nonumber\\
    &\leq&(2\ez)^\beta\,
    \hat{\epsilon}^{\frac{7}{4}-\frac{|\alpha|}{4}-\beta}
    \,\xi^{M+N}\;,
    \eeqn
This follows from the fact that for $|p|\leq\puppbd$,
$$\|\piop\|\;\leq\;\frac{4}{3}\;,$$
the induction hypotheses on $\derp\partial_\op^\alpha w_{M,N}$,
and $|\derp\DHf[z]|\leq O(g\,\epsilon)$.

Likewise, for (~\ref{indstepwMNmainprp3}) and the same cases (i),
(ii) as above,
\eqn\nm\;\derp^\beta
    \partial_X\hat{w}_{M,N}\;\nm
    &\leq&\rho\,\|\piop\|^2\,\nm\derp^\beta
    \partial_X w_{M,N}\nm\,+\,\ez^\beta\,
    \,
    C\,\epsilon^{\frac{3}{2}}\,
    \xi^{M+N}\nonumber\\
    &\leq&(2\ez)^\beta\,
    \,\left(\frac{8}{9}\,+\,
    O(\epsilon^{\frac{1}{4}})\right)\,
    \epsilon^{\frac{5}{4}}
    \,\xi^{M+N}\nonumber\\
    &\leq&(2\ez)^\beta\,
    \,\left(\frac{8}{9}\,+\,
    O(\epsilon^{\frac{1}{4}})\right)\,
    \epsilon^{\frac{5}{4}}\,\xi^{M+N}\nonumber\\
    &\leq&(2\ez)^\beta\,
    \hat{\epsilon}^{\frac{5}{4}}\,\xi^{M+N}\eeqn
is obtained in the same manner.
\\

(VI) Finally, analyticity of $\hat{w}_{M,N}$ in $\P_f$ follows
from the analyticity in $\P_f$ of $w_{M,N}$, cf. the induction
hypothesis, and of $\Delta w_{M,N}$, cf. Lemma
{~\ref{deropalphaDeltwMNlemma1}}. \qed

{\bf Proof of Theorem {~\ref{indstepmainthm3}}.} Theorem
{~\ref{indstepmainthm3}} is a summary of the statements of the
propositions of this chapters. \qed

\chapter{THE RENORMALIZATION GROUP FLOW AND PROOF
OF THE MAIN THEOREMS} \label{RGfloweqssect3}

The main insights gained in the two previous   chapters can be
summarized as follows. There is a point
\eqn\h^{(0)}\;\in\;\Polyd_{2\epsilon_0,\xi}\;,\eeqn whose
associated effective hamiltonian $H[\h^{(0)}]$ in $\Bound(P_1\Hp)$
is, in the Feshbach sense, isospectral to the model hamiltonian
$\Hps$.

Iteration of $\ren$ produces an orbit $\{\h^{(n)}\}_{n\in\N_0}$ in
$\Polyd_{2\epsilon_0,\xi}$ which emanates from $\h^{(0)}$.
Moreover, it establishes a map
\eqnn\ren\;:\;\Polyd_{\epsilon,\xi}&\longrightarrow&
    \Polyd_{\hat{\epsilon},\xi} \;,\eeqnn
where \eqnn\epsilon\,,\,\hat{\epsilon}\;\leq\;2\epsilon_0\;
,\eeqnn and where $\hat{\epsilon}\leq\epsilon$ is the renormalized
value for $\epsilon$. $\xi$ is a fixed constant, while $\epsilon$
possesses the r\^ole of the running coupling constant.

In the systems considered in \cite{bfs1,bfs2}, the interactions
are irrelevant, such that the sequence $\{\epsilon_n\}_{n\in\N_0}$
(with $\hat{\epsilon}_n=\epsilon_{n+1}$) tends to zero. Hence,
there is a sequence of nested polydiscs
$$\cdots\supseteq\;\Polyd_{\epsilon_n,\xi}\;\supseteq\;
    \Polyd_{\epsilon_{n+1},\xi}\;\supseteq\;\cdots$$
which converges, in the topology induced by the norm on
$\Polyd_{2\epsilon_0,\xi}$ defined in Chapter
{~\ref{rgdefsection}}, to a 1-dimensional polydisc that
parametrizes non-interacting effective hamiltonians. In
particular, this shows that $\ren$ is a contraction on
$\Polyd_{2\epsilon_0,\xi}$, and  it is proved in \cite{bfs1,bfs2}
by the Banach fixed point theorem that the 1-dimensional limiting
element $\Polyd_{0,\xi}$ is the unique fixed point set.

The interactions within the model hamiltonian $\Hps$ are also
irrelevant, but only so due to the introduction of the artificial
infrared regularization at $\ssig>0$. The limiting element
$\Polyd_{0,\xi}$ in this case is 2-dimensional, and parametrizes
operators of the form $\zeta_1 H_f+\zeta_2 \Ppar$, with
$\zeta_i\in\R$. In the limit $\ssig\rightarrow0$, where no
regularization is included, the corresponding sequence
$\{\epsilon_n\}_{n\in\N_0}$ does not tend to zero (unless $|p|=0$,
as we will see), since the interaction is strongly marginal. There
are very strong reasons to expect that
$\h^{(\infty)}:=\lim_{n\rightarrow\infty}\h^{(n)}$ exists, and
that the set of all such limiting points cover a 3-dimensional
center manifold in the polydisc $\Polyd_{2\epsilon_0,\xi}$.
However, the proof of these statements requires dynamical systems
theory on Banach spaces, and can presumably not be established by
a straightforward application of the Banach fixed point theorem.
Clarification of these matters is beyond the scope of the present
work.
\\

\section{Solution of the flow equations}

The cumulative result of the previous two chapters is the system
of flow equations generated by $\ren$ on the polydisc
$\Polyd_{\epsilon_0,\xi}\subset\Hspace$, given by Theorem
{~\ref{indstepmainthm3}}, together with the initial condition
$\h^{(0)}\in \Polyd_{\epsilon_0,\xi}$, provided by Theorem
{~\ref{firstdecstepmainthm3}}. The orbit under $\ren$ that
emanates from $\h^{(0)}$ is analytically controlled by the flow of
the parameters $\epsilon$ and  $a[z]$.
\\

\begin{thm}\label{definitergflowthm}
Let $\rho=\frac{1}{2}$, and
$$\Ns\;=\;\left\lceil\frac{\log\ssig}{\log\rho}
    \right\rceil\;.$$
Let $\{\h^{(n)}\}\in\Polyd_{\epsilon_0,\xi}$ denote the orbit of
$\h^{(0)}$ obtained from iterating the renormalization map, that
is,
$$\h^{(n+1)}\;=\;\ren[\h^{(n)}]\;\;,\;\;{\rm for}\;n\,\geq\,0
    \;,$$
and let the initial spectral parameter in $\h^{(0)}$ satisfy
$$z_0\;\in\; [\Eb-g,\Eb]\;.$$
Then, there is a constant $\puppbd\in[\puppbdnum,1)$, such that
for all $p$ with $|p|\,\leq\,\puppbd$, the running coupling
constant $\epsilon_n$ is given by
\eqn\epsilon_n\;=\;\left\{\begin{aligned}\max
    \left\{\left(\epfrac\right)^n\epsilon_0\,,\,
    \ezfac\,(1+\sqez)\,|p|\,\epsilon_0\,\right\}
    \hspace{1cm}& n\leq\Ns\\
    \left(\epfrac\right)^{n-\Ns}\,
    \epsilon_{\Ns}\hspace{3cm}&n>\Ns
    \;,\end{aligned}\right.\; \eeqn
such that $\epsilon_n\leq2\epsilon_0$ for all $n$. The coefficient
$a_n[z_n]$ of the operator $\Ppar$ in $T_{lin}[\h^{(n)}]$
satisfies
\eqn\left|\,\derp^\beta\left(a_n[z_n]\,+\,|p|\right)\,\right|&
    \leq&\sqez
    \; \label{partialzanznthm3}\eeqn
for $\beta=0,1$, hence
\eqn a_n[z_n]\;=\;-\,|p|(1\,+\,O\left(\sqez\right))
    \label{derpbetaanminpauxform3}\eeqn
for all $n\in\N_0$.
\end{thm}

\prf We have
\eqn\epsilon_n\;\leq\;\max
    \left\{\left(\epfrac\right)^n\epsilon_0\,,\,
    \ezfac\,(1+\sqez)\,|p|\,\epsilon_0\,\right\}\;,\eeqn
which follows straightforwardly from
\eqnn\hat{\epsilon}\;=\;\left\{
    \begin{aligned}
    \max\left\lbrace\epfrac\;\epsilon\;,\;
    \ezfac\,(1+\sqez)\,|p|\,\epsilon_0\right\rbrace
    &\;\;\;\;{\rm if }\;\sigma\leq1\\
    \epfrac\;\epsilon\hspace{2cm}&\;\;\;\;{\rm if }\;
    \sigma>1\;\;,\;\end{aligned}\right.\eeqnn
as stated in Theorem {~\ref{indstepmainthm3}}.

The statements about $a_n[z_n]$ have also been proved there. \qed

\section{Proof of the main theorems} \label{Proofmainthmssect3}

Using the results of the operator-theoretic renormalization group
analysis for the orbit $\{\h^{(n)}\}$, Theorems {~\ref{mainthm1}}
and {~\ref{mainthm2}} from Chapter {~\ref{statmainthmsect}} can
now be proved. Let us briefly recall their statements.
\\

\subsection{ Theorem {~\ref{mainthm1}}} It asserts that
for every arbitrarily small, but fixed value of the infrared
cutoff $\ssig>0$, there exists a constant
$\puppbd\in[\puppbdnum,1)$, such that for all $p$ with
$|p|\,\leq\,\puppbd$, the physical hamiltonian $\Hps$ possesses an
eigenvalue at the bottom of its spectrum, bordering to absolutely
continuous spectrum, which corresponds to a unique eigenvector
$\Omgrd\in\Hp$. Under the normalization condition
$$\left\langle\,\vac\,,\,\Omgrd\,\right\rangle\;=\;1\;,$$
there are constants $c,C>0$ and $c',C'>0$, such that
$$c\,\exp\left[
    c'\,g^2\,|p|^2\,|\log\ssig|\right]\;\leq\;
    \left\|\,\Omgrd\,\right\|^2\;\leq\;C\,\exp\left[
    C'\,g^2\,|p|^2\,|\log\ssig|\right]\;.$$
Thus, $\Omgrd\in\Hp$ for all $\ssig\geq0$ if and only if $|p|=0$.
\\

\subsection{Theorem {~\ref{mainthm2}}}
It asserts that the model hamiltonian $H_p(\ssig)$ possesses a
ground state eigenvalue $\Egrd$ at the bottom of its spectrum,
bordering to absolutely continuous spectrum. It is at least of
class $C^2$ with respect to $|p|$, and \eqnn\derp^\beta
    \left(\Egrd\;-\;\frac{|p|^2}{2}\right)\;
    \leq\;O(\epsilon_0 )\;\eeqnn
holds for $\beta=0,1$, {\bf uniformly} in $\ssig\geq0$.
Furthermore,
$$\,\derp^2\left(\Egrd\,-\,\frac{|p|^2}{2}\right)\;\leq\;0\;,$$
that is, the renormalized mass of the infraparticle is bounded
from below by the bare mass of the charged particle, for all
$\ssig\geq0$. This theorem thus controls smoothness properties of
the ground state energy under removal of the infrared
regularization.
\\

\section{Proof of Theorem {~\ref{mainthm1}}}

We have seen in the previous chapters that the ratio
\eqnn\dHfFQQ_n[z_n]\;=\;\frac{\left\langle \,
    \partial_{H_f}\Fn[z]
    \right\rangle_{\vac}}{\left\langle \,
    \Qns[z]\,\Qn[z]
    \right\rangle_{\vac}}\; \eeqnn
is given by
$$\dHfFQQ_n[z_n]\;=\;1\,+\,O(\epsilon_n^2)\;,$$
where $\lim_{n\rightarrow\infty}\epsilon_n=0$. Since
$$\Omgrd\;=\;s-\lim_{n\rightarrow\infty}\Qn[z]\vac\;,$$
it thus follows that
\eqn\left\|\Omgrd\right\|^2&=&
    \lim_{n\rightarrow\infty}\left\langle\,
    \partial_{H_f}\Fn[z]
    \right\rangle_{\vac}\nonumber\\
    &=&\left(
    \prod_{i=1}^{\Ns}(1+\DHf^{(i)}[E_{0,i}])\right)
    \left(
    \prod_{i=\Ns+1}^{\infty}(1+\DHf^{(i)}[E_{0,i}])\right)
    \nonumber\\
    &\leq&\exp\left[\sum_{i=1}^{\Ns} \DHf^{(i)}[E_{0,i}] \right]
    \exp\left[\sum_{i=\Ns+1}^{\infty}\DHf^{(i)}[E_{0,i}] \right]
    \;,\eeqn
while
$$\left\langle\vac\,,\,\Omgrd\right\rangle_{\vac}\;=\;1\;,$$
because
\eqnn\left\langle\vac\,,\,\Qn[z]\vac\right\rangle_{\vac}
    &=&\left\langle\vac\,,\,\vac\right\rangle_{\vac}\\
    &=&1\eeqnn
for all $n$.

The upper bound in the assertion is easy to obtain. It follows
from
\eqn\left\|\,\Omega_0\,\right\|\;\leq\;\exp\left(
    O\left(\epsilon_n^{2}\,\xi^2\right)\,
    \sum_{j=0}^{\Ns}1\right)\,
    \exp\left[\sum_{j=\Ns+1}^{\infty}
    \left(\epfrac\right)^j
    \,\epsilon_0 \right]\;,\eeqn
since
\eqn\epsilon_n\;\leq\;\max\left\{\left(\epfrac\right)^n
    \,\epsilon_0\,,\,\ezfac\,(1+\sqez)\,|p|\,
    \epsilon_0\,\right\}\;,\eeqn
as has been proved in the renormalization group analysis.

To prove the lower bound, we observe that
\eqnn\prod_{i=1}^{\Ns}(1+\DHf^{(i)}[E_{0,i}])
    &\geq&
    \exp\left[\frac{1}{20}\,g^2\,|p|^2\,\Ns\right]\,\times
    \nonumber\\
    &&\;\times\,\exp\left[
    \sum_{j=0}^{\Ns}\left(\lTnl^{(j)}
    \,O(\epsilon_j^2\xi^2)
    \,+\,O\left(\epsilon_j^{\frac{5}{2} }\,\xi^2
    \right)\right)\right]\;.\eeqnn
Since
$$\epsilon_n\;\leq\;\max\left\{\left(\epfrac\right)^n
    \epsilon_0\,,\,\ezfac\,(1+\sqez)\,g\,|p|\right\}\;,$$
for $n\leq\Ns$, we have
$$\epsilon_n^\gamma\;\leq\;\left(\epfrac\right)^{n\gamma}
    \epsilon_0^\gamma\,+\,
    \left(\ezfac\,(1+\sqez)\,g\,|p|\right)^\gamma\;$$
for all $\gamma>0$, and
\eqn\frac{1}{20}\,|p|^2\,g^2\,-\,
    O\left((g\,|p|)^{\frac{5}{2} }\right)\;
    \geq\;\frac{|p|^2\,g^2}{22}\eeqn
for $g$ sufficiently small. Furthermore,
\eqnn\prod_{i=\Ns+1}^{\infty}
    (1+\DHf^{(i)}[E_{0,i}])&\geq&
    \exp\left[-\,\sum_{j=\Ns+1}^{\infty}
    \left(\epfrac\right)^j
    \,\epsilon_0 \right]\\
    &=&1\,+\,O(\epsilon_0)\;.\eeqnn
Therefore,
\eqn\left\|\,\Omega_0\,\right\|&\geq&(1+O(\epsilon_0))\,
    \exp\left[\frac{1}{22}\,g^2\,|p|^2 \,\Ns\right]
    \,\times\nonumber\\
    &&\hspace{1cm}\times\,
    \exp\left[O\left(\epsilon_0^{\frac{5}{2} }\,\xi^2\right)
    \sum_{j=0}^{\Ns}
    \left(\epfrac\right)^{j}
    \right]\,\times\nonumber\\
    &&\hspace{1cm}\times\,
    \exp\left[O\left(\epsilon_0^{2}\,\xi^2\right)
    \sum_{j=0}^{\Ns }
    \left(\frac{1}{2}\right)^{j}\lTnl^{(0)}\right]
    \nonumber\\
    &\geq&\frac{1}{2}\,
    \exp\left[\frac{1}{22}\,g^2\,|p|^2 \,\Ns\right]
    \;,\eeqn
since $\lTnl^{(j)}=\lTnl^{(0)}\rho^j$, where $\rho=\frac{1}{2}$.
With
\eqnn\Ns&=&\left\lceil\frac{\log\ssig}{\log\rho}\right\rceil\\
    &\geq&\frac{|\log\ssig|}{2}\;,\eeqnn
the assertion of Theorem {~\ref{mainthm1}} follows. \qed

$\;$

\section{Proof of Theorem {~\ref{mainthm2}}}

Let us consider once more the eigenvalue equation for the ground
state of $\Hps$,
\eqn\left(\Hps-\Egrd\right)\Omgrd
    \;=\;0\;,\eeqn
where
\eqn\Omgrd\;=\;s-\lim_{n\rightarrow\infty}
    Q_{\ch_n,\tau_n}[\Hps-E_0]\,\vac\;.
    \eeqn
Thus, taking two derivatives with respect to $|p|$,
\eqn\derp^2\Egrd&=&\lim_{n\rightarrow\infty}
    \frac{\left\langle \Qn^\sharp
    [E_0]\,,\,\left(\derp^2
    \Hps\right)\,\Qn[E_0]\,\right\rangle_{\vac}}
    {\left\langle \Qn^\sharp
    [E_0]\,
    \Qn[E_0]\right\rangle_{\vac}}
    \nonumber\\
    &&-\;2\,\lim_{n\rightarrow\infty}
    \frac{\left\langle\;(\derp \Qn^\sharp
    [E_0]\,
    \left(\Hps-E_0\right)\;
    \derp \Qn[E_0]\right\rangle_{\vac}}
    {\left\langle \Qn^\sharp
    [E_0]\,
    \Qn[E_0]\right\rangle_{\vac}}\nonumber\\
    &=&1\;-\;2\,\lim_{n\rightarrow\infty}
    \frac{\left\langle\;(\derp \Qn^\sharp
    [E_0]\,
    \left(\Hps-E_0\right)\;
    \derp \Qn[E_0]\right\rangle_{\vac}}
    {\left\langle \Qn^\sharp
    [E_0]\,
    \Qn[E_0]\right\rangle_{\vac}}\;,
    \label{secderE0prime1}\eeqn
where
$$\Qn^{(\sharp)}[E_0]\;\equiv\;Q_{\ch_{\rho^n},\tau_n}^{(\sharp)}
    [\Hps-E_0]$$
are maps between $P_{\rho^n}\Hp$ and $\Hp$. We have here used that
$\derp^2\Hps=1$.

>From Propositions {~\ref{derpanznprp4444}} and
{~\ref{hatazminazestkappaprp}}, we know that
\eqn\derp
    a_n[z_n]&=&-\,
    \frac{\derp\dHfFQQ_n[z[z_n]]}{\dHfFQQ_n[z[z_n]]}
    \,a_n[z_n] \,+\,\dHfFQQ_n[z[z_n]]^{-1}\nonumber\\
    &&-\,2\,\dHfFQQ_n[z[z_n]]^{-1}\,\frac{\left\langle
    (\derp\Qns[z[z_n]])\left(\partial_{\Ppar} \Hps
    \right)\derp\Qn[z[z_n]]
    \right\rangle_{\vac}}{\left\langle
    \Qns[z[z_n]]\Qn[z[z_n]]
    \right\rangle_{\vac}}\nonumber\\
    &&+\,O\left(\epsilon_0^2\,\rho^{n}\,n^2\right)\;.
    \eeqn
Thus,
\eqn\derp^2 E_0&=&\lim_{n\rightarrow\infty}
    \left(\dHfFQQ_n[z[z_n]]\derp
    a_n[z_n]\,+\,(\derp\dHfFQQ_n[z[z_n]])a_n[z_n]\right)
    \eeqn
(since $O\left(\epsilon_0^2\,\rho^{n}\,n^2\right)=o(n)$, with
$\rho=\frac{1}{2}$).

Theorem {~\ref{indstepmainthm3}} implies that
\eqn\left|\derp^\beta (a_n[z_n]+|p|)\right|\,,\,
    \left|\derp \dHfFQQ_n[z[z_n]]\right|\;\leq\;\sqrt{\epsilon_0}
    \; \eeqn
holds for all $n$. Thus,
\eqnn\derp^2 E_0&=&1\,+\,O(\sqrt{\epsilon}_0)\\
    &=&1\,+\,O(g^{\frac{1}{6}})\;,\eeqnn
as claimed. \qed

\chapter{APPENDIX: PROOFS RELATED TO THE FIRST DECIMATION STEP}
\label{firstdecappendix}

In this appendix, we summarize proofs of lemmata used for the
analytical control of the first Feshbach decimation step in
Chapter {~\ref{firstdecchap}}.

\section{Proof of Lemma {~\ref{elhighenrelW1W2bounds1}}}
Clearly, $H_f$ is invertible on $\bar{P}_1\;\Hp$. Let us first
consider the case of $W_1[\op]$. We recall that
\eqn|\;w_{M,N}[\op;k,\lambda]\;|\;\leq\;
    c\,g\,(|p|+|\P_f|)\;\chi_{\Lambda}[|k|]\eeqn
if $M+N=1$. Thus, picking any $\phi\in\Hp$, Cauchy-Schwarz gives
\eqn\left\|\,\bar{R}_0^{\frac{1}{2}}\bch_1 W_{0,1}\bch_1\,
    \bar{R}_0^{\frac{1}{2}}\,
    \phi\,\right\|&\leq&c\,g\,
    \left\|\,\bP1\frac{|p|+|\P_f|}{\sqrt{(1-|p|)H_f+
    \frac{1}{2}|\P_f|^2}}
    \bP1\,\right\|\,\times\nonumber\\
    &&\hspace{1cm}\times\,\sum_\lambda\,\int
    d^3k\,\frac{\chi_{\Lambda}[|k|]}{\sqrt{|k|}}\,
    \left\|\,a_\lambda(k)\,\phi'\;\right\|
    \nonumber\\
    &\leq&c\,g\,\left(\frac{|p|}{1-|p|}+2\right)\,
    \sum_\lambda\,\int
    d^3k\,\frac{\chi_{\Lambda}[|k|]}{\sqrt{|k|}}
    \,
    \left\|\;a_\lambda(k)\,\phi'\;\right\|
    \nonumber\\
    &\leq&c\,g\,\left(\frac{|p|}{1-|p|}+2\right)\,
    \left(\,\sup_{\lambda}\int d^3 k\;|k|^{-2}\;
    \chi_{\Lambda}[|k|]\,
    \right)^{\frac{1}{2}}\,\times\nonumber\\
    &&\hspace{1cm}\times\,
    \left(\,\sum_\lambda\,\int d^3 k\,\left\langle\,\phi'\,,
    \,a^*_\lambda(k)\,|k|\,
    a_\lambda(k)\,\phi'\,
    \right\rangle\,\right)^{\frac{1}{2}}\;,
    \eeqn
with $\phi'\equiv \bar{R}_0^{\frac{1}{2}}\,\phi$, and where we
have used
$$|\bar{R}_0|\;\leq\;
    \left((1-|p|)H_f+\frac{1}{2}|\P_f|^2\right)^{-1}
    \;$$
on $\bar{P}_1\Hp$. Thus, we obtain the bound
\eqn\left\|\,\bar{R}_0^{\frac{1}{2}} \bch_1W_{0,1}\bch_1\,
    \bar{R}_0^{\frac{1}{2}}\,
    \phi\,\right\|&\leq&c'\,g\,\left(\frac{|p|}{1-|p|}+2\right)\,
    \left\|\;H_f^{\frac{1}{2}}\,((1-|p|)H_f)^{-\frac{1}{2}}\,\phi\,
    \right\|\nonumber\\
    &\leq&c'\,g\,\left(\frac{|p|}{1-|p|}+2\right)\,
    \frac{1}{\sqrt{1-|p|}}\,
    \left\|\,\phi\,
    \right\|\nonumber\\
    &\leq&c'\,g\,\|\phi\|\;,\eeqn
since $|p|\leq \puppbd$. The case for $W_{1,0}$ is identical.

Next, let us consider $W_2[\op]$. We recall that if $M+N=2$,
\eqn\nm\;w_{M,N}[\op;k,\lambda,\tilde{k},\tilde{\lambda}]
    \;\nm\;\leq\;c\,g^2\,\chi_{\Lambda}[|k|]\;
    \chi_{\Lambda}[|\tilde{k}|]\;.\eeqn
Picking any $\phi,\psi\in\Hp$, it is clear that for $M=N=1$,
\eqn\left|\;\left\langle\;\psi\;,\;W_{1,1}[\op]\;
    \phi\;\right\rangle\;\right|
    &\leq&c\,g^2\,\sum_{\lambda,\tilde{\lambda}}\;\int
    \frac{d^3k}{\sqrt{|k|}}\,
    \frac{d^3\tilde{k}}{\sqrt{|\tilde{k}|}}
    \,\chi_{\Lambda}[|k|]\;
    \chi_{\Lambda}[|\tilde{k}|]\;\nonumber\\
    &&\hspace{1cm}\times\,
    \left\|\,a_\lambda(k)\,\psi\,\left\|\,
    \left\|\,a_{\tilde{\lambda}}(\tilde{k})\,
    \phi\,\right\|\right.\right.\nonumber\\
    &\leq&c'\,g^2
    \left\|\;H_f^{\frac{1}{2}}\;\psi\;\right\|\;
    \left\|\;H_f^{\frac{1}{2}}\;\phi\;\right\|\;
    \label{psichiW11estaux3}\eeqn
by the Cauchy-Schwarz inequality. Using
\eqn\|\bar{R}_0\,H_f\|\,\leq\,C\;,
    \label{barR0Hfboundaux3}\eeqn
this implies that
\eqn\left\|\,\bar{R}_0^{\frac{1}{2}}\,\bch_1W_{1,1}[\op]\bch_1
    \,\bar{R}_0^{\frac{1}{2}}\,
    \right\|&\leq&c'\,C\,g^{2}\;.\eeqn

For $M=0$ and $N=2$, we find
\eqn&&\left\|\;(H_f+\Lambda+1)^{-\frac{1}{2}}\;
    W_{0,2}[\op]\;\phi\;\right\|^2\nonumber\\
    &\leq&\left(\;\sum_{\lambda,\tilde{\lambda}}\;\int d^3k\;
    d^3\tilde{k}\;|k|^{-\frac{1}{2}}\;
    |\tilde{k}|^{-\frac{1}{2}}\;
    \left\|\;w_{0,2}[\op;k,\lambda,\tilde{k},\tilde{\lambda}]\;
    \right\|\;\times\right.\nonumber\\
    &&\hspace{3.5cm}\times\;\left.\left\|\;(H_f+\Lambda+1)^{-\frac{1}{2}}\;
    a_\lambda(k)\;
    \;a_{\tilde{\lambda}}(\tilde{k})\;\phi\;\right\|\right)^2\nonumber\\
    &\leq&c\,g^2\,
    \left(\;\int d^3k\;d^3\tilde{k}\;|k|^{-\frac{1}{2}}\;
    |\tilde{k}|^{-\frac{1}{2}}\;\chi_{\Lambda}[|k|]
    \;\chi_{\Lambda}[|\tilde{k}|]\;\right)\;\times
    \nonumber\\
    &&\hspace{1cm}\times\;\sum_{\lambda,\tilde{\lambda}}\;
    \int d^3k \;d^3\tilde{k}\;|k|\;|\tilde{k}|\;\chi_{\Lambda}[|k|]\;
    \times\nonumber\\
    &&\hspace{1.5cm}\times\;
    \left\langle\;\phi\;,\;a^*_{\tilde{\lambda}}(\tilde{k})\;
    a^*_\lambda(k)\;(H_f+\Lambda+1)^{-1}\;
    a_\lambda(k)\;
    \;a_{\tilde{\lambda}}(\tilde{k})\;\phi\;\right\rangle\;
    \;\nonumber\\
    &\leq&c'\,g^2\,
    \left\|\;H_f^{\frac{1}{2}}\;\phi\;\right\|^2\;,\eeqn
again by using the Cauchy-Schwarz inequality, and \eqn
&&\sum_\lambda\;\int
     d^3k\;|k|\;\chi_{\Lambda}[|k|]\;a^*_\lambda(k)\;
    (H_f+\Lambda+1)^{-1}\;a_\lambda(k)\nonumber\\
    &=&
    \int d^3k\;|k|\;\chi_{\Lambda}[|k|]\;a^*_\lambda(k)\;
    a_\lambda(k)\;(H_f+\Lambda-|k|+1)^{-1}\nonumber\\
    &\leq&H_f\;(H_f+1)^{-1}\nonumber\\
    &\leq&1\;.\eeqn
>From \eqn\left\|\;H_f^{-\frac{1}{2}}\;\bP1\;
    (H_f+\Lambda+1)^{\frac{1}{2}}\;\right\|\;\leq\;
    \sqrt{\Lambda +2}\;\eeqn
and (~\ref{barR0Hfboundaux3}), we arrive at the desired bound
\eqn\left\|\,\bar{R}_0^{\frac{1}{2}}\,\bch_1W_{0,2}\bch_1\,
    \bar{R}_0^{\frac{1}{2}}\,\right\|&\leq&c'\,
    g^{2}\;.\eeqn
The case of $W_{2,0}[\op]$ is treated in the same fashion.

Collecting the above bounds, this proves Lemma
{~\ref{elhighenrelW1W2bounds1}}. \qed

\section{Proof  of Lemma {~\ref{elhighenrelW1W2bounds2}}}

Again using the bound \eqn\nm\,w_{M,N}\,\nm\;\leq\;
    c\,g\,\chi_{\Lambda}[|k|]\eeqn
for $M+N=1$, and picking any $\phi\in\Hp$, the Cauchy-Schwarz
inequality yields
\eqn\left\|\,\bar{R}_0^{\frac{1}{2}}
    \bch_1W_{0,1}[\op]\,P_1\,
    \phi\,\right\|&\leq&c\,g\,
    \left\|\,\bP1\frac{|p|+|\P_f|}{\sqrt{(1-|p|)H_f+
    \frac{1}{2}|\P_f|^2}}
    \bP1\,\right\|\,\times\nonumber\\
    &&\hspace{1cm}\times\,\sum_\lambda\,\int
    d^3k\,\frac{\chi_{\Lambda}[|k|]}{\sqrt{|k|}}\,
    \left\|\,a_\lambda(k)\,P_1\,\phi\;\right\|
    \nonumber\\
    &\leq&c\,g\,\left(\frac{|p|}{1-|p|}+2\right)\,
    \sum_\lambda\,\int
    d^3k\,\frac{\chi_{\Lambda}[|k|]}{\sqrt{|k|}}
    \,
    \left\|\;a_\lambda(k)\,P_1\,\phi\;\right\|
    \nonumber\\
    &\leq&c\,g\,\left(\frac{|p|}{1-|p|}+2\right)\,
    \left(\,\sup_{\lambda}\int d^3 k\;|k|^{-2}\;
    \chi_{\Lambda}[|k|]\,
    \right)^{\frac{1}{2}}\,\times\nonumber\\
    &&\hspace{1cm}\times\,
    \left(\,\sum_\lambda\,\int d^3 k\,\left\langle\,P_1\,\phi\,,
    \,a^*_\lambda(k)\,|k|\,
    a_\lambda(k)\,P_1\,\phi\,
    \right\rangle\,\right)^{\frac{1}{2}}\;,
    \eeqn
which is bounded by
$$c'\,g\,\|\phi\|\;,$$
as asserted, since $\|H_f^{\frac{1}{2}}P_1\phi\|\leq\|\phi\|$. The
case for $W_{1,0}$ is identical.

Next, let us consider $W_2[\op]$. Recalling
(~\ref{psichiW11estaux3}),  that is,
$$\left|\;\left\langle\,\psi\,,\,W_{1,1}[\op]\,
    \phi\;\right\rangle\;\right|
    \,\leq\,c'\,g^2\,\|H_f^{\frac{1}{2}}\psi\|\,
    \|H_f^{\frac{1}{2}}\chi\|\;, $$
we have \eqn\left|\,\left\langle\,\psi\,,\,
    \bar{R}_0^{\frac{1}{2}}\,\bch_1W_{1,1}\,P_1
    \phi\right\rangle
    \right|&\leq&c'\, g^{2}\,\|\psi\|\,\|\phi\|\;,\eeqn
using  (~\ref{barR0Hfboundaux3}).

For $M=0$ and $N=2$, we recall
$$\left\|\;(H_f+\Lambda+1)^{-\frac{1}{2}}\;
    W_{0,2}[\op]\;\phi\;\right\|\,\leq\,c'\,g^2
    \left\|\;H_f^{\frac{1}{2}}\;\phi\;\right\|\;,$$
and find
\eqn\left\|\,\bar{R}_0^{\frac{1}{2}}\,\bch_1W_{0,2}[\op]\,
    P_1\,\phi\,\right\|&\leq&
    c'\,g^{2}\|\phi\|\;,\eeqn
again using (~\ref{barR0Hfboundaux3}). The case of $W_{2,0}[\op]$
is treated in the same fashion.

This proves Lemma {~\ref{elhighenrelW1W2bounds2}}. \qed

\section{Proof of Lemma
{~\ref{firststepderopalphaDeltwMNlemma1}}} We first prove the
following  bounds on the Hilbert space $P_1\Hp$.
\\

\begin{lm}
\label{derWmnpqauxlemma3} Under the assumptions of Lemma
{~\ref{firststepderopalphaDeltwMNlemma1}}, there exists a constant
$C<\infty$, such that \eqnn\nm\,\bar{R}_0^{\frac{1}{2}}\,
    \derp^\beta\,\bcl(\partial_{\op}^\alpha\,W^{m,n}_{p,q})\bcl\,
    \bar{R}_0^{\frac{1}{2}}\,\nm&\leq&
    C \,\epsilon_0^{m+n+p+q}\,\xi^{m+n}
    (2\sqrt{\pi}\xi)^{p+q}\\
    \nm\,\bar{R}_0^{\frac{1}{2}}\,\bcl
    (\derp^\beta\,\partial_{\op}^\alpha\,
    W^{m,n}_{p,q})\,
    \ch_1\,\nm&\leq&
    C \,\epsilon_0^{m+n+p+q}\,\xi^{m+n}
    (2\sqrt{\pi}\xi)^{p+q}\;\eeqnn
for $|\alpha|\leq2$, $\beta=0,1$, and
\eqnn\nm\,\bar{R}_0^{\frac{1}{2}}\,\bcl
    (\derp^\beta\,\partial_{X}\,W^{m,n}_{p,q})\bcl\,
    \bar{R}_0^{\frac{1}{2}}\,\nm&\leq&
    C \,\epsilon_0^{m+n+p+q}\,\xi^{m+n}
    (2\sqrt{\pi}\xi)^{p+q}\\
    \nm\,\bar{R}_0^{\frac{1}{2}}\,
    (\derp^\beta\,\partial_{X}\,W^{m,n}_{p,q})\,
    \ch_1\,\nm&\leq&
    C \,\epsilon_0^{m+n+p+q}\,\xi^{m+n}
    (2\sqrt{\pi}\xi)^{p+q}\;\eeqnn
on the Hilbert space $P_1\Hp$, for $m+n+p+q\leq2$, and $X=|k_i|$,
$|\tilde{k}_j|$ or $z$.
\end{lm}

\prf The asserted bounds are obtained in the precise same manner
as in the cases of Lemma {~\ref{elhighenrelW1W2bounds1}} and Lemma
{~\ref{elhighenrelW1W2bounds2}}. The only difference now are the
bounds \eqnn|\,\derp^\beta\,\partial_{Y}^\alpha\,w_{M,N}\,|
    \;\leq\;c\,g^{M+N}\,(1+|\P_f|)\,\prod_{r=1}^{M}
    \chi_{[0,\Lambda]}(|k_r|)
    \,\prod_{s=1}^{N}\chi_{[0,\Lambda]}(|\tilde{k}_s|)\,\eeqnn
for $M+N\leq2$, and $Y$ denoting either $\op$ (with
$|\alpha|\leq2$) or $X$ (with $|\alpha|\leq1$), instead of those
on $\nm w_{M,N}\nm$. Furthermore, we recall that
$\epsilon_0\,\xi=2g$. \qed

\begin{lm}\label{mathcalELfirststepauxlm3}
Let us assume that there is a constant $C$ independent of
$\epsilon$, such that
$$\|\partial_{|p|}^{\beta'}
    \partial^{\alpha'}_Y\,(\bcl\bar{R}_0\bcl)\|\;,\;
    \|\partial_{|p|}^{\beta'}
    \partial^{\alpha'}_Y\,\piop\|\;\leq\;
    C\;$$
for any $0\leq\beta'\leq1$, and $0\leq|\alpha'|\leq2$. The
operator
\begin{gather}\mathcal{E}_{L}
    \left[\{\underline{m},\underline{n},\underline{p},\underline{q}\};
    z;\op;K^{(M,N)}\right]
    \;=\;
    \left\langle\,\piop W_{p_1,q_1}^{m_1,n_1}\bcl
    \bar{R}_0\bcl\cdots W^{m_l,n_l}_{p_L,q_L}\piop\,
    \right\rangle_{\op,sym}\;,\nonumber\end{gather}
with $\underline{m}=(m_1,\dots,m_L)$, etc., and $M=\sum m_i$,
$N=\sum n_i$, satisfies \eqnn\nm\,\partial_{|p|}^\beta
    \partial^\alpha_{\op}\,
    \mathcal{E}_{L}\,\nm\;
    \leq\;c_{\alpha,\beta}\,L^{|\alpha|+\beta}\;
    \epsilon_0^{L}\,C^L\,
    \xi^{M+N}\,
    (2\sqrt{\pi}\xi)^{\sum(p_i+q_i)}\;.\eeqnn
for $|\alpha|\leq2$, $\beta=0,1$, and
\eqnn\nm\,\partial_{|p|}^\beta
    \partial_X\,
    \mathcal{E}_{L}\,\nm\;
    \leq\;c_{X,\beta}\,L^{1+\beta}\;
    \epsilon_0^{L}\,C^L\,
    \xi^{M+N}\,
    (2\sqrt{\pi}\xi)^{\sum(p_i+q_i)}\;.\eeqnn
The constants $c_{X,\beta}$ and $c_{\alpha,\beta}$ do not depend
on $\epsilon,\xi$. Furthermore, $\mathcal{E}_{L}$ is {\bf
analytic} in $\op$ for all $M+N\geq0$.
\end{lm}

\prf $\mathcal{E}_{L}$ and the symbol
$\langle\cdot\rangle_{\op,sym}$ were defined below
(~\ref{mthcalELdef3}).

$\mathcal{E}_L$ is the $\langle\cdot\rangle_{\op,sym}$-expectation
value of $L+(L-1)<2L$ operators ($L$ interaction operators
$W_{p_i,q_i}^{m_i,n_i}$ and $L-1$ free resolvents $\bar{R}_0$).
Thus, by the product rule, $\partial_{|p|}^\beta
\partial^\alpha_{\op}\mathcal{E}_{L}$ is the sum of
$O(L^{|\alpha|+\beta})$ vacuum expectation values.

Derivatives of $\bcl\bar{R}_0\bcl$, $\piop$ can be estimated by
$$\|\partial_{|p|}^{\beta'}
    \partial^{\alpha'}_Y\,(\bcl\bar{R}_0\bcl)\|\;,\;
    \|\partial_{|p|}^{\beta'}
    \partial^{\alpha'}_Y\,\piop\|\;\leq\;
    C\;,$$
where the constant only depends on
$|\alpha'|\leq|\alpha|,\beta'\leq\beta$. Derivatives of
$W_{p_i,q_i}^{m_i,n_i}$ are controlled by Lemma
{~\ref{derWmnpqauxlemma3}}.

In each of the $O(L^{|\alpha|+\beta})$ terms obtained from
applying $\partial_{|p|}^\beta\partial^\alpha_{\op}$ to
$\mathcal{E}_L$, there are at most $L-1+|\alpha|+\beta$
occurrences of $\bar{R}_0$, and two of $\piop$, hence an overall
factor
$$C^{L+1+|\alpha|+\beta}$$
in the bound. Furthermore, from each $W_{p_i,q_i}^{m_i,n_i}$ or
its derivative, there emerges a factor
$$O(\epsilon_0)(2\sqrt{\pi}\xi)^{p_i+q_i}\xi^{m_i+n_i}\;.$$
Thus, there is an overall factor
$$O(\epsilon_0^{L})(2\sqrt{\pi}\xi)^{\sum(p_i+q_i)}
    \xi^{\sum(m_i+n_i)}$$
stemming from the interaction operators.

Since all implicit constants in these estimates only depend on
$\alpha$ and $\beta$, there is a constant $c_{\alpha,\beta}$, such
that
\eqnn\nm\,\partial_{|p|}^\beta
    \partial^\alpha_{\op}\,
    \mathcal{E}_{L}\,\nm&\leq&
    c_{\alpha,\beta}\,L^{|\alpha|+\beta}\;
    \epsilon_0^{L}\,C^L\,\times\\
    &&\times\;\xi^{M+N}\,
    (2\sqrt{\pi}\xi)^{\sum(p_i+q_i)}\;.\eeqnn
The asserted bound on $\nm\,\partial_{|p|}^\beta\partial_X\,
\mathcal{E}_{L}\,\nm$ are proved in the same manner.

Because every $\bar{R}_0$ and every $W_{p_i,q_i}^{m_i,n_i}$ is
analytic in $\P_f$, $\mathcal{E}_L$ also is.

For a more detailed exposition of the intermediate steps in the
derivation of these bounds, cf. \cite{bfs1,bfs2}. \qed

\noindent{\bf Proof of Lemma
{~\ref{firststepderopalphaDeltwMNlemma1}}.} Let $X$ denote $|k|$
or any component $\op^r$ of $\op=(H_f,\Ppar,\Pperp)$. Then, using
Lemma {~\ref{mathcalELfirststepauxlm3}},
\eqn\left\|\,\derp^\beta\,\partial_{\op}^\alpha\,\Delta
      w_{M,N}\;\right\|
      &\leq&c_{\alpha,\beta}\,\xi^{M+N}\,
      \sum_{L=2}^\infty\,L^{|\alpha|}\,
      \epsilon_0^{L}\,C^L\;\times
      \nonumber\\
      &&\times
      \sum_{\stackrel{1\leq M_i+N_i\leq 2}{i=1,\dots,L}}
      \delta_{M,\sum_{i=1}^L m_i}\delta_{N,\sum_{i=1}^L n_i}
      \;\times
      \nonumber\\
      &&\times \;\;
      \prod_{i=1}^L
      \left(\begin{array}{c}m_i+p_i\\p_i\end{array}\right)
      \left(\begin{array}{c}n_i+q_i\\q_i\end{array}\right)
      (2\sqrt{\pi}\xi)^{\sum(p_i+q_i)}\;
      \nonumber\\
      &\leq&c_{\alpha,\beta}\,
      \xi^{M+N}\,\sum_{L=2}^\infty \,4^L\,L^{|\alpha|}\,
      \epsilon_0^{L}\,C^L\,
      \;\times
      \nonumber\\
      &&\times \;\;
      \sum_{\stackrel{p_i+q_i\geq0}{i=1,\dots,L}}
      (2\sqrt{\pi}\xi)^{\sum(p_i+q_i)}
      \nonumber\\
      &\leq&C_{\alpha,\beta}\;
      \epsilon_0^{2}
      \,\xi^{M+N}\;
      \label{deroprDeltvMNphysham}\eeqn
for a constant $C_{\alpha,\beta}$ that is independent of
$\epsilon_0,\xi$. The combinatorial factors have here been
estimated by
$$\left(\begin{array}{c}m+p\\p
    \end{array}\right)\leq 2\;,$$
since $m+p\leq2$, and likewise for $m\leftrightarrow n$,
$p\leftrightarrow q$.

The bound
$$\nm\,\partial_{|p|}^\beta\partial_X\,
     \Delta w_{M,N}\,\nm\;\leq\;C_{X,\beta}\;
      \epsilon_0^{2}\,\xi^{M+N}$$
of the assertion follows in the same manner.

According to the previous lemma, $\mathcal{E}_{L}$ is analytic in
$\op$ for all $L$. Thus, $\Delta w_{M,N}$ is given by a series of
analytic functions of $\op$. In particular, the above results
imply that the defining series for $\partial_{\op}^\alpha\Delta
w_{M,N}$ are, for $|\alpha|\leq1$, uniformly convergent with
respect to $\nm\cdot\nm$. Thus, $\Delta w_{M,N}$ is analytic in
$\op$.

This proves Lemma {~\ref{firststepderopalphaDeltwMNlemma1}}. \qed

\chapter{APPENDIX: PROOFS RELATED TO THE INDUCTION STEP}
\label{iteratstepappendsect}

In this appendix, we give the proofs that are missing in Chapter
{~\ref{analiterstepchap}}.

\section{Proof of Proposition {~\ref{deropalphatildwMNprp1}}}

\begin{lm}
\label{derWmnpqauxindsteplemma3} Let $0\leq|\alpha|\leq2$,
$\beta=0,1$, and $X=|k_i|$, $|\tilde{k}_j|$ or  $z$, corresponding
to the assumptions of Lemma {~\ref{deropalphatildwMNprp1}},
\eqn\nm\,
    \derp^\beta\,\partial_{\op}^\alpha\,W^{m,n}_{p,q}\,
    \nm&\leq&(2\ez)^\beta \,
    \epsilon^{\frac{7}{4}-\frac{|\alpha|}{4}-\beta}\,
    (2\sqrt{\pi}\xi)^{p+q}\,\xi^{m+n}\nonumber\\
    \nm\,
    \derp^\beta\,\partial_{X}\,W^{m,n}_{p,q}\,
    \nm&\leq&(2\ez)^\beta \,
    \epsilon^{\frac{3}{2}-\beta}\,
    (2\sqrt{\pi}\xi)^{p+q}\,\xi^{m+n}\eeqn
on the Hilbert space $P_1\Hp$.
\end{lm}

\prf This follows directly from
$$\nm\,\derp^\beta\,\partial_{\op}^\alpha\,w_{M,N}\,\nm
    \,\leq\,(2\ez)^\beta \,
    \epsilon^{\frac{7}{4}-\frac{|\alpha|}{4}-\beta}\,\xi^{M+N}$$
and
$$\nm\,\derp^\beta\,\partial_{X}\,w_{M,N}\,\nm
    \,\leq\,(2\ez)^\beta \,
    \epsilon^{\frac{3}{2}-\beta}\,\xi^{M+N}$$
implied by the induction hypothesis. \qed

\begin{lm}\label{derexpvallemma3}
Let $0\leq|\alpha|\leq2$, $\beta=0,1$, and $X=|k_i|$,
$|\tilde{k}_j|$ or  $z$, corresponding to the assumptions of
Proposition {~\ref{deropalphatildwMNprp1}},
\begin{gather}\mathcal{E}_{L}
    \left[\{\underline{m},\underline{n},\underline{p},\underline{q}\};
    z;\op;K^{(M,N)}\right]
    \;=\;
    \left\langle\,\piop W_{p_1,q_1}^{m_1,n_1}\bch
    \bar{R}_0 \bch\cdots W^{m_l,n_l}_{p_L,q_L}
    \piop\,
    \right\rangle_{\op,sym}\;,\nonumber\end{gather}
with $\underline{m}=(m_1,\dots,m_L)$, etc., and $M=\sum m_i$,
$N=\sum n_i$, satisfies \eqnn\nm\,\partial_{|p|}^\beta
    \partial^\alpha_{\op}\,
    \mathcal{E}_{L}\,\nm\;
    \leq\;\ez^\beta \,
    c_{\alpha,\beta}\,L^{|\alpha|+\beta}\;
    \epsilon^{L-\beta}\,
    \rho^{1-L-|\alpha|-\beta}\;\times\nonumber\\
    \times\;\xi^{M+N}\,
    (2\sqrt{\pi}\xi)^{\sum(p_i+q_i)}\;.\eeqnn
and \eqnn\nm\,\partial_{|p|}^\beta
    \partial_X\,
    \mathcal{E}_{L}\,\nm\;
    \leq\;\ez^\beta \,
    c_{X,\beta}\,L^{1+\beta}\;
    \epsilon^{L-\beta}\,
    \rho^{L-\beta}\;\times\nonumber\\
    \times\;\xi^{M+N}\,
    (2\sqrt{\pi}\xi)^{\sum(p_i+q_i)}\;.\eeqnn
The constants $c_{\alpha,\beta}$ and $c_{X,\beta}$ are independent
of $\epsilon,\xi$. Furthermore, $\mathcal{E}_{L}$ is {\bf
analytic} in $\op$ for all $M+N\geq0$.
\end{lm}

\prf The definition of the symbol $\langle\cdot\rangle_{\op,sym}$
is given below (~\ref{mthcalELdef3}). By induction hypothesis, $T$
(and thus $\bar{R}_0[z]$) and $w_{M,N}$ are analytic in $\op$.
Thus, $\mathcal{E}_{L}$ is analytic in $\op$.

>From the product rule, it is first of all clear that there
are $O(L^{|\alpha|+\beta})$ terms in total, because
$\mathcal{E}_L$ is the $\langle\cdot\rangle_{\op,sym}$-expectation
value of a product of less than $2L$ operators.

Derivatives of $\bcr\bar{R}_0\bcr$ and $\piop$ can be estimated by
$$\|\partial_{|p|}^{\beta'}
    \partial^{\alpha'}_X\,(\bcr\bar{R}_0\bcr)\|\;,\;
    \|\partial_{|p|}^{\beta'}
    \partial^{\alpha'}_X\,\piop\|\;\leq\;
    C_\rho^{1+|\alpha|+\beta'}\;,$$
where the constant $C_\rho$ only depend on $\rho$, while
derivatives of $W_{p_i,q_i}^{m_i,n_i}$ can be estimated by Lemma
{~\ref{derWmnpqauxlemma3}}.

In each of the $O(L^{|\alpha|+\beta})$ terms obtained from
applying $\partial_{|p|}^\beta\partial^\alpha_X$ to
$\mathcal{E}_L$, there are at most $L-1+|\alpha|+\beta$
occurrences of $\bar{R}_0$, and two of $\piop$, hence an overall
factor $C_\rho^{L+1+|\alpha|+\beta} $ in the bound. Furthermore,
the induction hypotheses, together with Lemma
{~\ref{derWmnpqauxindsteplemma3}}, imply that
$$\nm\,\partial_{|p|}^{\beta_i}
    \partial^{\alpha'}_X\,W_{p_i,q_i}^{m_i,n_i}\,\nm\;
    \leq\;(2\ez)^{\beta_i} \,
    \epsilon^{\frac{7}{4}-\frac{|\alpha'|}{4}-\beta}
    (2\sqrt{\pi}\xi)^{p_i+q_i}
    \xi^{m_i+n_i}\;,$$
where $\beta_i=1$ if a derivative with respect to $|p|$ is taken,
and $\beta_i=0$ otherwise. Thus, there is an overall factor
$$O(\ez^\beta\epsilon^{L-\beta})(2\sqrt{\pi}\xi)^{\sum(p_i+q_i)}
    \xi^{\sum(m_i+n_i)}$$
due to the interaction operators.

Since all implicit constants in these estimates only depend on
$\alpha$ and $\beta$, there is a constant $c_{\alpha,\beta}$, such
that \eqn\nm\,\partial_{|p|}^\beta
    \partial^\alpha_{\op}\,
    \mathcal{E}_{L}\,\nm&\leq&\ez^\beta \,
    c_{\alpha,\beta}\,L^{|\alpha|+\beta}\;\times\nonumber\\
    &&\hspace{1cm}\times\;
    \epsilon^{L-\beta}\,
    \xi^{M+N}\,
    (2\sqrt{\pi}\xi)^{\sum(p_i+q_i)}\;,\eeqn
which proves the claim.

The estimate \eqnn\nm\,\partial_{|p|}^\beta
    \partial_X\,
    \mathcal{E}_{L}\,\nm\;
    \leq\;\ez^\beta \,
    c_{X,\beta}\,L^{1+\beta}\;
    \epsilon^{L-\beta}\,
    C_\rho^{L}\;\times\nonumber\\
    \times\;\xi^{M+N}\,
    (2\sqrt{\pi}\xi)^{\sum(p_i+q_i)}\;\eeqnn
is proved in the same way.

Furthermore, for every $L\geq2$, $\mathcal{E}_{L}$ is  analytic in
$\P_f$, since $\bar{R}_0$ and $W_{p_i,q_i}^{m_i,n_i}$ are.

We refer the interested reader to \cite{bfs1,bfs2} for a more
explicit exposition of some of the intermediate steps. \qed

\noindent{\bf Proof of Lemma {~\ref{deropalphaDeltwMNlemma1}}.}
For $0\leq|\alpha|\leq2$, \eqn\nm\;\partial_{|p|}^\beta\,
    \partial^\alpha_{\op}\,\Delta
    w_{M,N}\;\nm
    &\leq&\sum_{L=2}^\infty
    \sum_{\stackrel{1\leq M_i+N_i}{i=1,\dots,L}}
    \delta_{M,\sum_{i=1}^L m_i}\delta_{N,\sum_{i=1}^L n_i}
    \;\times\nonumber\\
    &&\hspace{1cm}\times\;
    \prod_{i=1}^L
    \left(\begin{array}{c}m_i+p_i\\p_i\end{array}\right)
    \left(\begin{array}{c}n_i+q_i\\q_i\end{array}\right)
    \,\left\|\,\partial_{|p|}^\beta\,
    \partial^\alpha_X\,\mathcal{E}_{L}\,
    \right\|\;\nonumber\\
    &\leq&(2\xi)^{M+N}\,c_{|\alpha|,\beta}\,
    \sum_{L=2}^\infty\;L^{|\alpha|}\;
    \epsilon^{L-\beta}
    \;\rho^{1-L-|\alpha|-\beta}\;\times\nonumber\\
    &&\hspace{1cm}\times\;\,\ez^{\beta}\,
    \sum_{\stackrel{p_i+q_i\geq0}{i=1,\dots,L}}
    2^{\sum (p_i+q_i)}\,
    (2\sqrt{\pi}\xi)^{\sum (p_i+q_i)}\nonumber\\
    &\leq&\ez^\beta \,C_{|\alpha|,\beta}\,
    \epsilon^{2-\beta}\,
    (2\xi)^{M+N}\nonumber\\
    &=&\ez^\beta \,C_{|\alpha|,\beta}\,
    \epsilon^{2-\beta}\,
    (2\xi)^{M+N}\;,\eeqn
using Lemma {~\ref{derexpvallemma3}}, and
$$\left(\begin{array}{c}m+p\\p\end{array}\right)\;
    \leq\;2^{m+p}\;,$$
while setting $\rho=\frac{1}{2}$. This is the first assertion of
Lemma {~\ref{deropalphaDeltwMNlemma1}}.

The assertion \eqn\nm\;\partial_{|p|}^\beta\,
    \partial_X\,\Delta
    w_{M,N}\;\nm
    &\leq&\ez^\beta \,C_{X,\beta}\,
    \epsilon^{2-\beta}\,
    (2\xi)^{M+N}\;, \eeqn
for $X=|k_i|$, $|\tilde{k}_j|$ or  $z$, is proved in the same way.

By the previous lemma, $\mathcal{E}_{L}$ is analytic in $\P_f$ for
all $L$. Thus, $\Delta w_{M,N}$ is given by a series of analytic
functions of $\P_f$. Since the first part of this proof showed
that the defining series for $\partial_{\P_f}^\alpha\Delta
w_{M,N}$ are, for $|\alpha|\leq1$, uniformly convergent with
respect to $\nm\cdot\nm$, $\Delta w_{M,N}$ is analytic in $\P_f$.
\qed

$\;$

\section*{Acknowledgements}

I am profoundly grateful to my thesis advisor Prof. J\"urg
Fr\"ohlich for his generosity, friendliness and support during the
completion of this work. I have learned and benefited immensely
from his knowledge, wisdom and penetrating insight.  I am most
grateful to Prof. Gian-Michele Graf for accepting the
coadvisorship for this thesis.

It is a pleasure to very warmly thank Prof. Volker Bach  for
illuminating and enjoyable discussions, for his interest in my
work and support. I also like to heartily thank Prof. Herbert
Spohn for highly interesting discussions, which taught me
important lessons about the class of problems dealt with in this
thesis, for his friendliness and support. I am also deeply
indebted to Profs. V. Bach, J. Fr\"ohlich and I. M. Sigal for
exposing me to unpublished notes which were highly relevant for
the present work, in particular those concerning the smooth
Feshbach map.

I am genuinely grateful to my wife Isabelle for her patience and
generosity, for her caring and support, and for being the lovely
person she is.

\end{document}